\documentclass[onecolumn]{elsart3p}

\usepackage{natbib}
\usepackage{epsfig}

\usepackage{amsfonts}
\usepackage{amssymb}
\usepackage{color}

\newcommand{\bA}{\mbox{\boldmath{$A$}}}
\newcommand{\bB}{\mbox{\boldmath{$B$}}}

\newcommand{\bp}{\mbox{\boldmath{$p$}}}
\newcommand{\bq}{\mbox{\boldmath{$q$}}}
\newcommand{\be}{\mbox{\boldmath{$e$}}}

\newcommand{\bV}{\mbox{\boldmath{$V$}}}

\newcommand{\bs}{\mbox{\boldmath{$s$}}}
\newcommand{\cS}{{\cal S}}

\newcommand{\bn}{\mbox{\boldmath{$n$}}}

\newcommand{\bbeta}{\mbox{\boldmath{$\beta$}}}
\newcommand{\etab}{\mbox{\boldmath{$\eta$}}}

\newcommand{\brho}{\mbox{\boldmath{$\rho$}}}
\newcommand{\bd}{\mbox{bd}}
\newcommand{\inter}{\mbox{int}}
\renewcommand{\thefootnote}{\arabic{footnote}}

\newcommand{\draftnote}[1]{}

\begin{document}

\begin{frontmatter}
\title{Evolutionary games on graphs}

\author[Szabo]{Gy\"orgy Szab\'o}
\ead{szabo@mfa.kfki.hu}
\author[Fath]{G\'abor F\'ath}
\ead{fath@szfki.hu}
\address[Szabo]{Research Institute for Technical Physics and Materials
Science, P.O. Box 49, H-1525 Budapest, Hungary}
\address[Fath]{Research Institute for Solid
State Physics and Optics, P.O. Box 49, H-1525 Budapest, Hungary}

\begin{abstract}
Game theory is one of the key paradigms behind many scientific disciplines from biology to behavioral sciences to economics. In its evolutionary form and especially when the interacting agents are linked in a specific social network the underlying solution concepts and methods are very similar to those applied in non-equilibrium statistical physics. This review gives a tutorial-type overview of the field for physicists. The first four sections introduce the necessary background in classical and evolutionary game theory from the basic definitions to the most important results. The fifth section surveys the topological complications implied by non-mean-field-type social network structures in general. The next three sections discuss in detail the dynamic behavior of three prominent classes of models: the Prisoner's Dilemma, the Rock-Scissors-Paper game, and Competing Associations. The major theme of the review is in what sense and how the graph structure of interactions can modify and enrich the picture of long term behavioral patterns emerging in evolutionary games.
\end{abstract}

\begin{keyword}
Game theory \sep graphs \sep networks \sep evolution
\PACS 02.50.Le \sep 89.65.-s \sep 87.23.Kg \sep 05.65.+b \sep 87.23.Ge
\end{keyword}
\end{frontmatter}

\tableofcontents

\section{Introduction}
\label{sec:intro}

Game theory is the unifying paradigm behind many scientific disciplines. It is a set of analytical tools and solution concepts, which provide explanatory and predicting power in interactive decision situations, when the aims, goals and preferences of the participating players are potentially in conflict. It has successful applications in such diverse fields as evolutionary biology and psychology, computer science and operations research, political science and military strategy, cultural anthropology, ethics and moral philosophy, and
economics. The cohesive force of the theory stems from its formal mathematical structure which allows the practitioners to abstract away the common strategic essence of the actual biological, social or economic situation. Game theory creates a unified framework of abstract models and metaphors, together with a consistent methodology, in which these problems can be recast and analyzed.

The appearance of game theory as an accepted physics research agenda is a relatively late event. It required the mutual reinforcing of two important factors: the opening of physics, especially statistical physics, towards new interdisciplinary research directions, and the sufficient maturity of game theory itself in the sense that it had started to tackle into complexity problems, where the competence and background experience of the physics community could become a valuable asset. Two new disciplines, socio- and econophysics were born, and the already existing field of biological physics got a new impetus with the clear mission to utilize the theoretical machinery of physics for making progress in questions whose investigation were traditionally connected to the social sciences, economics, or biology, and were formulated to a large extent using classical and evolutionary game theory \citep{sigmund_93,ball_04,nowak_06}. The purpose of this review is to present the fruits of this interdisciplinary collaboration in one specifically important area, namely in the case when non-cooperative games are played by agents whose connectivity pattern (social network) is characterized by a nontrivial graph structure.

The birth of game theory is usually dated to the seminal book of \citet{neumann_44}. This book was indeed the first comprehensive treatise with a wide enough scope. Note however, that as for most scientific theories, game theory also had forerunners much earlier on. The French economist Augustin Cournot solved a quantity choice problem under duopoly using some restricted version of the Nash equilibrium concept as early as 1838. His theory was generalized to price rivalry in 1883 by Joseph Bertrand. Cooperative game theory concepts appeared already in 1881 in a work of Ysidro Edgeworth. The concept of a mixed strategy and the minimax solution for two person games were developed originally by Emile Borel. The first theorem of game theory was proven in 1913 by E.\ Zermelo about the strict determinacy in chess. A particularly detailed account of early (and later) history of game theory is Paul Walker's chronology of game theory available on the Web \citep{walker_web95} or William Poundstone's book \citep{poundstone_92}. You can also consult \citet{gambarelli_td04}.

A very important milestone in the theory is John Nash's invention of a strategic equilibrium concept for non-cooperative games \citep{nash_pnas50}. The \emph{Nash equilibrium} of a game is a profile of strategies such that no player has a unilateral incentive to deviate from this by choosing another strategy. In other words, in a Nash equilibrium the strategies form ``best responses" to each other. The Nash equilibrium can be considered as an extension of von Neumann's minimax solution for non-zero-sum games. Beside defining the equilibrium concept, Nash also gave a proof of its existence under rather general assumptions. The Nash equilibrium, and its later refinements, constitute the ``solution" of the game, i.e., our best prediction for the outcome in the given non-cooperative decision situation.

One of the most intriguing aspects of the Nash equilibrium is that it is not necessarily \emph{efficient} in terms of the aggregate social welfare. There are many model situations, like the Prisoners Dilemma or the Tragedy of the Commons, where the Nash equilibrium could be obviously amended by a central planner. Without such  supreme control, however, such efficient outcomes are made unstable by the individual incentives of the players. The only stable solution is the Nash equilibrium which is inefficient. One of the most important tasks of game theory is to provide guidelines (normative insight) how to resolve such social dilemmas, and provide an explanation how microscopic (individual) agent-agent interactions without a central planner may still generate a (spontaneous) aggregate cooperation towards a more efficient outcome in many real-life situations.

Classical (rational) game theory is based upon a number of severe assumptions about the structure of a game. Some of these assumptions were systematically released during the history of the theory in order to push further its limits. Game theory assumes that agents (players) have well defined and consistent goals and preferences which can be described by a \emph{utility function}. The utility is the measure of satisfaction the player derives from a certain outcome of the game, and the player's goal is to maximize her utility. Maximization (or minimization) principles abound in science. It is, however, worth enlightening a very important point here: the maximization problem of game theory differs from that of physics. In a physical theory the standard situation is to have a single function (say, a Hamiltonian or a thermodynamic potential) whose extremum condition characterizes the whole system. In game theory the number of functions to maximize is typically as much as the number of interacting agents. While physics tries to optimize in a (sometimes rugged but) \emph{fixed} landscape, the agents of game theory continuously restructure the landscape for each other in pursuit of their selfish individual optimum.\footnote{A certain class of games, the so-called \emph{potential games}, can be recast into the form of a single function optimization problem. But this is an exception rather than a rule.}

Another key assumption in the classical theory is that players are \emph{perfectly rational} (hyper-rational), and this is \emph{common knowledge}. Rationality, however, seems to be an ill-defined concept. There are extreme opinions arguing that the notion of perfect rationality is not more then pure tautology: rational behavior is the one which complies with the directives of game theory, which in turn is based on the assumption of rationality. It is certainly not by chance that a central recurrent theme in the history of game theory is how to define rationality. In fact, any working definition of rationality is a negative definition, not telling us what rational agents do, but rather what they do not. For example, the usual minimal definition states that rational players do not play strictly dominated strategies \citep{aumann_92,gibbons_92}. Paradoxically, the straightforward application of this definition seems to preclude cooperation in games involving social dilemmas like a (finitely repeated) Prisoners Dilemma or Public Good games, whereas cooperation do occur in real social situations. Another problem is that in many games low level notions of rationality enable several, theoretically permitted outcomes of the game. Some of these are obviously more successful predictions then others in real-life situations. The answer of the classical theory for these shortcomings was to refine the concept of rationality and equivalently the concept of the strategic equilibrium. 

The post-Nash history of game theory is mostly the history of such \emph{refinements}. The Nash equilibrium concept seems to have enough predicting power in static games with complete information. The two mayor streams of extensions are toward dynamic games and games with incomplete information.
Dynamic games are the ones where the timing of decision making plays a role. In these games the simple Nash equilibrium concept would allow outcomes which are based on non-credible threats or promises. In order to exclude these spurious equilibria Selten has introduced the concept of a \emph{subgame perfect Nash equilibrium}, which requires Nash-type optimality in all possible subgames \citep{selten_zgs65}. Incomplete information, on the other hand, means that the players' available strategy sets and associated payoffs (utility) are not common knowledge.\footnote{Incomplete information differs from the similar concept of \emph{imperfect information}. The latter refers to the case when some of the history of the game is unknown to the players at the time of decision making. For example Chess is a game with perfect information because players know the whole previous history of the game, whereas the Prisoners dilemma is a game with imperfect information due to the simultaneity of the players' decisions. Nevertheless, both are games with complete information.}
In order to handle games with incomplete information the theory requires that the players hold beliefs about the unknown parameters and these believes are consistent (rational) is some properly defined sense. This has led to the concept of \emph{Bayesian Nash equilibrium} for static games and to  \emph{perfect Bayesian equilibrium} or \emph{sequential equilibrium} in dynamic games \citep{fudenberg_91,gibbons_92}. Many other refinements (Pareto efficiency, risk dominance, focal outcome, etc.) with lesser domain of applicability have been proposed to provide guidelines for equilibrium selection in the case of multiple Nash equilibria \citep{harsanyi_88}. Other refinements, like \emph{trembling hand perfection}, still within the classical framework, opened up the way for eroding the assumption of perfect rationality. However, this program has only reached its full potential by the general acceptance of \emph{bounded rationality} in the framework of evolutionary game theory.

Despite the undoubted success of classical game theory the paradigm has soon confronted its limitations. In many specific cases further progress seemed to rely upon the relaxation of some of the key assumptions. A typical example where rational game theory seems to give an inadequate answer is the ``backward induction paradox" related to repeated (iterated) social dilemmas like the Repeated Prisoner's Dilemma. According to game theory the only subgame perfect Nash equilibrium in the finitely repeated game is the one determined by backward induction, i.e., when both players defect in all rounds. Nevertheless, cooperation is frequently observed in real-life psycho-economic experiments. This result either suggests that the abstract Prisoner's Dilemma game is not the right model for the situation or that the players do not fulfill all the premises. Indeed, there is good reason to believe that many realistic problems, in which the effect of an agent's action depends on what other agents do are far more complex that perfect rationality of the players could be postulated \citep{conlisk_jel96}. The standard deductive reasoning looses its appeal when agents have non-negligible cognitive limitations, there is a cost of gathering information about possible outcomes and payoffs, the agents do not have consistent preferences, or the common knowledge of the players' rationality fails to hold. A possible way out is \emph{inductive reasoning}, i.e., a trial-and-error approach, in which agents continuously form hypotheses about their environment, build strategies accordingly, observe their performance in practice, and verify or discard their assumptions based on empirical success rates. In this approach the outcome (solution) of a problem is determined by the evolving mental state (mental representation) of the constituting agents. Mind necessarily becomes an endogenous dynamic variable of the model. This kind of bounded rationality may explain that in many situations people respond instinctively, play according to heuristic rules and social norms rather then adopting the strategies indicated by rational game theory.

Bounded rationality becomes a natural concept when the goal of the theory is to understand animal behavior. Individuals in an animal population do not make conscious decisions about strategy, even though the incentive structure of the underlying formal game they ``play" is identical to the ones discussed under the assumption of perfect rationality in classical game theory. In most cases the applied strategies are genetically coded and maintained during the whole life-cycle, the strategy space is constrained (e.g., mixed strategies may be excluded), or strategy adoption or change is severely restricted by biologically predetermined learning rules or mutation rates. The success of a strategy applied is measured by biological \emph{fitness}, which is usually related with reproductive success.

Evolutionary game theory is an extension of the classical paradigm towards bounded rationality. There is however, another aspect of the theory which was swept under the rug in the classical approach, but gets special emphasis in the evolutionary version, namely dynamics. Dynamical issues were mostly neglected classically, because the assumption of perfect rationality made such questions irrelevant. Full deductive rationality allows the players to derive and construct the equilibrium solution instantaneously. In this spirit, when dynamic methods were still applied, like Brown's fictitious play \citep{ brown_51}, they only served as a technical aid for deriving the equilibrium. Bounded rationality, on the other hand, is inseparable from dynamics. Contrary to perfect rationality, bounded rationality is always defined in a positive way, postulating what boundedly rational agents do. These \emph{behavioral rules} are dynamic rules, specifying how much of the game's earlier history is taken into consideration (memory), how long agents would think ahead (short-sightedness, myope), how they search for available strategies (search space), how they switch for more successful ones (adaptive learning), and what all these mean at the population level in terms of frequencies of strategies.

The idea of bounded rationality has the most obvious relevance in biology. It is not too surprising that early applications of the evolutionary perspective of game theory appeared in the biology literature. It is customary to cite R. A. Fisher's analysis on the equality of the sex ratio \citep{fisher_30} as one of the initial works with such ideas, and R.\ C.\ Lewontin's early paper which was probably the first to make a formal connection between evolution and game theory \citep{lewontin_jtb61}. However, the real onset of the theory can be dated to two seminal books in the early 80s: J. Maynard Smith's ``Evolution and the Theory of Games" \citep{maynard_82}, which introduced the concept of \emph{evolutionary stable strategies}, and R. Axelrod's ``The Evolution of Cooperation" \citep{axelrod_84}, which opened up the field for economics and the social sciences. Whereas biologists have used game theory to understand and predict certain outcomes of organic evolution and animal behavior, the social sciences community welcomed the method as a tool to understand social learning and ``cultural evolution", a notion referring to changes in human beliefs, values, behavioral patterns and social norms.

There is a \emph{static} and a \emph{dynamic} perspective of evolutionary game theory. Maynard Smith's definition of the evolutionary stability of a Nash equilibrium is a static concepts which does not require solving time-dependent dynamic equations. In simple terms evolutionary stability means that a rare mutant cannot successfully invade the population. The condition for evolutionary stability can be checked directly without incurring complex dynamic issues.
The dynamic perspective, on the other hand, operates by explicitly postulating dynamical rules. These rules can be prescribed as deterministic rules at the population level for the rate of change of strategy frequencies or as microscopic stochastic rules at the agent level (agent-based dynamics). Since bounded rationality may have different forms, there are many different dynamical rules one can consider. The most appropriate dynamics depends on the specificity of the actual biological or socio-economical situation under study. In biological applications the \emph{Replicator dynamics} is the most natural choice, which can be derived by assuming that payoffs are directly related to reproductive success. Socio-economic applications may require other adjustment or learning rules. Both the static and dynamic perspective of evolutionary game theory provide a basis for \emph{equilibrium selection} when the classical form of the game has multiple Nash equilibria.

Once the dynamics is specified the major concern is the long run behavior of the system: fixed points, cycles, and their stability, chaos, etc., and the connection between static concept (Nash equilibrium, evolutionary stability) and dynamic predictions. The connection is far from being trivial, but at least for normal form games with a huge class of ``reasonable" population-level dynamics the \emph{Folk theorem of evolutionary game theory} holds, asserting that stable rest points are Nash equilibria. Moreover, in games with only two strategies evolutionary stability is practically equivalent to dynamic stability. In general, however, it turns out that a static, equilibrium-based analysis is insufficient to provide enough insight into the long-run behavior of payoff maximizing agents with general adaptive learning dynamics \citep{hofbauer_bams03}. Dynamic rules of bounded rationality should not necessarily reproduce perfect rationality results.

As was argued above, the mission of evolutionary game theory was to remedy three key deficiencies of the classical theory: (1) bounded rationality, (2) the lack of dynamics, and (3) equilibrium selection in the case of multiple Nash equilibria. Although this mission was accomplished rather successfully, there was a series of weaknesses remaining. Evolutionary game theory in its early form considered population dynamics on the aggregate level. The state variables whose dynamics are followed are variables averaged over the population such as the relative strategy abundances. Behavioral rules, on the other hand, control the system on the microscopic, agent level. Agent decisions are frequently asynchronous, discrete and may contain stochastic elements. Moreover, agents may have different individual preferences, payoffs, strategy options (heterogeneity in agent types) or be locally connected to well-defined other agents (structural heterogeneity).

In large populations these microscopic fluctuations usually average out and produce smooth macroscopic behavior for the aggregate quantities. Even though the underlying microscopic rules can be rather different, there is a wide class of models where the standard population level dynamics, e.g., the replicator dynamics, can indeed be microscopically justified. In these situations the mean-field analysis, assuming an infinite, homogeneous population with unbiased random matching, can provide a good qualitative description. In other cases, however, the emerging aggregate level behavior may easily differ even qualitatively from the naive mean-field analysis. Things can go awry especially when the population is largely heterogeneous in agent types and/or when the topology of the interaction graph is nontrivial.

Although the importance of heterogeneity and structural issues was recognized long time ago \citep{follmer_jme74}, the systematic investigation of these questions is still in the forefront of research. The challenge is high, because heterogeneity in both agent types and connectivity structure breaks down the symmetry of agents, and thus requires a dramatic change of perspective in the description of the system from the aggregate level to the agent level. The resulting huge increase in the relevant system variables makes most standard analytical techniques, operating with differential equations, fixed points, etc., largely inapplicable. What remains is \emph{agent based modeling}, meaning extensive numerical simulations and analytical techniques going beyond the traditional mean-field level. Although we fully acknowledge the relevance and significance of the type heterogeneity problem, this Review will mostly concentrate on structural issues, i.e., we will assume that the population we consider consists of identical players (or at least the number of different player roles is small), and the players' only asymmetry stems from their unequal interaction neighborhoods.

It is well-known that real-life interaction networks can possess a rather complex topology, which is far from the traditional mean-field case. On one hand, there is a large class of situations where the interaction graph is determined by the geographical location of the participating agents. Biological games are good examples. The typical structure is two-dimensional. It can be modeled by a regular two-dimensional (2D) lattice or a graph with nodes in 2D and an exponentially decaying probability of distant links. On the other hand, games, motivated by economic or social situations, are typically played on scale free or small word networks, which have rather peculiar statistical properties \citep{albert_rmp02}. Of course, a fundamental geographic embedding cannot be ruled out either. Hierarchical structures are also possible with several levels. In many cases the inter-agent connectivity is not rigid but can continuously evolve in time.


In the simplest spatial evolutionary games agents are located on the sites of a lattice and play repeatedly with their neighbors. Individual income arises from two-person games played with neighbors, thereby the total income depends on the distribution of strategies within the neighborhood. From time to time agents are allowed to modify their strategies in order to increase their utility. Following the basic Darwinian selection idea, in many models the agents adopt (learn) one of the neighboring strategies that has provided a higher income. Similar models are widely and fruitfully used in different areas of science to determine macroscopic behavior from microscopic interactions. Apparently, many aspects of these models seem to be similar to many-particle systems, that is, one can observe different phases and phase transitions when the model parameters are tuned. The striking analogies inspired many physicists to contribute to the deeper understanding of the field by successfully adopting approaches and tools from statistical physics.

Evolutionary games can also exhibit behaviors which do not appear in typical equilibrium physical systems. These aspects require the methods of non-equilibrium statistical physics, where such complications have been investigated for a long time. 
In evolutionary games the interactions are frequently asymmetric, the time-reversal symmetry is broken for the microscopic steps, and many different (evolutionarily) stable states can coexist by forming frozen or self-organizing patterns.

In spatial models the short-range interactions limit the number of agents who can affect the behavior of a given player in finding her best solution. This process can be disturbed fundamentally if the number of possible strategies exceeds the number of neighbors. Such a situation can favor the formation of different strategy associations that can be considered as complex agents with proper spatio-temporal structure, and whose competition will determine the final stationary state. In other words, spatial evolutionary games provide a mathematical framework for studying the emergence of structural complexity that characterizes living material.

Very recently the research of evolutionary games has interfered with the extensive investigation of networks, because the actual social networks characterizing human interactions possess highly nontrivial topological properties. The first results clearly demonstrated that the topological features of these networks can influence significantly their behavior. In many cases ``games on graphs`` differ qualitatively from their counterparts defined in a well-mixed (mean-field) population.
The thorough mathematical investigation of these phenomena requires an extension of the traditional tools of non-equilibrium statistical physics. Evolutionary games lead to dynamical regimes much richer and subtle than those attainable with traditional equilibrium or non-equilibrium statistical physics models. We need revolutionary new concepts and methods for the characterization of the emerging complex, self-organizing patterns.


The setup of this review is as follows. The next three Sections summarize the basic concepts and methods of rational and evolutionary game theory. Our aim was to cut the material into a digestible size, and give a tutorial type introduction to the field, which is traditionally outside the standard curriculum of physicists. Admittedly, many highly relevant and interesting aspects have been left out. The focus is on non-cooperative matrix games (normal form games) with complete information. These are the games whose network extensions have received the most attention in the literature so far.
Section \ref{sec:noc} is devoted to the structure of realistic social networks on which the games can be played. Sections \ref{sec:epd} to \ref{sec:compass} review the dynamic properties of three prominent families of games: the Prisoner's Dilemma, the Rock-Scissors-Paper game, and Competing Associations. These games show a number of interesting phenomena, which occur due to the topological peculiarities of the underlying social network, and nicely illustrate the need to go beyond the mean-field approximation for a quantitative description.
We discuss open questions and provide an outlook for future research areas in Sec.\ \ref{sec:co}. There are three Appendices: the first gives a concise list of the most important games discussed in the paper, the second summarizes noteworthy strategies, and the third gives a detailed introduction to the Generalized Mean-field Approximation (Cluster Approximation) widely used in the main text.

\section{Rational game theory} 
\label{sec:rgt}

The goal of Sections \ref{sec:rgt} to \ref{sec:eg-abd} is to provide a concise summary of the necessary definitions, concepts and methods of rational (classical) and evolutionary game theory, which can serve as a background knowledge in later sections. Most of the material presented here is treated in much more detail in standard textbooks like \citet{fudenberg_91,gibbons_92,hofbauer_98,weibull_95,samuelson_97,gintis_00,cressman_03} and in a recent review by \citet{hofbauer_bams03}.

\subsection{Games, payoffs, strategies}
\label{sec:gps}

A game is an abstract formulation of an interactive decision situation with possibly conflicting interests. The \emph{normal (strategic) form representation} of a game specifies (1) the players of the game, (2) their feasible actions (to be called pure strategies), and (3) the payoffs received by them for each possible combination of actions (the action or strategy profile) that could be chosen by the players. Let $n=1,\dots,N$ denote the players; ${\cal S}_n=\{e_{n1},e_{n2},\dots e_{nQ}\}$ the set of pure strategies available to player $n$, with $s_n\in \cS_n$ an arbitrary element of this set; $(s_1,\dots,s_N)$ a given strategy profile of all players; and $u_n(s_1,\dots,s_N)$ player $n$'s payoff function (utility function), i.e., the measure of her satisfaction if the strategy profile $(s_1,\dots,s_N)$ gets realized. Such a game can be denoted as $G=\{\cS_1,\dots,\cS_N; u_1,\dots,u_N\}$.\footnote{Note that this definition only refers to static, one-shot games with simultaneous decisions of the players, and complete information. The more general case (dynamic games or games with incomplete information) is usually given in the \emph{extensive form representation} \citep{fudenberg_91,cressman_03}, which, in addition to the above list, also specifies: (4) when each player has the move, (5) what are the choice alternatives at each move, and (6) what information is available for the player at the moment of the move. Extensive form games can also be cast in normal form, but this may imply a substantial loss of information. We do not consider extensive form games in this review.}

In the case when there are only two players $n=1,2$ and the set of available strategies is discrete, $\cS_1=\{e_1,\dots,e_Q\}$, $\cS_2=\{f_1,\dots,f_R\}$ it is customary to write the game in a bi-matrix form $G=(\bA,\bB^T)$, which is a shorthand for the payoff table\footnote{It is customary to define $\bB$ as the transpose of what appears in the table.}
\begin{equation}\label{payofftable}
\begin{tabular}{lc|cccc}
  & & & \multicolumn{3}{c}{Player 2} \\
  & & & $f_1$ & $\cdots$  & $f_R$ \\[0.5ex] \hline
  & & &       &           &       \\[-2.0ex]
  &$e_1$&  & $(A_{11},B_{11}^T)$ & $\cdots$ & $(A_{1R},B_{1R}^T)$ \\
Player 1\phantom{a} & $\vdots$ &  &  $\vdots$ & $\ddots$ & $\vdots$  \\
  &$e_Q$&  & $(A_{Q1},B_{Q1}^T)$ & $\cdots$ &  $(A_{QR},B_{QR}^T)$  \\
\end{tabular}
\end{equation}
Here the matrix $A_{ij}=u_1(e_i,f_j)$ [resp., $B_{ij}^T=u_2(e_i,f_j)$] denotes Player 1's [resp., Player 2's] payoff for the strategy profile $(e_i,f_j)$.\footnote{Not all normal form games can be written conveniently in a matrix form. The Cournot and Bertrand games are simple examples \citep{gibbons_92,marsili_pa97}, where the strategy space is not discrete, $\cS_n=[0,\infty)$, and payoffs should be treated in functional form. This is, however, only a minute technical difference, the solution concepts and methods remain unchanged.}

Two-player normal-form games can be \emph{symmetric} (so called \emph{matrix games}) and \emph{asymmetric} (\emph{bi-matrix games}), with symmetry referring to the roles of players. For a symmetric game players are identical in all possible respects, they possess identical strategy options and payoffs, i.e., necessarily $R=Q$ and $\bB=\bA$. It does not matter for a player whether she should play in the role of Player 1 or Player 2.
A symmetric normal form game is fully characterized by the single payoff matrix $\bA$, and we can formally write $G=(\bA,\bA^T)=\bA$. The Hawk-Dove game (see Appendix \ref{sec:app:g} for a detailed description of games mentioned in the text), the Coordination game or the Prisoner's Dilemma are all symmetric games.

Player asymmetry, on the other hand, is often an important feature of the game like in male-female, buyer-seller, owner-intruder, or sender-receiver interactions. For asymmetric games (bi-matrix games) $\bB\ne\bA$, thus both matrices should be specified to define the game, $G=(\bA,\bB^T)$. It makes a difference whether the player is called to act in the role of Player 1 or Player 2.

Sometimes the players change roles frequently during interactions like in the Sender-Receiver game of communication, and a symmetrized version of the underlying asymmetric game is considered. The players can act as Player 1 with probability $p$ and as Player 2 with probability $1-p$. These games are called \emph{role games} \citep{hofbauer_98}. When $p=1/2$ the game is symmetric in the higher dimensional strategy space $\cS_1\times\cS_2$, formed by the pairs of the elementary strategies.

Symmetric games, whose payoff matrix is symmetric $\bA=\bA^T$ are called \emph{doubly symmetric}, and can be denoted as $G=(\bA,\bA)\equiv \bA$. These games belong to the so-called \emph{partnership} or \emph{potential} games. See Section \ref{sec:pzg} for a discussion. The asymmetric game $G=(\bA,-\bA)$ is called a \emph{zero-sum game}. This is the class where the existence of an equilibrium was first proved in the form of the Minimax Theorem \citep{neumann_ma28}.

The strategies that label the payoff matrices are \emph{pure strategies}. In many games, however, the players can also play \emph{mixed strategies}, which are probability distributions over pure strategies. {Poker} is a good example, where (good) players play according to a mixed strategy over the feasible actions of ``bluffing" and ``no bluffing". We can assume that players possess a randomizing device which can be utilized in decision situations. Playing a mixed strategy means that in each decision instance the player come up with one of her feasible actions  with a certain pre-assigned probability.
Each mixed strategy corresponds to a point $\bp$ of the \emph{mixed strategy simplex}
\begin{equation}\label{eq:simplex}
    \Delta_Q = \left\{ \bp=(p_1,\dots,p_Q)\in \mathbb{R}^Q :
    p_q\ge 0, \sum_{q=1}^Q p_q = 1  \right\},
\end{equation}
whose corners are the pure strategies.
In the two-player case of Eq.\ (\ref{payofftable}) a strategy profile is the pair $(\bp,\bq)$ with $\bp\in \Delta_Q, \bq\in \Delta_R$, and the expected payoffs of player 1 and 2 can be expressed as
\begin{equation}
    u_1(\bp,\bq)=\bp\cdot\bA\bq, \qquad u_2(\bp,\bq)=\bp\cdot\bB^T\bq=\bq\cdot\bB\bp.
\end{equation}
Note that in some games mixed strategies may not be allowed.

The timing of a normal form game is as follows: (1) players independently, but simultaneously choose one of their feasible actions (i.e., without knowing the co-players' choices), (2) players receive payoffs according to the action profile realized.

\subsection{Nash equilibrium and social dilemmas}

Classical game theory is based on two key assumptions: (1) perfect rationality of the players, and (2) that this is common knowledge. \emph{Perfect rationality} means that the players have well-defined payoff functions, and they are fully aware of their own and the opponents' strategy options and payoff values. They have no cognitive limitations in deducing the best possible way of playing whatever the complexity of the game is. In this sense computation is costless and instantaneous. Players are capable of correctly assessing missing information (in games with incomplete information) and process new information revealed by the play of the opponent (in dynamic games) in terms of probability distributions. \emph{Common knowledge}, on the other hand, implies that beyond the fact that all players are rational, they all know that all players are rational, and that all players know that all players know that all are rational, etc., ad infinitum \citep{fudenberg_91,gibbons_92}.

A strategy $s_n$ of player $n$ is \emph{strictly dominated} by a (pure or mixed) strategy $s_n'$, if for each strategy profile $s_{-n}=(s_1,\dots,s_{n-1},s_{n+1},\dots,s_N)$ of the co-players, player $n$ is always better off playing $s_n'$ than $s_n$,
\begin{equation}\label{eq:strict_dominance}
    \forall\,s_{-n}\,: \quad  u_n^{}(s_n',s_{-n}^{}) > u_n^{}(s_n^{},s_{-n}^{}).
\end{equation}
According to the standard minimal definition of rationality \citep{aumann_92}, rational players do not play strictly dominated strategies. Thus strictly dominated strategies can be iteratively eliminated from the problem. In some cases, like the Prisoner's Dilemma, only one strategy profile survives this procedure, which is then the perfect rationality ``solution" of the game. It is more common, however, that \emph{iterated elimination of strictly dominated strategies} do not solve the game, because either there are no strictly dominated strategies at all, or more than one profiles survive.

A stronger solution concept, which is applicable for all games, is the concept of a Nash equilibrium. A strategy profile $s^*=(s_1^*,\dots,s_N^*)$ of a game is said to be a \emph{Nash equilibrium} (NE), iff
\begin{equation}\label{eq:NE}
    \forall\,n, \forall\, s_n^{}\ne s_n^* : \quad
    u_n^{}(s_n^*,s_{-n}^*) \ge u_n^{}(s_n^{},s_{-n}^*).
\end{equation}
In other terms, each agent's strategy $s_n^*$ is a \emph{best response} (BR) to the strategies of the co-players,
\begin{equation}\label{eq:NEalt}
    \forall\,n\,:\quad  s_n^* = \textrm{BR}(s_{-n}^*),
\end{equation}
where
\begin{equation}\label{eq:BRdef}
    \textrm{BR}(s_{-n})\equiv {\rm argmax}_{s_n} u_n(s_n,s_{-n}).
\end{equation}

When the inequality above is strict, $s^*$ is called a \emph{strict Nash equilibrium}. The NE condition assures that no player has a unilateral incentive to deviate and play another strategy, because, given the others' choices, there is no way she could be better off. One of the most fundamental results of classical game theory is \emph{Nash's theorem} \citep{nash_pnas50}, which asserts that in normal-form games with a finite number of players and a finite number of pure strategies there exists at least one NE, possibly involving mixed strategies. The proof is based on Kakutani's fixed point theorem. An NE is called a \emph{symmetric (Nash) equilibrium} if all agents play the same strategy in the equilibrium profile.

The Nash equilibrium is a stability concept, but only in a rather restricted sense: stable against single-agent (i.e., unilateral) changes of the equilibrium profile. It does not speak about what could happen if more than one agent changed their strategies at the same time. In this latter case there are two possibilities which classify NEs into two categories. It may happen that there exists a suitable collective strategy change that increases some players' payoff while not decreasing all others'. Clearly then the original NE was \emph{inefficient} [sometimes called a \emph{deficient equilibrium} \citep{ rapoport_gs66}], and in theory can be emended by the new strategy profile. In all other cases the NE is such that any collective strategy change makes at least one player worse off (or, in the degenerate case, all payoffs remain the same). These NEs are called \emph{Pareto efficient}, and there is no obvious way for improvement.

Pareto efficiency can be used as an additional criterion (a so-called refinement) to the NE concept to provide \emph{equilibrium selection} in cases when the NE concept alone would provide more than one solution to the game like in some Coordination problems. For example, if a game has two NEs,  the first is Pareto efficient, the second is not, then we can say that the \emph{refined strategic equilibrium concept} (in this case the Nash equilibrium concept plus Pareto efficiency) predicts that the outcome of the game is the first NE. Preferring Pareto-efficient equilibria to deficient equilibria becomes an inherent part of the definition of rationality.

It may well occur that the game has a single NE, which is, however, not Pareto efficient, and thus the social welfare (the sum of the individual utilities) is not maximized in equilibrium. Two archetypical examples are the Prisoner's Dilemma and the Tragedy of the Commons (see later in Sec.\ \ref{sec:mpg}). Such situations are called \emph{social dilemmas}, and their analysis, avoidance or possible resolution is one of the most fundamental issues of economics and social sciences.

Another refinement concept which could serve as a guideline for equilibrium selection is \emph{risk dominance} \citep{harsanyi_88}. A strategy $s_n'$ risk dominates another strategy $s_n$, if the former has higher expected payoff against an opponent playing all his feasible strategies with equal probability. So playing strategy $s_n$ would have higher risk, if the opponent were, for some reason, irrational, or if were unable to decide between more than one equally appealing NEs. The concept of risk dominance is the precursor to stochastic stability to be discussed in the sequel. Note that Pareto efficiency and risk dominance may well give contradictory advise.

\subsection{Potential and zero-sum games}
\label{sec:pzg}

In general the number of utility functions to consider in a game equals to the number of players. However, there are two special classes where a single function is enough to characterize the strategic incentives of all players: potential games and zero-sum games. The existence of these single functions make the analysis more transparent, and as we will see, the methods of statistical physics directly applicable.

By definition \citep{Monderer_geb96}, if there exists a function $V=V(s_1,s_2,\dots,s_N)$ such that for each player $n=1,\dots,N$ the utility function differences satisfy
\begin{equation}\label{potential}
    u_n(s_n^\prime;s_{-n})-u_n(s_n;s_{-n}) = V(s_n^\prime;s_{-n})-V(s_n;s_{-n}),
\end{equation}
then the game is called a \emph{potential game} with $V$ being its \emph{potential}. If the potential exists it can be thought of as a single, fixed landscape, common for all players, in which they try to reach its maximum. In the case of the two-player game in Eq.\ (\ref{payofftable}) $V$ is a $Q\times R$ matrix, which represents a two-dimensional discretized landscape. The concept can be trivially extended to more players, in which case the dimension of the landscape is the number of players. For games with continuous strategy space (e.g., mixed strategies) the finite differences in Eq.\ (\ref{potential}) can be replaced by partial derivatives and the landscape is continuous.

For potential games the existence of a Nash equilibrium, even in pure strategies, is trivial if the strategy space (the landscape) is compact: the global maximum of $V$, which then necessarily exists, is a pure strategy NE.

For a two-player game, $G=(\bA,\bB^T)$, it is easy to formulate a simple sufficient condition for the existence of a potential. If the payoffs are equal for the two players in all strategy configurations, i.e., $\bA=\bB^T=\bV$, then $\bV$ is a potential, as can be checked directly. These games are also called ``partnership games" or ``games with common interests" \citep{Monderer_geb96,hofbauer_98}. If the game is symmetric, i.e., $\bA=\bB$ this condition implies that the potential is a symmetric matrix $\bV^T=\bV$.

There is, however, an even wider class of games for which a potential can be defined. Indeed, notice that Nash equilibria are invariant under certain rescaling of the game. Specifically, the game $G'=(\bA',\bB^{\prime T})$ is said to be \emph{Nash-equivalent} to $G=(\bA,\bB^T)$, and denoted $G\sim G'$, if there exist constants $\alpha,\beta>0$ and $c_r$, $d_q$ such that
\begin{equation}\label{eq:equivalence}
    A_{qr}^\prime = \alpha\, A_{qr} + c_{r}; \qquad
    B_{rq}^\prime = \beta\, B_{rq} + d_{q}.
\end{equation}
According to this, the payoffs can be freely multiplied, and arbitrary constants can be added to columns of Player 1's payoff and the rows of Player 2's payoff -- the Nash equilibria of $G$ and $G'$ are the same.
If there exists a $Q\times R$ matrix $\bV$ such that $(\bA,\bB^T)\sim (\bV,\bV)$, the game is called a \emph{rescaled potential game} \citep{hofbauer_98}.

The other class of games with a single strategy landscape is \emph{(rescaled) zero-sum games}, $(\bA,\bB^T)\sim (\bV,-\bV)$. Zero-sum games are only defined for two players. Whereas for potential games the players try to maximize $V$ along their associated dimensions, for zero-sum games Player 1 is a maximizer and Player 2 is a minimizer of $\bV$ along their respective strategy spaces. The existence of a Nash equilibrium in mixed strategies $(\bp,\bq)\in \Delta_Q\times\Delta_R$ follows from Nash's theorem, but in fact it was proved earlier by von Neumann.

Von Neumann's \emph{Minimax Theorem} asserts \citep{neumann_ma28} that each zero-sum game can be associated with a \emph{value} $v$, the optimal outcome. The payoffs $v$ and $-v$, respectively, are the best the two players can achieve in the game if both are rational. Denoting player 1's expected payoff by $u(\bp,\bq)=\bp \cdot\bV \bq$, the value $v$ satisfies
\begin{equation}\label{minimax}
    v= \max_{\bp} \min_{\bq}\, u(\bp,\bq) = \min_{\bq} \max_{\bp}\, u(\bp,\bq),
\end{equation}
i.e., the two extremization steps can be exchanged. The minimax theorem also provides a straightforward algorithm how to solve zero-sum games (\emph{minimax algorithm}), with direct extension (at least in theory) to dynamic zero-sum games such as Chess or Go.

\subsection{NEs in two-player matrix games}
\label{sec:netpmg}

A general two-player, two-strategy symmetric game is defined by the matrix
\begin{equation}\label{eq:general2x2}
{\bf A}=\left( \matrix{a & b \cr
                       c & d \cr}\right).
\end{equation}
where $a,b,c,d$ are real parameters. Such a game is Nash-equivalent to the rescaled game
\begin{equation}\label{Aprime}
{\bf A}'=\left( \matrix{\cos\phi & 0 \cr
                       0 & \sin\phi \cr}\right), \quad \tan\phi=\frac{d-b}{a-c}.
\end{equation}
Let $p_1$ characterize Player 1's mixed strategy $\bp=(p_1,1-p_1)$, and $q_1$ Player 2's mixed strategy $\bq=(q_1,1-q_1)$. By introducing the notation
\begin{equation}
    r=      \frac{1}{1+\displaystyle{\frac{a-c}{d-b}}} =
            \frac{1}{1+\cot\phi},
\end{equation}
the following classes can be distinguished as a function of the single parameter $\phi$:

\begin{description}
\item[Coordination Class] [$0<\phi<\pi/2$, i.e., $a-c>0, d-b>0$]. There are two strict, pure strategy NEs, and one non-strict, mixed strategy NE:
\begin{eqnarray}
  \textrm{NE}_1:& p_1^*=q_1^*=1; 
  \nonumber\\
  \textrm{NE}_2:& p_1^*=q_1^*=0; 
  \\
  \textrm{NE}_3:& p_1^*=q_1^*=r; 
  \nonumber
\end{eqnarray}
All the NEs are symmetric. The prototypical example is the Coordination game (see Appendix \ref{app:g:co} for details).
\item[Anti-Coordination Class] [$\pi<\phi<3\pi/2$, i.e., $a-c<0, d-b<0$] There are two strict, pure strategy NEs, and one non-strict, mixed strategy NE:
\begin{eqnarray}\label{eq:anti-coord}
  \textrm{NE}_1:& p_1^*=1,q_1^*=0; \nonumber\\
  \textrm{NE}_2:& p_1^*=0,q_1^*=1; \\
  \textrm{NE}_3:& p_1^*=q_1^*=r; \nonumber
\end{eqnarray}
$\textrm{NE}_1$ and $\textrm{NE}_2$ are asymmetric, $\textrm{NE}_3$ is a symmetric equilibrium. The most important games belonging to this class are the Hawk-Dove, the Chicken, and the Snowdrift games (see Appendix \ref{sec:app:g}).
\item[Pure Dominance Class] [$\pi/2<\phi<\pi$ or $3\pi/2<\phi<2\pi$, i.e., $(a-c)(d-b)\le 0$]. One of the pure strategies is strictly dominated. There is only one Nash equilibrium, which is pure, strict, and symmetric:
\begin{equation}
\textrm{NE}_1:\; \left\{\begin{array}{cc}
                             p_1^*=q_1^*=0 & \mbox{if $\pi/2<\phi<\pi$} \\
                             p_1^*=q_1^*=1 & \mbox{if $3\pi/2<\phi<2\pi$} \\
                            \end{array}\right.
\end{equation}
\end{description}
The best example is the Prisoner's Dilemma (see Appendix \ref{app:g:pd} and later sections).

\begin{figure}[t]
\centerline{\epsfig{file=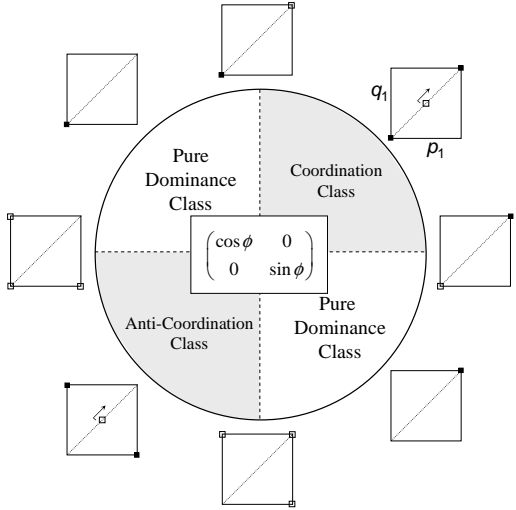,width=7.cm}}
\caption{\label{fig:Fig_2x2NE} Nash equilibria in $2\times 2$ matrix games $\bA'$, defined in Eq.\ (\ref{Aprime}), classified by the angle $\phi$. Filled boxes denote strict NE, empty boxes non-strict NE in the strategy space $p_1$ vs $q_1$. The $\mapsto$ shows the direction of motion of the mixed strategy NE for increasing $\phi$.}
\end{figure}

On the borderline between the different classes games are degenerate and a new equilibrium appears or disappears. At these points the appearing/disappearing NE is always non-strict. Figure \ref{fig:Fig_2x2NE} depicts the phase diagram.

The classification above is based on the number and type of Nash equilibria. Note that this is only a minimal classification, since these classes can be further divided using other properties.
For instance, the Nash-equivalency relation does not keep Pareto efficiency invariant. The latter is a separate issue, which does not depend directly on the combined parameter $\phi$.
In the Coordination Class, unless $a=d$, only one of the three NEs is Pareto efficient: if $a>d$ it is $\textrm{NE}_1$, if $a<d$ it is $\textrm{NE}_2$. $\textrm{NE}_3$ is never Pareto efficient.\draftnote{Check these!}
In the Anti-Coordination Class $\textrm{NE}_1$ and $\textrm{NE}_2$ are Pareto efficient, 
$\textrm{NE}_3$ is not.
The only NE of the Pure Dominance Class is Pareto efficient when $d\ge a$ for $\pi/2<\phi<\pi$, and when $d\le a$ for $3\pi/2<\phi<2\pi$. Otherwise the NE is deficient (social dilemma).


What can we say when the strategy space contains more than two pure strategies, $Q>2$? It is hard to give a complete classification as the number of the possible classes increases exponentially with $Q$ \citep{broom_mb00}. A classification based on the replicator dynamics (see later) is available for $Q=3$ \citep{bomze_bc83,bomze_bc95,zeeman_80}.

What we surely know is that for any finite game which allows mixed strategies, Nash's theorem \citep{nash_pnas50} assures that there is at least one Nash equilibrium, possibly in mixed strategies. Nash's theorem is only an existence theorem, and in fact the typical number of Nash equilibria in matrix games increases rapidly as the number of pure strategies $Q$ increases.

It is customary to distinguish interior and boundary NEs with respect to the strategy simplex. An \emph{interior NE} $\bp_{\rm int}^*\in \inter\Delta_Q$ is a mixed strategy equilibrium with $0<p_i^*<1$ for all $i=1,\dots,Q$. A \emph{boundary NE} $\bp^*\in \bd\Delta_Q$ is a mixed strategy equilibrium in which at least one of the pure strategies has zero weight, i.e., $\exists i$ such that $p_i^*=0$. These solutions are situated on the $Q-2$ dimensional boundary $\bd\Delta_Q$ of the $Q-1$ dimensional simplex $\Delta_Q$. Pure strategy NEs are sometimes called "corner solutions". As a special case of Nash's theorem it can be shown \citep{cheng_proc04} that a finite symmetric game always possesses at least one \emph{symmetric equilibrium}, possibly in mixed strategies.

Whereas for $Q=2$ we always have a pure strategy equilibrium, this is no longer true for $Q>2$. A typical example is the Rock-Scissors-Paper game whose only NE is mixed.

In case when the payoff matrix is regular, i.e., $\det \bA\ne 0$, there can be at most one interior NE. An interior NE in a matrix game is necessarily symmetric. If this exists it necessarily satisfies \citep{ bishop_jtb78}
\begin{equation}\label{eq:interiorNE}
    \bp_{\rm int}^* = \frac{1}{\cal N} \bA^{-1} \mathbf{1},
\end{equation}
where $\mathbf{1}$ is the $Q$-dimensional vector with all elements equals $1$, and $N$ is a normalization factor to assure $\Sigma_i [\bp_{\rm int}^*]_i=1$. 
The possible location of the boundary NEs, if they exist, can be calculated similarly by restricting (projecting) ${\bf A}$ to the appropriate boundary manifold under consideration.

When the payoff matrix is singular $\det \bA= 0$ the game may have an extended set of NEs, a so-called \emph{NE component}. An example for this will be given later in Section \ref{sec:es}.


In case of \emph{asymmetric games} (bi-matrix games) the two players have different strategy sets and different payoffs. An asymmetric Nash equilibrium is now a pair of strategies. Each component of such a pair is a best response to the other component. The classification of bi-matrix games is more complicated even for two possible strategies, $Q=R=2$. The standard reference is \citet{ rapoport_gs66} who define three major classes based on the number of players having a dominant strategy, and then define altogether fourteen different sub-classes within. Note, however, that they only classify bi-matrices with ordinal payoffs, i.e., when the four elements of the 2x2 matrix are the numbers 1,2,3 and 4 (the rank of the utility).

A different taxonomy based on the replicator dynamics phase portraits is given by \citet{ cressman_03}.

\subsection{Multi-player games}
\label{sec:mpg}

There are many socially and economically important examples where the number of decision makers involved is greater than two. Although sometimes these situations can be modeled as repeated play of simple pair interactions, there are many cases where the most fundamental unit of the game is irreducibly of multi-player nature. These games cannot be cast in a matrix or bi-matrix form. Still the basic solution concept is the same: when played by rational agents the outcome should be a Nash equilibrium where no player has an incentive to deviate unilaterally.

We briefly mention one example here: the Tragedy of the Commons \citep{ gibbons_92}. This abstract games exemplifies what has been one of the major concerns of political philosophy and economic thinking since at least David Hume in the 18th century \citep{hardin_s68}: \emph{without central planning and global control, private incentives would lead to overutilization of public resources and insufficient contribution to public goods.} Clearly, models of this kind provide valuable insight into deep socioeconomic problems like pollution, deforestation, mining, fishing, climate control, or environment protection, just to mention a few.

The Tragedy of the Commons is a simultaneous move game with $N$ players (farmers). The strategy for farmer $n$ is to choose a non-negative number $g_n\in [0,\infty)$, the number of goats, assumed to be a real number for simplicity, to graze on the village green. One goat implies a constant cost $c$, and a benefit $v(G)$ to the farmer, which is a function of the total number of goats $G=g_1+\dots +g_N$ grazing on the green. However, grass is a scarce resource. If there are many goats on the green the amount of available grass decreases. Goats become undernourished and their value decreases. We assume that $v'(G)<0$ and $v''(G)<0$. The farmer's payoff is
\begin{equation}
    u_n = [v(G) - c]\, g_n.
\end{equation}
Rational game theory predicts that the emerging collective behavior is a Nash equilibrium with choices $(g_1^*,\dots,g_N^*)$ from which no farmer has a unilateral incentive to deviate. The first order condition for the coupled maximization problem $\partial u_n/\partial g_n=0$ leads to
\begin{equation}
    v(g_n+G_{-n}^*) + g_n v'(g_n+G_{-n}^*)-c = 0
\end{equation}
where $G_{-n}=g_1+\dots+g_{n-1}+g_{n+1}+\dots+g_N$. Summing over the first order conditions for all players we get an equation for the equilibrium number of total goats, $G^*$,
\begin{equation}\label{eq:ToC1}
    v(G^*) + \frac{1}{N} G^* v'(G^*)-c=0,
\end{equation}
Given the function $v$, this can be solved (numerically) to obtain $G^*$.

Note, however, that the social welfare would be maximized by another value $G^{**}$, which maximizes the total profit $G v(G)-c G$ for $G$, i.e., satisfies the first order condition (note the missing $1/N$ factor)
\begin{equation}\label{eq:ToC2}
    v(G^{**}) + G^{**} v'(G^{**})-c=0.
\end{equation}
Comparing Eqs.\ (\ref{eq:ToC1}) and (\ref{eq:ToC2}), we find that $G^*>G^{**}$, that is too many goats are grazed in the Nash equilibrium compared to the social optimum. The resource is overutilized. The NE is not Pareto efficient (it could be easily improved by a central planner or a law fixing the total number of goats at $G^{**}$), which makes the problem a social dilemma.

The same kind of inefficiency, and the possible ways to overcome it, is studied actively by economists in Public Good game experiments \citep{ledyard_95}. See Appendix \ref{app:g:pg} for more details on Public Good games.


\subsection{Repeated games}

Most of the inter-agent interactions that can be modeled by abstract games are not one-shot relationships but occur repeatedly on a regular basis. When a one-shot game is played between the same rational players iteratively, a single instance of this series cannot be singled out and treated separately. The whole series should be analyzed as one big ``supergame". What a player does early on can effect what others choose to do later on.

Assume that the same game $G$ (the so called \emph{stage game}) is played a number of times $T$. $G$ can be the Prisoner's Dilemma or any matrix or bi-matrix game discussed so far. The set of feasible actions and payoffs in the stage game at period $t$ ($t=1,\dots,T$) are independent of $t$ and of the former history of the game. This does not mean, however, that actions themselves should be chosen independent of time and history. When $G$ is played at period $t$, all the game history thus far is common knowledge. We will denote this repeated game as $G(T)$ and distinguish \emph{finitely repeated games} $T<\infty$ and \emph{infinitely repeated games} $T=\infty$. For finitely repeated games the total payoff is simply the sum of the stage game payoffs
\begin{equation}\label{eq:nodiscount}
    U = \sum_{t=1}^T u_t.
\end{equation}

For infinitely repeated games \emph{discounting} should be introduced to avoid infinite utilities. Discounting is a regularization method, which is, however, not a pure mathematical trick but reflects real economic factors. On one hand, discounting takes into account the probability $p$ that the repetition of the game may terminate at any period. On the other hand, it represents the fact that the current value of a future income is less than its nominal value. If the interest rate of the market for one period is denoted $r$, the overall \emph{discount factor}, representing both effects reads $\delta=(1-p)/(1+r)$ \citep{gibbons_92}. Using this, the \emph{present value} of the infinite series of payoffs $u_1,u_2,\dots$ is\footnote{Although it seems reasonable to introduce a similar discounting for finitely repeated games, too, that would have no qualitative effect there.}
\begin{equation}\label{eq:discount}
    U = \sum_{t=1}^\infty \delta^{t-1} u_t.
\end{equation}
In the following we will denote a finitely repeated game based on $G$ as $G(T)$ and an infinitely repeated game as $G(\infty,\delta)$, $0<\delta<1$, and consider $U$ in Eq.\ (\ref{eq:nodiscount}), resp. Eq.\ (\ref{eq:discount}), as the payoff of the repeated game (supergame).\\

\leftline{\sf Strategies as algorithms}\medskip

In one-shot static games with complete information like those we have considered in earlier subsections a strategy is simply an \emph{action} a player can choose. For repeated games (and also for other kinds of dynamic games) the concept of a strategy becomes more complex. In these games \emph{a player's strategy is a complete plan of action, specifying a feasible action in any contingency in which the player may be called upon to act}. The number of possible contingencies in a repeated game is the number of possible histories the game could have produced thus far. Thus the strategy is in fact a mathematical algorithm which determines the output (the action to take in period $t$) as a function of the input (the actual time $t$ and the history of the actions of all players up to period $t-1$). In case of perfect rationality there is no limit on this algorithm. However, bounded rationality may restrict feasible strategies to those that do not require too long memories or too complex algorithms.

To illustrate how fast the number of feasible strategies for $G(T)$ increases, consider a symmetric, $N$-player stage game $G$ with action space $\cS=\{e_1,e_2,\dots,e_Q\}$. Let $M$ denote the length of the memory required by the strategy algorithm. $M=1$ if the algorithm only needs to know the actions of the players in the last round, $M=2$ if knowledge of the last two rounds is required, etc. The maximum possible memory length at period $T$ is $T-1$. Given the player's memory, a strategy is a function ${\cal Q}^{N M}\to {\cal Q}$. There are altogether $Q^{Q^{N M}}$ length-$M$ strategies. Notice, however, that beyond depending on the game history a strategy may also depend on the time $t$ itself. For instance, actions may depend whether the stage game is played on a weekday or on the weekend. In addition to the beginning of the game, in finitely repeated games the end of the game can also get special treatment, and strategies can depend on the remaining time $T-t$. As it will be shown shortly, this possibility can have a crucial effect on the outcome of the game.

In case of the Iterated Prisoner's Dilemma games some standard strategies like ``AllC" ($M=0$) ``Tit-for-Tat" (TFT, $M=1$) or ``Pavlov" ($M=1$) are finite-memory strategies, others like ``Contrite TFT" \citep{boerlijst_jtb97,panchanathan_n04} depends as input on the whole action history. However, the need for such a long-term memory can sometimes be traded off for some additional state variables (``image score", ``standing", etc.) \citep{panchanathan_n04,nowak_n05}. For instance, these variables can code for the ``reputation" of agents, which can be utilized in decision making, especially when the game is such that opponents are chosen randomly from the population in each round. Of course, then a strategy should involve a rule for updating these state variables too. Sometimes these strategies can be conveniently expressed as Finite State Machines (Finite State Automata) \citep{binmore_jet92,lindgren_proc97}.

The definition of a strategy should also prescribe how to handle errors (noise) if these have finite possibility to occur. It is customary to distinguish two kinds of errors: \emph{implementation} (``trembling hand") errors and \emph{perception} errors. The first refers to unintended ``wrong" actions like playing $D$ accidentally in the Prisoner's Dilemma when $C$ was intended. These errors usually become common knowledge for players. Perception errors, on the other hand, arise from events when the action was correct as intended, but one of the players (usually the opponent) interpreted it as another action. In this case players end up keeping a different track record of the game history.\\

\leftline{\sf Subgame perfect Nash equilibrium}\medskip

What is the prediction of rational game theory for the outcome of a repeated game? As always the outcome should correspond to a Nash equilibrium: no player can have a unilateral incentive to change its strategy (in the supergame sense), since this would induce immediate deviation from that profile. However, not all NEs are equally plausible outcomes in a dynamic game such as a repeated game. There are ones which are based on ``non-credible threats and promises" \citep{fudenberg_91,gibbons_92}. A stronger concept than Nash equilibrium is needed to exclude these spurious NEs. \emph{Subgame perfection} introduced by \citet{selten_zgs65} is a widely accepted criterion to solve this problem. Subgame perfect Nash equilibria are those that pass a credibility test.

A \emph{subgame} of a repeated game is a subseries of the whole series that starts at period $t\ge 1$ and ends at the last period $T$. However, there are many subgames starting at $t$, one for each possible history of the (whole) game before $t$. Thus subgames are labeled by the starting period $t$ and the history of the game before $t$. Thus when a subgame is reached in the game the players know the history of play. By definition an NE of the game is \emph{subgame perfect} if it is an NE in all subgames \citep{selten_zgs65}.

One of the key results of rational game theory that there is no cooperation in the Finitely Repeated Prisoner's Dilemma. In fact the theorem is more general and states that \emph{if a stage game $G$ has a single NE then the unique subgame perfect NE of $G(T)$ is the one, in which the NE of $G$ is played in every period}. The proof is by \emph{backward induction}. In the last round of a Prisoner's Dilemma rational players defect since the incentive structure of the game is the same as that of $G$. There is no room for expecting future rewards for cooperation or punishment for defection. Expecting defection in the last period they consider the next-to-last period, and find that the effective payoff matrix in terms of the remaining undecided $T-1$ period actions is Nash equivalent to that of $G$ (the expected payoffs from the last round constitutes a constant shift for this payoff matrix). The incentive structure is again similar to that of $G$ and they defect. Considering the next-to-next-to-last round gives similar result, etc. The induction process can be continued backward till the first period showing that rational players defect in all rounds.\\

\leftline{\sf The backward induction paradox}\medskip

There is substantial evidence from experimental economics showing that human players do cooperate in repeated interactions, especially in early periods when the end of the game is still far away. Moreover, even in situations which are seemingly best described as  single-shot Prisoner's Dilemmas cooperation is not infrequent.
The unease of rational game theory to cope with these facts is sometimes referred to as the \emph{backward induction paradox} \citep{pettit_jp89}. Similarly, contrary to rational game theory predictions, altruistic behavior is common in Public Good experiments, and ``fair behavior" appears frequently in the Dictator game or in the Ultimatum game \citep{forsythe_geb94,camerer_jep95}.

Game theory has worked out a number of possible answers to these criticisms: players are rational but the actual game is not what it seems to be; players are not fully rational, or rationality is not common knowledge; etc. As for the first, it seems at least possible that
the evolutionary advantages of cooperative behavior have exerted a selective pressure to make humans hardwired for cooperation \citep{fehr_n03}. This would imply a genetic predisposition for the player's utility function to take into account the well-being of the co-player too. The pure ``monetary" payoff is not the ultimate utility to consider in the analysis. Social psychology distinguishes various player types from altruistic to cooperative to competitive individuals. In the simplest setup the player's overall utility function is $U_n=\alpha u_n+\beta u_m$, i.e., is a linear combination of his own monetary payoff $u_n$ and the opponent's monetary payoff $u_m$, where the signs of the coefficients $\alpha$ and $\beta$ depend on the player's type \citep{brosig_jebo02,weibull_bookch04}. Clearly such a dependence may redefine the incentive structure of the game, and may transform a would-be social dilemma setup into another game, where cooperation is, in fact, rational. Also in many cases the experiment, which is intended to be a one-shot or a finitely repeated game is in fact perceived by the players -- despite all contrary efforts by the experimenter -- as being part of a larger, practically infinite game (say, the players' everyday life in its full complexity) where other issues like future threats or promises, or reputation matter, and strategies observed in the experiment are in fact rational in this supergame sense.

Contrary to what is predicted for finitely repeated games, \emph{infinitely} repeated games can behave differently. Indeed, it was found that cooperation can be rational in infinitely repeated games. A collection of theorems formalizing this result is usually referred to as the Folk Theorem.\footnote{The name "Folk Theorem" refers to the fact that some of these results were common knowledge already in the 1950s, even though no one had published them. Although later more specific theorems were proven and published, the name remained.} General results by \citet{friedman_res71} and later by \citet{fudenberg_e86} imply that if the discount factor is sufficiently close to 1, i.e., if players are sufficiently patient, Pareto efficient subgame perfect Nash equilibria (among many other possible equilibria) can be reached in an infinitely repeated game $G(\infty,\delta)$. In case of the Prisoner's Dilemma, for instance, this means that cooperation can be rational if the game is infinite and $\delta$ is sufficiently large.

When restricted to the Prisoner's Dilemma the proof is based on considering unforgiving (``trigger") strategies like ``Grim Trigger". Grim Trigger starts by cooperating and cooperates until the opponent defects. From that point on it always defects. The strategy profile of both players playing Grim Trigger is a subgame perfect NE provided that $\delta$ is larger than some well defined lower bound. See e.g.\ Gibbons' book \citep{gibbons_92} for a readable account. Clearly, cooperation in this case stems from the fear from an infinitely long punishment phase following defection, and the temptation to defect in a given period is suppressed by the threatening cumulative loss from this punishment. When the discount factor is small the threat diminishes, and the cooperative behavior evaporates.

Another interesting way to explain cooperation in finitely repeated social dilemmas is to assume that players are not fully rational or that their rationality is not common knowledge. Since cooperation of boundedly rational agents is the major theme of subsequent sections, we only discuss here the second possibility, that is when all players are rational but this is not common knowledge. Each player knows that she is rational, but perceives a chance that the co-players are not. This information asymmetry makes the game a different game, namely a finitely repeated game based on a stage game with \emph{incomplete information}. As was shown in a seminal paper by \citet{kreps_jet82}, such an information asymmetry is able to rescue rational game theory, and assure cooperation even in the \emph{finitely} repeated Prisoner's Dilemma, provided that the game is long enough or the chance of playing with a non-rational co-player is high enough, and the stage game payoffs satisfy some simple inequalities. The crux of the proof is that there is an incentive for rational players to mimic non-rational players and thus create a \emph{reputation} (of being non-rational, i.e., cooperative) for which the other's best reply is cooperation -- at least sufficiently far from the last stage.

\section{Evolutionary games: population dynamics}
\label{sec:eg-pd}

\emph{Evolutionary game theory} is the theory of dynamic adaptation and learning in (infinitely) repeated games played by boundedly rational agents. Although nowadays evolutionary game theory is understood as an intrinsically dynamic theory, it originally started in the 70s as a novel \emph{static} refinement concept for Nash equilibria. Keeping the historical order, we will first discuss the concept of an \emph{evolutionarily stable strategy} (ESS) and its extensions. The ESS concept investigates the stability of the equilibrium under rare mutations, and it does not require the specification of the actual underlying game dynamics. The discussion of the various possible evolutionary dynamics, their behavior and relation to the static concepts will follow afterwards.

In most of this section we restrict our attention to so-called ``population games". Due to a number of simplifying assumptions related to the type and number of agents and their interaction network, these models give rise to a relatively simple, \emph{aggregate level} description. We delegate the analysis of games with a finite number of players or with more complex social structure into later sections. There statistical fluctuations, stemming from the microscopic dynamics or from the structure of the social neighborhood cannot be neglected. These latter models are more complex, and usually necessitate a lower level, so-called \emph{agent-based}, analysis.

\subsection{Population games}

A \emph{mean-field} or \emph{population game} is defined by the underlying two-player stage game, the set of feasible strategies (usually mixed strategies are not allowed or are strongly restricted), and the heuristic updating mechanism for the individual strategies (update rules). The definition tacitly implies the following simplifying assumptions:
\begin{enumerate}
\item The number of boundedly rational agents is very large, $N\to\infty$;
\item All agents are equivalent and have identical payoff matrices (symmetric games), or agents form two different but internally homogeneous groups for the two roles (asymmetric games);\footnote{In the asymmetric case population games are sometimes called \emph{two-population games} to emphasize that there is a separate population for each possible role in the game.}\draftnote{Only discuss one-pop games here!}
\item In each stage game (round) agents are randomly matched with equal probability (symmetric games), or agents in one group are randomly matched with agents in the other group (asymmetric games), thus the social network is the simplest possible;
\item Strategy updates are \emph{rare} in comparison with the frequency of playing, so that the update can be based on the average success rate of a strategy.
\item All agents use the same strategy update rule.
\item Agents are myopic, i.e., their discount factor is small, $\delta\to 0$.
\end{enumerate}
In population games the fluctuations, arising from the randomness of the matching procedure, playing mixed strategies (if allowed), or from stochastic update rules, average out and can be neglected. For this reason \emph{population games comprise the mean-field level of evolutionary game theory}.

These simplifications allow us to characterize the overall behavioral profile of the population, and hence the game dynamics, by a restricted number of state variables.\\

\leftline{\sf Matrix games}\medskip

Let us consider first a symmetric game $G=(\bA,\bA^T)$. Assume that there are $N$ players, $n=1,\dots,N$, each playing a pure strategy $\bs_n$ from the discrete set of feasible strategies $\bs_n\in\cS=\{\be_1,\be_2,\dots,\be_Q\}$. It is convenient to think about strategy $\be_i$ as a $Q$ component unit vector, whose $i$th component is 1, the other components are zero. Thus $\be_i$ points to the $i$th corner of the strategy simplex $\Delta_Q$.

Let $N_i$ denote the number of players playing strategy $\be_i$. At any moment of time, the state of the population can be characterized by the relative frequencies (abundances, concentrations) of the different strategies, i.e., by the $Q$ dimensional state vector $\brho$:
\begin{equation}
    \brho=\frac{1}{N}\sum_{n=1}^N \bs_n=\sum_{i=1}^Q \frac{N_i}{N} \be_i
\end{equation}
Clearly, $\sum_i \rho_i = \sum_i N_i/N = 1$, i.e., $\brho \in \Delta_Q$. Note that $\brho$ can take any point on the simplex $\Delta_Q$ even though players are only allowed to play pure strategies (the corners of the simplex) in the game. Saying it differently, the average strategy $\brho$ may not be in the set of feasible strategies for a single player.

Payoffs can be expressed as a function of the strategy frequencies. The (normalized) expected payoff of player $n$ playing strategy $\bs_n$ in the population is
\begin{equation}\label{ui}
    u_n(\bs_n,\brho) = {1\over N}\sum_{m=1}^N \bs_n \cdot \bA \bs_m = \bs_n \cdot \bA \brho,
\end{equation}
where the sum is over all players $m$ of the population except $n$, but this omission is negligible in the infinite population limit.\footnote{This formulation assumes that the underlying game is a two-player matrix game. In a more general setup (non-matrix games, multi-player games) it is possible that the utility of a player cannot be written in a bilinear form containing her strategy and the mean strategy of the population as in Eq.\ (\ref{ui}), but is a more general nonlinear function, e.g., $u=\bs\cdot f(\bar{\brho})$ with $f$ nonlinear.} Equation (\ref{ui}) implies that for a given player the unified effect of the other players of the population looks as if she would play against a single \emph{representative agent} who plays the population's average strategy as a mixed strategy.

Despite the formal similarity, $\brho$ is not a valid mixed strategy in the game, but the population average of pure strategies. It is not obvious at first sight how to cope with real mixed strategies when they are allowed in the stage game. Indeed, in case of an unrestricted $S\times S$ matrix game, there are $S$ pure strategies and an infinite number of mixed strategies: each point on the simplex $\Delta_S$ represents a possible mixed strategy of the game. When all these mixed strategies are allowed, the state vector $\brho$ is rather a ``state functional" over $\Delta_S$. However, in many biological and economic applications the nature of the problem forbids mixed strategies. In this case the number of different strategy types in the population is simply $Q=S$.

Even if mixed strategies are not forbidden in theory, it is frequently enough to consider one or two well-prepared mixed strategies in addition to the pure ones, for example, to challenge evolutionary stability (see later). In this case the number of types $Q$ ($>S$) can remain finite. The general recipe is to convert the original game which has mixed strategies to a new (effective) game which only has pure strategies. Assuming that the underlying stage game has $S$ pure strategies, and the construction of the game allows, in addition, a number $R$ of well-defined mixed strategies as combinations of these pure ones, we can always define an effective population game with $Q=S+R$ strategies, which are then treated on equal footing. The new game is associated with an \emph{effective payoff matrix} of dimension $Q\times Q$, and from that point on, mixed strategies (as linear combinations of these $Q$ strategies) are formally forbidden.

As a possible example \citep{cressman_03}, consider population games based upon the $S=2$ matrix game in Eq.\ (\ref{eq:general2x2}). If the population game only allows the two pure strategies $\be_1=(1,0)^T$ and $\be_2=(0,1)^T$, the payoff matrix is that of Eq.\ (\ref{eq:general2x2}), and the state vector, representing the frequencies of $\be_1$ and $\be_2$ agents, lies in $\Delta_2$. However, in the case when some mixed strategies are also allowed such as, for example, a third (mixed) strategy $\be_3=(1/2,1/2)^T$, i.e., playing $\be_1$ with probability $1/2$ and $\be_2$ with $1/2$, the effective population game becomes a game with $Q=3$. The effective payoff matrix for the three strategies $\be_1,\be_2$, and $\be_3$ now reads
\begin{equation}\label{eq:example1}
{\bf A}=\left(
\begin{array}{ccc}
    a  & b  & \displaystyle{\frac{a+b}{2}}  \\[2.ex]
    c  & d  & \displaystyle{\frac{c+d}{2}}  \\[2.ex]
    \displaystyle{\frac{a+c}{2}}  & \displaystyle{\frac{b+d}{2}} &
    \displaystyle{\frac{a+b+c+d}{4}}
\end{array}
\right).
\end{equation}

It is also rather straightforward to formally extend the concept of a Nash equilibrium, which was only defined so far on the agent level, to the aggregate level, where only strategy frequencies are available. A \emph{state} of the population $\brho^*\in\Delta_Q$ is called a \emph{Nash equilibrium of the population game}, iff
\begin{equation}\label{eq:NEstate}
    \brho^*\cdot \bA \brho^* \ge \brho\cdot \bA \brho^*\quad
    \forall\, \brho\in\Delta_Q.
\end{equation}
Note that NEs of a population game are always \emph{symmetric} equilibria of the two-player stage game. Asymmetric equilibria of the stage game like those appearing in the Anti-Coordination Class in Eq.\ (\ref{eq:anti-coord}) cannot be interpreted at the population level.

It is not trivial that the definition in Eq.\ (\ref{eq:NEstate}) is equivalent to the agent-level (microscopic) definition of an NE, i.e., given the average strategy of the population, no player has a unilateral incentive to change her strategy. We delegate this question to Sec.\ \ref{sec:es}, where we discuss evolutionary stability.\\

\leftline{\sf Bi-matrix games}\medskip

\begin{figure}[t]
\centerline{\epsfig{file=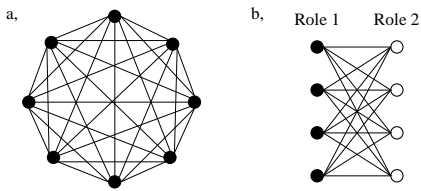,width=6.cm}}
\caption{\label{fig:population} Schematic connectivity structure for (a) population (symmetric) and (b) two-population (asymmetric) games. Black and white dots represent players in different roles.}
\end{figure}

In the case of asymmetric games (bi-matrix games) $G=(\bA,\bB^T)$ the situation is somewhat more complicated. Predators only play with Prey, Owners only with Intrudes, Buyers only with Sellers. The simplest model along the above assumptions is a \emph{two-population game}, one population for each possible kind of players. In this case random matching means that a certain player of one population is randomly connected with another player from the other population, but never with a fellow agent in the same population (see Fig.\ \ref{fig:population} for an illustration). The state of the system is characterized by two state vectors: $\brho\in\Delta_Q$ for the first population and $\etab\in\Delta_{R}$ for the second population. Obviously, symmetrized versions of asymmetric games (role games) can be formulated as one-population games.

For two-population games the Nash equilibrium is a pair of state vectors $(\brho^*,\etab^*)$ such that
\begin{eqnarray}
    \brho^*\cdot \bA \etab^* \ge \brho\cdot \bA \etab^*\quad
        \forall\, \brho\in\Delta_Q, \nonumber\\
    \etab^*\cdot \bB \brho^* \ge \etab\cdot \bB \brho^*\quad
        \forall\, \etab\in\Delta_R.
\end{eqnarray}

\subsection{Evolutionary stability}
\label{sec:es}

A central problem of evolutionary game theory is the stability and robustness of strategy profiles in a population. Indeed, according to the theory of ``punctuated equilibrium" \citep{gould_n93} (biological) evolution is not a continuous process, but is characterized by abrupt transition events of speciation and extinction of relatively short duration, which separate long periods of relative tranquility and stability. On one hand, punctuated equilibrium theory explains missing fossil records related to ``intermediate" species and the documented coexistence of freshly branched species, on the other hand it implies that most of the biological structure and behavioral pattern we observe around us, and aim to explain, is likely to possess a high degree of stability. In this sense a solution provided by a game theory model can only be a plausible solution if it is ``evolutionarily" stable.\\

\leftline{\sf Evolutionarily stable strategies}\medskip

The first concept of evolutionary stability was formulated by \citet{maynard_n73} in the context of symmetric population games. An \emph{evolutionarily stable strategy} (ESS) is a strategy that, when used by an entire population, is immune against invasion by a minority of mutants playing a different strategy. Players playing according to an ESS fares better than mutants in the population and thus in the long run outcompete and expel invaders. An ESS persists as the dominant strategy over evolutionary timescales, so strategies observed in the real world are typically ESSs. The ESS concept is relevant when the \emph{mutation} (biology) or \emph{experimentation} (economics) rate is low. In general evolutionary stability implies an \emph{invasion barrier}, i.e., an ESS can only resist invasion until mutants reach a finite critical frequency in the population.
The ESS concept does not contain any reference to the actual game dynamics, and thus it is a ``static" concept. The only assumption it requires is that a strategy performing better must have a higher replication (growth) rate.

\begin{figure}[ht]
\centerline{\epsfig{file=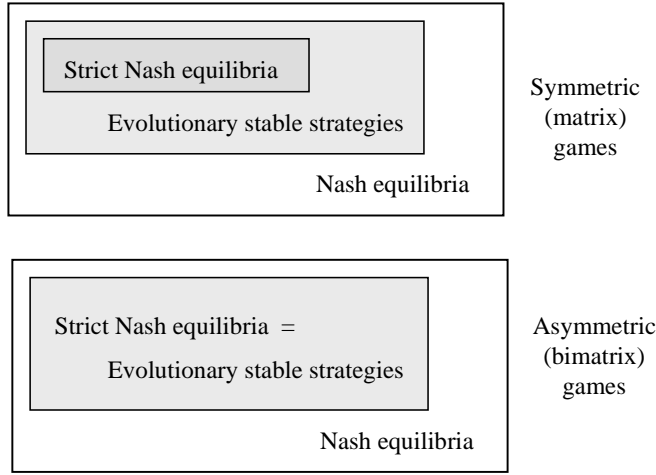,width=9.cm}}
\caption{\label{fig:ESS} Relation between evolutionary stability (ESS) and symmetric Nash equilibria for matrix and bi-matrix games.}
\end{figure}

In order to formulate evolutionary stability, consider a matrix game with $S$ pure strategies and all possible mixed strategies composed from these pure strategies, and a population in which the majority, $1-\epsilon$ part, of the players play the incumbent strategy $\bp^*\in\Delta_S$, and a minority, $\epsilon$ part, plays a mutant strategy $\bp\in\Delta_S$. Both $\bp^*$ and $\bp$ can be mixed strategies. The average strategy of the population, which determines individual payoffs in a population (mean-field) game, is $\brho=(1-\epsilon)\bp^*+\epsilon\bp \in\Delta_S$. The strategy $\bp^*$ is an ESS, iff for all $\epsilon>0$ smaller than some appropriate invasion barrier $\bar{\epsilon}$, and for any feasible mutant strategies $\bp\in\Delta_S$, the incumbent strategy $\bp^*$ performs strictly better in the mixed population than the mutant strategy $\bp$,
\begin{equation}
    u(\bp^*,\brho) > u(\bp,\brho),
\end{equation}
i.e.,
\begin{equation}\label{eq:ESScond}
    \bp^*\cdot \bA \brho > \bp\cdot \bA \brho.
\end{equation}
If the inequality is not strict, $\bp^*$ is usually called a \emph{weak} ESS. The condition in Eq.\ (\ref{eq:ESScond}) takes the explicit form
\begin{equation}
    \bp^*\cdot \bA \big[(1-\epsilon)\bp^*+\epsilon\bp\big] >
    \bp\cdot \bA \big[(1-\epsilon)\bp^*+\epsilon\bp\big],
\end{equation}
which can be rewritten as
\begin{equation}
    (1-\epsilon)( \bp^*\cdot \bA \bp^* -\bp\cdot \bA \bp^* )
    +\epsilon ( \bp^*\cdot \bA \bp -\bp\cdot \bA \bp ) > 0.
\end{equation}
It is thus clear that $\bp^*$ is an ESS, iff two conditions are satisfied:\\
(1) \emph{NE condition:}
\begin{equation}\label{eq:ESS-i}
    \bp^*\cdot \bA \bp^* \ge \bp\cdot \bA \bp^*\quad \mbox{for all $\bp\in\Delta_N$},
\end{equation}
(2) \emph{stability condition:}
\begin{eqnarray}\label{eq:ESS-ii}
    &&\mbox{if $\bp\ne \bp^*$ and $\bp^*\cdot \bA \bp^* = \bp\cdot \bA \bp^*$,}\nonumber\\
    &&\mbox{then $\bp^*\cdot \bA \bp >\bp\cdot \bA \bp$}.
\end{eqnarray}
According to (1) $\bp^*$ should be a symmetric Nash equilibrium of the stage game, hence the definition in Eq.\ (\ref{eq:NEstate}), and according to (2) if it is a non-strict NE, then $\bp^*$ should fare better against $\bp$ than $\bp$ against itself.
Clearly, all strict symmetric NEs are ESSs, and all ESSs are symmetric NEs, but not the other way around [see the upper panel in Fig.\ \ref{fig:ESS}]. Most notably not all NEs are evolutionary stable, thus the ESS concept provides a means for equilibrium selection. However, a game may have several ESSs or no ESS at all.

As an illustration consider the Hawk-Dove game defined in Appendix \ref{app:g:hd}. The stage game belongs to the Anti-Coordination Class in Eq.\ (\ref{eq:anti-coord}) and has a symmetric mixed strategy equilibrium:
\begin{equation}\label{HDGps}
    \bp^* = \left( \begin{array}{c}
                   V/C \\
                   1-V/C
                   \end{array}  \right), \\
\end{equation}
which is an ESS. Indeed, we find
\begin{equation}
    \forall \bp\in\Delta_2: \quad \bp\cdot \bA \bp^* = \frac{V}{2}-\frac{V^2}{2C}
\end{equation}
independent of $\bp$. Thus Eq.\ (\ref{eq:ESS-i}) is satisfied for all $\bp$ with equality. Hence we should check the stability condition Eq.\ (\ref{eq:ESS-ii}). Parameterizing $\bp$ as  $\bp=(p,1-p)^T$, we find
\begin{equation}\label{HDG2}
    \bp^*\cdot \bA \bp - \bp\cdot \bA \bp = \frac{(V-pC)^2}{2C}>0,\quad  \forall\, p\ne \frac{V}{C},
\end{equation}
thus Eq.\ (\ref{eq:ESS-ii}) is indeed satisfied, and $\bp^*$ is an ESS. It is in fact the only ESS of the Hawk-Dove game.

An important theorem [see, e.g., \citet{hofbauer_98} or \citet{cressman_03}] says that a strategy $\bp^*$ is an ESS, iff it is \emph{locally superior}, i.e., $\bp^*$ has a neighborhood in $\Delta_S$ such that for all $\bp\ne\bp^*$ in this neighborhood
\begin{equation}\label{locsup}
    \bp^*\cdot \bA \bp >\bp\cdot \bA \bp.
\end{equation}
In the Hawk-Dove game example, as is shown by Eq.\ (\ref{HDG2}), local superiority holds in the whole strategy space, but this is not necessary in other games.\\

\leftline{\sf Evolutionarily stable states and sets}\medskip

Even if a game has an evolutionarily stable strategy in theory, this strategy may not be feasible in practice.
For instance, Hawk and Dove behavior may be genetically coded, and this coding may not allow for a mixed strategy. Therefore, when the Hawk-Dove game is played as a population game in which only pure strategies are allowed, there would be no feasible ESS. Neither pure Hawk nor pure Dove, the only individual strategies allowed, are evolutionarily stable. There is, however, a stable composition of the population with strategy frequencies $\brho^*=\bp^*$ given by Eq.\ (\ref{HDGps}). This state satisfies both the ESS conditions Eqs.\ (\ref{eq:ESS-i}) and (\ref{eq:ESS-ii}) (with $\bp$ substituted everywhere by $\brho$) and the equivalent local superiority condition Eq.\ (\ref{locsup}). It is customary to call such a state an \emph{evolutionarily stable state} (also abbreviated traditionally as ESS), which is the direct extension of the ESS concept to population games with only pure strategies \citep{ hofbauer_98}.

The extension of the ESS concept to composite states of a population would only have a sense, if these states had similarly strong stability properties as evolutionarily stable strategies. As we will see later in Sec.\ \ref{sec:rd}, this is almost the case, but there are some complications. Because of the extension, the ESS concept loses some of its strength. Although it remains true that ESSs are dynamically stable, the converse loses validity: a state should not necessarily be an ESS for being dynamically stable and thus persisting for evolutionary times.

Up to now we have discussed the properties of ESSs, but how to find an ESS? One possibility is to use the Bishop-Cannings Theorem \citep{bishop_jtb78} which provides a necessary condition for an ESS: If $\brho^*$ is a mixed evolutionarily stable strategy with support $I=\{\be_1,\be_2,\dots,\be_k\}$, i.e., $\brho^*=\sum_{i=1}^k \rho_i \be_i$ with $\rho_i>0$ $\forall\, 1\le i\le k$, then
\begin{equation}\label{eq:Biship-Cannings}
    u(\be_1;\brho^*) = u(\be_2;\brho^*) = \dots = u(\be_k;\brho^*) =u(\brho^*;\brho^*),
\end{equation}
which leads to Eq.\ (\ref{eq:interiorNE}), the condition of an interior NE, when payoffs are linear. Bishop and Canning have also proved that if $\brho^*$ is an ESS with support $I$ and $\etab^*\ne \brho^*$ is another ESS with support $J$, then neither support set can contain the other. Consequently, if a matrix game has an \emph{interior} ESS, than it is the only ESS.

The ESS concept has a further extension towards \emph{evolutionarily stable sets} (ESset), which becomes useful in games with singular payoff matrices. An ESset is an NE component (a connected set of NEs) which is stable in the following sense: no mutant can spread when the incumbent is using a strategy in the ESset, and mutants using a strategy not belonging to the ESset are driven out \citep{balkenborg_95b}. In fact each point in the ESset is \emph{neutrally stable}, i.e., when the incumbent and the mutant are also elements of the set they fare just equally in the population.

\begin{figure}[t]
\centerline{\epsfig{file=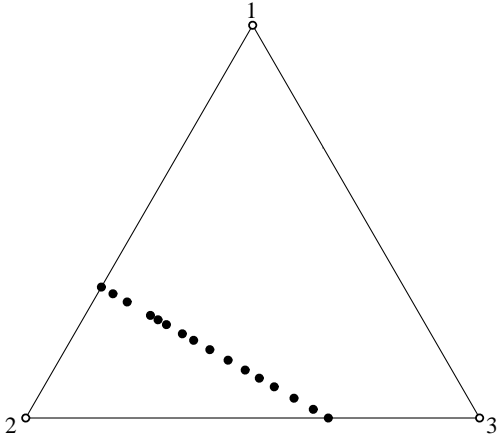,width=7cm}}
\caption{\label{fig:ESset} An ESset (line along the black dots) for the degenerate Hawk-Dove game Eq.\ (\ref{eq:example1}) with $a=-2,b=c=0,d=-1$.}
\end{figure}

As a possible example consider again the game defined by the effective payoff matrix Eq.\ (\ref{eq:example1}). If the basic 2-strategy game has an interior NE, $\bp^*=(p_1^*,p_2^*)^T$, the 3-strategy game has a boundary NE $(p_1^*,p_2^*,0)^T$. Then the line segment $E=\{\brho\in\Delta_3: \rho_1+\rho_3/2=p_1^*\}$ is an NE component (see Fig.\ \ref{fig:ESset}). It is an ESset, iff the 2-strategy game belongs to the Anti-Coordination (Hawk-Dove) Class \citep{cressman_03}.\\

\leftline{\sf Bi-matrix games}\medskip

The ESS concept was originally invented for symmetric games, and later extended for asymmetric games. In many respects the situation is easier in the latter. Bi-matrix games are two-population games, and thus a mutant appearing in one of the populations never plays with fellow mutants in her own population. In technical terms this means that the condition (2) in Eq.\ (\ref{eq:ESS-ii}) is irrelevant. Condition (1) is the Nash equilibrium condition. When the NE is not strict there exists a mutant strategy in at least one of the populations which is just as fit as the incumbent. These mutants are not driven out although they cannot spread either. This kind of neutral stability is usually called \emph{drift} in biology: the composition of the population changes by pure chance. Strict NEs, on the other hand, are always evolutionarily stable and as was shown by \citet{selten_jtb80}: \emph{In asymmetric games a strategy pair is an ESS iff it is a strictNE} (see the lower panel in Fig.\ \ref{fig:ESS}).

Since interior NEs are never strict it follows that asymmetric games can only have ESSs on $\bd (\Delta_Q\times\Delta_R)$. Again the number of ESSs can be high and there are games, e.g., the asymmetric, two-population version of the Rock-Scissors-Paper game, where there is no ESS.

Similarly to the ESset concept for symmetric games, the concept of \emph{strict equilibrium sets} (SEset) can be introduced \citep{cressman_03}. An SEset $F\in \Delta_Q\times\Delta_R$ is a set of NE strategy pairs such that $(\brho,\etab^*)\in F$ whenever $\brho\cdot \bA\etab^* = \brho^*\cdot \bA\etab^*$ and $(\brho^*,\etab)\in F$ whenever $\etab \cdot\bB \brho^*  = \etab^*\cdot \bB\brho^*$ for some $(\brho^*,\etab^*)\in F$. Again, mutants outside the SEset are driven out from the population, while those within the SEset cannot spread. When mutations appear randomly, the composition of the two populations drifts within the SEset.

\subsection{Replicator dynamics}
\label{sec:rd}

A model in evolutionary game theory is made complete by postulating the game dynamics, i.e., the rules that discribe the update of strategies in the population. Depending on the actual problem, different kinds of dynamics can be appropriate. The game dynamics can be continuous or discrete, deterministic or stochastic, and within these major categories a large number of different rules can be formulated depending on the situation under investigation.

On the macroscopic level, by far the most studied continuous evolutionary dynamics is the \emph{replicator dynamics}. It was introduced originally by \citet{taylor_mb78}, and it has exceptional status in the models of biological evolution. On the phenomenological level the replicator dynamics can be postulated directly by the reasonable assumption that the per capita growth rate $\dot \rho_i/\rho_i$ of a given strategy type is proportional to the fitness difference
\begin{equation}
    \frac{\dot \rho_i}{\rho_i} = \mbox{fitness of type $i$} - \mbox{average fitness}.
\end{equation}
The fitness is the individual's evolutionary success, i.e., in the game theory context the payoff of the game.
In population games the fitness of strategy $i$ is $(\bA\brho)_i$, whereas the average fitness is $\brho\cdot\bA\brho$. This leads to the equation
\begin{equation}\label{eq:replicator}
    \dot \rho_i = \rho_i \big( (\bA\brho)_i-\brho\cdot\bA\brho \big).
\end{equation}
which is usually called the \emph{Taylor form} of the replicator equation.

Under slightly different assumptions the replicator equation takes the form
\begin{equation}\label{eq:replicatorMS}
    \dot \rho_i = \rho_i \frac{ (\bA\brho)_i-\brho\cdot\bA\brho }{\brho\cdot\bA\brho},
\end{equation}
which is the so-called \emph{Maynard Smith form} (or adjusted replicator equation). In this the driving force is the relative fitness difference. Note that the denominator is only a rescaling of the flow velocity. For both forms the simplex $\Delta_Q$ is invariant and such are all of its faces: if the initial condition does not contain a certain strategy $i$, i.e., $\rho_i(t=0)=0$ for some $i$, then it remains $\rho_i(t)=0$ for all $t$. The replicator dynamics does not invent new strategies, and as such it is the prototype of a wider class of dynamics called \emph{non-innovative dynamics}. Both forms of the replicator equations can be deduced rigorously from microscopic assumptions (see later).

The replicator dynamics lends the notion of evolutionary stability, originally introduced as a static concept, an explicit dynamic meaning. From the dynamical point of view, fixed points (rest points, stationary states) of the replicator equation play a distinguished role since a population which happens to be exactly in the fixed point at $t=0$ remains there forever. Stability against fluctuations is, however, a crucial issue. It is customary to introduce the following classification of fixed points:\smallskip\\
(a) A fixed point $\brho^*$ is \emph{stable} (also called \emph{Lyapunov stable}) if for all open neighborhoods $U$ of $\brho^*$ (whatever small it is) there is another open neighborhood $O\subseteq U$ such that any trajectory initially inside $O$ remains inside $U$.\smallskip\\
(b) The fixed point $\brho^*$ is \emph{unstable} if it is not stable.\smallskip\\
(c) The fixed point $\brho^*$ is \emph{attractive} if there exists an open neighborhood $U$ of $\brho^*$ such that all trajectory initially in $U$ converges to $\brho^*$. The maximum possible $U$ is called the \emph{basin of attraction} of $\brho^*$.\smallskip\\
(d) The fixed point $\brho^*$ is \emph{asymptotically stable} (also called \emph{attractor}) if it is stable and attractive.\footnote{Note that a fixed point $\brho^*$ can be attractive without being stable. This is the case if there are trajectories which start close to $\brho^*$, but take a large excursion before converging to $\brho^*$ (see an example later in Sec.\ \ref{sec:pd3s}). Being attractive is not enough to classify as an attractor, stability is also required.} In general a fixed point is \emph{globally asymptotically stable} if its basin of attraction encompasses the whole space. In the case of the replicator dynamics where $\bd\Delta_Q$ is invariant, a state is called globally asymptotically stable if its basin of attraction contains $\inter\Delta_Q$.\smallskip

These definition can be trivially extended from fixed points to invariant sets such as a set of fixed points or limit cycles, etc.\\

\leftline{\sf Dynamic vs. evolutionary stability}\medskip

What is the connection between dynamic stability and evolutionary stability? Unfortunately, the two concepts do not perfectly overlap. The actual relationship is best summarized in the form of two collections of theorems: one relating to Nash equilibria, the other to evolutionary stability.
As for Nash equilibria in matrix games under the replicator dynamics the \emph{Folk Theorem of Evolutionary Game Theory} asserts:\smallskip\\
(a) \emph{NEs are rest points.}\smallskip\\
(b) \emph{Strict NEs are attractors.}\smallskip\\
(c) \emph{If an interior orbit converges to $\bp^*$, then it is an NE.}\smallskip\\
(d) \emph{If a rest point is stable then it is an NE.}\smallskip\\
See \citet{hofbauer_98,cressman_03,hofbauer_bams03} for detailed discussions and proofs.

None of the converse statements hold in general. For (a) there can be rest points which are not NEs. These can only be situated on $\bd \Delta_Q$. It is easy to show that a boundary fixed point is only an NE iff its transversal eigenvalues (associated with eigenvectors of the Jacobian transverse to the boundary) are all non-positive \citep{hofbauer_bams03}. An interior fixed point is trivially an NE.

\begin{figure}[ht]
\centerline{\epsfig{file=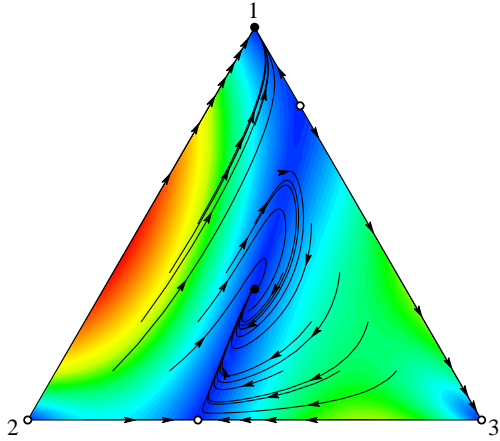,width=7cm}}
\caption{\label{fig:flow7.12}Flow pattern for the game ${\bf A}_{7_1}$ in Eq.\ \ref{eq:HS7.12a}. Red (blue) colors indicate fast (slow) flow. Black (white) circles are stable (unstable) rest points. Figure made by the game dynamics simulation program ``Dynamo" \citep{sandholm_dynamo}.}
\end{figure}

As for (b), not all attractors are strict NEs. As an example consider the matrix \citep{hofbauer_98,zeeman_80}
\begin{equation}\label{eq:HS7.12a}
{\bf A}_{7_1}=\left( \begin{array}{ccc}
                        0 & 6 & -4\cr
                       -3 & 0 & 5\cr
                       -1 & 3 & 0
                       \end{array}\right),
\end{equation}
with the associated flow diagram in Fig.\ \ref{fig:flow7.12}. The rest point $[1/3,1/3,1/3]$ is an attractor (the eigenvalues have negative real parts), but being in the interior of $\Delta_3$ it is a non-strict NE.

\begin{figure}[ht]
\centerline{\epsfig{file=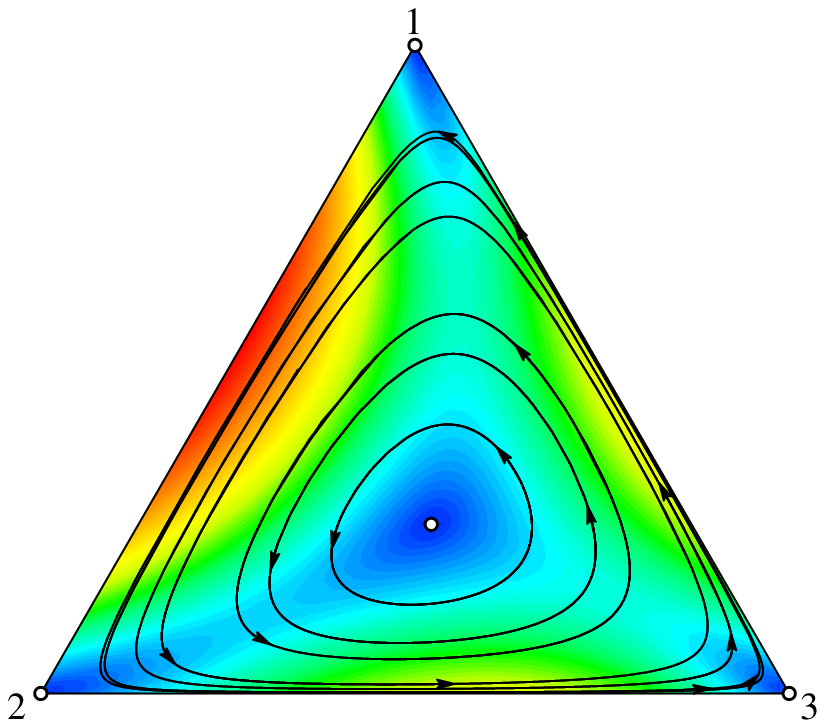,width=5cm}}
\centerline{\epsfig{file=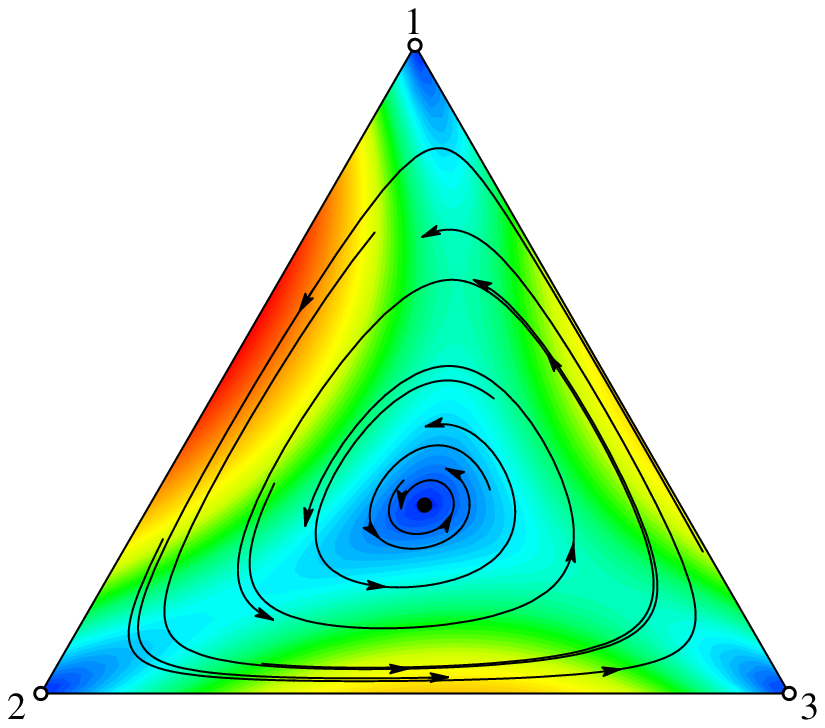,width=5cm} \epsfig{file=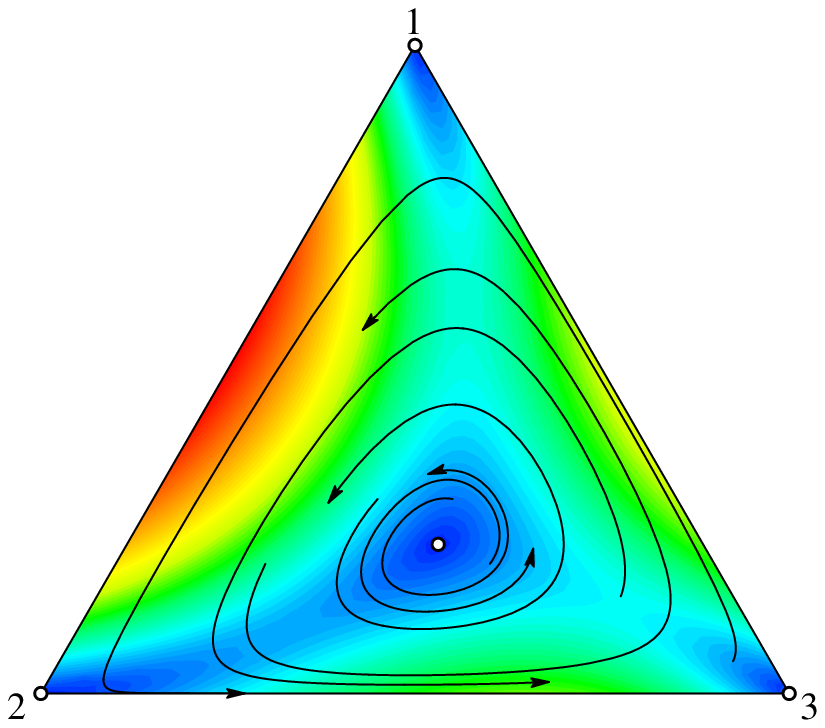,width=5cm}}
\caption{\label{fig:flalpha}Trajectories predicted by Eq.~(\ref{eq:Cr263}) for $\alpha=5.5$, $6.5$, and $7.5$ (from left to right). Pluses indicate the internal rest points. Figures made by the ``Dynamo" program \citet{sandholm_dynamo}.}
\end{figure}

It is easy to construct examples in which a point is an NE, but it is not a limit point of an interior orbit. Thus the converse of (c) does not hold. For an example of dynamically unstable Nash equilibria consider the following family of games \citep{cressman_03}
\begin{equation}\label{eq:Cr263}
{\bf A}=\left( \begin{array}{ccc}
                        0 & 2-\alpha & 4\cr
                       6 & 0 & -4\cr
                       -2 & 8-\alpha & 0
                       \end{array}\right).
\end{equation}
These games are generalized Rock-Scissors-Paper games which have the cyclic dominance property for $2<\alpha<8$. Even beyond this interval, for $-16<\alpha<14$, there is a unique symmetric interior NE, $\bp^*=[28-2\alpha,20,16+\alpha]/(64-\alpha)$, which is a rest point of the replicator dynamics (see Fig.\ \ref{fig:flalpha}). This rest point is attractive for $-16<\alpha<6.5$ but repulsive for $6.5<\alpha<14$.

The main results for the subtle relation between the replicator dynamics and evolutionary stability can be summarized as follows \citep{hofbauer_98,cressman_03,hofbauer_bams03}:\smallskip\\
(a) \emph{ESSs are attractors.}\smallskip\\
(b) \emph{interior ESSs are global attractors.} \smallskip\\ 
(c) \emph{For potential games a fixed point is an ESS iff it is an attractor.}\smallskip\\
(d) \emph{For $2\times 2$ matrix games a fixed point is an ESS iff it is an attractor.}\smallskip\\
Again, the converses of (a) and (b) do not necessarily hold. To show a counterexample where an attractor is not an ESS consider again the payoff matrix $\bA_{7_1}$ in Eq.\ (\ref{eq:HS7.12a}) and Fig.\ \ref{fig:flow7.12}. The only ESS in the game is $\be_1=[1,0,0]$ on $\bd\Delta_3$. The NE $[1/3,1/3,1/3]$ is an attractor but not evolutionarily stable. Recall that the Bishop-Cannings theorem forbids the occurrence of interior and boundary ESSs at the same time.

Another example is Eq.\ (\ref{eq:Cr263}), where the fixed point $\bp^*$ is only an ESS for $-16<\alpha<3.5$ At $\alpha=3.5$ it is neutrally evolutionarily stable. For $3.5<\alpha<6.5$ the fixed point is dynamically stable but not evolutionarily stable.
\\

\leftline{\sf Classification of matrix game phase portraits}\medskip

\begin{figure}[t]
\centerline{\epsfig{file=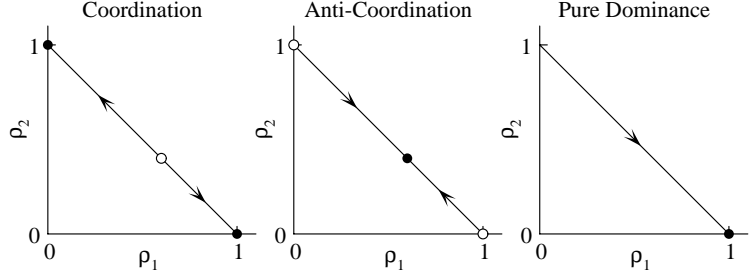,width=10cm}}
\caption{\label{fig:delta2} Classification of two-strategy (symmetric) matrix games based on the replicator dynamics flow. Full circles are stable, open circles are unstable NEs.}
\end{figure}

The complete classification of possible replicator phase portraits for two-strategy games, where $\Delta_2$ is one-dimensional, is easy and goes along with the classification of Nash equilibria discussed in Sec.\ \ref{sec:netpmg}. The three possible classes are shown in Fig.\ \ref{fig:delta2}. In the Coordination Class the mixed strategy NE is an unstable fixed point of the replicator dynamics. Trajectories converge to one or the other pure strategy NEs, which are both ESSs (cf.\ theorem (d) above). For the Anti-Coordination Class the interior (symmetric) NE is the global attractor thus it is the unique ESS. In case of the Pure Dominance Class there is one pure strategy ESS, which is again a global attractor.

\begin{figure}[ht]
\centerline{\epsfig{file=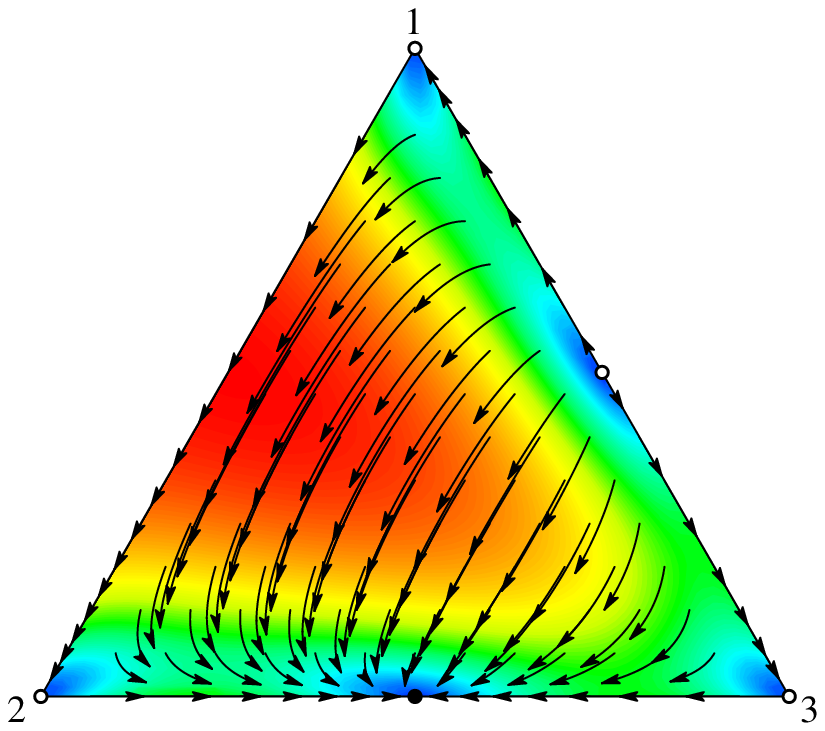,width=5cm}
\epsfig{file=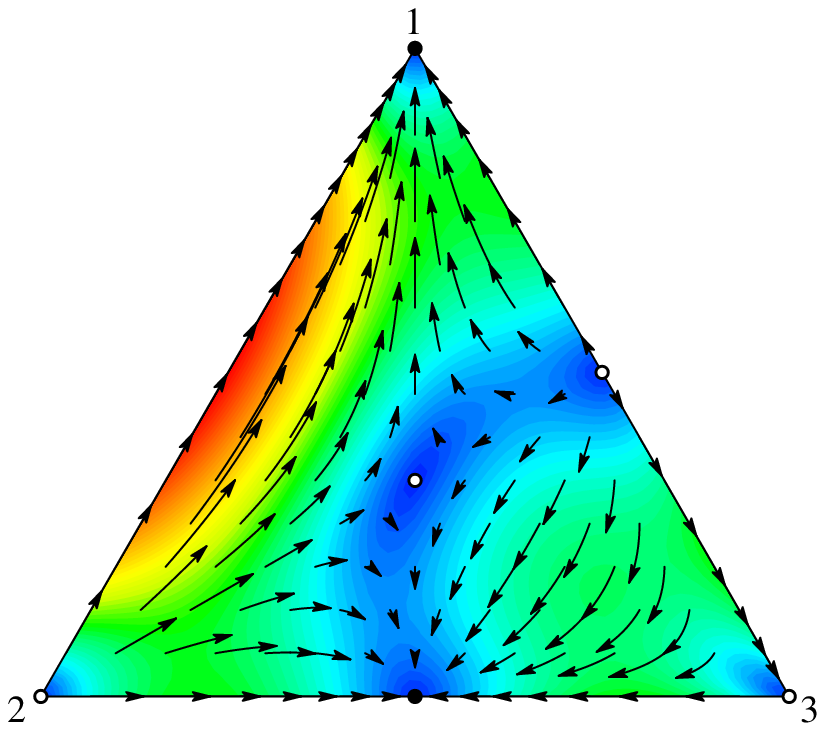,width=5cm}}
\centerline{\epsfig{file=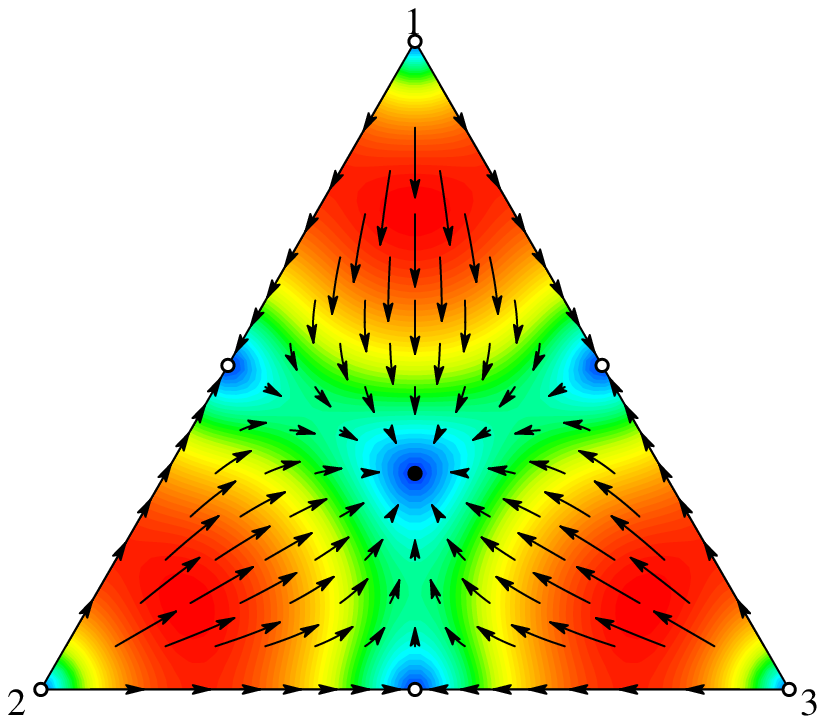,width=5cm}
\epsfig{file=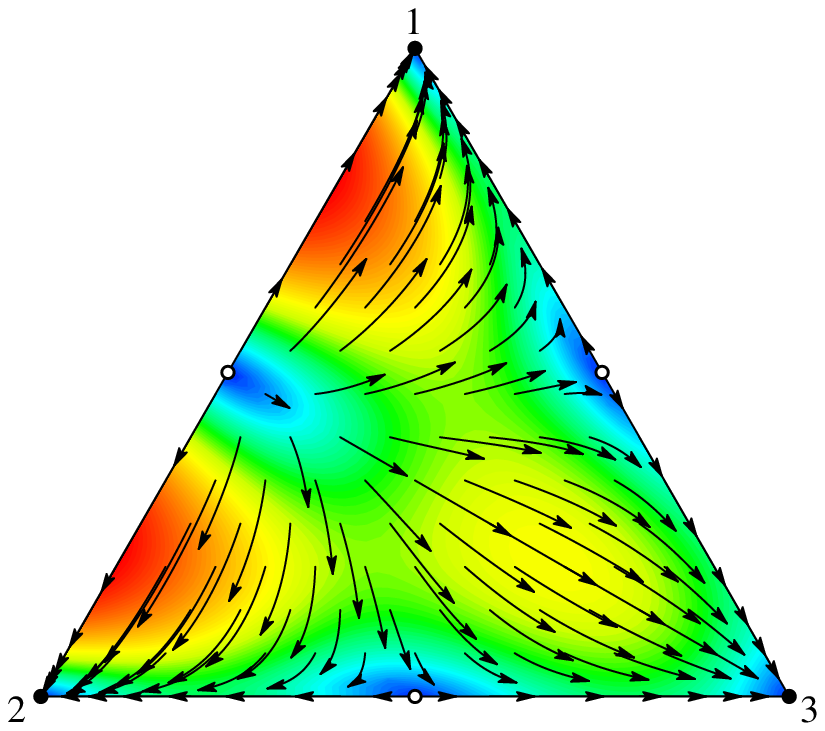,width=5cm}}
\caption{\label{fig:3portraits} Some possible flow diagrams for three-strategy games. Upper left: Zeeman's code $8$; upper right: $7_2$; lower left: $10_1$; lower right: $-10_2$. Red (blue) colors indicate fast (slow) flow. Black (white) circles are stable (unstable) rest points. Figures made by the simulation program ``Dynamo" \citep{sandholm_dynamo}.}
\end{figure}

A similar complete classification in the case of three-strategy matrix games is more tedious \citep{hofbauer_bams03}. In the generic case there are 19 different phase portraits for the replicator dynamics \citep{zeeman_80}, and this number rises to about fifty when degenerate cases are included \citep{bomze_bc83,bomze_bc95}.
In general Zeeman's classification is based on counting the Nash equilibria: there can be at most one interior NE, three NEs on the boundary faces, and three pure strategy NEs. An NE can be stable, unstable or a saddle point, but topological rules only allow a limited combinations of these. Four possibilities from the 19,
\begin{eqnarray}\label{eq:3portraits}
&&{\bf A}_8=\left( \begin{array}{ccc}
                        0 & -1 & -1\cr
                       1 & 0 & 1\cr
                       -1 & 1 & 0
                       \end{array}\right),\quad
{\bf A}_{7_2}=\left( \begin{array}{ccc}
                        0 & 1 & -1\cr
                       -1 & 0 & 1\cr
                       -1 & 1 & 0
                       \end{array}\right), \nonumber\\
&&{\bf A}_{10_1}=\left( \begin{array}{ccc}
                        0 & 1 & 1\cr
                       1 & 0 & 1\cr
                       1 & 1 & 0
                       \end{array}\right),\quad
{\bf A}_{-10_2}=\left( \begin{array}{ccc}
                        0 & -3 & -1\cr
                       -3 & 0 & -1\cr
                       -1 & -1 & 0
                       \end{array}\right),\nonumber
\end{eqnarray}
are shown in Fig.\ \ref{fig:3portraits} (other classes have been show in Figs.\ \ref{fig:flow7.12} and \ref{fig:flalpha}). The subscript of $\bA$ denotes the Zeeman classification code \citep{ zeeman_80}. In degenerate cases extended NE components can appear.

An important result is that no \emph{isolated} limit cycles exist \citep{zeeman_80}. Limit cycles can exist in degenerate cases but then they occur in non-isolated families like for the Rock-Scissor-Paper game in Eq.\ (\ref{eq:Cr263}) with $\alpha=6.5$ as shown in Fig.\ \ref{fig:flalpha}.

Isolated limit cycles can exist in four-strategy games (and above), and there numerical simulations also indicate the possible presence of chaotic attractors \citep{hofbauer_bams03}. A complete taxonomy of all possible behaviors (topologically distinct phase portraits) for more than three strategies seems rather hopeless and has not yet been given.\\

\leftline{\sf Classification of bi-matrix game phase portraits}\medskip

The generalization of the replicator dynamics for bi-matrix games is rather straightforward
\begin{eqnarray}\label{eq:bireplicator}
    \dot \rho_i &=& \rho_i \big[ (\bA\etab)_i-\brho\cdot\bA\etab \big], \nonumber\\
    \dot \eta_i &=& \eta_i \big[ (\bB\brho)_i-\etab\cdot\bB\brho \big].
\end{eqnarray}
where $\brho$ and $\etab$ denote the state of the two populations, respectively. (The Maynard Smith form in Eq.\ (\ref{eq:replicatorMS}) can be generalized similarly.) The relation between dynamic and evolutionary stability is somewhat easier for bi-matrix than for matrix games. We have already seen that ESSs are strict NEs and vice versa. As a set-wise extension of bi-matrix evolutionary stability we have introduced SESets. It turns out that these concepts are exactly equivalent to asymptotic stability:
\emph{In bi-matrix games a set of rest points of the bi-matrix replicator dynamics Eq.\ (\ref{eq:bireplicator}) is an attractor if and only if it is a SESet} \citep{cressman_03}.

It can also be proven that the bi-matrix replicator flow is incompressible, thus there can be no interior attractor. Consequently, ESSs, if any, are necessarily pure strategies in bi-matrix games.

\begin{figure}[ht]
\centerline{\epsfig{file=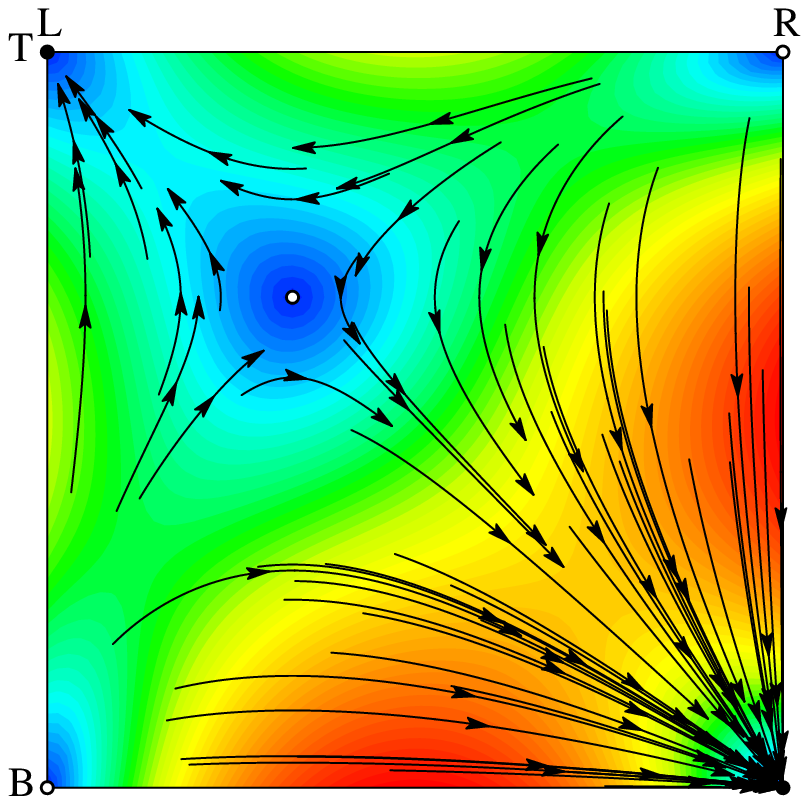,width=5cm}
\epsfig{file=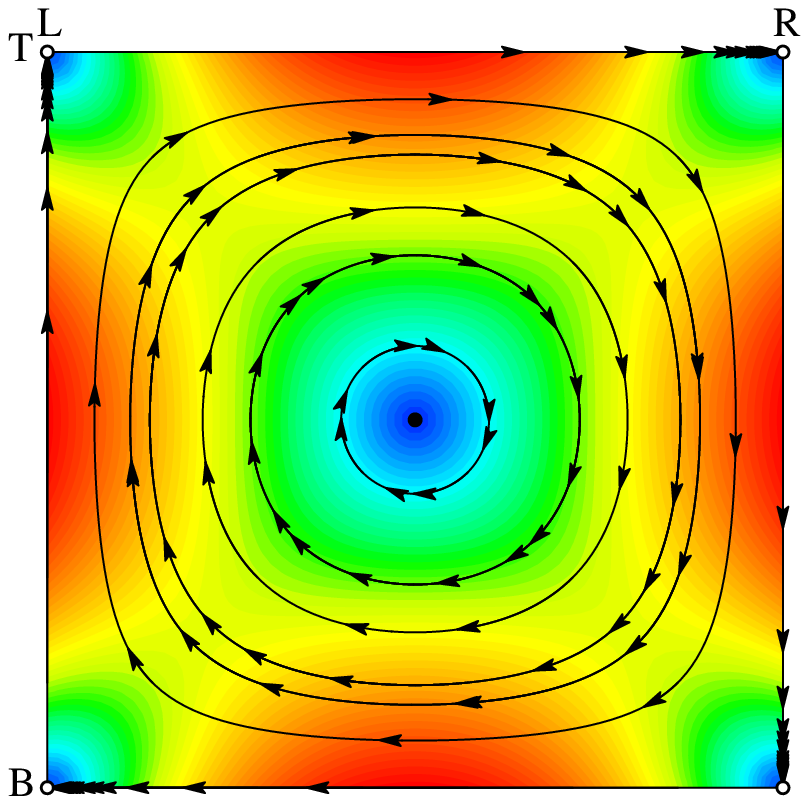,width=5cm}}
\centerline{\epsfig{file=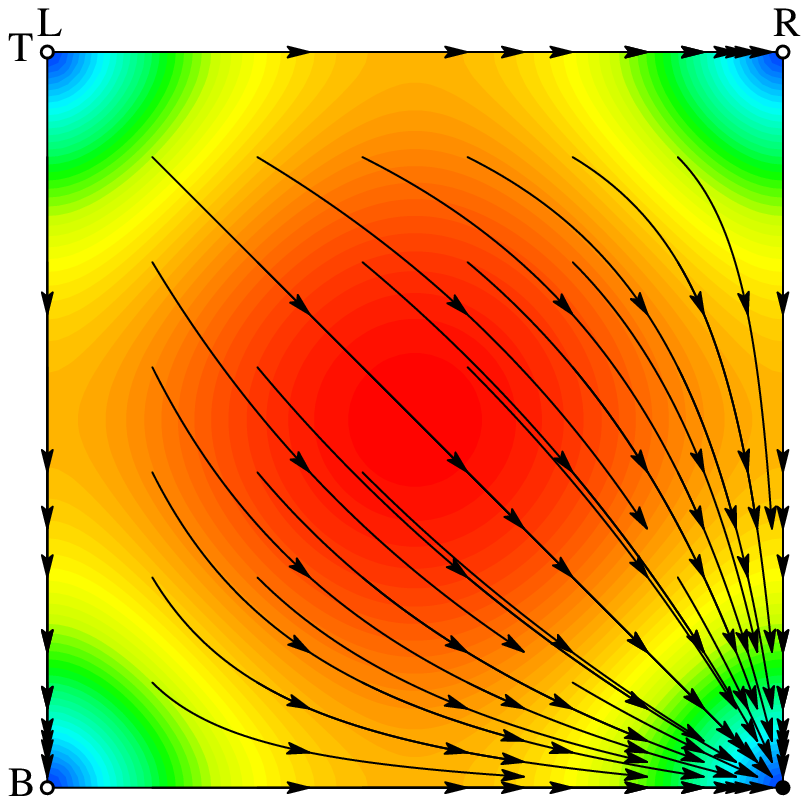,width=5cm}
\epsfig{file=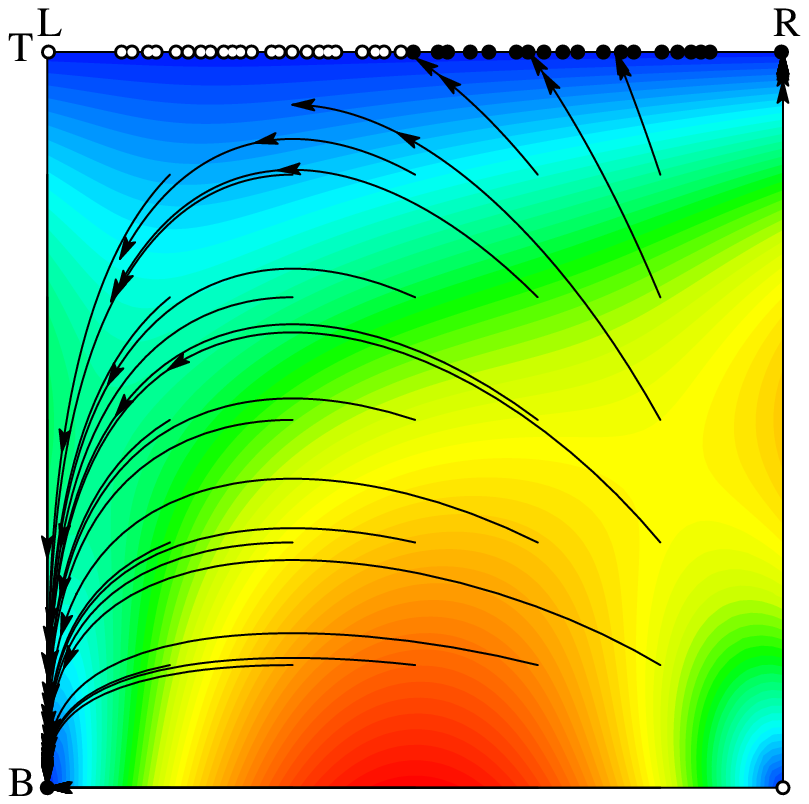,width=5cm}}
\caption{\label{fig:bimatclass} Archetypes of bi-matrix phase portraits in two-population games with the replicator dynamics. The players have two options: T or B for the row player, and L or R for the column player. Upper left: Saddle Class, upper right: Center Class, lower left: Corner Class, lower right: a degenerate case with $\alpha=0$. Red (blue) colors indicate fast (slow) flow. Black (white) circles are stable (unstable) rest points. Figures made by the simulation program ``Dynamo" \citep{sandholm_dynamo}. }
\end{figure}

As an example consider the Owner-Intruder game, the bi-matrix (two-population) version of the traditional Hawk-Dove game. All players in this game have a tag, they are either Owners or Intruders (of, say, a territory). Both Owners and Intruders and can play two strategies: Hawk or Dove. The payoff matrix is equal to that of the Hawk-Dove game [see Eq.\ (\ref{eq:HDG})], but the social connectivity is such that Owners only play with Intruders and vice versa, thus it is a two-population game with a connectivity structure shown in Fig.\ \ref{fig:population}(b). In fact this is the simplest possible deviation from mean-field connectivity in the Hawk-Dove game, and as is shown by the flow diagram in the upper left part of Fig.\ \ref{fig:bimatclass} this has strong consequences. On the $(\rho,\eta)$ plane, where $\rho$ ($\eta$) is the ratio of Owners (Intruders) playing the Hawk strategy, the interior fixed point, which was an attractor for the original Hawk-Dove game, becomes now a saddle point. Losing stability it is no longer an ESS. Instead the replicator dynamics flows towards two possible pure strategy pairs: either all Owners play Hawk and all Intruders play Dove, i.e., $(\rho^*,\eta^*)=(1,0)$ or vice versa, i.e., $(\rho^*,\eta^*)=(0,1)$. These strategy pairs, and only these, are evolutionarily stable.

It is possible to give a complete classification of the different phase portraits of bi-matrix games where both players have two possible pure strategies \citep{cressman_03}. Since the replicator dynamics is invariant for adding constants to the columns of the payoff matrices and interchanging strategies the bi-matrix can be parameterized as
\begin{equation}\label{eq:HS7.12}
(\bA,\bB^T)=\left( \begin{array}{cc}
                        (a,\alpha)\; & (0,0) \cr
                       (0,0)\; & (d,\delta)
                       \end{array}\right).
\end{equation}
An interior NE exists if and only if $ad>0$ and $\alpha\delta>0$, and reads
\begin{equation}
    (\brho_1^*,\etab_1^*) = \left( \frac{\delta}{\alpha+\delta},\frac{d}{a+d} \right).
\end{equation}
There are three generic cases:
\begin{description}
\item[Saddle Class] [$a,d,\alpha,\delta>0$]. The interior NE is a saddle point singularity. It is not an ESS. There are two pure strategy attractors in opposite corners (both pairs are possible), which are ESSs. The bi-matrix Coordination game and the Owner-Intruder game belong to this class. See the upper left panel of Fig.\ \ref{fig:bimatclass} for an illustration.
\item[Center Class] [$a,d>0$, $\alpha,\delta<0$] There is only one NE, the interior NE. It is a center-type singularity. There are no ESSs. On the boundary the trajectory form a heteroclinic cycle. The trajectories do not converge to an NE. The Buyer-Seller game or the Matching Pennies are typical examples. See the upper right part of Fig.\ \ref{fig:bimatclass}.
\item[Corner Class] [$a<0<d$, $\delta>0$, $\alpha\ne 0$] There is no interior NE. All trajectories converge to a unique ESS in the corner. The Prisoner's Dilemma game when played as a two-population game is in this class. See the lower left part of Fig.\ \ref{fig:bimatclass}.
\end{description}

In addition to the three generic cases there are \emph{degenerate cases} when one or more of the four parameters are zero. None of these contains an interior singularity, but many of these contain an extended NE component which may be, but not necessarily, a SESet. We refer the interested reader to \citet{cressman_03} for details. Practically interesting games like the Centipede game of Length Two or the Chain Store game belongs to these degenerate cases. The latter is defined by the payoff matrix
\begin{equation}\label{eq:chainstore}
(\bA,\bB^T)=\left( \begin{array}{cc}
                        (0,4)\; & (0,4) \cr
                       (2,2)\; & (-4,-4)
                       \end{array}\right),
\end{equation}
and the associated replicator flow diagram with a SESet is illustrated in the lower right panel of Fig.\ \ref{fig:bimatclass}.

\subsection{Other game dynamics} 

A large number of different population-level dynamics is discussed in the game theory literature. These can be either derived rigorously from microscopic (i.e., agent-based) strategy update rules in the large population limit (see Sec.\ \ref{sec:mve}), or they are simply posited as the starting point of the analysis on the aggregate (population, macro) level. Many of these share important properties with the replicator dynamics, others behave quite differently. An excellent recent review on the various game dynamics is \citet{ hofbauer_bams03}.

There are two important dimensions along which macro (aggregate) dynamics can be classified: (1) being innovative or non-innovative, and (2) being payoff monotone or non-monotone. \emph{Innovative} strategies have the potential to introduce new strategies not currently present in the population, or revive formerly extinct ones. Non-innovative dynamics maintain (sometimes diminish) the support of the strategy space. Payoff monotonicity, on the other hand, refers to the relative speed of the change of strategy frequencies. A dynamics is \emph{payoff monotone} if for any two strategies $i$ and $j$,
\begin{equation}
    \frac{\dot{\rho}_i}{\rho_i} > \frac{\dot{\rho}_{j}}{\rho_{j}} \Longleftrightarrow
    u_i(\brho) > u_{j}(\brho),
\end{equation}
i.e., iff at any time strategies with higher payoff proliferate more rapidly than those with lower payoff. By these definitions, the Replicator dynamics is non-innovative and payoff monotone. Although some dynamics may not share all the qualitative properties of the replicator dynamics, it can be shown that when a dynamic is payoff monotone, the Folk theorem, discussed above, remains valid \citep{hofbauer_bams03}.\\

\leftline{\sf Best Response Dynamics}\medskip

A typical innovative dynamics is the \emph{Best Response dynamics}: in each moment a small fraction of the population updates her strategy, and chooses her best response (BR) to the current state $\brho$ of the system, leading to the dynamical equation
\begin{equation}\label{eq:BRdyn}
    \dot{\brho} = BR(\brho)-\brho.
\end{equation}
Usually the game is such that in a large domain of the strategy simplex the best response is a unique (and hence pure) strategy $\bbeta$. Then the solution of Eq.\ (\ref{eq:BRdyn}) is a linear orbit
\begin{equation}\label{eq:BRdynsol}
    \brho(t) = (1-e^{-t})\bbeta + e^{-t}\brho_0,
\end{equation}
where $\brho_0$ is the aggregate state of the population at $t=0$. The solution tends towards $\bbeta$, but this is only valid up to a finite time $t'$, when the actual best response suddenly jumps to another strategy $\bbeta'$. From this time on the solution is another linear orbit tending towards $\bbeta'$, again up to another singular point, and so on. The overall trajectory is composed of linear segments.

\begin{figure}[ht]
\centerline{\epsfig{file=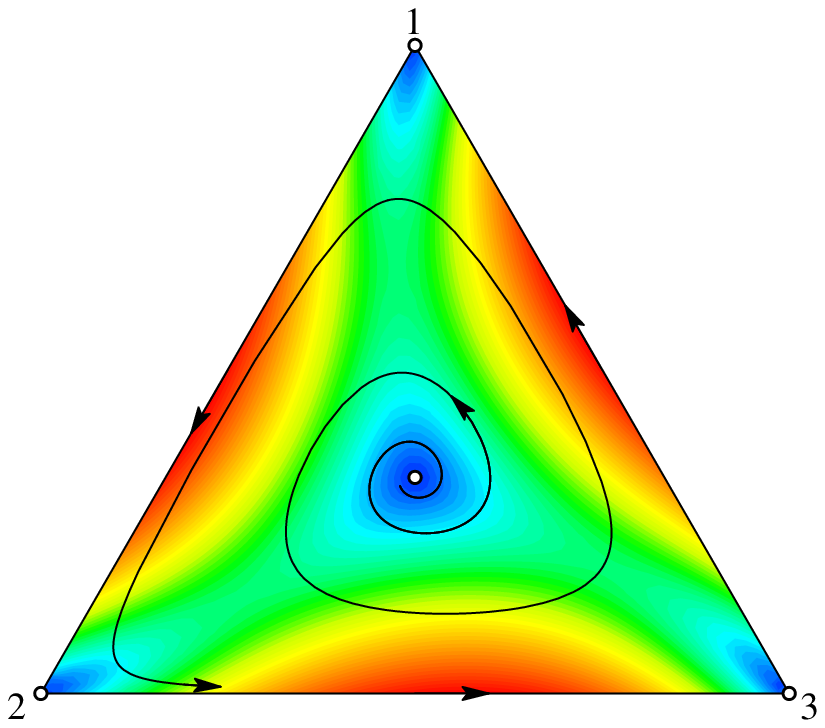,width=5cm}}
\centerline{\epsfig{file=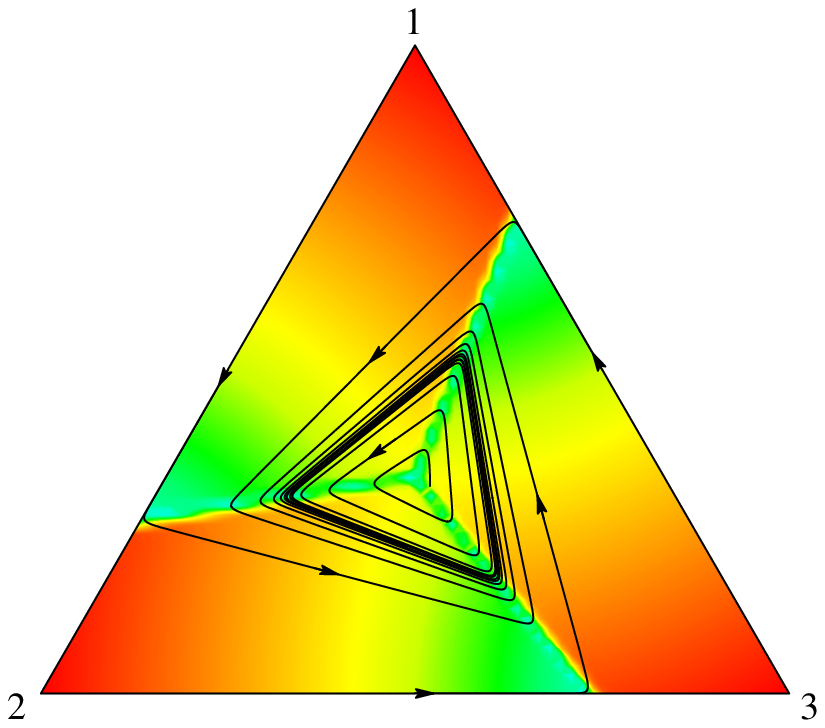,width=5cm} \epsfig{file=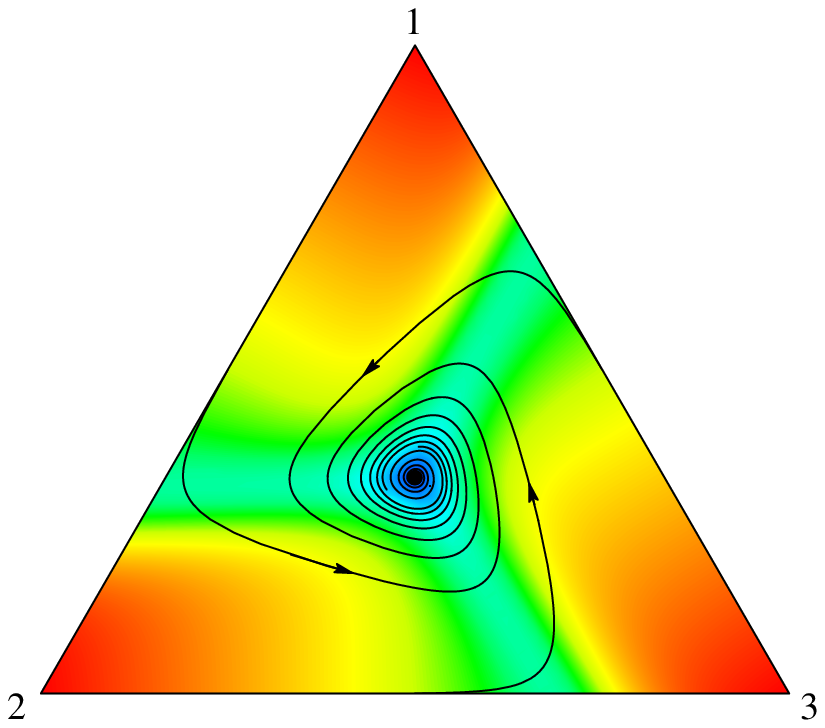,width=5cm}}
\caption{\label{fig:BRdyn}Trajectories predicted by Eqs.\ (\ref{eq:replicator}), (\ref{eq:BRdyn}) and (\ref{eq:logit}) for the game Eq.\ (\ref{eq:Ho_example}) with $b=0.55$. Upper panel: Replicator equation with a repulsive fixed point; lower left: Best Response dynamics with a stable limit cycle; lower right: Logit dynamics with $K=0.1$ showing an attractive fixed point.}
\end{figure}

The Best Response dynamics can produce qualitatively different behavior from the Replicator equation. As an example \citep{ hofbauer_bams03}, consider the trajectories for the Rock-Scissors-Paper-type game
\begin{equation}\label{eq:Ho_example}
{\bf A}=\left( \begin{array}{ccc}
                        0 & -1 & b \cr
                        b & 0  & -1\cr
                       -1 & b  & 0
                       \end{array}\right),
\end{equation}
with $b=0.55$, as depicted in Fig.\ \ref{fig:BRdyn}. This shows the appearance of a limit cycle, the so-called Shapley triangle \citep{shapley_ams64} for the Best Response dynamics, but not for the Replicator equation.\\

\leftline{\sf Logit Dynamics}\medskip

A generalization of the Best Response dynamics for bounded rationality is the \emph{Logit dynamics} (Smoothed Best Response), defined as \citep{fudenberg_98}
\begin{equation}\label{eq:logit}
    \dot{\rho}_i = \frac{\exp{[u_i(\brho)/K]}}{\sum_j \exp{[u_j(\brho)/K]}} - \rho_i.
\end{equation}
In the limit when the noise parameter $K\to 0$, we get back the Best Response equation (\ref{eq:BRdyn}). Again, the Logit dynamics may produce rather different qualitative results than other dynamics. For the game in Eq.\ (\ref{eq:Ho_example}) with $b=0.55$ the system goes through a Hopf bifurcation when $K$ is varied. There is a critical value $K_c\approx 0.07$ below which the interior fixed point is unstable and a limit cycle develops, whereas above $K_c$ the fixed point becomes attractive (see Fig. \ref{fig:BRdyn}).\\

\leftline{\sf Adaptive Dynamics}\medskip

All dynamics discussed so far model how strategies compete with each other on shorter time scales. Given the initial composition of the population a possible outcome is that one of the strategies drives out all others and prevails in the system. However, when this strategy is not an ESS it is unstable against the attack of some superior mutants (not present originally in the population). \emph{Adaptive Dynamics} \citep{metz_96,dieckmann_jmb96,geritz_prl97,nowak_s04} models the situation when the mutation rate is low and possible mutants only differ slightly from residents. Under these conditions the dynamic behavior is governed by short transient periods when a successful mutant conquers the system, separated by long periods of relative tranquility with unsuccessful invasion attempts by inferior mutations. By each successful invasion event, whose details are not modeled, the defining parameters of the prevailing strategy change a little. The process can be modeled on longer time scales as a smooth dynamics in strategy space. It is usually assumed that the flow points into the direction of the most favorable local strategy, i.e., the actual monomorphic strategy $s(t)$ of the population satisfies
\begin{equation}\label{eq:adaptivedyn}
    \dot s = \left. \frac{\partial u(q,s)}{\partial q}\right|_{q=s}
\end{equation}
where now $u(q,s)$ is the payoff of a $q$-strategist in a homogeneous population of $s$-strategists.
Adaptive dynamics may not converge to a fixed point (limit cycles are possible), and even if it converges, the attractor may not be an ESS \citep{nowak_s04}. Counterintuitively, Adaptive Dynamics can lead to a fitness minimum, where a monomorphic population becomes unstable and split into two, providing a theory of evolutionary branching and speciation \citep{geritz_prl97,nowak_s04}.

\section{Evolutionary games: agent-based dynamics}
\label{sec:eg-abd}

What we used in the former section for the description of the state of the population was a \emph{population-level} (\emph{macro-level, aggregate-level}) description. This specified the state of the population and its dynamic evolution in terms of a small number of strategy frequencies. Such a macroscopic description is adequate, if the social network is mean-field-like and the number of agents is very large. However, when these premises are not fulfilled, a more detailed, lower-level analysis is required. Such a lower-level approach is usually termed ``agent-based", since on this level the basic units of the theory are the individual agents themselves. The agent-level dynamics of the system is usually defined by \emph{strategy update rules}, which describe how the agents perceive their surrounding environment, what information they acquire, what believes and expectations they form from former experience, and how this all translates into strategy updates during the game. These rules can mimic genetically coded Darwinian selection or boundedly rational human learning, both affected by possible mistakes. When games are played on a graph, the update rule may not only concern a strategy change alone, but a reorganization of the agent's local network structure, too.

These update rules can also be viewed as ``meta-strategies", as they represent strategies about strategies. The distinction between strategies and meta-strategies is somewhat arbitrary, and is only  justified if there is a hierarchic relation between them. This is usually assumed, i.e., the given update rule is utilized much more rarely in the repeated game than the basic stage-game strategies. A strategy update has low probability during the game. \citet{roca_prl06} have discussed the significant consequences when varying the ratio of time scales between payoff refreshment and strategy update. Also, while players can use different strategies in the stage games, we usually posit that all players use the same update rule in the population. Note that these assumptions may not be justified in certain situations.

There is a huge variety of microscopic update rules defined and applied in the game theory literature. There is no general law of nature which would dictate such behavioral rules: even though we call these rules ``microscopic", they, in fact, emerge as simplified phenomenological rules of an even more fundamental layer of mechanisms describing the operation of the human psyche. The actual choice of the update rule depends very much on the concrete problem under consideration.

Strategy updates in the population may be synchronized or randomly sequential in the social graph. Some of these rules are generically stochastic, some others are deterministic, sometimes with small stochastic components representing random mutations (experimentation). Furthermore, there exist many different ways how the strategy change is determined by the local environment. In many cases the strategy choice for a given player depends on the payoff differences of her payoff and her neighbors' payoffs. This difference may be determined by a one-shot game between the confronting players (see, e.g., the spatial Rock-Scissors-Paper games), by a summation of the stage game payoffs over all neighbors, or perhaps these summed payoffs are accumulated over some time with a weighing factor decreasing with the time elapsed. Usually the rules are myopic, i.e., optimization is based on the current state of the population without anticipating possible future alterations. We will mostly concentrate on memoryless (Markovian) systems, where the evolutionary rule is determined by the current payoffs. A well-known exception, which we do not discuss here, is the Minority game, which was covered in ample detail in two recent books by \citet{challet_04} and by \citet{coolen_05}.

In the following we consider a system with equivalent players distributed on the site of a lattice (or graph). The player at site $x$ follows one of the $Q$ pure strategies characterized by a set of $Q$-component
unit vectors,
\begin{equation}
\label{eq:state}
{\bf s}_x=\left( \matrix{1 \cr \vdots \cr 0
\cr}\right)\, , \; \cdots \, , \, \left( \matrix{0 \cr \vdots \cr
1 \cr}\right)\ .
\end{equation}

The income of player $x$ comes from the same symmetric two-person stage game with her neighbors. In this case her total income can be expressed as
\begin{equation}
\label{eq:pox}
U_x=\sum_{y \in \Omega_x} {\bf s}_x \cdot {{\bf A}} {\bf s}_y,
\end{equation}
where the summation runs over agent x's neighbors $y\in\Omega_x$ defined by the connectivity structure.

For any strategy distribution $\{ {\bf s}_x \}$ the total income of the system is given as
\begin{eqnarray}
\label{eq:tpo}
U &=& \sum_x U_x  = \sum_{x,y \in \Omega_x} {\bf s}_x \cdot{{\bf A}} {\bf s}_y \;.
\end{eqnarray}
Notice that for potential games ($A_{ij}=A_{ji}$) this formula is analogous to the (negative) energy of an Ising-type model. In the simplest case ($Q=2$), if $A_{ij} = - \delta_{ij}$ then $U$ is equivalent to the Hamiltonian of the ferromagnetic Ising model where spins positioned on a lattice can point upward or downward. Within the lattice gas formalism the up and down spins are equivalent to occupied and empty sites, respectively. These models are widely used to describe ordering processes in  two-component solid solutions \citep{kittel_04}. The generalization for higher $Q$ is straightforward.  The energy of the $Q$-state Potts model can be reproduced by a $Q \times Q$ unit matrix \citep{wu_rmp82}. Consequently, for certain types of dynamical rules, evolutionary potential games become equivalent to many-particle systems, whose investigations by the tools of statistical physics are very successful [for a textbook see e.g. \citet{chandler_87}].

The exploration of the possible dynamic update rules has resulted in an extremely wide variety of models, which cannot be surveyed completely. In the following, we focus our attention on some of the most relevant and/or popular rules appearing in the literature.

\subsection{Synchronized update}

Strategy revision opportunities can arrive synchronously or asynchronously for the agents. In spatial models the difference between synchronous and asynchronous update is enhanced, because these processes yield fundamentally different spatio-temporal patterns and general behavior \citep{huberman_pnas93}.

In \emph{synchronous update} the whole population updates simultaneously in discrete time steps, giving rise to a discrete-time dynamics on the macro level. This is the kind of update used in cellular automata. Synchronous update is applied, for instance, in biological models, when generations are clearly discernible in time, or when seasonal effects synchronize metabolic and reproductive processes. Synchronous update may be relevant for some other systems, as well, where  appropriate time delays enforce neighboring players to modify their strategy with respect to the same current time surrounding.

Cellular automata can well represent those evolutionary games, where the players are located on the sites of a lattice. The generalization to arbitrary networks \citep{abramson_pre01,masuda_pla03,duran_pd05} is straightforward and not detailed here. At discrete time steps ($t=0, 1, 2, \ldots$) each player refreshes her strategy simultaneously according to a deterministic rule, depending on the state of the neighborhood. For evolutionary games this rule is usually determined by the payoffs in Eq.\ (\ref{eq:pox}). For example, in the model suggested by \citet{nowak_n92b} each player adopts the strategy of those neighbors (including herself) who achieved the highest income in the last round.

Spatio-temporal patterns (or behaviors) occurring in these cellular automata can be classified into four distinct classes \citep{wolfram_rmp83,wolfram_pd84,wolfram_02}. In Class 1 the evolution leads exactly to the same uniform final pattern (called frequently a ``fixed point" or an ``absorbing state") from almost all initial states. In Class 2 the system can develop into many different states built up from a certain set of simple local structures that either remain unchanged or repeat themselves after a few steps (limit cycles). The behavior becomes more complicated in Class 3, where the time-dependent patterns exhibits random elements. Finally, in Class 4 the time-dependent patterns involve high complexity (mixture of some order and randomness) and certain features of these nested patterns exhibit power law behavior. The best known example belonging to this universality class is the Game of Life invented by John Conway. This two-dimensional cellular automaton has a lot of localized structures (called animals that can blink, move, collide, come to life, etc.) as discussed in detail by \citet{gardner_sa70} and \citet{sigmund_93}. \citet{killingback_jtb98} have shown that the game theoretical construction suggested by \citet{nowak_n92b} exhibits complex dynamics with long range correlations between states in both time and space. This is a characteristic feature in Class 4.

A cellular automaton rule defines the new strategy for every player as a function of the strategy distribution in her neighborhood. For the Nowak-May model the spatio-temporal behavior remains qualitatively the same within a given range of payoff parameters, while larger variations of these parameters can modify substantially the behavior and can even put it into another class. In Section \ref{sec:spdsu}, we will demonstrate explicitly the consecutive transitions occurring in the model.


Cellular automaton rules may involve stochastic elements. In this case, while preserving synchronicity, the update rule defines the probability to take one of the possible states for each site. In general, stochastic rules destroy states belonging to Classes 2 and 4. This extension can also cause additional non-equilibrium phase transitions \citep{kinzel_zpb85}.

Agent heterogeneity can also be included. In the stochastic cellular automaton model suggested by \citet{lim_pre02} each player has an acceptance level $\delta_x$ chosen randomly from a region $-\Delta < \delta_x < \Delta$ at her birth, and the best strategy of the neighborhood is accepted if $U_x < \delta_x + \max (U_y)$, $y \in \Omega_x$. This type of stochasticity can largely affect the behavior.

There is a direct way to make a cellular automaton stochastic \citep{mukherji_n96,tomochi_pre02}. If the prescription of the local deterministic rule is accepted at each site with probability $\mu$, while the site remains in the previous state with probability $1-\mu$, the parameter $\mu$ characterizes the degree of synchronization. This update rule reproduces the original cellular automaton for $\mu=1$, whereas for $\mu<1$ it leads to stochastic update. This type of stochastic cellular automata allows us to study the continuous transition from deterministic synchronized update to random sequential evolution (limit $\mu\to 0$).

The effects of weak noise, $1- \mu << 1$, have been studied from different points of view. \citet{mukherji_n96} and \citet{abramson_pre01} demonstrated that a small amount of noise can prevent the system to fall into a "frozen" state characteristic of Class 2. The sparsely occurring "errors" are capable of initiating avalanches, which transform the system into another meta-stable state. Under some conditions the size distribution of these avalanches exhibits power-law behavior, at least for some parameter regions, as reported by \citet{lim_pre02,holme_pre03}. Details of the frozen patterns and the avalanches are discussed in detail by \citet{zimmermann_pre05}.

\subsection{Random sequential update}
\label{sec:rsu}

The other basic update mechanism is \emph{asynchronous update}. In many real social systems the players modify their strategies independently of each other. For these systems random sequential (asynchronous) update gives a more appropriate description. One possibility is that in each time step one agent at random is selected from the population. Thus the probability per time of updating a given agent is $\lambda=1/N$. Alternatively, each agent may possess an independent ``Poisson clock", which ``rings" for update according to a Poisson process at rate $\lambda$. These assumptions assure that the simultaneous update of more than one agents has zero probability, and thus in each moment the macroscopic state of the population can only change a little. In the infinite population limit asynchronous update leads to smooth, continuous dynamics as we will see in Sec.\ \ref{sec:mve}.  Asynchronous strategy revisions are  appropriate for overlapping generations in the Darwinian selection context. Also they are more typical in economics applications.

In the case of random sequential update the central object of the agent-level description is the \emph{individual transition rate} $w(s\to s')$ which denotes the conditional probability per unit time that an agent, given the opportunity to update, flips from strategy $s$ to $s'$. Clearly, this should satisfy the sum rule
\begin{equation}\label{sumrule}
    \sum_{s'} w(s\to s') = \lambda
\end{equation}
($s'=s$ included in the sum).
In population games the individual transition rate only depends on the macroscopic state of the population $w(s\to s')=w(s\to s';\brho)$. In network games the rate depends on the other agents' actual state in the neighborhood, including their current strategies or payoffs $w(s_x\to s_x')=w(s_x\to s_x';\{s_y,u_y\}_{y\in\Omega_x})$. In theory the transition probabilities could also depend explicitly on time (the round of the game), or on the complex history of the game, etc., but we usually disregard these possibilities.

\subsection{Microscopic update rules}
\label{sec:mur}

In the following we enlist some of the most representative microscopic rules, based on replication, imitation, and learning. 
\\

\leftline{\sf Mutation and experimentation}\medskip

The simplest case is when the individual transition probability is independent of the state of the other agents. Strategy change arises due to intrinsic mechanisms with no influence from the rest of the population. The prototypical example is \emph{mutation} in biology. The similar mechanism is often called \emph{experimentation} in economics contexts. It is customary to assume that mutation probabilities are payoff-independent constants
\begin{equation}\label{mutation}
    w(s\to s') = c_{ss'}.
\end{equation}
such that the sum rule in Eq.\ (\ref{sumrule}) is satisfied.

If mutation is the only dynamic mechanism without any other selective force, it leads to spontaneous \emph{genetic drift} (random walk) in the state space. If the population is finite there is a finite probability that a strategy becomes extinct by pure chance \citep{drossel_ap01}. Drift usually has less relevance in large populations [for exceptions see, e.g., \citet{traulsen_pre06b}], but becomes an important issue in the case of new mutants. Initially mutant agents are small in numbers, and even if the mutation is beneficial, the probability of fixation (i.e., reaching a macroscopic number in the population) is less that one, because the mutation may be lost by random drift.\\

\leftline{\sf Imitation}\medskip

Imitation processes form a wide class of microscopic update rules. The essence of imitation is that the agent who has the opportunity to revise her strategy takes over the strategy of one of the fellow players with some probability. The strategy of the fellow player remains intact; it only plays a catalyzing role. Imitation cannot introduce a new strategy which is not yet played in the population; this rule is \emph{non-innovative}.

Imitation processes can differ in two respects: whom to imitate and with what probability. The standard procedure is to choose the agent to imitate at random from the neighborhood. In the mean-field case this means a random partner from the whole population. The imitation probability may depend on the information available for the agent. The rule can be different if only the strategies used by the neighbors are known, or if both the strategies and their resulting last-round (or accumulated) payoffs are available for inspection. In the first case the agent should decide only knowing her own payoff. Rules in this case are similar in spirit to the Win-Stay-Lose-Shift rules that we discuss later on.

In the case when both strategies and payoff can be compared, one of the simplest imitation rules is \emph{Imitate if Better}. Agent $x$ with strategy $s_x$ takes over the strategy of another agent $y$, chosen randomly from $x$'s neighborhood $\Omega_x$, iff $y$'s strategy has yielded higher payoff, otherwise the original strategy is maintained. If we denote the set of neighbors of agent $x$ who play strategy $s$ by $\Omega_x(s)\subseteq \Omega_x$, the individual transition rate for $s'_x\ne s_x$ can be written as
\begin{equation}\label{imit_better}
    w(s_x\to s_x') = \frac{\lambda}{|\Omega_x|} \sum_{y\in \Omega_x(s_x')} \theta[U_y-U_x],
\end{equation}
where $\theta$ is the Heaviside function, $\lambda>0$ is an arbitrary constant, and $|\Omega_x|$ is the number of neighbors. In the mean-field case (population game) this simplifies to \begin{equation}\label{imit_betterMF}
    w(s_x\to s_x') = \lambda\, \rho_{s_x'} \theta[U(s_x')-U(s_x)].
\end{equation}

Imitation rules are more realistic if they take into consideration the actual payoff differences between the original and the imitated strategies. Along this line, an update rule with nice dynamical properties, as we will see, is Schlag's \emph{Proportional Imitation} \citep{helbing_incoll98,schlag_jet98,schlag_jme99}. In this case another agent's strategy in the neighborhood is imitated with a probability proportional to the payoff difference, provided that the new payoff is higher than the old one:
\begin{equation}\label{prop_imit}
    w(s_x\to s_x') =  \frac{\lambda}{|\Omega_x|} \sum_{y\in \Omega_x(s_x')} \max[U_y-U_x,0].
\end{equation}
Imitation only occurs if the target strategy is more successful, and in this case its rate goes linearly with the payoff difference. For later convenience we give again the mean-field rates
\begin{equation}\label{prop_imitMF}
    w(s_x\to s_x') =  \lambda\, \rho_{s_x'} \max[U(s_x')-U(s_x),0].
\end{equation}

Proportional imitation does not allow for an inferior strategy to replace a more successful one. Update rules which forbid this are usually called \emph{payoff monotone}. However, payoff monotonicity is frequently broken in case of bounded rationality, and a strategy may be imitated with some finite probability even if it has produced lower payoff in earlier rounds. A possible general form of \emph{Smoothed Imitation} is
\begin{equation}\label{gen_imit}
    w(s_x\to s_x') =  \frac{\lambda}{|\Omega_x|} \sum_{y\in \Omega_x(s_x')} g(U_y-U_x)
\end{equation}
where $g$ is a monotonically increasing smoothing function, for instance,
\begin{equation}\label{gen_imit2}
    g(\Delta u)=\frac{1}{1+\exp(-\Delta u/K)}.
\end{equation}
where $K$ can measure the extent of noise.

Imitation rules can be generalized to the case when the entire neighborhood is monitored simultaneously and plays a collective catalyzing role. Denoting by $\bar\Omega_x=\{ x,\Omega_x \}$ the neighborhood which includes agent $x$ as well, a possible form used by \citet{nowak_ijbc94,nowak_pnas94} and their followers (e.g., \citep{alonsosanz_ijbc01b,masuda_pla03}) for the Prisoner's Dilemma is
\begin{equation} \label{eq:nbmscak}
    w(s_x\to s_x') = \lambda\, \frac{ \sum_{y\in \bar\Omega_x(s_x')} g(U_y) }
                         { \sum_{y\in \bar\Omega_x} g(U_y) }.
\end{equation}
where again $g(U)$ is an arbitrary positive smoothing function, and the sum in the numerator is limited to agents in the neighborhood who pursue the target strategy $s_x'$. \citet{nowak_ijbc94,nowak_pnas94} considered $g(z)=z^k$ ($z>0$). This choice reproduces the deterministic rule Imitate the Best (in the neighborhood) in the limit $k \to \infty$. Most of their analysis, however, focused on the linear version $k=1$. When $k<\infty$ this rule is not payoff monotone, but assures in general that strategies which perform better on average in the neighborhood have higher chances to be imitated.

All types of imitation rules have the possibility to reach a homogeneous state sooner or later when the system size is finite. Once the homogeneous state is reached the system remains there forever, that is, evolution stops. Homogeneous strategy distributions are absorbing states for imitative behavior. The time to reach one of the absorbing states, however, increases very rapidly with the system size. With a small mutation rate added, the homogeneous states cease to remain steady states, and the dynamical process becomes ergodic.
\\

\leftline{\sf Moran process}\medskip

Although imitation seems to be a conscious decision process, the underlying mathematical structure may emerge in dynamic processes which have nothing to do with higher level consciousness. Consider, for instance, the \emph{Moran process}, which is biology's prototype dynamic update rule for asexual (haploid) replication \citep{moran_62,nowak_n04}. At each time step, one individual is chosen for
reproduction with a probability proportional to its fitness. (In the game theory context fitness can be the payoff or a non-negative, monotonic function of the payoff.) An identical offspring (clone) is produced, which replaces another individual. In the mean-field case this latter is chosen randomly from the population. The population size $N$ remains constant. Update in the population is asynchronous.

The overall result of the Moran process is that one individual of a constant population forgoes her strategy, and instead takes over the strategy of another agent with a probability proportional to the relative abundance of that strategy. The Moran process can be viewed as a kind of imitation.
Formally, in the mean-field case, the individual transition rates read
\begin{equation}\label{Moran}
    w(s_x\to s_x') = \lambda\, \rho_{s_x'} \frac{U(s_x')}{\bar{U}},
\end{equation}
where $\bar{U}=\sum_s \rho_s U(s)$ is the average fitness of the population. The right hand side is simply the frequency of $s'$-strategists, weighted by their relative fitness. This is formally the rule Eq.\ (\ref{eq:nbmscak}) with $g(z)=z$.

The Moran process is investigated recently by \citet{nowak_n04,taylor_bmb04,wild_prslb04,antal_bmb06,traulsen_pre06c} to determine the size-dependence of the relaxation time and the extinction probability characterizing how the system reaches a homogeneous absorbing state via random fluctuations. An explicit mean-field description in the form of Fokker-Planck equation was derived and studied by \citep{traulsen_prl05,traulsen_pre06a}.
\\

\leftline{\sf Better and Best Response}\medskip

Imitation dynamics, including the Moran process, considered thus far are all \emph{non-innovative} dynamics, which cannot introduce new strategies into the population. If a strategy becomes extinct, a non-innovative rule cannot revive it. In contrast with this, dynamics which can introduce new strategies are called \emph{innovative}.

One of the most important innovative dynamics is the \emph{Best Response} rule. This assumes that agents, when they get a revision opportunity, adopt their best possible strategy (best response) to the current strategy distribution of the neighborhood. The Best Response rule is myopic, agents have no memory and do not worry about the long run consequences of their strategy choice. Nevertheless, best response require more cognitive abilities than imitation: (1) the agent should be aware of the distribution of the co-players' strategies, and (2) should fully know the conceivable strategy options. Note that these may not be realistic assumptions in complicated real-life situations.

Denoting by $s_{-x}$ the current strategy profile of the population in the neighborhood of agent $x$, and by $\textrm{BR}(s_{-x})$ the set of the agent's best replies to $s_{-x}$, the Best Response rule can be formally written as
\begin{equation}\label{eq:BRrule}
    w(s_x\to s_x')\sim \left\{%
    \begin{array}{ll}
    \lambda/|\textrm{BR}| & \hbox{if $s_x'\in \textrm{BR}(s_{-x})$,} \\
    0 & \hbox{if $s_x'\notin \textrm{BR}(s_{-x})$} \\
    \end{array}%
    \right.
\end{equation}
where $|\textrm{BR}|$ is the number of possible best replies (there may be more than one).

The player may not be able to assess the distribution of strategies in the population (neighborhood), but may remember the ones confronted with in earlier rounds. Such setup leads to \emph{Fictitious Play}, a prototype dynamic rule studied already in the 50s \citep{brown_51}, in which best response is given to the overall empirical distribution of former rounds.

Bounded rationality may also imply that the agent is only able to consider a limited set of opportunities and optimize within this set (\emph{Better Response}). For instance, in the case of continuous strategy spaces, taking a small step in the direction of the local payoff gradient is called \emph{Gradient Dynamics}. The new strategy adopted is
\begin{equation}
    s_x' = s_x + \frac{\partial U(s_x,s_{-x})}{\partial s_x} \Delta s_x.
\end{equation}
which improves the player's payoff by a small amount.

In many cases it is reasonable to assume that the strategy update is a stochastic process, and instead of a sharp Best Response a smoothed rule is posited. In \emph{Smoothed Best Response} the transition
rates are usually written as
\begin{equation}\label{eq:SBRrule}
    w(s_x\to s_x') = \lambda \frac{g[U(s_x';s_{-x})]}{\sum_{s_x''\in S_x} g\left[U(s_x'';s_{-x})\right] }
\end{equation}
where $g$ is a monotonically increasing positive smoothing function assuring that better strategies have more chance to be adopted. A typical choice, as will be discussed in the next Section, is when $g$ takes the Boltzman form
\begin{equation}\label{SBR}
    w(s_x\to s_x') = \lambda\frac{\exp[U(s_x';s_{-x})/K]}{\sum_{s_x''\in S_x}
    \exp[U(s_x'';s_{-x})/K] }.
\end{equation}
This rule is usually called the ``Logit rule" or the ``Log-linear rule" in the game theory literature \citep{blume_geb93}.

Smoothed Best Response is not payoff monotone, and is very similar in form to smoothed imitation in Eq.\ (\ref{eq:nbmscak}). The major difference is the extent of the strategy space: imitation only considers strategies actually played in the social neighborhood, whereas the variants of the best response rules challenge the agent's all feasible strategies, even if they are not played in the population.

Smoothed Best Response can describe a myopic player who is capable of estimating the variation of her own payoff upon a strategy change while assuming that the neighbors' strategies remain unchanged. Such a player always adopts (rejects) the advantageous (disadvantageous) strategy in the limit $K \to 0$ but makes a boundedly rational (noisy) decision with some probability of mistaking when $K>0$.
The rule in Eq.\ (\ref{SBR}) may also model a situation in which players are rational but payoffs have a small hidden idiosyncratic random component or when gathering perfect information about decision alternatives is costly \citep{ blume_geb03,blume_igtr03,helbing_td96}. All these features together can be handled mathematically by the ``temperature" parameter $K$.

Such dynamical rules were used by \citet{ebel_pre02} in the consideration of the Prisoner's Dilemma, by \citet{sysiaho_epjb05} who studied an evolutionary Snow-drift game, and by \citet{blume_geb93,blume_98,blume_geb03,blume_igtr03} in Coordination games. \draftnote{Igaz ez?}
\\

\leftline{\sf Win-Stay-Lose-Shift}\medskip

When the co-players' payoffs are unobservable, the player should decide about a new strategy in the knowledge of her own payoffs realized earlier in the game. She may have an ``aspiration level": when her last round payoff (or payoffs averaged over a number of rounds) is below this aspiration level, she shifts over to a new strategy, otherwise she stays with the original one. The new (target) strategy can be imitated from the neighborhood or chosen randomly from the agent's own strategy set.

This is the philosophy behind the \emph{Win-Stay-Lose-Shift} rules. These rules have been frequently discussed in connection with the Prisoner's Dilemma, where there are four possible payoff values $S<P<R<T$ in a stage game (see Appendix \ref{app:g:pd}). If the aspiration level is set between the second and third payoff values, and only the last round payoff is considered (no averaging), we obtain the so-called ``Pavlov rule" [also called ``Pavlov strategy", when considered as an elementary (non-meta) strategy]. The individual transition rate can be written as
\begin{equation}
    w(s_x\to \bar{s}_x) =\lambda\, \Theta(a-U_x);\qquad P<a<R,
\end{equation}
where $a$ is the aspiration level, $s_x=C,D$ and $\bar{s}_x=D,C$ (the opposite of $s$), respectively. Pavlov is known to beat Tit-for-Tat in a noisy environment \citep{nowak_n93}. In the spatial versions the Pavlov rule was used by \citet{fort_jsm05}, who observed that the size distribution of $C$ (or $D$) strategy clusters follows power-law scaling.

Other aspiration levels, with or without averaging, define other variants of the Win-Stay-Lose-Shift class of (meta-)\-strategies. The set of these rules can be extended by allowing a dynamic change of the aspiration level, as it appears frequently in human and animal examples \citep{colman_95}. These strategies involve a way how the aspiration level changes step by step knowing the previous payoffs. For example, the so-called \emph{Yesterday strategy} repeats the previous action
if, and only if, the payoff is at least as good as in the previous round \citep{posch_prslb99}. Other versions of these strategies are discussed in \citep{posch_jtb99}.\\

\subsection{From micro to macro dynamics in population games}
\label{sec:mve}

The individual transition rates introduced in the former section can be used to formulate the  dynamics in terms of a master equation. This is especially convenient in the mean-field case, i.e., for population games. For random sequential update in each infinitesimal time interval there is at most one agent who considers a strategy change. When the change is accepted, the old strategy $i$ is replaced by a new strategy $j$, thereby decreasing the number of players pursuing $i$ by one, and increasing the number of players pursuing $j$ by one in the population.  These processes are called \emph{$Q$-type birth-death Markov processes} \citep{blume_98,gardiner_04}, where $Q$ is the number of different strategies in the game.

Let $n_i$ denote the number of $i$-strategists in the population, and
\begin{equation}
    \bn=\{n_1,n_2,\dots,n_Q\}, \quad \sum_i n_i = N,
\end{equation}
the macroscopic configuration of the system. We introduce $\bn^{(jk)}$ as a shorthand for the configuration \citep{helbing_td96,helbing_incoll98}
\begin{equation}
    \bn^{(jk)}=\{n_1,n_2,\dots,n_j-1,\dots,n_k+1,\dots,n_Q\},
\end{equation}
which differs from $\bn$ by the elementary process of changing one $j$-strategist into a $k$-strategist.
The time-dependent probability density over configurations, $P(\bn,t)$, satisfies the master equation
\begin{equation}\label{mastereq}
    \frac{dP(\bn,t)}{dt} = \sum_{\bn'} \left[ P(\bn',t) W(\bn'\to\bn) - P(\bn,t) W(\bn\to\bn') \right].
\end{equation}
where $W(\bn\to\bn')$ is the \emph{configurational transition rate}. The first (second) term represents the inflow (outflow) into (from) configuration $\bn$. This form assumes that the update rule is a Markov process with no memory. [For a generalized master equation with a memory effect see, e.g., \citet{helbing_incoll98}].

The configurational transition rates can be calculated from the individual transition rates. Notice that $W(\bn\to\bn')$ is only nonzero if $\bn'=\bn^{(jk)}$ with some $j$ and $k$, and in this case it is proportional to the number of $j$ strategists $n_j$ in the population. Thus we can write the configurational rates as
\begin{equation}\label{Wdelta}
    W(\bn\to\bn') = \sum_{j,k} n_j\, w(j\to k;\bn)\, \delta_{\bn',\bn^{(jk)}}.
\end{equation}

The master equation in Eq.\ (\ref{mastereq}) describes the dynamics of the configurational probabilities. Knowing the actual state of the population in a given time means that the probability distribution $P(\bn,t)$ is extremely sharp (delta function). This initial probability density becomes wider and wider as time elapses. When the population is large, this widening is slow enough such that a deterministic approximation can give a satisfactory description. The temporal trajectory of the mean strategy densities, i.e., the state vector $\brho\in\Delta_Q$ obeys a deterministic, continuous-time, first-order ordinary differential equation \citep{helbing_td96,helbing_incoll98,benaim_e03}.

In order to show this, we define the (time-dependent) average value of a quantity $f$ as $\langle f\rangle = \sum_{\bn} f(\bn) P(\bn,t)$. Following \citet{helbing_incoll98}, we express the time derivative of $\langle n_i\rangle$ from the master equation as
\begin{eqnarray}
    \frac{d\langle n_i \rangle}{dt} =
    \sum_{\bn,\bn'} (n_i^\prime-n_{i}) W(\bn\to\bn') P(\bn,t).
\end{eqnarray}
Using Eq.\ (\ref{Wdelta}) we get
\begin{eqnarray}\label{det1}
    \frac{d\langle n_i \rangle}{dt} &=&
    \sum_{\bn,j,k} \left( n^{(jk)}_i-n_i \right) n_j w(j\to k;\bn) P(\bn,t) \nonumber\\
    &=&
    \sum_{\bn,j} \left[ n_j w(j\to i;\bn) -n_i w(i\to j;\bn)  \right] P(\bn,t) \nonumber\\
\end{eqnarray}
where we have used that $n^{(jk)}_i-n_i=\delta_{ik}-\delta_{ij}$.

Equation (\ref{det1}) is exact. However, to get a closed equation which only contains mean values an approximation should be made. When the probability distribution is narrow enough, we can write
\begin{equation}
    \langle n_i w(i\to j;\bn)\rangle \approx \langle n_i\rangle\, w(i\to j;\langle\bn\rangle).
\end{equation}
This approximation leads to the \emph{approximate mean value equation} for the strategy frequencies $\rho_i(t)=\langle n_i \rangle/N$ \citep{weidlich_pr91,helbing_incoll98,helbing_td96,benaim_e03}:
\begin{eqnarray}\label{amveq}
    \frac{d\rho_i(t)}{dt} =
    \sum_{j} \left[\rho_j(t) w(j\to i;\brho) -\rho_i(t) w(i\to j;\brho)\right]. 
\end{eqnarray}

This equation can be used to derive the macroscopic dynamic equations from the microscopic update rules. Take, for instance, Proportional Imitation in Eq.\ (\ref{prop_imitMF}). Equation (\ref{amveq}) then gives
\begin{eqnarray}\label{amveq2}
    \frac{d\rho_i(t)}{dt} =
    \sum_{j\in {\cal S}_{-}} \rho_j\rho_i(u_i-u_j) -
    \sum_{j\in {\cal S}_{+}} \rho_i\rho_j(u_j-u_i)
\end{eqnarray}
where ${\cal S}_{+}$ (${\cal S}_{-}$) denotes the set of strategies superior (inferior) to strategy $i$. It can be readily checked that this leads to
\begin{eqnarray}\label{amveq3}
    \frac{d\rho_i(t)}{dt} = \lambda \rho_i(u_i-\bar{u})\qquad  \bar{u}=\sum_j \rho_j u_j
\end{eqnarray}
where $\bar{u}$ is the average payoff in the population. This is exactly the replicator dynamics in the Taylor-Jonker form, Eq.\ (\ref{eq:replicator}).

If the microscopic rule is chosen to be the Moran process, Eq.\ (\ref{Moran}), the approximate mean value equation leads instead to the Maynard-Smith form of the replicator equation, Eq.\ (\ref{eq:replicatorMS}). Similarly, the Best Response rule in Eq.\ (\ref{eq:BRrule}) or the Logit rule in Eq.\ (\ref{SBR}) lead to the corresponding macroscopic counterparts in Eqs.\ (\ref{eq:BRdyn}) and (\ref{eq:logit}).

Since Eq.\ (\ref{amveq}) is only approximate, the actual behavior in a large but finite population will eventually deviate from the solution of Eq.\ (\ref{amveq}). Going beyond the approximate mean value equation requires a Taylor expansion of the right hand side of Eq.\ (\ref{det1}), and leads to an infinite hierarchy of coupled equations involving higher moments. This hierarchy should be truncated by an appropriate decoupling scheme. The simplest (second order) approximation beyond Eq.\ (\ref{amveq}) yields two coupled equations: one for the mean and one for the covariance $\langle n_i n_j\rangle_t$ \citep{weidlich_pr91,helbing_incoll98,helbing_td96} [for an alternative approach see \citet{binmore_geb95}].

\subsection{Potential games and the kinetic Ising model}
\label{sec:pgkim}

The kinetic Ising model, introduced by \citet{glauber_jmp63}, exemplifies how a static spin model can be extended by a dynamic rule such that the dynamics drives the system towards the thermodynamic equilibrium characterized by Boltzmann probabilities.

\begin{figure}[ht]
\centerline{\epsfig{file=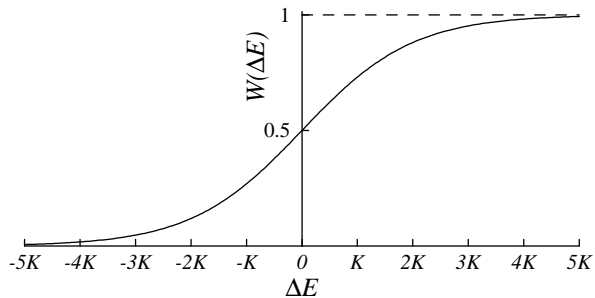,width=8cm}}
\caption{\label{fig:fddf}Transition probability (\ref{eq:glauber}) as a function of $\Delta E$ provides a smooth transition from 0 to 1 within a region comparable to $K$.}
\end{figure}

In Glauber dynamics the following steps are repeated: (1) choose a site (player) $x$ at random; (2) choose a possible spin state (strategy) $s_x^{\prime}$ at random; 3) change the configuration $\bs=\{s_x,s_{-x}\}$ to $\bs'=\{s_x',s_{-x}\}$ with a probability
\begin{equation}
    \label{eq:glauber}
    W(\bs \to \bs' ) = {1 \over 1+ \exp (-\Delta E(\bs,\bs')/K) },
\end{equation}
where $\Delta E(\bs,\bs') = E(\bs') - E(\bs)$ is the change in energy of the system with respect to the initial and final spin configurations. The parameter $K$ is the ``temperature" which measures the extent of noise. \citet{traulsen_jtb07,traulsen_jtb07b,perc_njp06a} have shown that the stochastic evaluation of payoffs can be interpreted as an increased temperature. Notice that $w(s_x \to s_x^{\prime}) \to 1$ (0) if $\Delta E >> K$ ($\Delta E << K$ ) as illustrated in Fig.~\ref{fig:fddf}. In the $K\to 0$ limit the dynamics becomes deterministic.

Glauber dynamics allows one-site spin flips with a probability depending on the energy difference $\Delta E$ between the final and initial states. The equilibrium distribution $P(\bs)$ of this stochastic process is known to satisfy the condition of detailed balance \citep{glauber_jmp63,kawasaki_72}
\begin{eqnarray}\label{detbal}
    W(\bs \to \bs^{\prime})\, P(\bs) =
    W(\bs^{\prime} \to \bs)\, P(\bs^\prime).
\end{eqnarray}
In equilibrium there is no net probability current flowing between any two microscopic states connected by the dynamics. If the detailed balance is satisfied, the equilibrium can be calculated trivially: independently of the initial state, the dynamics drives the system towards a limit distribution in which the probability of a given spin configuration $\bs$ is given by the Boltzmann distribution at temperature $K$,
\begin{equation}
    \label{eq:boltzmanndf}
    P( \bs)={ \exp (-E( \bs )/K) \over \sum_{\bs^\prime} \exp (-E(\bs^\prime) /K)} \;.
\end{equation}

The equilibrium distribution in Eq.\ (\ref{eq:boltzmanndf}) can be reached by other transition rates as well. The only requirement following from the detailed balance condition Eq.\ (\ref{detbal}) is that the ratio of forward and backward transitions should be
\begin{equation}
    \label{eq:glauberratio}
    \frac{W(\bs \to \bs^{\prime})}{W(\bs^{\prime} \to \bs)} = \exp (-\Delta E(\bs,\bs')/K).
\end{equation}
Equation (\ref{eq:glauber}) corresponds to one of the standard choices with single spin flips, where the sum of the forward and backward transition probabilities equal to 1, but this choice is largely arbitrary. Alternative dynamics may even allow two spin exchanges \citep{kawasaki_72}.

We remind the readers that Glauber dynamics favors elementary processes reducing the total energy in the model, while the contact with the heat reservoir provides an average energy depending on the temperature $K$. In other words, this dynamics drives the system into a (maximally disordered) macroscopic state where the entropy
\begin{equation}
    \label{eq:entropy}
    S = -\sum_{ \bs} P( \bs) \ln P( \bs);,
\end{equation}
reaches its maximum, if the average energy is fixed. The mathematical details of this approach are well described in many books [see e.g., \citet{chandler_87,haken_88}].

Glauber dynamics, as introduced above, defines a simple stochastic rule whose limit distribution (equilibrium) turns out to be the Boltzmann distribution. In the following we will show that the Logit rule, as defined in Eq.\ (\ref{SBR}), is equivalent to Glauber dynamics when the game has a potential \citep{blume_geb93,blume_geb95,blume_98}. In order to prove this we have to show two things: the ratio of forward and backward transitions satisfy Eq.\ (\ref{eq:glauberratio}) with a suitable energy function $E$, and the detailed balance condition Eq.\ (\ref{detbal}) is indeed satisfied.

The first is easy: for the Logit rule, Eq.\ (\ref{SBR}), the ratio of forward and backward transitions reads
\begin{equation}
    \label{eq:SBRratio}
    \frac{W(\bs \to \bs^{\prime})}{W(\bs^{\prime} \to \bs)} =
    \frac{w(s_x \to s_x^{\prime})}{w(s_x^{\prime} \to s_x)} = \exp (\Delta U_x/K).
%
%
\end{equation}
If the game has a potential $V$ then Eq.\ (\ref{potential}) assures that
\begin{eqnarray}\label{potential2}
    \Delta U_x &=& U_x(s_x^\prime,s_{-x})-U_x(s_x,s_{-x})\nonumber\\
               &=& V(s_x^\prime,s_{-x})-V(s_x,s_{-x}).
\end{eqnarray}
Thus if we define the ``energy function" as $E(\{s_x\})=-V(\{s_x\})$ then Eq.\ (\ref{eq:glauberratio}) is formally satisfied.

As for the detailed balance requirement what we have to check is \emph{Kolmogorov's Criterion} \citep{freidlin_84,blume_98,kelly_79}: detailed balance is satisfied if and only if the product of forward transition probabilities equals to the product of backward transition probabilities along each possible closed loop ${\cal L}: \bs^{(1)}\to \bs^{(2)}\to \dots\to\bs^{(L)}\to \bs^{(1)}$ in the state space, i.e.,
\begin{equation}
    R_{\cal L} = \quad\frac{W(\bs^{(1)} \to \bs^{(2)})}{W(\bs^{(2)} \to \bs^{(1)})}\,
    \frac{W(\bs^{(2)} \to \bs^{(3)})}{W(\bs^{(3)} \to \bs^{(2)})}\times \dots \frac{W(\bs^{(L)} \to \bs^{(1)})}{W(\bs^{(1)} \to \bs^{(L)})} = 1.
\end{equation}
Fortunately, it suffices to check the criterion for one-agent three-cycles and two-agent four-cycles: each closed loop can be built up from these elementary building blocks.

For one-agent three-cycles ${\cal L}_3: s_x\to s_x'\to s_x''\to s_x$ (with no change in the other agents' strategies) the Logit rule satisfies trivially the criterion:
\begin{eqnarray}
    R_{{\cal L}_3}=&&
    \exp\left(\frac{U_x(s_x',s_{-x})-U_x(s_x,s_{-x})}{K}\right)
    \exp\left(\frac{U_x(s_x'',s_{-x})-U_x(s_x',s_{-x})}{K}\right) \times \nonumber \\
    && \exp\left(\frac{U_x(s_x,s_{-x})-U_x(s_x'',s_{-x})}{K}\right)  =1,
\end{eqnarray}
even if the game has no potential. However, the existence of a potential is essential for the elementary four-cycle, ${\cal L}_4: (s_x,s_y)\to (s_x',s_y)\to (s_x',s_y')\to (s_x,s_y')\to (s_x,s_y)$ (with no change in the other agents' strategies). In this case
\begin{eqnarray}
 R_{{\cal L}_4}=&&
 \exp \left\{\frac{1}{K}
    \Big[ U_x(s_x',s_y)-U_x(s_x,s_y) +U_y(s_x',s_y')-U_y(s_x',s_y)\Big] \right\} \times \nonumber \\
  && \exp \left\{\frac{1}{K}
    \Big[ U_x(s_x,s_y')-U_x(s_x',s_y') +U_y(s_x,s_y)-U_y(s_x,s_y') \Big]
    \right\}
\end{eqnarray}
where the notation was simplified by omitting the reference to other, fixed-strategy players, $U_x(s_x,s_y)\equiv U_x(s_x,s_y; s_{-\{x,y\}})$. If the game has a potential, and thus Eq.\ (\ref{potential2}) is satisfied, this can be written as
\begin{eqnarray}
    R_{{\cal L}_4}= &&
\exp \left\{\frac{1}{K}
    \Big[ V(s_x',s_y)-V(s_x,s_y) +V(s_x',s_y')-V(s_x',s_y)\Big]
    \right\} \times \nonumber \\
&&\exp \left\{\frac{1}{K}
    \Big[ V(s_x,s_y')-V(s_x',s_y') +V(s_x,s_y)-V(s_x,s_y')
    \Big] \right\} = 1.
\end{eqnarray}

Kolmogorov's criterion is satisfied since all terms in the exponent cancel. Thus we have shown that for potential games the Logit rule is equivalent to Glauber dynamics, consequently the limit distribution is the Boltzmann distribution at temperature $K$ with energy function $-V$,
\begin{equation}
    \label{eq:boltzmanndf2}
    P( \bs)={ \exp (V( \bs )/K) \over \sum_{\bs^\prime} \exp (V(\bs^\prime) /K)} \;.
\end{equation}

The above equivalence with kinetic Ising-type models (Glauber dynamics) opens a direct way for the application of statistical physics methods in this class of evolutionary games.
For example, the mutual reinforcement effect between neighboring (connected) agents can be considered as an attractive interaction in case of Coordination games. A typical example is when agents should select between two competing technologies like LINUX and WINDOWS \citep{lee_res00} or VHS and BETAMAX \citep{ helbing_td96}. When the connectivity structure is characterized by a square lattice and the individual strategy update is defined by the Logit rule (Glauber dynamics) then the stationary states of the corresponding model are well described by the thermodynamic behavior of the Ising model, which exhibits a critical phase transition in the absence of an external magnetic field \citep{stanley_71}. Moreover, the ordering processes (including the decay of metastable states, nucleation, and domain growth) are also well investigated for Ising-type models \citep{binder_prb74,bray_ap94}. The behavior of potential games is similar to that of the equivalent Ising-type model.

The above analogy can also be useful in games which do not strictly have a potential, but which are close to a potential game that is equivalent to a well-known physical system. For example a small parameter $\varepsilon$ can characterize the symmetry-breaking of the payoff matrix ($A_{ij}-A_{ji}=\pm 2 \varepsilon$ or $0$) as happens when considering the combination of a three-state Potts model with the Rock-Scissors-Paper game \citep{szolnoki_pre05a}. One can study how the cyclic dominance ($\varepsilon>0$) can prevent the formation of long-range order described by the three-state Potts model ($\varepsilon=0$) below the critical temperature. A similar idea was used for the so-called ``driven lattice gases" suggested by \citet{katz_prb83,katz_jsp84}, who studied the effect of an external electric field (inducing particle transport through the system) on the ordering process controlled by  nearest-neighbor interactions in Ising-type lattice gas models [for a review see \citet{schmittmann_95,marro_99}].

\subsection{Stochastic stability}

Although stochastic update rules are essential at the level of individual decision making, the fluctuations average out and produce smooth, deterministic population dynamics on the aggregate level, when the population is infinitely large and all pairs are connected. In the case of finite populations, however, the deterministic approximation discussed in Section \ref{sec:mve} is no longer exact, and usually it only provides acceptable prediction for the short run behavior \citep{benaim_e03,reichenbach_pre06}. The long-run analysis requires the calculation of the full stochastic (limit) distribution. Even if the basic micro-dynamics is deterministic, a small stochastic noise, originating from mutations, random errors in strategy implementation, or deliberate experimentation with new strategies may play a crucial role. It turns out that predictions based on the asymptotic behavior of the deterministic approximation may not always be relevant for the long run behavior of noisy systems. The stability of dynamic equilibria with respect to pertinent perturbations becomes a fundamental question. The concept of noise tolerance can be introduced as a novel refinement criterium to provide additional guideline for realistic equilibrium selection.

The analysis of stochastic perturbations goes back to \citet{foster_tpb90} who define the notion of \emph{stochastic stability}. \emph{A stochastically stable set is a set of strategy profiles, which appear with nonzero probability in the limit distribution of the stochastic evolutionary process,  when the noise level becomes infinitesimally small.} As such, stochastically stable sets may contain states which are not Nash equilibria such as limit cycles. Nevertheless, for many important games this set contains a single Nash equilibrium. For instance, in the case of 2x2 Coordination games, where the game has two asymptotically stable Nash equilibria, only one of these, the \emph{risk-dominant} \draftnote{Is it defined?}
equilibrium proves to be stochastically stable. The connection between stochastic stability and risk dominance was found to hold under rather wide assumptions on the dynamic rule \citep{kandori_e93,young_e93,blume_98,blume_geb03}. Stochastically stable equilibria are sometimes called "long-run equilibria" in the literature.

Following \citet{kandori_e93}, we illustrate the concept of stochastic stability by a simple, symmetric, two-strategy Coordination game
\begin{equation}
\begin{tabular}{c|ccc}
              & Hare & &Stag     \\[0.5ex] \hline
Hare          & a=4      & &b=3   \\
Stag          & c=0     & &d=5   \\
\end{tabular}
\end{equation}
played as a population game by a finite number $N$ of players. We can think about this game as the Stag Hunt game, where a population of hunters learn to coordinate on hunting for Stags or for Hares. (See Appendix \ref{app:g:sh} for a detailed description). Coordinating on Stag is Pareto optimal, but choosing Hare is less risky if the hunter is unsure of the coplayer's choice. As for a definite micro-dynamics, we assume a deterministic Best Response rule with asynchronous update, perturbed by a small probability $\varepsilon$ that a player, when it is her turn to act, makes an erroneous choice (noise).

Let $N_S=N_S(t)$ denote the number of Stag hunters in the population. Without mutations the game has two Nash equilibria: (1) everybody hunts for Stag (``AllStag", i.e., $N_S=N$), and (2) everybody hunts for Hare (``AllHare", i.e., $N_S=0$). Both are evolutionarily stable. There is also a third, mixed strategy equilibrium with $N_S^*=2N/3$, which is not evolutionarily stable. For the deterministic dynamics the first two NEs are stable fixed points, the third is an unstable fixed point playing the role of a separatrix. When $N_S(t)<N_S^*$ ($N_S(t)>N_S^*$) each agent's best response is to hunt for Stag (Hare).  Depending on the initial condition one of the fixed point is readily reached by the deterministic rule.

\begin{figure}[t]
\centerline{\epsfig{file=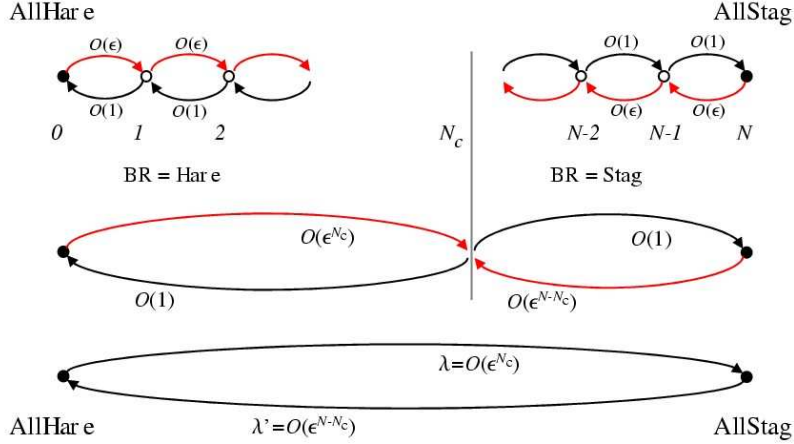,width=11cm}}
\caption{\label{fig:staghunt} The Stag Hunt game with uniform noise. The $N$-state birth-death Markov process can be approximated by a two-state process in the small noise limit.}
\end{figure}

When the noise rate is finite, $\varepsilon> 0$, we should calculate a probability distribution  over the state space, which satisfies the appropriate master equation. The process is a one-type, $N$-state birth-death Markov process, which satisfies detailed balance, and has a unique stationary distribution $P_\varepsilon(N_S)$. It is intuitively clear that in the $\varepsilon\to 0$ limit the probability will cluster on the two deterministic fixed points, and thus the most important features of the solution can be obtained without detailed calculations. When $\varepsilon$ is small we can ignore all intermediate states and estimate an effective transition rate between the two stable fixed points \citep{kandori_e93}.  Reaching the border of the basin of attraction from AllHare requires $N_S^*$ subsequent mutations, whose probability is ${\cal O}(\varepsilon^{N_S^*})$ (see Fig.\ \ref{fig:staghunt}). From here the other fixed point can be reached by a rate ${\cal O}(1)$. Thus the effective transition rate from AllHare to AllStag is $\lambda={\cal O}(\varepsilon^{N_S^*})={\cal O}(\varepsilon^{2N/3})$. A similar calculation gives the effective rate for the inverse process $\lambda'={\cal O}(\varepsilon^{N-N_S^*})={\cal O}(\varepsilon^{N/3})$. Using this approximation, the stationary distribution over the two states AllHare and AllStag (neglecting all the intermediate states) becomes
\begin{equation}
    P_{\varepsilon} \approx \left(
    \frac{\lambda'}{\lambda+\lambda'}\, ,  \frac{\lambda}{\lambda+\lambda'} \right)
    \mathop{\longrightarrow}_{\varepsilon\to 0}\;
    (1, 0)
\end{equation}
We can see that in this example as $\varepsilon\to 0$ the probability weight clusters on the state AllHare, showing that only this NE is stochastically stable. Of course, this qualitative result would remain true if we calculated the full limit distribution of this stochastic problem. 

Even though the system flips occasionally into AllStag by a series of appropriate mutations, it spends most of the time near AllHare. If we look at the system at a random time we can almost surely find it in the state (more precisely, in the domain of attraction) of AllHare. The deterministic approximation based on Eq.\ (\ref{amveq}), which predicts AllStag as an equally possible solution depending on the initial condition, is misleading in the long run. It should be noted, however, that when the noise level is low or the population is large there is a huge time needed to take the system out from a non-stochastically stable configuration, which clearly limits the practical relevance of the concept of stochastic stability.

In the example above we have found that only AllHare is stochastically stable. AllStag, which is otherwise Pareto optimal is not stochastically stable. Of course, this finding depends on the actual parameters in the payoff matrix. In general (still keeping the game to be a Coordination game), we find that the stochastic stability of AllHare holds iff
\begin{equation}\label{riskdominant}
    a-c \ge d-b.
\end{equation}
This is exactly the condition that AllHare is risk-dominant, exemplifying the general connection between risk dominance and stochastic stability in this class of games \citep{kandori_e93,young_e93}. When Eq.\ (\ref{riskdominant}) loses validity, the other NE, AllStag, becomes risk-dominant and, at the same time, stochastically stable.

Stochastic stability probes the limit distribution of the stochastic process in the zero noise limit for a given finite population. Another interesting question arises if the noise level is low but fixed, and the size of the population is taken to infinity. In this limit any given configuration (strategy profile) will have zero probability, but the distribution may concentrate on one Nash configuration and its small perturbations, where most players play the same strategy. Such a Nash configuration is called \emph{ensemble stable} \citep{miekisz_jsp04,miekisz_jpa04,miekisz_pa04}. The concepts of stochastic stability and ensemble stability may not necessarily coincide \citep{miekisz_jsp04,miekisz_jpa04,miekisz_pa04}.

\section{The structure of social graphs}
\label{sec:noc}

In realistic multi-player systems players do not interact with all other players. In these situations the connectivity between players is given by two graphs as suggested recently by \citet{ohtsuki_prl07,ohtsuki_jtb07b}, where the nodes represent players, and the edges, connecting the nodes, refer to the connection between the corresponding players. The first graph defines how the connected players play a game to gain some income. The second graph describes the learning (strategy adoption) mechanism. In a more general formalism the edges of graphs may have two weight factors characterizing the strength of influence along both directions. These extensions allow us to study the effects of preferential learning \citep{lieberman_n05,wu_cpl06,guan_epl06} that can involve an asymmetry in the teaching-learning activity between the connected players \citep{kim_pre02,szolnoki_epl07}. The behavior of evolutionary games is strongly affected by the underlying social structure.

Henceforth our analysis concentrates on systems, where both the game theoretical interaction and the learning mechanism are based on the same network of connectivity, and the weight of an edge is unity. The theory of graphs [see textbooks, e.g., \citet{bollobas_85,bollobas_98}] gives a mathematical background for the structural analysis. In the last decades an extensive research has focused on the different random networks [for detailed survey see \citet{amaral_pnas00,albert_rmp02,dorogovtsev_03,newman_siamr03,boccaletti_pr06}] that can be considered as potential structures of connectivity. The investigation of evolutionary games on these structures has lead to the exploration of new phenomena and raised a number of interesting questions.

The connectivity structure can be characterized by a number of topological properties. In the following we assume that the corresponding graph is connected, i.e., there is at least one path along edges between any two sites. Graph theory defines the degree $z_x$ of site $x$ as the number of neighbors (co-players) connected to $x$. For random graphs the \emph{degree distribution} $f(z)$ determines the probability of finding exactly $z$ neighbors for a player. The degree 
is uniform, i.e., $f(z)=\delta(z-z_0)$, for \emph{regular structures}\footnote{A graph is called regular, if the number of links $z_x$, emanating from node $x$, is the same $z_x=z$ for all $x$.}
(e.g., for lattices), and $f(z) \propto z^{-\gamma}$ (typically $2 < \gamma < 3$) for the so-called \emph{scale-free graphs}. These latter exhibit sites with extremely large number of neighbors. Real networks can have statistical properties in between these extreme limits. It is customary to classify structures according to their degree distributions \citep{amaral_pnas00} as 1) \emph{scale-free networks}  (e.g., the world-wide web) having a power-law tail for large $z$, 2) \emph{truncated scale-free networks} (e.g., the network of movie actors) with an extended power-law behavior up to some large $z$ cutoff, and 3) \emph{single-scale networks} (e.g., some friendship networks) with an exponential or Gaussian dependence in the whole regime with a characteristic degree. For other specific real-life examples see, e.g., \citet{amaral_pnas00} and \citet{boccaletti_pr06}.

In graph theory a \emph{clique} means a complete subgraph in which all pairs are linked together. The \emph{clustering coefficient} characterizes the "cliquishness" of the closest environment of a site. More precisely, the clustering coefficient ${\cal C}_x$ of site $x$ is the proportion of actual edges among the sites within its neighborhood to the number of potential edges that could possibly exist among them. In a different context the clustering coefficient characterizes the fraction of possible triangles (two edges with a shared site) which are in fact triangles (three-site cliques). The distribution of $\cal{C}$ can be also introduced but usually only its average value $\bar{\cal{C}}$ is considered. In many cases the percolation of the overlapping triangles (or cliques with more site) seems to be the crucial feature responsible for interesting phenomena \citep{palla_n05}.

\begin{figure}[ht]
\centerline{\epsfig{file=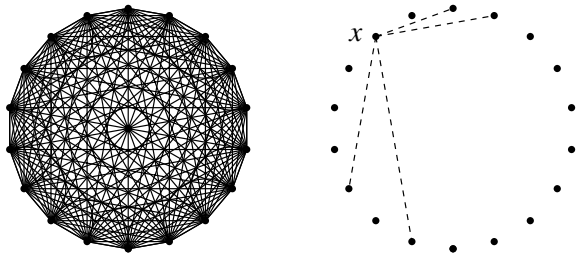,width=8cm}}
\caption{\label{fig:mfstruc}Connectivity structures for which the mean-field approach is valid in the limit $N \to \infty$. On the right hand side the dashed lines indicate temporary connections to   co-players chosen at random for a given time.}
\end{figure}

Now we briefly survey some prototypical connectivity structures, which have been investigated actively in recent years. First we have to emphasize that a mean-field-type behavior occurs for two drastically different graph structures as $N \to \infty$ (see Fig.~\ref{fig:mfstruc}). On one hand, the mean-field approximation is exact by definition for those systems where all the pairs are linked, that is, the corresponding graph is complete. On the other hand, similar behavior arises if each player's income comes from games with only $z$ temporary players who are chosen randomly in every step. In this latter setup there is no correlation between partners from one stage to the next.

\subsection{Lattices}

For spatial models the fixed interaction network is defined by the sites of a lattice and the edges between those pair whose distance does not exceed a given value. The most frequently used structure is the square lattice with von Neumann neighborhood (including connections between nearest neighbor sites, $z=4$) and Moore neighborhood (with connections between nearest and next-nearest neighbors, $z=8$). Many  transitions from one behavior to a distinctly different one depend on the dimensionality of the embedding space, therefore the considerations can be extended to $d$-dimensional hyper-cubic structures too. When using periodic boundary conditions these systems become translation invariant, and the spatial distribution of strategies can be investigated by  mathematical tools developed in solid state theory and non-equilibrium statistical physics (see e.g., Appendix \ref{app:gmfa}).

\begin{figure}[th]
\centerline{\epsfig{file=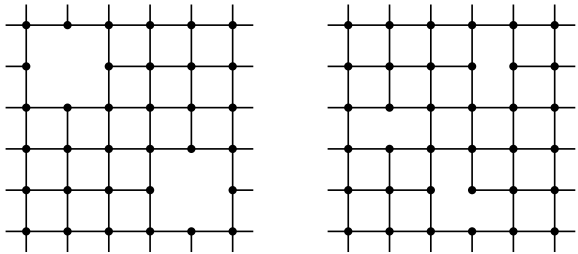,width=8cm}}
\caption{\label{fig:dillat}Two types of diluted lattices. Left: randomly chosen sites are removed together with the corresponding edges. Right: randomly chosen edges are removed.}
\end{figure}

In many cases the regular lattice only provides an initial structure for the creation of more realistic social networks. For example, diluted lattices (see Fig.~\ref{fig:dillat}) can be used to study what happens if a portion $q$ of players and/or interactions are removed at random \citep{nowak_ijbc94}. The resultant connectivity structure is inhomogeneous and we cannot use  analytical methods assuming translation invariance. A systematic numerical analysis in the limit $q \to 0$, however, can give us a picture about the effect of these types of defects.

\begin{figure}[th]
\centerline{\epsfig{file=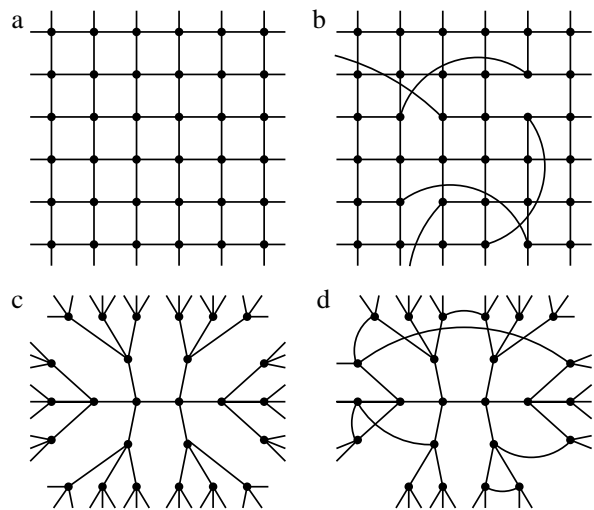,width=8cm}}
\caption{\label{fig:stages4} Different regular connectivity structures
where each player has four neighbors: (a) square lattice,
(b) regular "small-world" network, (c) Bethe lattice (or tree-like structure) and (d) random regular graph. }
\end{figure}

For many (partially) random connectivity structures more than one topological features change simultaneously, and the individual effects of these are mixed up. The clarification of the role a feature may play requires their separation and independent tuning. Figure~\ref{fig:stages4} shows some regular connectivity structures (for $z=4$), whose topology is strikingly different, meanwhile the clustering coefficients are all zero (or vanishing in the limit $N \to \infty$.

\subsection{Small worlds}

In Figure~\ref{fig:stages4} a \emph{regular small-world network} is created from a square lattice by randomly rewiring a fraction $q$ of connections in a way that conserve the degree for each site. Random links reduce drastically the average distance $\bar{l}$ between randomly chosen pair of sites, producing a small world phenomenon characteristic to many social networks \citep{milgram_pt67}. In the limit $q \to 0$ the depicted structure is equivalent to the square lattice. If all connections are replaced ($q=1$) the rewiring process yields the well-investigated \emph{random regular graph} \citep{wormald_inc99}. Consequently, the set of these structures provides a continuous structural transition from the square lattice to random regular graphs.

In random regular graphs the concentration of short loops vanishes as $N\to\infty$ \citep{wormald_jctb81}, therefore these structures become locally similar to a tree. In other words, the local structure is similar to the one characterizing the fictitious Bethe lattice (existing in the limit $N\to\infty$), on which analytical calculations can be performed thanks to translation invariance. Hence the results of simulations on large random regular graphs can be compared with analytical predictions (e.g., the pair approximation) on the Bethe lattice, at least for those systems where the effect of large loops is negligible.

Spatial lattices are full of short loops whose number decreases during the rewiring process. There exists a wide range of $q$, however, where the two main features of small-world structures are present simultaneously \citep{watts_n98}. These properties are (1) the small average distance $\bar{l}$ between two randomly chosen sites, and (2) the high clustering coefficient $\bar{\cal{C}}$, i.e., the high density of connections within the neighborhood of a site. Evidently, the average distance increases with $N$ in a way that depends on the topology. For the sake of comparison, on a $d$-dimensional lattice $\bar{l} \approx N^{1/d}$; on a complete (fully connected) graph $\bar{l}=1$; and on random graphs $\bar{l} \approx \ln N / \ln \bar{z}$ where $\bar{z}$ denotes the average degree. When applying the method of random rewiring the small average distance ($\bar{l} \approx \ln N / \ln \bar{z}$) can be achieved for a surprisingly low portion of random links ($q>0.01$).

\begin{figure}[ht]
\centerline{\epsfig{file=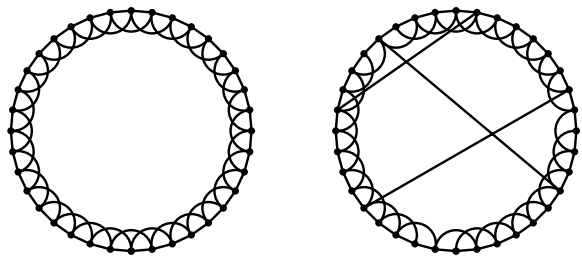,width=8cm}}
\caption{\label{fig:wssw}Small-world structure (right) is created from
a ring (left) with the nearest- and next-nearest neighbors. Random links are substituted for a portion $q$ of the original links as suggested by \citet{watts_n98}.}
\end{figure}

For the structures in Fig.~\ref{fig:stages4} the topological character of the graph cannot be characterized by the clustering coefficient, because $\bar{\cal{C}}=0$ (in the limit $N \to \infty$) as mentioned above. In the original small-world model suggested by \citet{watts_n98} the initial structure is a one-dimensional lattice with periodic boundary conditions, where each site is connected to its $z$ (even) nearest neighbors as shown in Fig.~\ref{fig:wssw}. During a rewiring process one of the ends of $q z N/2$ bonds is shifted to another site chosen at random. The final structure has sufficiently high clustering coefficient and short average distances within a wide range of $q$. Further versions of small-world networks were suggested by \citet{newman_pre99}. In the simplest structures random links are added to a lattice.

In the initial structure (left graph in Fig.~\ref{fig:wssw}) the average clustering coefficient $\bar{\cal{C}}=1/2$ for $z=4$, and its value decreases linearly with $q$ (if $q << 1$) and tends to a value of order $1/N$ as $q \to 1$. The analysis of these types of graphs becomes interesting for those evolutionary games, where the clustering or percolation of the overlapping cliques play a crucial role (see Sect. \ref{sec:pdsocnet}).

\subsection{Scale-free graphs}

In the above inhomogeneous connectivity structures the degree distribution $f(z)$ has a sharp peak around $\bar{z}$ and the occurrence of sites with $z>>\bar{z}$ is unlikely. There exists, however, a lot of real networks in nature (e.g., the internet, the network of acquaintance, collaborations, metabolic reactions, and many other biological networks), where the presence of sites with large $z$ is essential and is due to fundamental processes.

\begin{figure}[ht]
\centerline{\epsfig{file=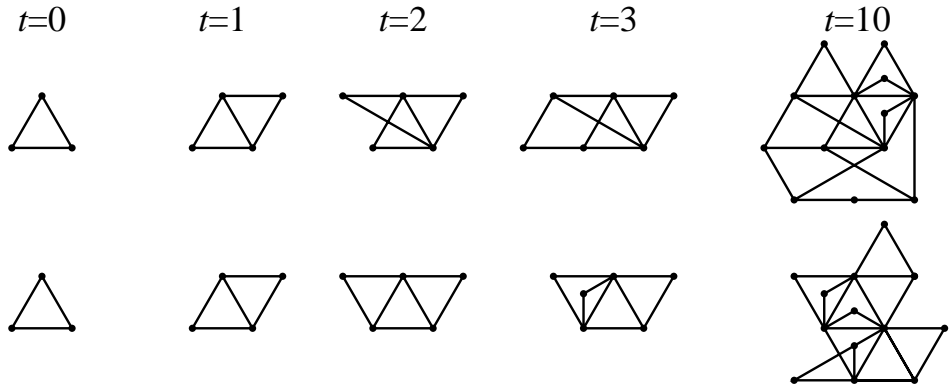,width=13cm}}
\caption{\label{fig:sfbadm}Creation of two scale-free networks with the same average degree ($\langle z \rangle =4$) as suggested by \citet{barabasi_s99} (top) and \citet{dorogovtsev_pre01} (bottom).}
\end{figure}

In recent years a lot of models have been developed to reproduce the main characteristics of these networks. Now we only discuss two procedures for growing networks exhibiting scale-free properties. Both procedures start with $k$ connected sites and for each step $t$ we add one site linked to $m$ (different) existing sites as demonstrated in Fig.~\ref{fig:sfbadm} for $k=3$ and $m=2$.

For the growth procedure suggested by \citet{barabasi_s99} a site with $m$ links is added to the system step by step, and the new site is preferentially linked to those sites that have large degrees already. To realize the "rich-gets-richer" phenomenon the new site is linked to the existing site $x$ with a probability depending on its degree
\begin{equation}
\Pi_x = {z_x \over \sum_y z_y} \;.
\label{eq:bamp}
\end{equation}
After $t$ steps this random graph has $N=3+t$ sites and $3+t m$ edges. For large $t$ (or $N$) the degree distribution exhibits a power law behavior within a wide range of $z$, i.e., $f(z) \approx 2 m^2 z^{- \gamma}$ where $\gamma =3$, because older sites increase their degree at the expense of the younger ones. The average connectivity stays at $\langle z \rangle=m$.

In fact, network growth with \emph{linear preferential attachment} naturally leads to power-law degree distributions. A more general family of growing networks can be introduced if the attachment probability $\Pi_x$ is modified as
\begin{equation}
\Pi_x = {g(z_x) \over \sum_y g(z_y)} \;,
\label{eq:mbamp}
\end{equation}
where $g(z)>0$ is an arbitrary function. The analytical calculations of \citet{krapivsky_prl00} demonstrated that \emph{nonlinear preferential attachment} destroys the power-law behavior. The scale-free nature of the growing network can only be achieved if the attachment probability is asymptotically linear, i.e.,  $\Pi_x \sim a z_x$ as $z \to \infty$. In this case the exponent $\gamma$ can be tuned to any value between 2 and $\infty$.


In the Barab\'asi-Albert model (as well as in its variants) the clustering coefficient vanishes as $N \to \infty$. On the contrary, $\bar{\cal C}$ remains finite in many real networks. \citet{dorogovtsev_pre01} have suggested another procedure to create growing graphs providing more realistic clustering coefficients. In this procedure one site is added to the system in each step in such a way that the new site is linked to both ends of a randomly chosen edge. The degree distribution of this structure is similar to those found in the Barab\'asi-Albert model. Apparently these two growth procedures yield very similar graphs for small $t$ as illustrated in Fig.~\ref{fig:sfbadm}. The algorithm suggested by \citet{dorogovtsev_pre01}, however, creates at least one triangle for each step, therefore it results in a finite clustering coefficient ($\bar{\cal C} \sim 0.5$) for large $N$.

For realistic networks the growing process is affected by the aging of vertices and also by the cost of linking or the limited capacity of a vertex \citep{amaral_pnas00}. The mentioned phenomena prevent the formation of scale-free degree distribution by yielding fast decrease in the probability for sufficiently large degrees. Evidently, many models of network growth were developed during the last years, and their investigation have already become an extensive area within statistical physics. Nowadays this research covers the investigation of time-dependent networks too.

\subsection{Evolving networks}

The simplest possible time-dependent connectivity structure is illustrated by the right hand plot in Fig.~\ref{fig:mfstruc}. Here co-players are chosen randomly with equal probability in each round, making the mean-field approximation valid in the limit $N \to \infty$. Partially random connectivity structures with some temporary co-players are convenient from a technical point of view. For example, if randomly chosen co-players are substituted in each round for a portion $q$ of the standard neighbors on the square lattice, then spatial symmetries are conserved on average, and the generalized mean-field technique discussed in Appendix~\ref{app:gmfa} can be safely applied. The increase of $q$ can induce a transition from spatial to mean-field-type behavior as it will be discussed later on.

Dynamically evolving networks in general are the subject of intensive research in many fields \citep{boccaletti_pr06,jackson_05,dutta_03}. Here we only focus on models where the growth and/or rewiring of the network is dictated by an evolutionary game played over the evolving social structure. Some early works on this subject include \citet{skyrms_pnas00} who investigated various dynamic models of network formation assuming reinforcement learning. Choosing partners to play with in the next round was assumed to depend on a probability distribution over possible co-players, which was updated according to achieved payoffs in the former round. Even models with simple constant reinforcement rules were able to demonstrate how structure can emerge from uniformity or, under different model parameters, how uniformity may develop from an initial structure. In more complicated games like the Stag Hunt game with possible discounting and/or noise the co-evolution of network and strategies can strongly depend on the relative speed of the two fundamental updating processes. It was found that when strategy evolution is faster than structural evolution there is a tendency to coordinate on the risk dominant equilibrium (hunting hare) as in the mean-field case. In the opposite limit of fast network evolution the payoff-dominant equilibrium outcome (hunting stag) becomes more likely. In between the two limits the population is typically split into disconnected clusters of stag or hare hunters which coexist in the society.

In a conceptually similar model \citet{zimmermann_00,zimmermann_01,santos_ploscb06} considered what happens if the links of a social graph can be removed and/or replaced by other links. They also assumed that strategies and structure evolve simultaneously with different rates. The (rare) cancelation of a link depended on the total payoff received by the given pair of players. Their model becomes interesting for the Prisoner's Dilemma when an unsatisfied defector breaks the link to neighboring defectors with some probability and looks for other randomly chosen partners. This mechanism conserves the number of links. In  models suggested by \citet{biely_cm05,pacheco_jtb06} the unsatisfied players are allowed to cancel links (to defectors), and the cooperators can create new links to one of the second neighbors suggested by the corresponding first neighbor. \citet{pacheco_prl06} have developed an elegant mean-field type approach assuming finite life times for the links and allowing preferential search for new links. The main predictions of these models will be discussed in Sec.~\ref{sec:pdevolnet}.

For many real systems the migration of players plays a crucial role. It is assumed that during a move in the graph the player conserves her original strategy but faces a new neighborhood. There are several ways how this feature can be built into a model. If the connectivity structure is defined by a lattice (or graph) then we can introduce empty sites where neighboring players can move to. Depending on the microscopic details this additional feature can either decrease or increase the frequency of cooperators in the spatial Prisoner's Dilemma \cite{vainstein_jtb07}. The effects of local migration can be studied by using a more convenient way: two neighboring players (chosen randomly) are allowed to exchange their positions. In Section \ref{sec:compass} we will consider the consequences of tuning the relative strength between imitation and site exchange (migration). In continuous space the random migration of players can be studied using reaction diffusion equations \citep{hofbauer_na97,wakano_mb06} which exhibit either traveling waves or self-organizing patterns.

\section{Prisoner's Dilemma}
\label{sec:epd}

By now the Prisoner's Dilemma has become a world-wide known paradigm for studying the emergence of cooperative (altruistic) behavior in communities consisting of selfish individuals. The name of this game as well as the traditional notation are explained in Appendix \ref{app:g:pd}.

In this two-player game the players have two options to choose from which are called \emph{defection} (D) and \emph{cooperation} (C). For mutual cooperation the players receive the highest total payoff shared equally. The highest individual payoff is reached by a defector against a cooperator. In the most interesting case the defector's extra income (related to the income for mutual cooperation), $T-R$, is less than the relative loss of the cooperator, $R-S$. According to classical (rational) game theory both players should play defection, since this is the Nash equilibrium. However, this outcome would provide them with the second worst income, thus creating the dilemma.

There are other Social Dilemmas \citet{macy_pnas02,skyrms_03} where mutual cooperation could provide the highest total income, although selfish individual reasoning often leads to other choices. For example, in the Snowdrift game (see Appendix \ref{app:g:sd}) it is better to choose the opposite of what the other player does; in the Stag Hunt game (see Appendix \ref{app:g:sh}) the player is better off doing whatever the co-player does. In the last years several authors (e.g., \citep{santos_pnas06,hauert_jtb06a,hauert_jtb06b}) have studied all these dilemmas in a unified framework by expanding the range of payoff parameters. In the following our discussion will focus on the Prisoner's Dilemma which is the most prominent example - other social dilemmas are less favorable to defection.

People face frequently the situation of Prisoner's Dilemmas in real life when they have to choose between to be selfish or altruistic, to keep the ethical norms or not, to work (study) hard or lazy, {\it etc}
\citep{alexander_87,axelrod_apsr86,binmore_94,trivers_85}. Disarmament and some business negotiations are also burdened with this type of dilemma. The traffic flow models and game theoretic concepts were combined by \citet{helbing_acs05,perc_njp07}. \citet{dugatkin_bs96} have shown that the  Prisoner's Dilemma can be recognized in the behavior of hermaphroditic fish alternately releasing
eggs and sperm, because the production of eggs implies a higher metabolic investment. The application of this game, however, is not restricted to human or animal societies. When considering the interactions between two bacteriophages, living and reproducing within infected cells,
\citet{turner_n99} have identified the defective and cooperative versions \citep{nowak_n99}. The investigation of two-dimensional models helps our understanding how altruistic behavior occurs in biofilms \citep{kreft_mb04,maclean_n06}. In a biochemical example \citep{pfeiffer_s01} the two different pathways of the adenosine triphosphate (ATP) production represent cooperators (low rate but high yield of ATP production) and defectors (high rate but low yield). Some authors speculate that in multicellular organisms cells can be considered as behaving cooperatively, except for tumor cells that shifted towards a more selfish behavior \citep{pfeiffer_tbs05}. Game theoretical methods may provide a promising new approach for understanding cancer \citep{gatenby_n03,frick_n03,axelrod_pnas06}.

Despite the prediction of classical game theory the emergence of cooperation can be observed in many naturally occurring Prisoner's Dilemma situations. In the subsequent sections we will discuss the possibilities and conditions how cooperative behavior can subsist in multi-agent models. During the last decades it turned out that the rate of cooperation is affected by the strategy set, the evolutionary rule, the payoffs, and the structure of connectivity.
For such a high degree of freedom the analysis cannot be complete. In many cases investigations are carried out for a payoff matrix with a limited number of variable parameters.

\subsection{Axelrod's Tournaments}
\label{sec:er}

In the late 1970's, a computer tournament were conducted by Robert \citet{axelrod_jcr80a,axelrod_jcr80b,axelrod_84}, whose interest in game theory arose essentially from a deep concern about international politics and especially the risk of nuclear war. Axelrod wanted to identify the conditions under which cooperation could emerge in a Prisoner's Dilemma, therefore he invited game theorists to submit strategies (in the form of computer programs) for playing an Iterated Prisoner's Dilemma game with the payoff matrix
\begin{equation}\label{eq:Axelpom}
\begin{tabular}{cccc}
  & $D\;\;\; C$ & & $D\;\;\; C$ \\[2pt]
  ${\bf A}\,= \, \matrix{D \cr C \cr }$ &
  $\left( \matrix{P & T \cr
                  S & R \cr }\right)$ &
  $\,=\, \matrix{D \cr C \cr }$ &
  $\left( \matrix{1 & 5 \cr
                  0 & 3 \cr }\right)$  \\
\end{tabular}
\end{equation}

The computer tournament was structured as a round robin game. Each player was paired with all other ones, then with its own twin (the same strategy) and with a random strategy choosing cooperation and defection with equal probability. Each iterated game between fixed opponents consisted of two hundred rounds, and the entire round robin tournament was repeated five times to improve the reliability of the scores. Within a tournament the players were allowed to take into account the history of the interactions as had developed thus far. All these rules had been announced well before the tournament.

The highest average score was achieved by the so-called "Tit-for-Tat" strategy developed by Anatol Rapoport. This was the simplest of all the fourteen strategies submitted. Tit-for-Tat is a strategy which cooperates on the first round, and thereafter repeats what the opponent has done on the previous move.

The main results of this computer tournament were published and people were soon invited to submit other strategies for a second tournament. The new strategies were developed in the knowledge of the fourteen old strategies, therefore a large portion of them would have performed substantially better than Tit-for-Tat in the environment of the first tournament. Surprisingly, Tit-for-Tat won the second tournament too.

Using this set of strategies some further systematic investigations were performed to examine the robustness of the results. For example, the tournaments were repeated with some modifications in the population of strategies. The most remarkable investigations are related to the adaption of a dynamical rule \citep{trivers_qrb71,dawkins_76,maynard_sa78} that mimics Darwinian selection. This evolutionary computer tournament was started with $Q$ strategies, which were represented by a portion of players in the limit $N \to \infty$. After a round robin game the score for each player was evaluated, and this quantity served as a fitness to determine the abundance of strategies in the subsequent generation. By repeating these steps one can determine the variations in the population of strategies step by step. In most of these simulations, the success of Tit-for-Tat was confirmed because the evolutionary process almost always ended up with a population of some mutually cooperating strategies prevailed by Tit-for-Tat.

Nevertheless, these investigations gave numerical evidence that there is no absolutely best strategy independent of the environment. The empirical success of Tit-for-Tat is related to some of its fundamental properties. Tit-for-Tat does not wish to exploit others, it tries to maximize the total payoff. On the other hand, Tit-for-Tat cannot be exploited (except for the first step), because it reciprocates defection (as a punishment) until the opponent unilaterally cooperates and thus compensates her by the largest income. One can think that Tit-for-Tat is a forgiving strategy because its choice of defection is determined by the last choice of its co-player only. A Tit-for-Tat strategy is identifiable easily, and its future behavior is predictable, therefore for Iterated Prisoner's Dilemmas the best strategy against Tit-for-Tat is to cooperate always. On the other hand, exploiting strategies are suppressed by Tit-for-Tat because of their mutual defection sequences. Consequently, in evolutionary games the presence of Tit-for-Tat promotes mutual cooperation in the whole community.

The main conclusions of these pioneering works were collected in a paper by \citet{axelrod_s81}, and many technical details were presented in the Axelrod's book \citep{axelrod_84}. The relevant message for people facing a Prisoner's Dilemma can be summarized as follows:

(1) Don't be envious.

(2) Don't be the first to defect.

(3) Reciprocate both cooperation and defection.

(4) Don't be too clever.

In short, the above game theoretical investigations suggest that people wishing to benefit from a Prisoner's Dilemma situation should follow a strategy similar to Tit-for-Tat. The application of this strategy (in repeated games) provides a way to avoid the ``Tragedy of the Community".

Anyway, if the payoff of a one-shot Prisoner's Dilemma game can be divided into small portions then it is usually very useful to transform the game into a Repeated Prisoner's Dilemma game with an uncertainty in the ending (because of the applicability of Tit-for-Tat). This approach has been used successfully in various political negotiations.

Beside the deduction of the above very important results, the evolutionary approach also gave insight into the mechanisms and processes resulting in the spontaneous emergence of cooperation. Here it is worth recalling that the results were concluded from a set of numerical investigations with a rather limited number of strategies (14 and 64 in the first and second tournaments), whose construction includes eventualities. For the purpose of a more quantitative and reliable analysis, we will briefly describe another approach, which is based on a well-defined, continuous set of strategies capable of representing a remarkably reach variety of interactions.

\subsection{Emergence of cooperation for stochastic reactive strategies}
\label{sec:pqstrat}

The success of the Tit-for-Tat strategy in Axelrod's computer tournament inspired \citet{nowak_amc89,nowak_jtb89,nowak_aam90} to study a set of \emph{stochastic reactive strategies}, where the choice of action in a given round is only affected by the opponent's behavior in the previous round. These mixed strategies are described by three parameters, $s=(u,p,q)$, where $u$ is the probability to choose cooperation in the first round, while $p$ and $q$ are the conditional probabilities to cooperate in later rounds, given that the opponent's previous choice was cooperation or defection, respectively. So, if a player using strategy $s=(u,p,q)$ is matched against an opponent with a strategy $s^{\prime}=(u^{\prime},p^{\prime},q^{\prime})$ then they will cooperate with probabilities
\begin{equation}
u_1=u \; \quad \mbox{and} \quad u_1^{\prime}=u^{\prime}
\label{eq:pdpqc1}
\end{equation}
in the first step, respectively. In the second step the probability of cooperation becomes
\begin{eqnarray}
u_2&=& p u^{\prime}+q (1-u^{\prime})  \; \nonumber \\
u_2^{\prime}&=& p^{\prime} u+q^{\prime} (1-u)  \;.
\label{eq:pdpqc2}
\end{eqnarray}
In subsequent steps the variation of these quantities is given by the recursive relation
\begin{eqnarray}
u_{n+2}&=& v u_{n}+ w  \;, \nonumber \\
u_{n+2}^{\prime}&=& v u_{n}^{\prime}+ w^{\prime} \;,
\label{eq:pdpqcn}
\end{eqnarray}
where
\begin{eqnarray}
v &=& (p-q)(p^{\prime}-q^{\prime})  \;, \nonumber \\
w &=& p q^{\prime}+q (1-q^{\prime}) \;, \nonumber \\
w^{\prime}&=& p^{\prime} q+q^{\prime}(1-q) \;. \nonumber
\label{eq:pdpqcnx}
\end{eqnarray}
According to this recursive process, if $|p-q|, |p^{\prime} - q^{\prime}| < 1$ then the probabilities of cooperation tend towards the stationary values,
\begin{eqnarray}
\bar{u}&=& {q+(p-q)q^{\prime} \over 1-(p-q)(p^{\prime}-q^{\prime})}  \;, \nonumber \\
\bar{u}^{\prime}&=& { q^{\prime}+(p^{\prime}-q^{\prime})q \over
 1-(p-q)(p^{\prime}-q^{\prime}) } \;,
\label{eq:pdpqcs}
\end{eqnarray}
independently of $u$ and $u^{\prime}$. In this case we can neglect the notation of $u$ and label strategies with $p$ and $q$ only, $s=s(p,q)$.

\begin{figure}[ht]
\centerline{\epsfig{file=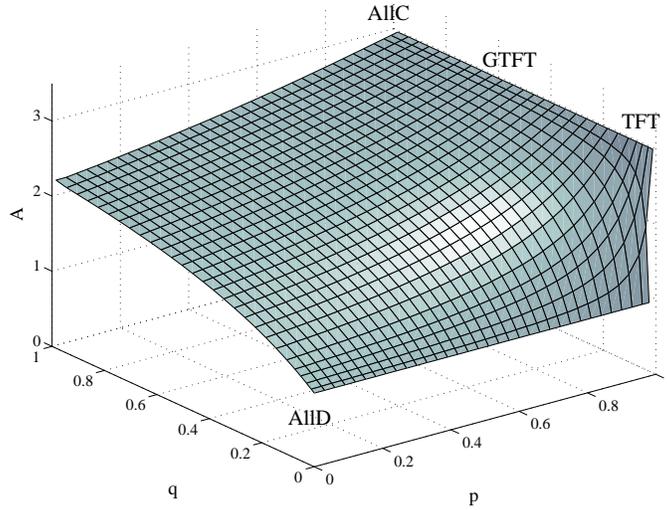,width=9cm}}
\caption{\label{fig:pqavpo}Average payoff if all players follow an $s=(p,q)$ strategy for $T=5$, $R=3$, $P=1$, and $S=0$. The function has a singularity at $s=(1,0)$ (TFT).}
\end{figure}

Using the expressions for $\bar{u}$ and $\bar{u}^{\prime}$ the expected income of strategy $s$ against $s^{\prime}$ reads
\begin{equation}
A(s,s^{\prime})=R \bar{u} \bar{u}^{\prime}+S \bar{u}(1-\bar{u}^{\prime})+T (1-\bar{u}) \bar{u}^{\prime} + P (1-\bar{u})(1-\bar{u}^{\prime}) \;,
\label{eq:pdpqapo}
\end{equation}
at least, when the stationary state is reached. For a homogeneous system consisting of strategy $s$ only, the average payoff obeys a simple form,
\begin{equation}
A(s,s)=P+(T+S-2P)\bar{u}+(R-S-T+P) \bar{u}^2 \;,
\label{eq:pdpqhapo}
\end{equation}
which has a maximum at $p=1$ (nice strategies cooperating mutually forever) as illustrated in Fig.~\ref{fig:pqavpo}. Notice that $A(s,s)$ becomes a non-analytical function at $s=(1,0)$ (Tit-for-Tat strategy), where the payoff depends on the initial probability of cooperations $u$ and $u^{\prime}$.

In their simulation \citet{nowak_n92a} used Axelrod's payoff matrix in Eq. (\ref{eq:Axelpom}) and the above stationary payoff functions. Their numerical investigation aimed to clarify the role of Tit-for-Tat (TFT) and Generous (Forgiving) Tit-for-Tat (GTFT) strategies (see Appendix B for a description). Deterministic Tit-for-Tat has a weakness when playing against itself in a noisy environment \citep{axelrod_84}: after any mistakes two (deterministic) Tit-for-Tat strategies would choose alternately to defect and to cooperate in opposite phases. This shortcoming is suppressed by Generous Tit-for-Tat, which, instead of retaliating defection, chooses cooperation against a previously defecting opponent with a nonzero probability $q$.

The evolutionary game suggested by \citet{nowak_n92a} started with $Q=100$ strategies, with $p$ and $q$ parameters chosen at random and with the same initial strategy concentrations. Now we will discuss this approach using strategies distributed equidistantly on the $p-q$ plane.

\begin{figure}[ht]
\centerline{\epsfig{file=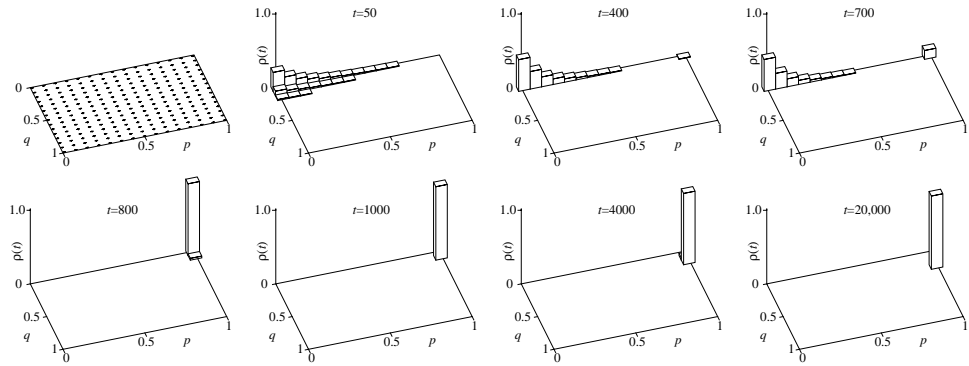,width=13cm}}
\caption{\label{fig:pqtsum8}Distribution of the relative strategy concentrations for different $t$ values illustrated in the plots. The height of columns indicates the relative concentrations for the corresponding strategies excepting those where $\rho_s(t)/\rho_s(0)<1$. In the upper-left plot the crosses show the distribution of ($Q=225$) strategies on the $p-q$ plane.}
\end{figure}

Consider an evolutionary Prisoner's Dilemma game with $Q=15^2=225$ stochastic reactive strategies $s_i=(p_i,q_i)$ with parameters $p_i=0.01+0.07 k_1$ and $q_i=0.01+0.07 k_2$ where $k_1= i\, (\mbox{mod} 15)$ and $k_2=i-15 k_1$ for $i=0, 1, \ldots , Q-1$. In this discretized strategy space,
$(0.01, 0.01)$ is the closest approximation of the strategy AllD (unconditional defection), and $(0.99,0.99)$ is that of Tit-for-Tat. Besides these, this space also includes several Generous Tit-for-Tat strategies, e.g., $(0.99,0.08)$, $(0.99,0.15)$, {\it etc}. Initially all strategies have the same concentration, $\rho_{s_i}(0)=1/Q$. In subsequent steps ($t=1,2,\ldots$) the concentration for each strategy is determined by its previous payoff,
\begin{equation}
\rho_{s_i}(t+1)={ \rho_{s_i}(t) E(s_i,t)
\over \sum_{s_j} \rho_{s_j}(t) E(s_j,t)}\;,
\label{eq:pdpqevol}
\end{equation}
where the total payoff for strategy $s_i$ is given as
\begin{equation}
E(s_i,t)= \sum_{s_j} \rho_{s_j}(t) A[s_i(t),s_j(t)] \nonumber;
\end{equation}
and $A[s_i(t),s_j(t)]$ is defined by Eq.\ (\ref{eq:pdpqapo}) for $T=5$, $R=3$, $P=1$, and $S=0$. Equation (\ref{eq:pdpqevol}) is the discretized form of the replicator equation in the Maynard Smith form in Eq.\ (\ref{eq:replicatorMS}). In the present system the payoffs are positive and the evolutionary rule yields exponential decrease in the concentration for the worst strategies. The evolutionary process is illustrated in Fig.~\ref{fig:pqtsum8}.

Figure \ref{fig:pqtsum8} shows that the strategy concentrations near $(0,0)$
grow rapidly due to the large income received from the exploited, and thus diminishing strategies near $(1,1)$. After a short transient process the system is dominated by defective strategies whose income tends towards the minimum, $P=1$. This phenomenon reminds us of the following ancient Chinese script (circa 1000 B.C.) as interpreted by \citet{wilhelm_50}:

\emph{"Here the climax of the darkening is reached. The dark power at first held so high a place that it could wound all who were on the side of good and of the light. But in the end it perishes of its own darkness, for evil must itself fall at the very moment when it has wholly overcome the good, and thus consumed the energy to which it owed its duration."}

\begin{figure}[ht]
\centerline{\epsfig{file=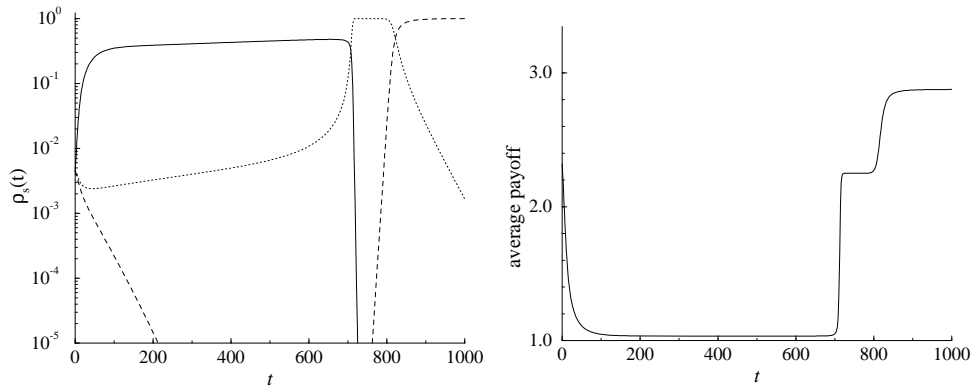,width=13cm}}
\caption{\label{fig:pdpqt2}Time dependence of the strategy concentrations (left) for three relevant strategies (solid line: $(0.01,0.01)$; dotted line: $(0.99,0.01)$); dashed line: $(0.99,0.08)$) playing relevant role until $t=1000$ and the variation of average payoff (right).}
\end{figure}

During the extinction process the few exceptions are the remnants of the Tit-for-Tat strategies near $(1,0)$ cooperating with each other. Consequently, after a suitable time the defector's income becomes smaller than the payoff of Tit-for-Tats, which grows up at the expense of defectors as shown in Fig.~\ref{fig:pdpqt2}.

The increase of the population of Tit-for-Tat is accompanied by a striking increase in the average payoff as demonstrated in Fig.~\ref{fig:pdpqt2}. However, the present stochastic version of Tit-for-Tat cannot provide the maximum possible total payoff forever. The evolutionary process does not terminate by the prevalence of Tit-for-Tat, because this state can be invaded by Generous Tit-for-Tat \citep{axelrod_s88,nowak_tpb90}. The left plot of Fig.~\ref{fig:pdpqt2} shows the nearly exponential growth of $(0.99,0.08)$ and the exponential decay of $(0.99,0.01)$ around $t\approx 750$.

\begin{figure}[ht]
\centerline{\epsfig{file=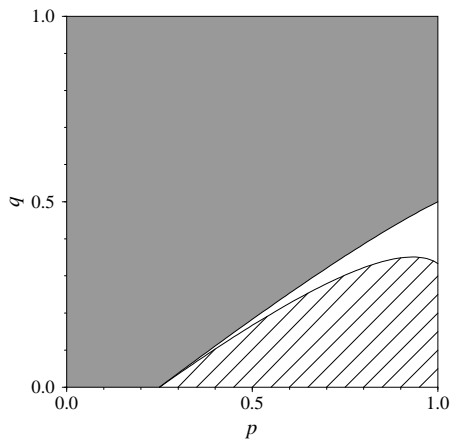,width=6cm}}
\caption{\label{fig:pqstab} Defectors $(0,0)$ can invade a homogeneous population of stochastic reactive strategies belonging to the gray territory. Within the hatched region $\dot{p}=\dot{q}>0$, while along the inside boundary $\dot{p}=\dot{q}=0$. The boundaries are evaluated for $T=5$, $S=3$, $P=1$, and $S=0$.}
\end{figure}

A stability analysis helps to understand the relevant features of this behavior. In Fig.~\ref{fig:pqstab} the gray territory indicates those strategies whose homogeneous population can be invaded by AllD. Evidently, one can also determine the set of strategies $s^{\prime}$ which are able to invade a homogeneous system with a given strategy $s$. The complete stability analysis of this system is given by \citet{nowak_aam90}.

\begin{figure}[ht]
\centerline{\epsfig{file=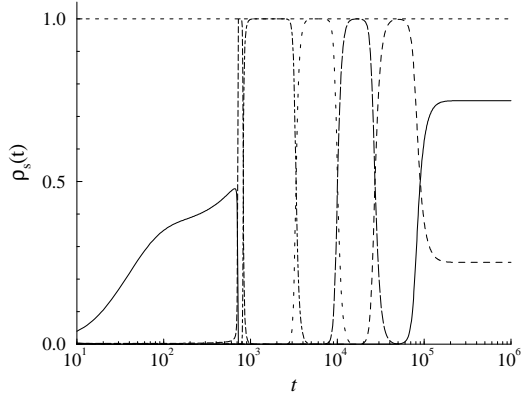,width=7cm}}
\caption{\label{fig:pdpqrgt}Time-dependence of the concentration of the strategies dominating in some time interval. From left to right the peaks are reached by the following $p$ and $q$ values: $(0.01,0.01)$, $(0.99,0.01)$, $(0.99,0.08)$, $(0.99,0.15)$, $(0.99,0.22)$, $(0.99,0.29)$, and $(0.99,0.36)$. The horizontal dotted line shows the maximum when only one strategy exists.}
\end{figure}

In biological applications it is natural to assume that mutant strategies only differ slightly from their parents. Thus \citet{nowak_aam90} assumed an Adaptive Dynamics, and in the spirit of Eq.\ (\ref{eq:adaptivedyn}) they introduced a vector field in the $p-q$ plane pointing towards the preferred direction of infinitesimal evolution,
\begin{equation}
\dot{p}= \left. {\partial A(s,s^{\prime}) \over \partial p}\right|_{s'=s},
\qquad 
\dot{q}= \left. {\partial A(s,s^{\prime}) \over \partial q}\right|_{s'=s}.
\label{eq:pqveloc}
\end{equation}
It is found that both quantities become zero along the same boundary separating positive and negative values. In Fig.~\ref{fig:pqstab} the hatched region shows those strategies where $\dot{p}, \dot{q}>0$. In the above simulation (see Fig.~\ref{fig:pqtsum8}) the noisy Tit-for-Tat strategy $(0.99,0.01)$ is within this region, therefore a homogeneous system develops into a state dominated by a more generous strategy with higher $q$ value. Thus, for the present set of strategies, the strategy $(0.99,0.01)$ will be dominated by $(0.99,0.08)$, that will be overcome by $(0.99,0.15)$, and so on, as illustrated by the variation of strategy frequencies in Fig.~\ref{fig:pdpqrgt}.

The evolutionary process stops when the average value of $q$ reaches the boundary defined by $\dot{p}=\dot{q}=0$. In the low noise limit, $p\to 1$, $\dot{p}$ and $\dot{q}$ vanishes at $q \simeq 1-(T-R)/(R-S)$ ($=1/3$ for the given payoffs). On the other hand, this state is invasion-proof against AllD until $q<(R-P)/(T-P)$ ($=1/2$ here). Thus the quantity
\begin{equation}
q \simeq \min{\left(1-{T-R \over R-S}, {R-P \over T-P} \right)}
\label{eq:pqmaxq}
\end{equation}
(for $p \simeq 1$) can be considered as the optimum of generosity (forgiveness) in the week noise limit \citep{molander_jcr85}. A rigorous discussion, including the detailed stability analysis of stochastic reactive strategies in a more general context is given in \citet{nowak_amc89,nowak_jtb90,nowak_tpb90}.

\subsection{Mean-field solutions}
\label{sec:pdmf}

Consider a situation where the players interact with a limited number of randomly chosen co-players within a large population. For the sake of simplicity, we assume that a portion $\rho$ of the population follows unconditional cooperation (AllC, to be denoted simply as C), and a portion $1-\rho$ plays unconditional defection (AllD, or simply D). No other strategies are allowed. This is the extreme limit where players have no memory of the game.

Each player's income comes from $z$ games with randomly chosen opponents. The average payoff for cooperators and defectors are given as
\begin{eqnarray}
    U_C &=& R z \rho + S z (1-\rho) \;, \nonumber \\
    U_D &=& T z \rho + P z (1-\rho) \;,
    \label{eq:pdmfpocd}
\end{eqnarray}
and the average payoff of the population is
\begin{equation}
    \bar{U}=\rho\, U_C + (1-\rho)\,U_D.
\end{equation}
With the replicator dynamics Eq.\ (\ref{eq:replicator}), the variation of $\rho$ satisfies the differential equation
\begin{equation}
    \dot{\rho} = \rho (U_C - \bar{U}) = \rho (1-\rho)(U_C - U_D) \;.
    \label{eq:pdmfrepd}
\end{equation}

In general, the macro-level dynamics satisfies the approximate mean value equation Eq.\ (\ref{amveq}), which in the present case reads
\begin{eqnarray}
    \dot{\rho} &=& (1-\rho)\, w(D \to C) -\rho\, w(C \to D) \;.
    \label{eq:pdmfrpd}
\end{eqnarray}
In the case of the Smoothed Imitation rule Eq.\ (\ref{gen_imit}) with (\ref{gen_imit2}), the transition rates are
\begin{eqnarray}
    w(C\to D) &=& (1-\rho)\, \frac{1}{1+ \exp{[(U_C-U_D)/K]}} \;, \nonumber \\
    w(D\to C) &=& \rho\, \frac{1}{1+ \exp{[(U_D-U_C)/K]}} \;,
\end{eqnarray}
and the dynamical equation becomes
\begin{eqnarray}
    \dot{\rho} &=& \rho (1-\rho) \tanh\left( {U_C-U_D \over 2 K} \right) \;.
    \label{eq:pdmfrpd2}
\end{eqnarray}

In both cases above, $\rho$ tends to 0 as $t \to \infty$ since $U_D > U_C$ for the Prisoner's Dilemma independently of the value of $z$ and $K$. This means that in the mean-field case cooperation cannot be sustained against defectors with imitative update rules. Notice furthermore, that both systems have two absorbing states, $\rho=0$ and $\rho=1$, where $\dot{\rho}$ vanishes.

For the above dynamical rules we assumed that players adopt one of the co-player's strategy with a probability depending on the payoff difference. The behavior is qualitatively different for innovative strategies such as Smoothed Best Response (Logit or Glauber dynamics), where players update strategies independently of the actual density of the target strategy. For two strategies Eq.\ (\ref{SBR}) gives
\begin{eqnarray}
    w(C\to D) &=& { 1 \over 1+ \exp{[(U_C-U_D)/K]}} \;, \nonumber \\
    w(D\to C) &=& { 1 \over 1+ \exp{[(U_D-U_C)/K]}} \;,
    \label{eq:pdmyop}
\end{eqnarray}
where $K$ characterizes the noise as before. In this case the dynamical equation becomes
\begin{eqnarray}
    \dot{\rho} &=& (1-\rho)\,w(D\to C)-\rho\, w(C\to D) \nonumber \\
     &=& w(D\to C)-\rho \;,
\label{eq:pdmfmyop}
\end{eqnarray}
and the corresponding stationary solution satisfies the implicit equation,
\begin{eqnarray} \label{eq:pdmyopst}
    \rho &=& w(D\to C) \\
         &=& { 1 \over 1+ \exp{\big(z[(T-R) \rho + (P-S)(1 - \rho)]/K\big)}} \;. \nonumber
\end{eqnarray}
In the limit $K \to 0$ this reproduces the Nash equilibrium $\rho = 0$, whereas the frequency of cooperators increases monotonously from $0$ to $1/2$ as $K$ increases from $0$ to $\infty$ (noise-sustained cooperation). In comparison with the algebraic decay of cooperation predicted by Eqs.\ (\ref{eq:pdmfrepd}) and (\ref{eq:pdmfrpd2}), Smoothed Best Response yields a faster (exponential) tendency towards the Nash equilibrium for $K=0$. On the other hand, for finite noise the absorbing solutions are missing.

The situation changes drastically when the players can follow Tit-for-Tat strategies or they are positioned on a lattice and the interaction is limited to their neighborhood as will be detailed in the next sections.

\subsection{Prisoner's Dilemma in finite populations}
\label{sec:pdfp}

Following the work by \cite{nowak_n04} we will discuss what happens in a population of $N$ players who can choose AllD or Tit-for-Tat (TFT) strategies, and play $n$ games with each other within a round. In this case the effective payoff matrix can be given as
\begin{equation}
\label{eq:NSTFPD}
    {\bf A}=\left( \matrix{a & b \cr
                           c & d \cr }\right)
\end{equation}
where $a=Pn$, $b=T+P(n-1)$, $c=S+P(n-1)$, and $d=Rn$. Notice that the original rank of order can be modified by $n$. Consequently, both AllD and TFT become strict Nash equilibria (and ESSs) if $n > (T-P)/(R-P)$. The deterministic replicator dynamics for infinite populations predicts that TFT is eliminated by natural selection if its initial frequency is less than a critical value $x^{\star}= (a-c)/(a-b-c+d)$ (otherwise the frequency of AllD tends to zero).

\cite{nowak_n04} have considered a stochastic (Moran) process when at each time step a player is chosen for reproduction with a probability proportional to her payoff and her offspring is substituted for a randomly chosen player. This stochastic process is investigated in the so-called weak-selection limit, when only a small part of payoff (fitness) comes from games, i.e.,
\begin{eqnarray} \label{eq:weakpo}
U_{D} &=& 1 - w + w [a(i-1) + b (N-i)]/(N-1) \;,\nonumber \\
U_{T} &=& 1-w + w [c i + d (N-i-1)]/(N-1) \;,
\end{eqnarray}
where the strength of selection is characterized by $w$ ($0 < w << 1$), and the number of AllD and TFT strategies are denoted by $i$ and $(N-i)$. This approach resembles the high temperature limit and seems to be adequate for biological systems where strategy changes have little effect on overall fitness \citep{ohta_pnas02}.

The elementary steps can increase or decrease the value of $i$ by 1 with probabilities $\mu_{i+}$ and  $\mu_{i-}$ given as
\begin{eqnarray} \label{eq:moranpr}
\mu_{i+} &=& {U_D i (N-i) \over N[ i  U_D +  (N-i) U_C]} \;,\nonumber \\
\mu_{i-} &=& {U_C i (N-i) \over N[ i  U_D +  (N-i) U_C]} \;.
\end{eqnarray}
This Moran process is equivalent to a random walk on sites $i=0, \ldots , N$. The evolution is stopped when the system reaches one the two absorbing states: $i=0$ and $i=N$. \cite{nowak_n04} have determined the fixation probability $\nu_T$ that a single TFT strategy will invade the whole population containing $(N-1)$ AllD strategies at the beginning. It is well known that $\nu_T=1/N$ if $w=0$. Thus selection favors TFT invading AllD if $\nu_T > 1/N$. For the limit of large $N$ and weak selection the analytical result predicts that TFT is favored if  $x^{\star} < 1/3$ (this is the so-called $1/3$ rule).

When investigating the same system \citet{antal_bmb06} have found that the average fixation time $t_{fix}$ is proportional to $N \ln N$ if the system has at least one ESS, and $t_{fix} \sim e^N/[\sqrt{N}(N-1)]$ if the two strategies coexist in the limit $N \to \infty$.

\subsection{Spatial Prisoner's Dilemma with synchronized update}
\label{sec:spdsu}

In 1992 Nowak and May introduced a spatial evolutionary game to demonstrate that local interactions within a spatial structure can maintain cooperative behavior indefinitely. In the first model they proposed \citep{nowak_n92b,nowak_ijbc93}, the players, located on the sites of a square lattice, could only follow two memoryless strategies: AllD (to be denoted simply as D) and AllC (or simply C), i.e., unconditional defection and unconditional cooperation, respectively. In most of their analysis they used a simplified, rescaled payoff matrix,
\begin{equation}\label{eq:weakPD}
    {\bf A}=\left( \matrix{P & T \cr
                           S & R \cr }\right)=
    \lim_{c\to -0}\left( \matrix{0 & b \cr
                                 c & 1 \cr }\right) =
    \left( \matrix{0 & b \cr
                                 0 & 1 \cr }\right),
\end{equation}
which only contains one free parameter $b$, but is expected to preserve the essence of the Prisoner's Dilemma.\footnote{The case $c=0$ is sometimes called a ``Weak" Prisoner's Dilemma, where not only $(D,D)$, but $(D,C)$ and $(C,D)$ are also Nash equilibria. Nevertheless, it was found \citep{nowak_n92b} that when played as a spatial game the weak version has the same qualitative properties as the typical version with $c<0$, at least when $|c|<<1$.}

In each round (time step) of this iterated game the players play a Prisoner's Dilemma with all their neighbors, and the payoffs are summed up. After this, each player imitates the strategy of those neighboring players (including herself) who has scored the highest payoff. Due to the deterministic, simultaneous update this evolutionary rule defines a cellular automaton. [For a recent survey about the rich behavior of cellular automata, see the book by \citet{wolfram_02}.]

Systematic investigations have been made for different types of neighborhoods. In most cases the authors considered the Moore neighborhood (involving nearest and next-nearest neighbors) with or without self-interactions. The computer simulations were started from random initial states (with approximately the same number of defectors and cooperators) or from symmetrical starting configurations. Wonderful sequence of patterns (kaleidoscopes, dynamic fractals) were found when the simulation was started from a symmetrical state (e.g., one defector in the sea of cooperators).

\begin{figure}[ht]
\centerline{\epsfig{file=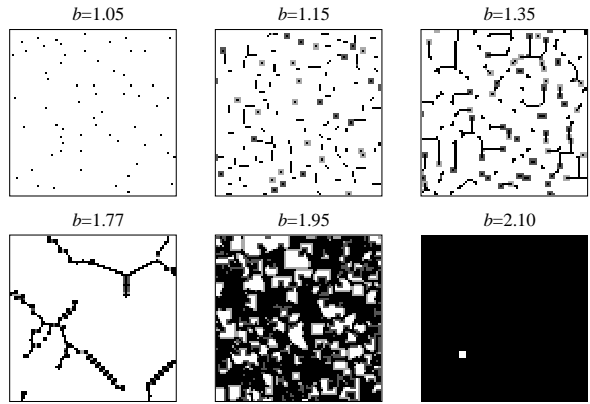,width=8cm}}
\caption{\label{fig:pdca6ss}Stationary $80 \times 80$ patterns indicating the distribution of defectors and cooperators on a square lattice with synchronized update (with self-, nearest-, and next-nearest-neighbor interactions) as suggested by \citet{nowak_n92b}. The simulations are started from random initial states with different values of $b$ as indicated. The black (white) pixels refer to defectors (cooperators) who have chosen defection (cooperation) in the previous step too. The dark (light) gray pixels indicate defectors (cooperators) who invaded the given site in the very last step.}
\end{figure}

Figure \ref{fig:pdca6ss} shows typical patterns appearing after a large number of time steps with random initial conditions. The underlying connectivity structure is a square lattice with periodic boundary condition. The players' income derives from nine games against nearest- and next-nearest neighbors and against the player's own strategy as opponent. These self-interactions favor cooperators, and become relevant in the formation of a number of specific patterns.

In these spatial models defectors receive one of the possible payoffs $U_D=0, b, \ldots, zb$ depending on the given configuration, whereas the cooperators' payoff is $U_C=0, 1, \ldots, z$ (or $U_C=1, \ldots, z+1$ if self-interaction is included), where $z$ denotes the number of neighbors.
Consequently, the parameter space for $b$ can be divided into small equivalency regions by break points, $b_{bp}=k_2/k_1$, where $k_1=2, 3, \ldots, z$ and $k_2=3, 4, \ldots, z$ [and $(z+1)$ for self-interaction]. Within these regions the cellular automaton rules are equivalent. Notice, however, that the strategy change at a given site depends on the actual configuration within a sufficiently large surroundings, including all neighbors of neighbors. For instance, for the Moore neighborhood this means a square block of $5 \times 5$ sites. As a result, for $1<b<9/8$ one can observe a frozen pattern of solitary defectors, whose typical spatial distribution is plotted in Fig.~\ref{fig:pdca6ss} for $b=1.05$.

In the absence of self-interactions the pattern of solitary defectors can be observed for $7/8 <b <1$. For $b>1$, however, solitary defectors always receive the highest score, therefore their strategy is adopted by their neighbors in the next step. In the step after this the payoff of the defecting offsprings is reduced by interacting with each other, and they flip back to cooperation. Consequently, for $1<b<7/5$ we find that the $3 \times 3$ block of defectors shrink back to their original one-site size. In the sea of cooperators there are isolated islands of defectors with fluctuating size, $1 \times 1 \leftrightarrow 3 \times 3$. These objects are easily recognizable in several snapshots of Fig.~\ref{fig:pdca6ss}. This phenomenon demonstrates how a strategy adoption (learning) mechanism for interacting players can prevent the spreading of defection (exploitation) in a spatial structure.

\begin{figure}[ht]
\centerline{\epsfig{file=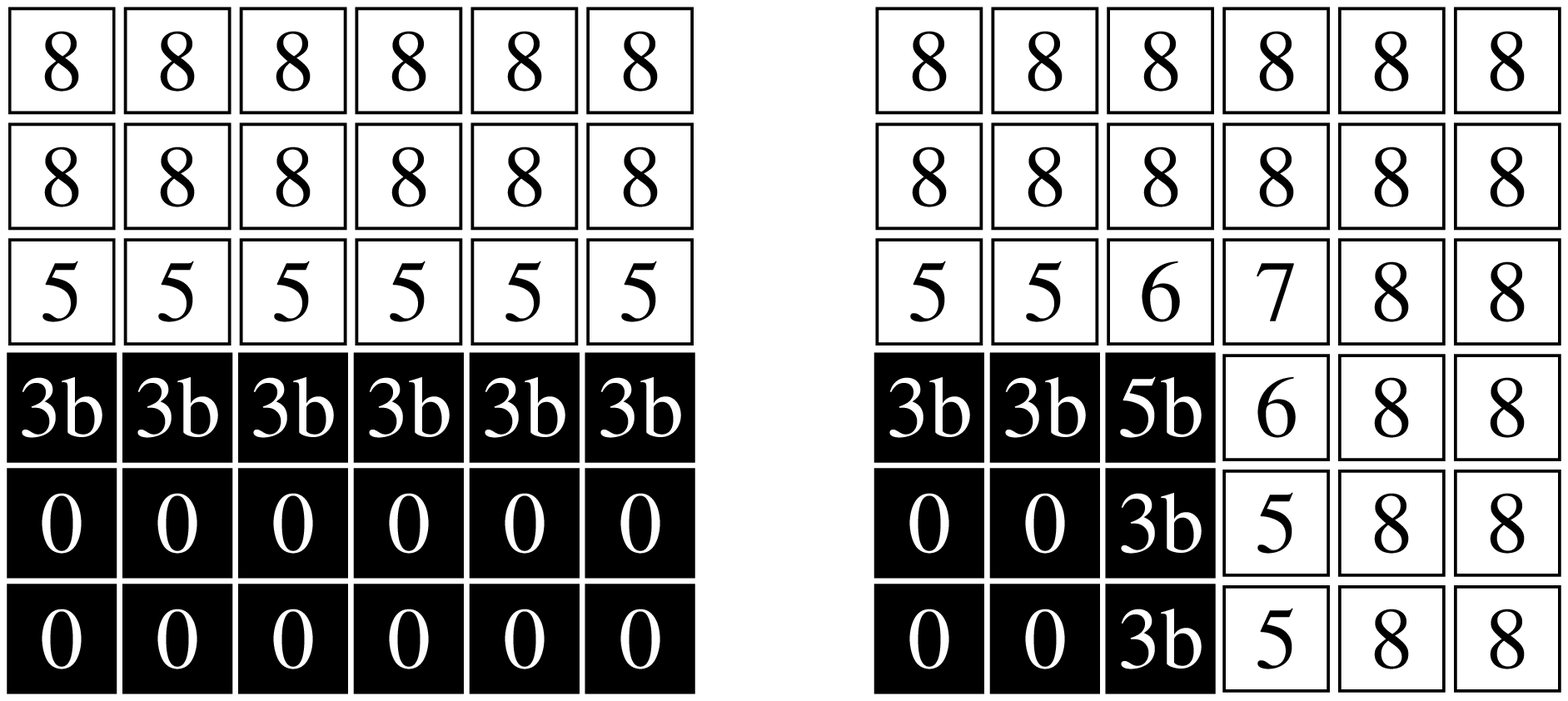,width=7.cm}}
\caption{\label{fig:podistrn} Local payoffs for two distributions of defectors (black boxes) and cooperators (white boxes) if $z=8$. The cellular automaton rules yield that cooperators invade defectors along the horizontal boundary for the left hand constellation if $b<5/3$. This horizontal boundary is stopped if $5/3 < b < 8/3$, whereas defector invasion occurs for $b > 8/3$.  Similar cooperator invasions can be observed along both the horizontal and vertical fronts for the right hand distribution, except for the corner, where three defectors survive if $7/5 < b < 5/3$.}
\end{figure}

In the formation of the patterns shown in Fig.~\ref{fig:pdca6ss}, the invasion of cooperators along  horizontal and vertical straight line fronts plays a crucial role. As demonstrated in Fig.\ \ref{fig:podistrn}, this process can be observed for $b<5/3$ (or $b<2$ in the presence of self-interactions). The invasion process stops when two fronts surround defectors who form a network (or fragments of lines) as illustrated in Fig.~\ref{fig:pdca6ss}. In some constellations the defectors' income becomes sufficiently high to be followed by some neighboring cooperators. However, in the subsequent steps, as described above, cooperators strike back, and this leads to local oscillations (limit cycles) as indicated by the grey pixels in the snapshots.

\begin{figure}[ht]
\centerline{\epsfig{file=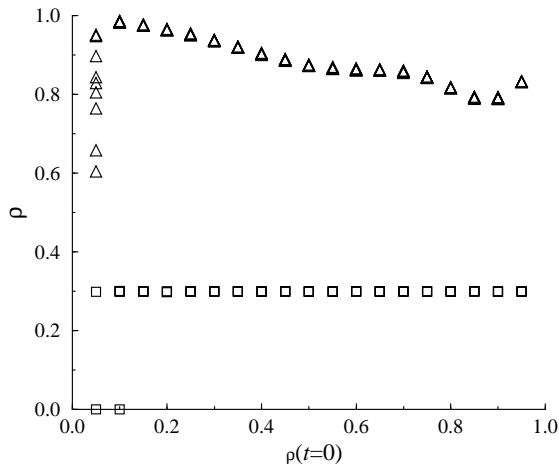,width=7.5cm}}
\caption{\label{fig:pdcarr}Monte Carlo data for the average frequency of cooperators as a function of initial cooperator's frequency for $b=1.35$ (triangles) and 1.65 (squares) in the absence of self-interaction.}
\end{figure}

There exists a region of $b$, namely $8/5 <b<5/3$ (or $9/5 < b < 2$ with self-interactions included), where consecutive defector invasions also occur, and the average ratio of cooperators and defectors is determined by the dynamical balance between these competing invasion processes. Within this region the average concentration of strategies becomes independent of the initial configuration, except for some possible pathologies. Otherwise, in other regions, the time-dependence of the strategy frequencies and the limit values depend on the initial configuration \citep{nowak_ijbc93,schweitzer_acs02}. Figure \ref{fig:pdcarr} compares two series of numerical results obtained by varying the initial frequency $\rho(t=0)$ of cooperators on a lattice consisting of $10^6$ sites. These simulations are repeated 10 times to demonstrate that the standard deviation of limit values increases drastically for low values of $\rho(t=0)$. The triangles represent typical behavior when the system evolves into a "frozen (blinking) pattern".

The visualization of the evolution makes clear that cooperators have a fair chance of survival if they form compact colonies. For random initial states the probability of finding compact cooperator colonies decreases very fast with their concentration. In this case the dynamics becomes very sensitive to the initial configuration. The last snapshot in Fig.~\ref{fig:pdca6ss} demonstrates that cooperators forming a $k \times l$ ($k,l >2$) rectangular block can remain alive even for $5/3 < b < 8/3$ (or $2 < b < 3$ for self-interactions), when cooperator invasions along the horizontal and vertical fronts are already blocked.

\begin{figure}[ht]
\centerline{\epsfig{file=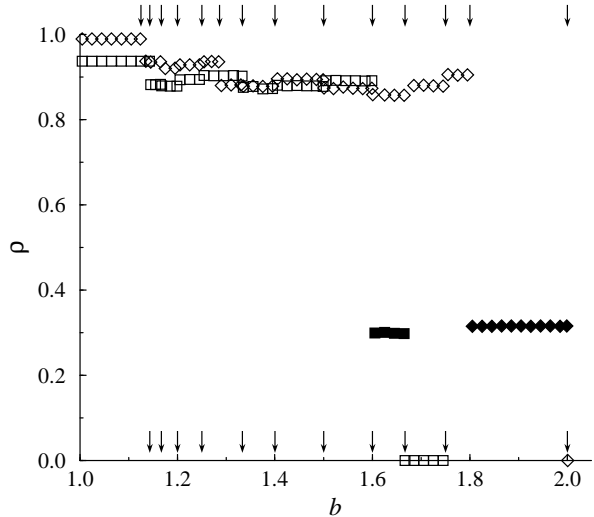,width=8cm}}
\caption{\label{fig:pdcamc89}Average (non-vanishing) frequency of cooperators as a function of $b$ in the model suggested by \citet{nowak_n92b}. Diamonds (squares) represent Monte Carlo data obtained for a Moore neighborhood on the square lattice in the presence (absence) of self-interactions. The simulations start from a random initial state with equal frequency of cooperators and defectors. Closed symbols indicate states, where the coexistence of cooperators and defectors is maintained by a dynamical balance between opposite invasion processes. Open symbols refer to frozen (blinking) patterns. The arrows at the bottom (top) indicate break points in the absence (presence) of self-interactions.}
\end{figure}

Keeping in mind that the stationary frequencies of strategies depend on the initial configuration, it is useful to compare the numerical results obtained in the presence or absence of self-interactions. As indicated in Fig.~\ref{fig:pdcamc89} the coexistence of cooperators and defectors is maintained for larger $b$ values in the case of self-interactions. These numerical data (obtained on a large box containing $10^6$ sites) have sufficiently low statistical error to demonstrate the occurrence of break points. We have to emphasize, that for frozen patterns the variation of $\rho(t=0)$ in the initial state can cause significant deviation for the data denoted by open symbols.

The results obtained via cellular automaton models have raised a lot of questions, and inspired people to develop a wide variety of models. For example, in subsequent papers \citet{nowak_ijbc94,nowak_pnas94} studied different spatial structures including triangular and cubic lattices and a random grid, whose sites were distributed randomly on a rectangular block with periodic boundary condition with a range of interaction limited by a given radius. It turned out that cooperation can be maintained in spatial models even for some randomness.

The rigorous comparison of all the results obtained under different conditions is difficult, because most of the models were developed without knowing all previous investigations published in different areas of science. The conditions for the coexistence of the $C$ and $D$ strategies and some quantitative features of these states on a square lattice were studied systematically for more general, two-parameter payoff matrices containing $b$ and $c$ parameters by \citet{lindgren_pd94,hauert_prslb01,schweitzer_acs02}. Evidently, on the two-dimensional parameter space the break points form crossing straight lines.

The cellular automaton model with nearest-neighbor interactions was studied by \citet{vainstein_pre01} on diluted square lattices (see the left hand structure in Fig.~\ref{fig:dillat}). If too many sites are removed the spatial structure fragments into small parts, on which cooperators and defectors survive separately from each other, and the final strategy frequencies depend on the initial state. The most interesting result was found for small $q$ values. The simulations have indicated that the fraction of cooperators increases with $q$ if $q<q_{th}\simeq 0.05$. In this case the sparsely removed sites can play the role of sterile defectors who block further spreading of defection.

Cellular automaton models on random networks were also studied by \citet{abramson_pre01}, \citet{masuda_pla03}, and \citet{duran_pd05}. These investigations highlighted some interesting and general features. It was observed that local irregularities in the small-world structures or random graph shrink the region of parameters where frozen patterns occur (Class 2 of cellular automata). On random graphs this effect may prevent the formation of frozen patterns if the the average value of connectivity $\bar{\cal C}$ is large enough \citep{duran_pd05}. Furthermore, the simulations also revealed at least two factors which can increase the frequency of cooperators. Both an inhomogeneity in the degree distribution and a sufficiently large clustering coefficient can support cooperation.

\citet{kim_pre02} studied a cellular automaton model on a small-world graph, where many players are allowed to adopt the strategy of an influential player. On this inhomogeneous structure the asymmetric role of the players can cause large fluctuations, triggered by the strategy flips of the influential site. Many aspects of this phenomenon will be discussed later on.

Huge efforts were focused on the extension of the strategy space. Most of these investigations studied what happens if the players are allowed to follow a finite subset of stochastic reactive strategies [see e.g., \citet{nowak_ijbc94,lindgren_pd94,grim_jtb95,brauchli_jtb99}]. It was shown that the addition of some Tit-for-Tat strategy to the standard $D$ and $C$ strategies supports the prevalence of cooperation in the whole range of payoff parameters. A similar result was earlier found by \citet{axelrod_84} for non-spatial games. \citet{nowak_ijbc94} demonstrated that the application of three cyclically dominating strategies leads to a self-organizing pattern, to be discussed later on within the context of the spatial Rock-Scissors-Paper game.

\citet{grim_jtb95,grim_bs96} studied a cellular automaton model with $D$, $C$, and several Generous Tit-for-Tat strategies, characterized by different values of the $q$ parameter. The chosen strategy set included those strategies that play a dominating role in Figs.~\ref{fig:pqtsum8}, \ref{fig:pdpqt2}, and \ref{fig:pdpqrgt}. Besides deterministic cellular automata Grim also studied the effects of mutants introduced in small clusters after each step. In both cases the time-dependence of the strategy population was very similar to those plotted in Figs.~\ref{fig:pdpqt2} and \ref{fig:pdpqrgt}. There was, however, an important difference: the spatial evolutionary process ended up with the dominance of a Generous Tit-for-Tat strategy, whose generosity parameter exceeded the optimum of the mean-field system by a factor of 2 \citep{molander_jcr85,nowak_n92a,grim_bs96}. This can be interpreted as an indication that spatial effects encourage greater generosity.

Within the family of spatial multi-strategy evolutionary Prisoner's Dilemma games the model introduced by \citet{killingback_prslb99} represents another remarkable cellular automaton. For this approach a player's strategy is characterized by an investment (incurring some cost for her) that will be beneficial for all her neighbors including herself. This formulation of the Prisoner's Dilemma is similar to Public Good games as described in Appendix \ref{app:g:pg}. The cellular automaton on a square lattice was started with (selfish) players having low investment. The authors demonstrated that these individuals can be invaded by mutants who have higher investment rate due to spatial effects.

Another direction towards which cellular automaton models were extended involves the stochastic cellular automata discussed briefly in Sec. \ref{sec:er}. The corresponding spatial Prisoner's Dilemma game was introduced and investigated by many authors, e.g., \citet{nowak_ijbc94,mukherji_n96,kirchkamp_jebo00,hauert_ijbc02}, and further references are given in the paper by \citet{schweitzer_lnems05}. The stochastic elements of the rules are capable of destroying local regularities, which appear in deterministic cellular automata belonging to Class 2 or 4. As a result, stochasticity modifies the phase boundary separating quantitatively different behaviors in the parameter space. In the papers by \citet{hauert_ijbc02} and \citet{schweitzer_lnems05} the reader can find phase diagrams obtained by numerical simulations. The main features caused by these types of randomness are similar to those observed for random sequential update. Anyway, if the strategy change suggested by the deterministic rule is realized with low probability then the corresponding dynamical system becomes equivalent to a model with random sequential update. In the next section we will show that random sequential update usually provides a more efficient and convenient framework for studying the effects of payoff parameters, noise, randomness, and different connectivity structures on the emergence of cooperation.

\subsection{Spatial Prisoner's Dilemma with random sequential update}
\label{sec:spdrsu}

In many cases random sequential update provides a more realistic approach as detailed by \citet{huberman_pnas93}. Evolution starts from a random initial state, and the following elementary strategy adoption steps are repeated: choose two neighboring players at random and the first player (at site $x$) adopts the second's strategy (at site $y$) with a probability $P(s_x \to s_y)$ depending on the payoff difference. This rule belongs to the general (smoothed) imitation rules defined by Eq.\ (\ref{gen_imit}). In this section we use Eq.\ (\ref{gen_imit2}) and write the transition probability as
\begin{equation}\label{eq:smoothedimit}
    P(s_x \to s_y)={1 \over 1 +\exp[(U_x-U_y)/K]}.
\end{equation}
Our principal goal is to study the effect of noise $K$.

\begin{figure}[ht]
\centerline{\epsfig{file=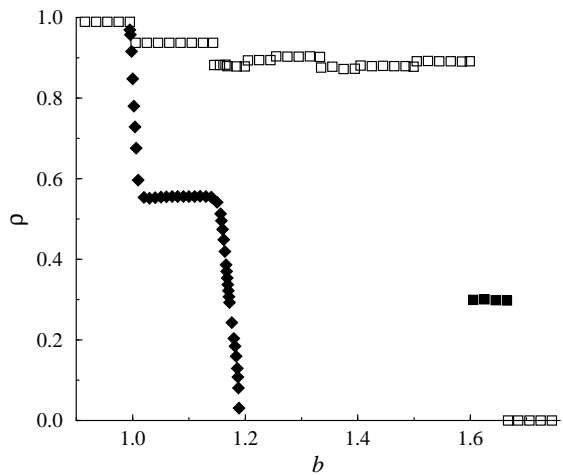,width=7.5cm}}
\caption{\label{fig:pdcamc} Average frequency of cooperators as a function of $b$, using synchronized update (squares) and random sequential updates (diamonds) for nearest- and next-nearest interactions. Squares are the same data as plotted in Fig.~\ref{fig:pdcamc89}, diamonds are obtained for $K=0.03$ within the coexistence region of $b$.}
\end{figure}

In order to quantify the differences between synchronized and random sequential update, in Fig.~\ref{fig:pdcamc} we compare two sets of numerical data obtained on square lattices with nearest- and next-nearest neighbor interactions ($z=8$) in the absence of self-interactions. First we have to emphasize that "frozen" (blinking) patterns cannot be observed for random sequential update. Instead, the system always develops into one of the homogeneous absorbing states with $\rho=0$ or $\rho=1$, or into a two-strategy co-existence state being independent of the initial state.

The most striking difference is that for noisy dynamics cooperation can only be maintained until a lower threshold value of $b$. As mentioned above, for synchronized update cooperation is strongly supported by the invasion of cooperators perpendicular to the horizontal or vertical interfaces. For random sequential update, however, the invasion fronts become irregular, favoring defectors. At the same time the noisy dynamics is capable to eliminate solitary defectors who appear for low values of $b$, as illustrated in Fig.~\ref{fig:pdcamc}.

\begin{figure}[ht]
\centerline{\epsfig{file=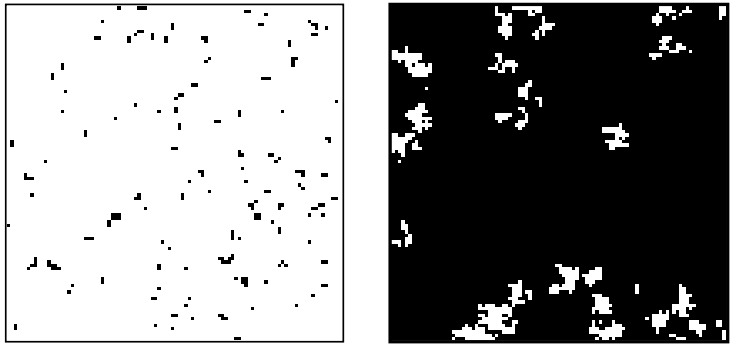,width=10cm}}
\caption{\label{fig:pdcd2}Distribution of defectors (black pixels) and cooperators (white areas) on a square lattice for $b=0.95$ (left) and $b=1.036$ (right) if $K=0.1$.}
\end{figure}

The left snapshot of Fig.~\ref{fig:pdcd2} shows that solitary defectors can create an offspring positioned on one of the neighboring sites for $7/8 < b < 1$ in the absence of self-interactions. As the parent and the offspring mutually reduce each other's income, one of them becomes extinct within a short time, and the survival is ready to create another offspring. Due to this mechanism defectors perform random walks on the lattice. These moving objects collide and one of them is likely to be destroyed. On the other hand, there is some probability that they split into two objects. In short, these processes are analogous to \emph{branching and coalescing random walks}, thoroughly investigated in non-equilibrium statistical physics. \citet{cardy_prl96,cardy_jsp98} have shown that these systems exhibit a non-equilibrium phase transition (more precisely an extinction process) when the parameters are tuned. There the transition belongs to the so-called \emph{directed percolation} universality class.

A very similar phenomenon can be observed at the extinction of cooperators in our spatial game. In the vicinity of the extinction point the cooperators form colonies (see right snapshot of Fig.~\ref{fig:pdcd2}) who also perform branching and coalescing random walks. Consequently, both extinction processes exhibit the same universal features \citep{szabo_pre98,chiappin_pre99}. However, the occurrence of these universal features of the extinction transition is not restricted to random sequential update. Similar properties have also been found in several stochastic cellular automata (synchronous update) \citep{domany_prl84,bidaux_pra89,jensen_pra91,wolfram_02}.

The universal properties of this type of non-equilibrium critical phase transitions have been intensively investigated. For a survey see the book by \citet{marro_99} or the review paper by \citet{hinrichsen_ap00}. The very robust features of this transition from the active phase $\rho > 0$ to the absorbing phase ($\rho = 0$) occur in many homogeneous spatial systems where the order parameter is scalar and the interactions are short ranged \citep{janssen_zpb81,grassberger_zpb82}. The simplest model exemplifying this transition is the \emph{contact process} proposed by \citet{harris_ap74} as a toy model to describe the spread of an epidemic. In his model healthy and infected objects are localized on a lattice. Infection spreads with some rate through nearest-neighbor contacts (in analogy to a learning mechanism), while infected sites recover at a unit rate. Due to its simplicity, the contact process allows for a very accurate numerical investigation of the universal properties of the critical transition.

In the vicinity of the extinction point (here $b_{cr}$), the frequency of the given strategy plays the role of the order parameter. In the limit $N \to \infty$ it vanishes as a power law,
\begin{equation}
    \rho \propto |b_{cr}-b|^{\beta},
    \label{eq:dpbeta}
\end{equation}
where $\beta = 0.583(4)$ in two dimensions, independently of many irrelevant details of the system. The algebraic decrease of $\rho$ is accompanied by an algebraic divergence of the fluctuations of $\rho$,
\begin{equation}
    \chi = N \left[ \langle \rho^2(t) \rangle - \langle \rho(t) \rangle^2  \right]
    \propto         |b_{cr}-b|^{-\gamma} \;,
    \label{eq:dpgamma}
\end{equation}
as well as of the correlation length
\begin{equation}
    \xi \propto |b_{cr}-b|^{-\nu_{\perp}}
\end{equation}
and the relaxation time
\begin{equation}
    \tau \propto |b_{cr}-b|^{-\nu_{\parallel}}.
\end{equation}
For two-dimensional systems the numerical values of the exponents are: $\gamma=0.35(1)$, $\nu_{\perp}=0.733(4)$, and $\nu_{\parallel}=1.295(6)$. Further universal properties and techniques to study these transitions are well described in \citet{marro_99} and \citet{hinrichsen_ap00} (and further references therein).

The above divergencies cause serious difficulties in numerical simulations. Within the critical region the accurate determination of the average frequencies requires large system sizes, $L>>\xi$, and simultaneously long thermalization and sampling times, $t_{th}, t_s >> \tau$. In typical Monte Carlo simulations the suitable accuracy for $\rho \simeq 0.01$ can be achieved by run times as long as several weeks on a PC.

The peculiarity of evolutionary games is that there may be two subsequent critical transitions. The numerical confirmation of the universal features becomes easier, if the spatial evolutionary game is  as simple as possible. The simplicity of the model also allows us to perform the generalized mean-field analyses, using larger and larger cluster sizes (for details see Appendix \ref{app:gmfa}). Therefore, in the rest of this section our attention will be focused on those systems, where the players only have four neighbors, and the evolution is governed by sequential strategy adoptions, as described at the beginning of this section.

\begin{figure}[ht]
\centerline{\epsfig{file=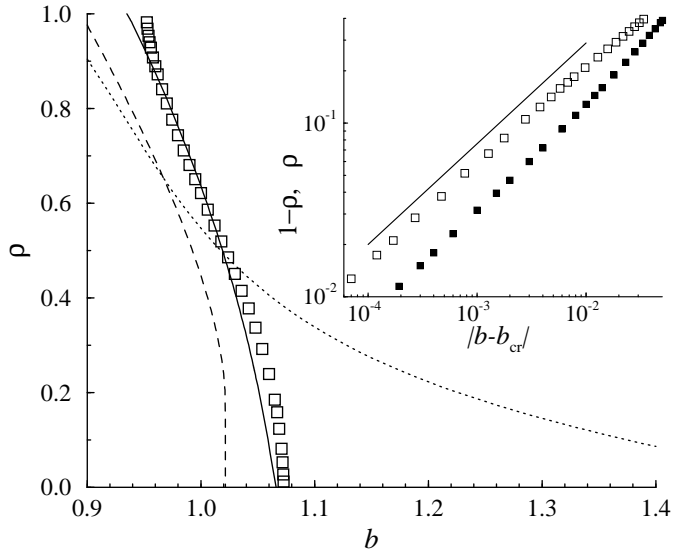,width=9cm}}
\caption{\label{fig:pdsqt04}Average frequency of cooperators as a function of $b$ on the square lattice for $K=0.4$. Monte Carlo results are denoted by squares. Solid, dashed, and dotted lines indicate the predictions of the generalized mean-field technique at the level of $3 \times 3$-, $2 \times 2$-, 2-site clusters. Inset shows the log-log plot of the order parameter $\Phi$ {\it vs.} $|b-b_{cr}|$, where $\Phi = 1-\rho$ or $\Phi = \rho$ in the vicinity of the first and second transition points. The solid line has the slope $\beta = 0.58$ characterizing directed percolation.}
\end{figure}

Figure~\ref{fig:pdsqt04} shows a typical variation of $\rho$, when we tune the value of $b$ within the region of coexistence for $K=0.4$. These data refer to two critical transitions at $b_{c1}=0.9520(1)$ and at $b_{c2}=1.07277(2)$. The log-log plot in the inset provides numerical evidence that both extinction processes belong to the directed percolation universality class.

In Fig.~\ref{fig:pdsqt04} the Monte Carlo data are compared with the results of the generalized mean-field technique. The large difference between the Monte Carlo data and the prediction of the pair approximation is not surprising, if we know that traditional mean-field theory (one-site approximation) predicts $\rho=0$ at $b>1$ for any $K$ as detailed above. As expected, the prediction improves gradually for larger and larger clusters. The slow convergence towards the exact (Monte Carlo) results for increasing cluster sizes highlights the importance of short range correlations related to the occurrence of solitary defectors and cooperator colonies. According to these results, the first continuous (linear) transition appears at $b=b_{c1}<1$. As this transition lies outside the parameter region of the Prisoner's Dilemma ($b>1$), we do not consider this in the sequel.

It is worth mentioning that the four-site approximation predicts a first-order phase transition, i.e., a sudden jump to $\rho=0$ at $b=b_{c2}$, while the more accurate nine-site approximation predicts linear extinction at a threshold value $b_{c2}^{(9s)}=1.0661(2)$, very close to the exact result. We have to emphasize that the generalized mean-field method is not able to reproduce the exact algebraic behavior for any finite cluster sizes, although its accuracy increases gradually. Despite this shortcoming, this technique can be used to estimate the region of parameters where strategy coexistence is expected.

\begin{figure}[ht]
\centerline{\epsfig{file=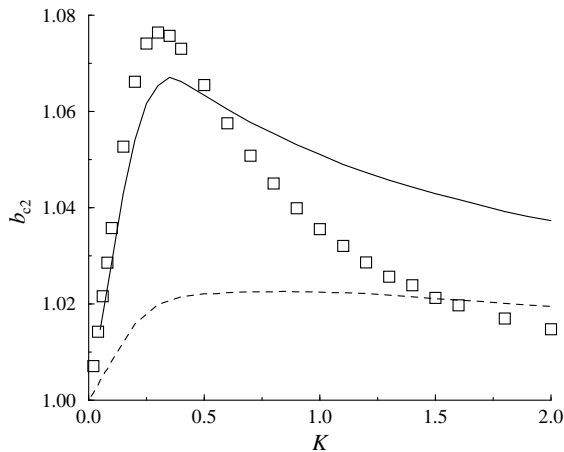,width=7.5cm}}
\caption{\label{fig:pdpdsq}Monte Carlo data (squares) for the critical value of $b$ as a function of $K$ on the square lattice. The solid and dashed lines denote predictions of the generalized mean-field technique at the levels of $3 \times 3$- and $2 \times 2$-site approximations.}
\end{figure}

By the above techniques one can determine the transition point $b_{c2}>1$ for different values of $K$. The remarkable feature of the results (see Fig.~\ref{fig:pdpdsq}) is the extinction of cooperators for $b>1$ in the limits when $K$ goes to $0$ or $\infty$ \citep{szabo_pre05}. In other words, there exists an optimum value of noise to maintain cooperation at the highest possible level. This means that noise plays a crucial role in the emergence of cooperation, at least in the given connectivity structure.

The prediction of the pair approximation cannot be illustrated in Fig.~\ref{fig:pdpdsq} because the corresponding data lie outside the plotted region. The most serious shortcoming of this approximation is that it predicts $\lim_{K \to 0} b_{c2}^{(2s)} = 2$. Notice, however, that the generalized mean-field approximations are capable to reproduce the correct result, $\lim_{K \to 0} b_{c2}^{\rm (MC)} = 1$, at higher levels.

\begin{figure}[ht]
\centerline{\epsfig{file=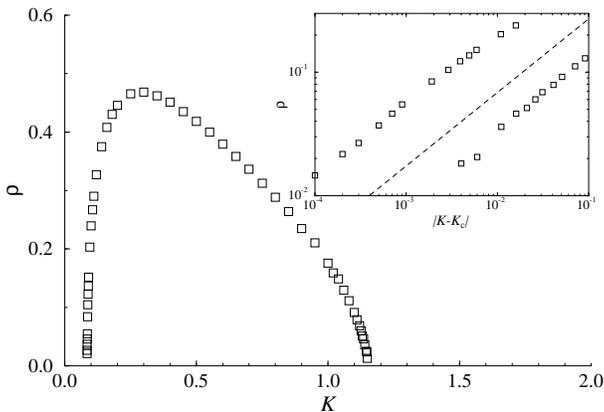,width=8cm}}
\caption{\label{fig:b103}Frequency of cooperators has a maximum when increasing $K$ for $b=1.03$. Inset shows the log-log plot of the frequency of cooperators {\it vs.} $|K-K_c|$ ($K_{c1}=0.0841(1)$ and $K_{c2}=1.151(1)$). The straight line indicates the power-law behavior characterizing directed percolation.}
\end{figure}

The occurrence of noise enhanced cooperation can be demonstrated more strikingly when we plot the frequency of cooperators against $K$ at a fixed $b<\max (b_{c2})$. As expected the results show two subsequent critical transitions belonging to the directed percolation universality class, as demonstrated in Fig.~\ref{fig:b103}.

Using a different approach, \citet{traulsen_prl04} also found that the learning mechanism (for weak external noise) can drive the system into a state, where the unfavored strategy appears with an enhanced frequency. They showed that the effect can be present in other games (e.g. the Matching Pennies) too. The frequency of the unfavored strategy reaches its maximum at intermediate noise intensities in a resonance-like manner. Similar phenomena, called \emph{stochastic resonance}, are widely studied in climatic change \citep{benzi_t82}, biology \citep{douglass_n93}, ecology \citep{blarer_el99}, and excitable media \citep{jung_prl95,gammaitoni_rmp98} when considering the response to a periodic external force. The present phenomenon, however, is more similar to \emph{coherence resonance} \citep{pikovsky_prl97,traulsen_prl04,perc_pre05,perc_njp06b} because no periodic force is involved here. In other words, the enhancement in the frequency of cooperators appears as a nonlinear response to purely noisy excitations.

The analogy becomes more striking when we also consider additional payoff fluctuations. In a recent paper \citet{perc_njp06a} studied an evolutionary Prisoner's Dilemma game with a noisy dynamical rule at very low value of $K$, and made the payoffs stochastic by adding some noise $\zeta$, chosen randomly in the range $-\sigma < \zeta < \sigma$. When varying $\sigma$ the numerical simulations indicated a resonance-like behavior in the frequency of cooperators. His curves, obtained for a somewhat different parametrization, are very similar to those plotted in Fig.~\ref{fig:b103}.

\begin{figure}[ht]
\centerline{\epsfig{file=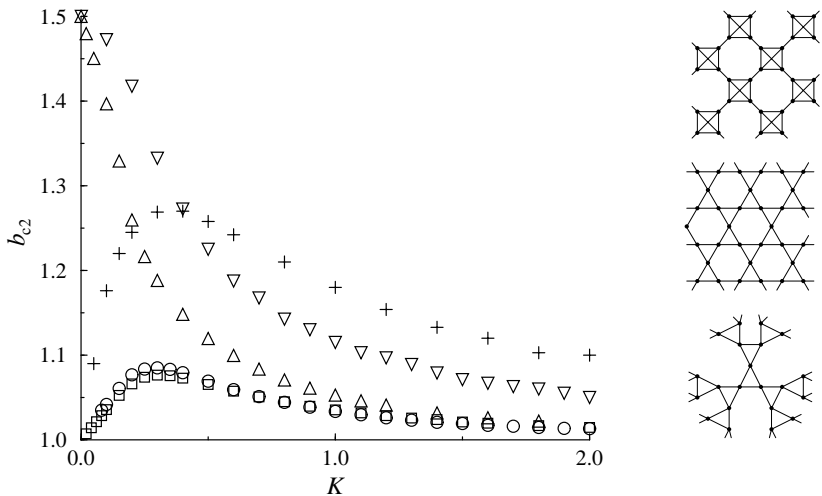,width=11cm}}
\caption{\label{fig:pdpd5}Critical values of $b$ {\it vs.} $K$ for five different connectivity structures characterized by $z=4$. Monte Carlo data are indicated by squares (square lattice), pluses (random regular graph or Bethe lattice), $\bigcirc$ (lattice of four-site cliques), $\bigtriangleup$ (kagome lattice), and $\bigtriangledown$ (regular graph of overlapping triangles). The last three structures are illustrated above from top to bottom on the right-hand side.}
\end{figure}

The disappearance of cooperation in the zero noise limit for the present topology is contrary to naive expectations that could be deduced from previous simulations \citep{nowak_ijbc94}. To further clarify the reasons, systematic investigations were made for different connectivity structures \citep{szabo_pre05}. According to the simulations, this evolutionary game exhibits qualitatively similar behavior if the connectivity is characterized by a simple square or cubic ($z=6$) lattice, a random regular graph (or Bethe lattice), or a square lattice of four site cliques (see the right hand side of Fig.~\ref{fig:pdpd5}). Notice that these structures are practically free of triangles, or the overlapping triangles form larger cliques that are not connected by triangles but by single links (see the top structure on the right in Fig.\ \ref{fig:pdpd5}).

In contrast with these, cooperation is maintained in the zero noise limit on two-dimensional triangular ($z=6$) and kagome ($z=4$) lattices, on the three-dimensional body centered cubic lattice ($z=8$) and on random regular graphs (or Bethe lattice) of one-site connected triangles (see the right hand side of Fig.~\ref{fig:pdpd5}). Similar behavior was found on the square lattice when second-neighbor interactions are also taken into account. The common topological feature of these structures is the percolation of the overlapping triangles over the whole system. The exploration of this topological feature of networks has begun very recently \citep{palla_n05,derenyi_prl05}, and for the evolutionary Prisoner's Dilemma it seems to be more important than the spatial character of the connectivity graph. In order to separate the dependence on the degree of nodes from other topological characteristics in Fig.~\ref{fig:pdpd5} we compare results obtained for structures with the same number of neighbors $z=4$. It is worth mentioning that these qualitative differences are also confirmed by the generalized mean-field technique for sufficiently large clusters \citep{szabo_pre05}.

In the light of the above results it is conjectured that for the given dynamical rule cooperation can be maintained in the zero noise limit only for those regular connectivity structures, where the overlapping triangles span an infinitely large portion of the system for $N \to \infty$. For low values of $K$ the highest frequency of cooperators is found on those random (non-spatial) structures, where two overlapping triangles have only one common site (see Fig.~\ref{fig:pdpd5}).

\begin{figure}[ht]
\centerline{\epsfig{file=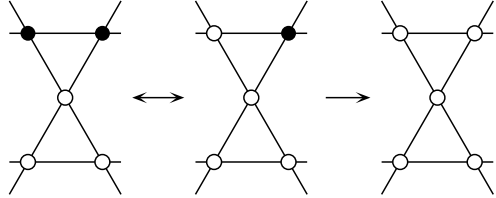,width=7cm}}
\caption{\label{fig:sprckag}Schematic illustration of the spreading of cooperators (white dots) within the sea of defectors (black dots) on a network built up from one-site overlapping triangles. These structures are locally similar to those plotted in the middle and bottom structures on the right-hand side of Fig.~\ref{fig:pdpd5}. Arrows show the direction of the most frequent transitions in the low noise limit.}
\end{figure}

The advantageous feature of these topologies \citep{vukov_pre06} can be explained most easily on those $z=4$ structures where the overlapping triangles have only one common site (the middle and bottom structures on the right hand side of Fig.~\ref{fig:pdpd5}). Let us assume that originally only one triangle is occupied by cooperators in the sea of defectors. Such a situation is plotted in Fig.~\ref{fig:sprckag}.

The payoff of cooperators within a triangle is 2, neighboring defectors receive $b$, and all the other defectors get 0. For this constellation the most likely process is that one of the neighboring defectors adopts the more successful cooperator strategy. As the state of the new cooperator is not stable, she can return to defection again within a short time, unless the other neighboring defector changes to cooperation too, provided that $b < 3/2$. In the latter case we obtain another triplet of cooperators. This is a stable configuration against the attack of neighboring defectors for low noise. Similar processes can occur at the tip of the branches, if the overlapping triangles form a tree. Consequently, the iteration of these events gives rise to a growing tree of overlapping cooperator triplets. The growth process, however, is blocked at defector sites separating two branches of the tree. This blocking process is missing if only one tree of cooperator triplets exists like on a tree-like structure. Nevertheless, it constraints the spreading of cooperators on spatial structures (e.g., on the kagome lattice). The blocking process controls the spreading of cooperators (and assure the coexistence of cooperation and defection) for both types of connectivity structures, if many trees of cooperator triplets are simultaneously present in the system. Evidently, more complicated analysis is required to clarify the role of different topological properties for $z > 4$, as well as for cases when the overlapping triangles have more than one common sites.

The above analysis also suggests some further interesting conclusions (conjectures). In the large $K$ limit, simulations show similar behavior for all the investigated spatial structures. The highest rate of cooperation occurs on random regular graphs, where the role of loops is negligible. In other words, for high noise the (quenched) randomness in the connectivity structure provides the highest rate of cooperation, at least, if the structure of connectivity is restricted to regular graphs with $z=4$.

\begin{figure}[ht]
\centerline{\epsfig{file=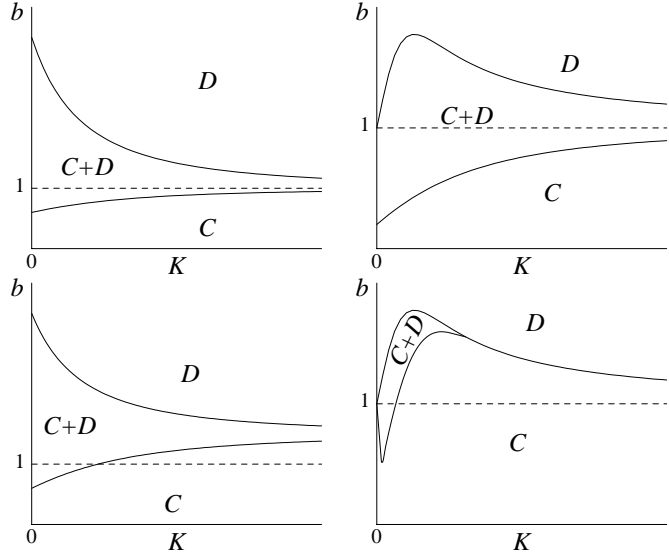,width=9.cm}}
\caption{\label{fig:spdspd} Possible schematic phase diagrams of the Prisoner's Dilemma on the $K$--$b$ plane. The curves show the phase transitions between the coexistence region ($C+D$) and homogeneous solutions ($C$ or $D$). The $K\to\infty$ limit reproduces the mean-field result $b_{c1}=b_{c2}=1$ for three plots, the exception represents a behavior occurring for some inhomogeneous strategy adoption rates.}
\end{figure}

According to preliminary results some typical phase diagrams on the $K$--$b$ plane are depicted in Fig.\ \ref{fig:spdspd}. The versions on the top have already been discussed above. The bottom left diagram is expected to occur on the kagome lattice if the strategy adoption probability from some players (with fixed position) is reduced by a suitable factor \citep{szolnoki_epl07}. The fourth type (bottom right) is characteristic to some one-dimensional systems.

\subsection{Two-strategy Prisoner's Dilemma game on a chain}
\label{sec:pd1d}

First we consider a system with players located on the sites $x$ of a one-dimensional lattice. The players follow one of the two strategies, $s_x=C$ or $D$, and their total utility comes from a Prisoner's Dilemma game with their left and right neighbors. If the evolution is governed by some imitative learning mechanism, which involves nearest-neighbors, then a strategy change can only occur at those pair of sites, where the neighbors follow different strategies. At these sites the maximum payoff of cooperators, $1$, is always lower than the defectors' minimum payoff $b$. Consequently, for any Darwinian selection rule cooperators should become extinct sooner or later.

In one-dimensional models the imitation rule leads to a domain growing process (where one of the strategies is favored) driving the system towards the homogeneous final state. \citet{nakamaru_jtb97} studied the competition between $D$ and Tit-for-Tat strategies under different evolutionary rules. It is demonstrated that the pair approximation predicts correctly the final state dependent on the evolutionary rule and initial strategy distribution.

Considering $C$ and $D$ strategies only, the behavior changes significantly if second neighbors are also taken into account. In this case the players have four neighbors and the results can be compared with those discussed above (see Fig.~\ref{fig:pdpd5}), provided we choose the same dynamical rule. Unfortunately, systematic investigations are hindered by extremely slow transient events. According to the preliminary results, the two subsequent phase transitions coincide, $b_{c1}=b_{c2}>1$, if the noise exceeds a threshold value, $K>K_{th}$. This means that evolution reaches a homogeneous absorbing state of cooperators (defectors) for $b<b_{c1}=b_{c2}$ ($b>b_{c1}=b_{c2}$). For lower noise levels, $K<K_{th}$, coexistence occurs in a finite region,
$b_{c1}(K) < b < b_{c2}(K)$. The simulations indicate that $b_{c2} \to 1$ if $K$ goes to zero. Shortly, for this connectivity structure the system represents another class of behavior, complementing those discussed in the previous section.

Very recently \citet{santos_prl05} have studied a similar evolutionary Prisoner's Dilemma game on the one-dimensional lattice varying the number $z$ of neighbors. In their model a randomly chosen player $x$ adopts the strategy of her neighbor $y$ (chosen at random) with a probability $(U_y-U_x)/U_0$ (where $U_0$ is a suitable normalization factor) if $U_y>U_x$. This is Schlag's Proportional Imitation rule, Eq.\ (\ref{prop_imit}).
It turns out that cooperators can survive if $4 \le z  \le z_{max} \approx 64$. More precisely, the range of $b$ with coexisting $C$ and $D$ strategies decreases when $z$ increases. Similar results are reported by \citet{ifti_jtb04} and \citet{tang_epjb06} who used a different dynamical rules. As expected, a mean-field type behavior governs the system for sufficiently large $z$.

The importance of one-dimensional systems lies in the fact that they act as starting structures for small-world graphs as suggested by \citet{watts_n98}, who substituted random links for a portion $q$ of bonds connecting nearest- and next-nearest neighbors on a one-dimensional lattice (see Fig.~\ref{fig:wssw}). The long-range connections generated by this rewiring process decrease the average distance between sites, and produce a small world phenomenon \citep{milgram_pt67}, which characterizes typical social networks.

\subsection{Prisoner's Dilemma on social networks}
\label{sec:pdsocnet}

The spatial Prisoner's Dilemma has been studied by many authors on different random networks. Some results for synchronized update were briefly discussed in Sec.~\ref{sec:spdsu}. In Fig.~\ref{fig:spdspd} we have compared the phase diagrams for regular graphs, i.e., lattices and random graphs where the number of neighbors is fixed. However, the assumption of a uniform degree is not realistic for most social networks, and these models do not necessarily give adequate predictions. In this section we focus our attention on graphs where the degree has a non-uniform scale-free distribution. The basic message is that degree heterogeneity can drastically enhance cooperation.

\begin{figure}[ht]
\centerline{\epsfig{file=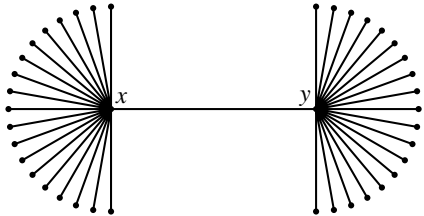,width=6cm}}
\caption{\label{fig:bossconf}A typical subgraph of a scale-free social network, where two connected central players are linked to many others having significantly less neighbors.}
\end{figure}

The emergence of cooperation around the largest hub was reported by \citet{pacheco_aipcp05}. To understand the effect of heterogeneity in the connectivity network, first we consider a simple situation, which represents a typical part of a scale-free network. We restrict our attention to the surroundings of two connected central players (called hubs), each linked to a large number of other, less connected players, as shown in Fig.~\ref{fig:bossconf}. Following \citet{santos_prslb06}, we wish to illustrate the basic mechanism how cooperation gets supported when a cooperator and a defector face each other in such a constellation.

For the sake of simplicity we assume that both the cooperator $x$ and the defector $y$ are linked to the same number $N_x=N_y$ of additional co-players ($xn$ and $yn$ players), who follow the $C$ and $D$ strategies with equal probability in the initial state. This random initial configuration provides the highest total income for the central defector. The central cooperator's total utility is lower, but exceeds the income of most of her neighbors, because she has much more connections. We can assume that the evolutionary rule is Smoothed Imitation defined by Eq.\ (\ref{eq:smoothedimit}), with the imitated player chosen randomly from the neighbors.
The influence of the neglected part of the whole system on $xn$ and $yn$ players is taken into consideration in a simplified manner. We assume no direct interaction with the rest of the population but posit that $xn$ and $yn$ players adopt a strategy from each other with probability $P_{\rm rsa}$ (rsa - random strategy adoption), because the strategy distribution in this subset is alike to that of the whole system.

\begin{figure}[ht]
\centerline{\epsfig{file=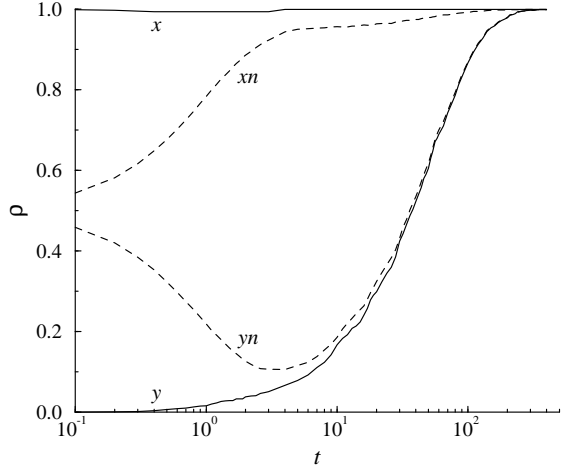,width=7.5cm}}
\caption{\label{fig:compboss}Frequency (probability) of cooperators as a function of time for the topology in Fig.\ \protect\ref{fig:bossconf} at the central sites $x$ and $y$, as well as at their neighbors ($xn$ and $yn$). Initially $s_x=C$ and $s_y=D$, and the neighboring players choose $C$ or $D$ with equal probability. The numerical results are averaged over 1000 runs for $b=1.5$, $P_{\rm rsa}=0.1$, and the number of neighbors $N_x=N_y=49$.}
\end{figure}

The result of a numerical simulation is shown in Fig.~\ref{fig:compboss}. At the beginning of the evolutionary process the frequency of cooperators increases (decreases) in the neighborhood of the central cooperator (defector), and this variation is accompanied by an increase (decrease) of the utility $U_x$ ($U_y$). Consequently, after some time the central cooperator will obtain the highest payoff and becomes the best player to be followed by others. A necessary condition for this scenario is
$N_x,N_y >> 1$ which can be satisfied in scale-free networks, ant that the random effect of the surroundings characterized by $P_{\rm rsa}$ should not be too strong. A weak asymmetry in the number of neighbors cannot modify this picture.

\begin{figure}[ht]
\centerline{\epsfig{file=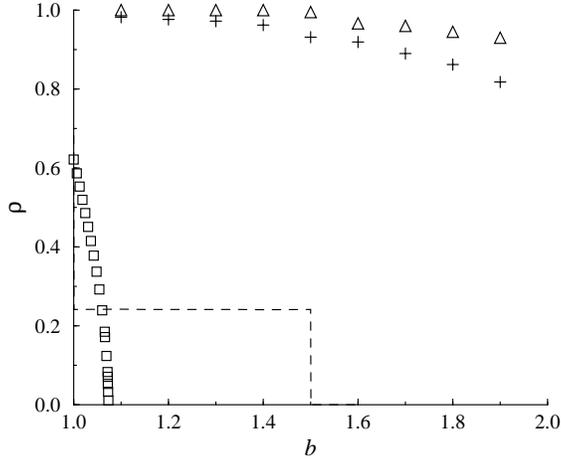,width=7.5cm}}
\caption{\label{fig:sfmccomp}Frequency of cooperators {\it vs.} $b$ for two different scale-free connectivity structures, where the average number of neighbors is $\langle z \rangle =4$ \citep{santos_prl05}. The Monte Carlo data are denoted by pluses (Barab\'asi-Albert model) and triangles (Dorogovtsev-Mendes-Samukhin model). Squares and the dashed line show $\rho$ on the square and kagome lattices, resp., at a noise level $K$ where $b_{cr}$ reaches its maximum.}
\end{figure}

This mechanism seems to work well for all scale-free networks, as was
reported by \citet{santos_prl05,santos_prslb06,santos_jeb06}, who observed a significant enhancement in the frequency of cooperators. The stationary frequency of cooperators were determined both for the \citet{barabasi_s99} and \citet{dorogovtsev_pre01} structures (see Fig.~\ref{fig:sfbadm}). Cooperation remains high in the whole region of $b$, especially for the Dorogovtsev-Mendes-Samukhin network (see Fig.~\ref{fig:sfmccomp}).
To illustrate the striking enhancement, these data are compared with the results obtained for two lattices at the optimum value of noise (i.e., $K=0.4$ on the square lattice and $K=0$ on the kagome lattice) in Fig.~\ref{fig:sfmccomp}.

\citet{santos_prslb06} found no qualitative change when using asynchronous update instead of the synchronous rule. The simulations clearly indicated that the highest frequency of cooperation occurs on sites with a large degree. Furthermore, contrary to what was observed for regular graphs, the frequency of cooperators increases with $\langle z \rangle$. Conversely, cooperative behavior evaporates if the edges connecting the hubs are artificially removed. Considering an evolutionary Prisoner's Dilemma game with synchronized update (without irrational strategy adoption) \citet{gomez_prl07} have found that cooperation remains unchanged in a vicinity of hubs. All these features support the conclusion that the appearance of connected hubs is responsible for enhanced cooperation in these structures.

Note that in these simulations an imitative adoption rule was applied, and the total utility was posited to be simply additive. These elements favor cooperative players with many neighbors. Robustness of the above mechanism is questionable for other type of dynamics like Best Response, or when the utility is non-additive. Indeed, using different evolutionary rules \citet{wu_cm05} have studied what happens on a scale-free structure if the strategy adoption is based on the normalized payoff (payoff divided by the number of co-players) rather than the total payoff. The simulations indicate that the huge increase in the cooperator frequency found in \citet{santos_prslb06} is largely suppressed.

A weaker, although well-detectable consequence of the inhomogeneous degree distribution was observed by \citet{santos_pre05}, when they compared the stationary frequency of cooperators on the classic (see Fig.~\ref{fig:wssw}) and regular version of the small-world structure suggested by \citet{watts_n98}. They found that on  regular structures (with $z=4$) the frequency of cooperators decreases (increases) below (above) a threshold value of $b$, when the ratio of rewired links is increased.

In some sense hubs can be considered as privileged players, whose interaction with the neighborhood is asymmetric. Such a feature was artificially built into a cellular automaton model by \citet{kim_pre02}, who  observed large drops in the level of cooperation after flipping an influential site from cooperation to defection. Very recently \citet{wu_cpl06}, \citet{ren_cm06a}, and \citet{szolnoki_epl07} have demonstrated numerically that inhomogeneities in the strategy adoption rate can also enhance cooperation. Besides inhomogeneities, the percolation of the overlapping triangles in the connectivity structure can also give an additional boost for cooperation. Notice that the highest frequency of cooperators were found for the Dorogovtsev-Mendes-Samukhin structure (see Fig.~\ref{fig:sfbadm}), where the construction of the growing graph guarantees the percolation of triangles.

All the above analysis is based on the payoff matrix given by
Eq.~(\ref{eq:weakPD}). In the last years another parametrization of
$\bf A$ is used frequently. Within this approach the cooperator
pays a cost $c$ for her opponent to receive a benefit $b$ ($0 < c <
b$). The defector pays nothing and does not distribute benefits.
The corresponding payoff matrix is:
\begin{equation}
\label{eq:costbenefitPD}
    {\bf A}=\left( \matrix{0 & b \cr
                           -c & b-c \cr }\right).
\end{equation}
Using this payoff matrix \citet{ohtsuki_n06} have studied the fixation of
solitary $C$ (and $D$) strategy on finite social networks for several evolutionary rules (e.g., ''death-birth'', ''birth-death'', and imitation updating). For example, for the elementary steps of ''death-birth'' updating a random player is chosen to die and subsequently the neighboring players compete for the empty site in proportion to their fitness. In the so-called weak-selection limit, defined by expression (\ref{eq:weakpo}),  the pair approximation can be performed analytically. The pair approximation predicts that the cooperator has on average one more cooperator neighbor than the defector, and the cooperators have a fixation probability larger than $1/N$ (i.e., $C$ is favored in comparison to the voter model dynamics) if the ratio of benefit to cost exceeds the average number of neighbors ($b/c > z$). The results of simulations (performed on different networks) show sufficiently good agreement with the prediction of pair approximation although this approximation neglects the role of loops in the connectivity structure. Evidently, the accuracy of pair approximation is improved as $z$ is increased. It is underlined that the mentioned simple rule seems to be similar to Hamilton's rule saying that kin selection can favor cooperation if $b/c > 1/r$, where the social relatedness $r$ can be measured by the inverse of the number of neighbors \citep{hamilton_jtb64a}. \citet{nowak_s06} has shown that the altruistic behavior emerges if the benefit-to-cost ratio ($b/c$) exceeds some threshold value for three other mechanisms called direct reciprocity, indirect reciprocity, and group selection.

All the above phenomena indicate that there are possibly many ways to influence and enhance cooperative behavior in social networks. Further investigations would be required to clarify the distinguished role of noise, appearance of different time scales, and inhomogeneities in these structures.

\subsection{Prisoner's Dilemma on evolving networks}
\label{sec:pdevolnet}

In real-life situations the preferential choice and refusal of partners play an important role in the emergence of cooperation. Several aspects of this process have been investigated by using different dynamics in the mathematical models \citep{ashlock_bs96,bala_em00,hanaki_ms06} and experimentally [see \citep{hauk_jcr01,coricelli_jcr04} and further references therein] for several years.

An interesting model where the social network and the individual strategies co-evolve was introduced by \citet{zimmermann_00,zimmermann_pre04}. The construction of the system starts with the creation of a random graph [as suggested by \citep{erdos_pmd59}], which serves as an initial connectivity structure with a prescribed average degree. The strategy of the players, positioned on this structure, can be either $C$ or $D$. In discrete time steps ($t=1,2,\ldots$) each player plays a one-shot Prisoner's Dilemma game with all her neighbors and the payoffs are summed up. Knowing the payoffs,  unsatisfied players (whose payoff is not the highest in their neighborhood) adopt the strategy of their best neighbors. This is the deterministic version of the Imitate the Best rule in Eq.\ (\ref{eq:nbmscak}).

It is assumed that after the imitation phase unsatisfied players who have become defectors (and only these players) break the link with the imitated defector with probability $p$. The broken link is replaced by a new link to a randomly chosen player (self-links and multiple links are forbidden). This network adaption rule conserves the number of links and allows cooperators to receive new links from the unsatisfied defectors. The simulation starts from a random initial strategy distribution, and the evolution of the network, whose intensity is characterized by $p$, is switched on after some thermalization time, typically 150 generations.

The numerical analysis for $N=10^4$ sites shows that evolution ends up either in an absorbing state with defectors only or in a state where cooperators and defectors form a frozen pattern. In the final cooperative state the defectors are isolated, and their frequency depends on the initial state due to the cellular automaton rule. It turns out that modifications in $\langle z \rangle$ can only cause a slight variation in the results. The adaptation of the network leads to fundamental changes in the connectivity structure: the number of high degree sites are enhanced significantly, and these sites (leaders) are dominantly occupied by cooperators. These structural changes are accompanied by a significant increase in the frequency of cooperators. The final concentration of cooperators is as high as found by \citet{santos_prl05} (see Fig.~\ref{fig:sfmccomp}) in the whole region of the payoff parameter $b$ (temptation to defect). It is observed, furthermore, that social crisis, i.e., large cascades in the evolution of both structure and frequency of cooperators propagate through the network, if a leader's state is changed suddenly \citep{zimmermann_pre05,eguiluz_ajs05}.

The final connectivity structure exhibits hierarchical characters. The formation of this structure is preceded by the cascades mentioned above, in particular, for high values of $b$. This phenomenon may be related to the competition of leaders \citep{santos_prslb06}, discussed in the previous section (see Figs.~\ref{fig:bossconf} and \ref{fig:compboss}). This type of network adaption rule could not cause a relevant increase in the clustering coefficient. In \citep{zimmermann_pre04,eguiluz_ajs05} the network adaption rule was altered to favor the choice of second neighbors (the neighbors of the $D$ neighbor who are companions in distress) with another probability $p$ when selecting a new partner. As expected, this modification resulted in a remarkable increase in the final clustering coefficient. A particularly high increase was found for large $b$ and small $p$ values.

Fundamentally different co-evolutionary dynamics were studied by \citet{biely_cm05}. The synchronized strategy adoption rule was extended by the possibility of cancellation and/or creation of new links with some probability. The number of removed and new links were limited by model parameters. In this model the players were capable to estimate the expected payoffs in subsequent steps by taking into account the possible variations in the neighborhood. In contrast to the previous model, here the number of connections did not remain fixed. The numerical simulations indicated the emergence of oscillations in some quantities (e.g., the frequency of cooperators and the number of links) within some region of the parameter space. For high frequency of the cooperators the network characteristics resemble those of scale-free graphs discussed above. The appearance of global oscillations reminds us of similar phenomena found for voluntary Prisoner's Dilemma games (see below).

\begin{figure}[ht]
\centerline{\epsfig{file=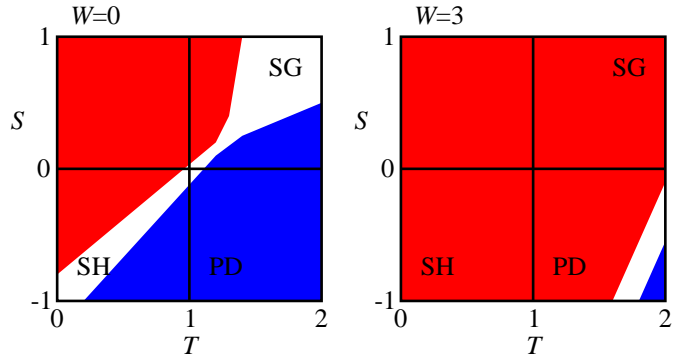,width=9cm}}
\caption{\label{fig:spl}Red areas indicate regions of the $S-T$ plane where cooperators prevail the system for two different values of $W$. Blue territories refer to defectors' dominance. The quadrants of Snowdrift, Stag Hunt, and Prisoner's Dilemma are denoted by the corresponding abbreviations \citep{santos_ploscb06}.}
\end{figure}

The above versions of co-evolutionary models are extended to involve stochastic elements (irrational choices) in the evolution of both the strategy distribution and the connectivity structure \citep{santos_ploscb06,pacheco_jtb06}. Such an adaptive individual behavior introduces two time scales ($\tau_l$ and $\tau_s$) associated with the typical lifetime of the links and individual strategies modified by imitation via pairwise comparison.  In the limit $W=\tau_s / \tau_l \to 0$ this model reproduces the results achieved on quenched structure. In the opposite limit ($W \to \infty$) the players readjust their connectivity structure for fixed strategy distribution. This latter phenomenon was investigated by \citet{pacheco_jtb06,pacheco_prl06} who assumed that the connections can disappear exponentially with a life time and new links are created between randomly chosen players with some rates. The characteristic death and birth rates of connections depend on the player's strategies and the corresponding parameters will determine the stationary frequency  of connections ($\Phi_{XY}$) between the $s-s^{\prime}$ strategy pairs ($s, s^{\prime} = D$ or $C$). In this case the slow evolution of the strategy frequencies can be described by a mean-field type equation based on a rescaled payoff matrix (${\bf A} \to {\bf A}^{\prime}$). In other words, under the mean-field conditions the players feel an effective payoff matrix with coefficients $A_{ss^{\prime}}^{\prime}=A_{ss^{\prime}} \Phi_{ss^{\prime}}$. Evidently, this transformation can change the ranking of payoff values. \citet{pacheco_jtb06,pacheco_prl06} have demonstrated that the Prisoner's Dilemma can be transformed into a new game favoring cooperation definitely when choosing a suitable dynamics of linking. The numerical analysis indicates that similar consequences can be observed even for $W=1$. Two schematic plots in Fig.~\ref{fig:spl} show the expansion of those region of parameters where cooperators dominates the system as $W$ is increased from 0 to 3.

In the multilevel selection model introduced by \citet{traulsen_pnas06} $N$ players are divided into $m$ groups. Individual incomes (fitness) come from games between members of the same group, i.e., the connectivity structure is a union of $m$ disjoint complete subgraphs (cliques). At each time step of the evolution a player is chosen randomly from the whole population with a probability proportional to her income. The player's offspring is added to the same group. The players in the given group are divided into two groups randomly with a probability $q$ if the group size reaches a critical value $n$. Otherwise one of the players is eliminated from this group (with a probability $1-q$). The number of groups is fixed by eliminating a randomly chosen group whenever a new group is created. Although the evolutionary dynamics is based on individual fitness the selection occurs at two levels, and favors the fastest reproduction for both the individuals and the groups. This model can be interpreted as a hierarchy of two types of Moran processes,  and has been analyzed analytically in the weak selection limit for low values of $q$. The model was extended by allowing migration between the groups with some probability. Using the payoff matrix (\ref{eq:costbenefitPD}) and considering the fixation probabilities \citet{traulsen_pnas06} have shown that cooperators are favored over defectors if the benefit-to-cost ratio exceeds a threshold value, $b/c > 1 + z + n/m$, where $z$ is the average number of migrants arising from a group during its lifetime.

Finally we call the reader's attention to the work by \citet{vainstein_jtb07} who surveyed the effect of player's migration on the measure of cooperation. The introduction of mobility drives the system towards the mean-field situation favoring the survival of defectors. Despite of this expectation Vainstein {\it et al}.\ have described several dynamical rules increasing the frequency of cooperators. Similar results were reported by \citet{aktipis_jtb04} who considered the effect of contingent movement of cooperators: the players walked away once a defection occurred in the previous step. In some sense, this ``win-stay, lose-move" strategy is analogous to voluntary participation which also enhance cooperation \citep{saijo_jet99,hauert_s02,szabo_pre02d,wu_pre05} as detailed in the subsequent section.

\subsection{Spatial Prisoner's Dilemma with three strategies}
\label{sec:pd3s}

In former sections we considered in detail the Iterated Prisoner's Dilemma game with two memoryless strategies, AllD (shortly $D$) and AllC (shortly $C$). Enlarging the strategy space to three or more strategies (now with memory) is a straightforward extension. The first numerical investigations in this genre focused on games with a finite subset of stochastic reactive strategies. As mentioned in Sec.\ref{sec:pqstrat} one can easily select three stochastic reactive strategies, for which cyclic dominance appears naturally as detailed by \citet{nowak_ijbc94,nowak_jtb89,nowak_amc89}. In the corresponding spatial games most of the evolutionary rules give rise to self-organizing patterns, similar to those characterizing spatial Rock-Scissors-Paper games that we discuss later in Sec.~\ref{sec:rsp}.

In the absence of cyclic dominance, a three-strategy spatial game usually develops into a state where one or two strategies die out. This situation can occur in a model which allows for $D$, $C$, and Tit-for-Tat (shortly $T$) strategies (see Appendix \ref{app:str:tft} for an introduction on Tit-for-Tat). In this case the $T$ strategy will invade the territory of $D$, and finally the evolution terminates in a mixture of $C$ and $T$ strategies, as was described by \citet{axelrod_84} in the non-spatial case. In most spatial models, for sufficiently high $b$ any external support of $C$ against $T$ yields cyclic dominance: $T$ invades $D$ invades $C$ invades $T$.

In real social systems a number of factors can favor $C$ against $T$. For example, permanent inspection, i.e., keeping track of the opponents moves introduces a cost which reduces the net income of $T$ strategists. Similar payoff reduction is found for Generous $T$ players (see again Appendix \ref{app:str:tft} for an introduction on Generous Tit-for-Tat). In some populations the $C$ strategy may be favored by education or by the appearance of new, unexperienced generations. These effects can be built into the payoff matrix \citep{imhof_pnas05} and/or can be handled via the introduction of an external force \citep{szabo_pre00a}. Independently of the technical details, these evolutionary games have some common features that we discuss here for the Voluntary Prisoner's Dilemma and later for the Rock-Scissors-Paper game.

The idea of a Voluntary Prisoner's Dilemma game comes from Voluntary Public Good games \citep{hauert_s02} or Stag Hunt games (see Appendix \ref{app:g:sh}), where the players are allowed to avoid exploitation. Namely, the players can refuse to participate in these multi-player games. Besides the traditional two strategies ($D$ and $C$) this option can be considered as a third strategy called ``Loner" (shortly $L$). The cyclic dominance between these strategies ($L$ invades $D$ invades $C$ invades $L$) is due to the average payoff of $L$. The experimental verification that this leads to a Rock-Scissors-Paper-like dynamics is reported in \citet{semmann_n03}.

In the spatial versions of Public Good games \citep{hauert_s02,szabo_prl02} the players income comes from multi-player games with their neighborhood. This makes the calculations very complicated. However, all the relevant properties of these models can be reserved when we reduce the five- (or nine-) site interactions into pairwise interactions. With this proviso, the model becomes equivalent to the Voluntary Prisoner's Dilemma characterized by the following payoff matrix:
\begin{equation}\label{eq:vpdpo}
\begin{tabular}{cc}
  &    $D\;\;\; C \;\;\; L $  \\[2pt]
  ${\bf A}\,= \, \matrix{D \cr C \cr L \cr}$ &
  $\left( \matrix{0 & b & \sigma \cr
                  0 & 1 & \sigma \cr
                  \sigma & \sigma & \sigma \cr }\right)$ \\
\end{tabular}
\end{equation}

where the loner and her opponent share a payoff $0< \sigma < 1$. Notice that the sub-matrix corresponding to the $D$ and $C$ strategies is the same we used previously in Eq.\ (\ref{eq:weakPD}). Similar payoff structure can arise for a job made in collaboration by two workers, if the collaboration itself involves a Prisoner's Dilemma. In this case, the workers can look for other independent jobs yielding lower income $\sigma$, if any of them declines the common work.

\begin{figure}[ht]
\centerline{\epsfig{file=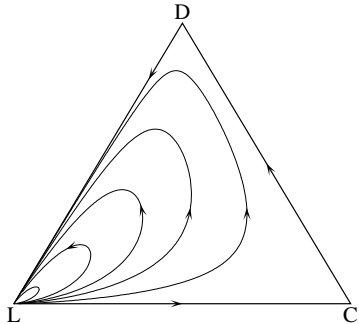,width=5cm}}
\caption{\label{fig:vpdmf}Trajectories of the Voluntary Prisoner's Dilemma game predicted by the traditional mean-field approximation for $\sigma=0.3$, $b=1.5$, and $K=0.1$.}
\end{figure}

First we consider the Voluntary Prisoner's Dilemma game on a square lattice with nearest neighbor interaction \citep{szabo_pre02d}. The evolutionary dynamics is assumed to be random sequential with a transition rate defined by Eq.\ (\ref{eq:smoothedimit}). According to the traditional mean-field approximation, the time-dependent strategy frequencies follow the typical trajectories shown in Fig.~\ref{fig:vpdmf}, that is, evolution tends towards the homogeneous loner state.\footnote{Notice that $L$ is attractive but unstable (see the relevant discussion in Sec.\ \ref{sec:rd}).} A similar behavior was found by \citet{hauert_s02}, when they considered a Voluntary Public Good game.

The simulations indicate a significantly different behavior on the square lattice. In the absence of $L$ strategies this model agrees with the system discussed in Sec.~\ref{sec:rsu}, where cooperators become extinct if $b > b_{cr}(K)$ ($\max (b_{cr}) \simeq 1.1$). Contrary to these, cooperators can survive for arbitrary $b$ in the presence of loners.

\begin{figure}[ht]
\centerline{\epsfig{file=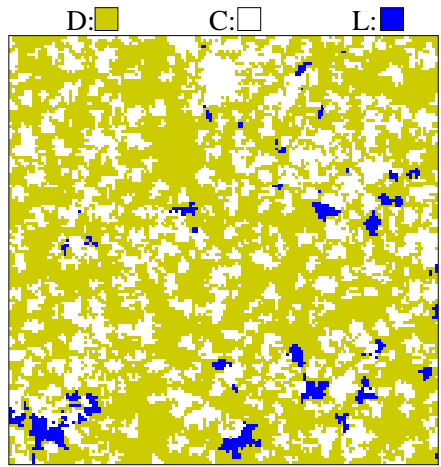,width=6cm}}
\caption{\label{fig:dclpattn}A typical distribution of the three strategies on the square lattice for the Voluntary Prisoner's Dilemma in the vicinity of the extinction of loners.}
\end{figure}

Figure~\ref{fig:dclpattn} shows the coexistence of all three strategies. Note that this is a dynamic image. It is perpetually in motion due to propagating interfaces, which reflect the cyclic dominance nature of the strategies. Nevertheless, the qualitative properties of this self-organizing pattern are largely unchanged for a given payoff matrix.

It is worth emphasizing that cyclic dominance is not directly built into the payoff matrix (\ref{eq:vpdpo}). Cyclic invasions appear at the corresponding boundaries, where the role of strategy adoption and clustering is important.

\begin{figure}[ht]
\centerline{\epsfig{file=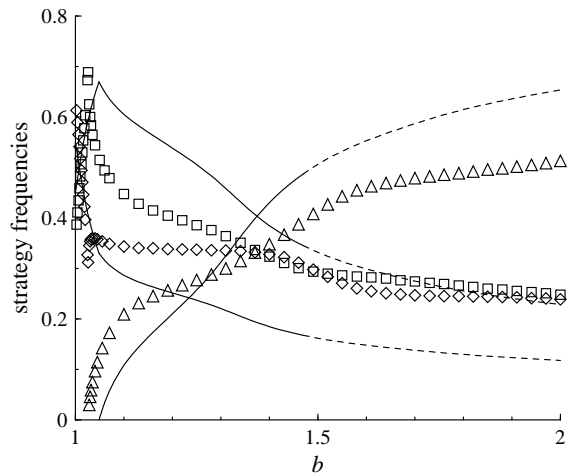,width=7.5cm}}
\caption{\label{fig:vpdsqmc}Stationary frequency of defectors (squares), cooperators (diamonds), and loners (triangles) {\it vs.} $b$ for the Voluntary Prisoner's Dilemma on the square lattice at $K=0.1$ and $\sigma = 0.3$. The predictions of the pair approximation are indicated by solid lines. Dashed lines refer to the region where this stationary solution becomes unstable.}
\end{figure}

The results of a systematic analysis are summarized in Fig.~\ref{fig:vpdsqmc} \citep{szabo_pre02d}. Notice that loners cannot survive for low values of $b$, where the defector frequency is too low to feed loners. Furthermore, the loner frequency increases monotonously with $b$, despite the fact that large $b$ only increases the payoff of defectors. This unusual behavior is related to the cyclic dominance as explained later in  Sec.~\ref{sec:rspur}.

As $b$ decreases the stationary frequency of loners vanishes algebraically at a critical point \citep{szabo_pre02d}. In fact, the extinction of loners belongs to the directed percolation universality class, although the corresponding absorbing state is an inhomogeneous and time-dependent one. However, on large spatial and temporal scales, the rate of birth and death becomes homogeneous on this background, and these are the relevant criteria for belonging to the robust universality class of directed percolation, as was discussed in detail by \citet{hinrichsen_ap00}.

The predictions of the pair approximation are also illustrated in Fig.~\ref{fig:vpdsqmc}. Contrary to the two-strategy model, here the agreement between the pair approximation and the Monte Carlo data is surprisingly good. Figure~\ref{fig:pdsqt04} nicely demonstrates that the two-strategy pair approximation can fail in those regions, where one of the strategy frequencies is low. This is, however, not the case here. For three-strategy solutions the accuracy of this approach becomes better. The stability analysis, however, reveals a serious shortcoming: the stationary solutions become unstable in the region of parameters indicated by dashed lines in Fig.~\ref{fig:vpdsqmc}. In this region the numerical analysis indicates the appearance of global oscillations in the frequencies of strategies.

Although the shortcomings of the generalized mean-filed approximation on the square lattice diminish when we go beyond the pair approximation and consider higher levels (larger clusters), it would be enlightening to find other connectivity structures on which the pair approximation becomes acceptable. Possible candidates are the Bethe lattice and random regular graphs for large $N$ (as discussed briefly in Sec.~\ref{sec:noc}), because this approach neglects correlations mediated by four-site loops. In other words, the prediction of the pair approximation is more reliable if loops are missing, as in the Bethe lattice.

\begin{figure}[thb]
\centerline{\epsfig{file=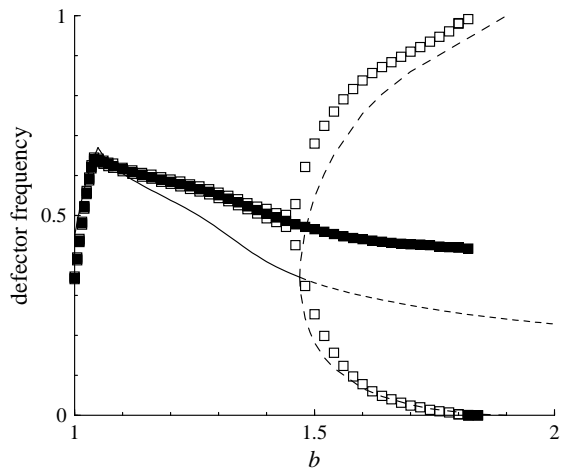,width=7.5cm}}
\caption{\label{fig:vpddmm}Monte Carlo results (closed squares) for the frequency of defectors as a function of $b$ on a random regular graph for $K=0.1$ and $\sigma = 0.3$. The average frequency of defectors are denoted by closed squares. Open squares indicate the maximum and minimum values taken during the simulation. Predictions of the pair approximation are shown by solid (stable) and dashed (unstable) lines.}
\end{figure}

The simulations \citep{szabo_pre02d} found a similar behavior to those on the square lattice (see Fig.~\ref{fig:vpdsqmc}) if $b$ remains below a threshold value $b_1$ depending on $K$. For higher $b$ global oscillations were observed, in nice agreement with the prediction of the pair approximation, as illustrated in Fig.~\ref{fig:vpddmm}. To avoid confusion in this plot we only illustrate the frequency of defectors, obtained on a random regular graph with $10^6$ sites. The plotted values (closed squares) are determined by averaging the defector's frequency over a long sampling time, typically $10^5$ Monte Carlo steps (MCS). During this time the maximum and minimum values of the defector's frequency were also determined (open squares). The predictions of the pair approximation for the same quantities can be easily determined, and they were compared with the Monte Carlo data in Fig.~\ref{fig:vpddmm}. Both methods predict that the "amplitude" of global oscillation increases with $b$ until a second threshold value $b_2$. If $b > b_2$, the strategy frequencies oscillate with a growing amplitude, and after some time one of them dies out. Finally the system develops into a homogeneous state. In most cases cooperators become extinct first, and then loners sweep out defectors.

Further investigations of this model were aimed to clarify the effect of partial randomness \citep{szabo_pre04b}. When tuning the portion of rewired links on a regular small-world network (see Fig.~\ref{fig:stages4}), the self-organizing spatial pattern transforms into a state exhibiting global oscillations. If the portion of rewired bonds exceeds
about ten percent ($q \gtrsim 0.1$) 
the small-world effect leads to a behavior similar to those found on random regular graphs. Using a different evolutionary rule \citet{wu_pre05} reported similar features on the small-world structure suggested by \citet{newman_pre99}.

In summary, the possibility of voluntary participation in spatial Prisoner's Dilemma games supports cooperation via a cyclic dominance mechanism, and leads to a self-organizing pattern. This spatio-temporal structure is destroyed by random links (or small-world effects) and synchronizes the local oscillations over the whole system. In some parameter regions the evolution is terminated in a homogeneous state of loners after some transient processes, during which the amplitude of the global oscillations increase. In this final state the interaction between the players vanishes, that is, they do not try to benefit from possible mutual cooperations. According to the classical mean-field calculation \citep{hauert_s02} an analogous behavior is expected for the connectivity structures shown in Fig.~\ref{fig:mfstruc}. These general properties are rather robust, similar behavior can be observed if the loner strategy is replaced by some Tit-for-Tat strategies in these evolutionary Prisoner's Dilemma games.

\subsection{Tag-based models}
\label{sec:pdtbm}

\emph{Kin selection} \citep{hamilton_jtb64a,hamilton_jtb64b} can support cooperation among individuals (relatives), who mutually help each other. This mechanism assumes that individuals are capable of identifying their relatives to be donated. In fact, tags characterizing a group of individuals can facilitate selective interactions, leading to growing aggregates in spatial systems \citep{holland_95}. Such a tag-matching mechanism was taken into account in the image scoring model introduced by \citet{nowak_n98}, and many other, so-called tag-based models.

In the context of the Prisoner's Dilemma a simple tag-based model was introduced by \citet{hales_lnai00}. In his model each player possesses a memoryless strategy, $D$ or $C$, and a tag, which can take a finite number of discrete values. A tag is a phenotypic marker like sex, age, cultural traits, language, etc., or any other signal that individuals can detect and distinguish \citep{robson_jtb90}. Agents play a Prisoner's Dilemma game with random opponents, provided that the co-player possesses the same tag. This mechanism partitions the social network into disjunct clusters with identical tags. There is no play between clusters. Nevertheless, mutation can take individuals from one tag cluster into another, or change her strategy with some probability. Players reproduce in proportion to their average payoff.

\citet{hales_lnai00} found that cooperative behavior flourishes in the model when the number of tags is large enough. Clusters with different tag values compete with each other in the reproductive step, and those that contain a large proportion of cooperators outperform other clusters. However, such clusters have a finite life time, since if mutations implant defectors into a tag cluster, they soon invade the cluster and shortly cause its extinction (within a cluster it is a mean-field game where defectors soon expel cooperators). The room is taken over by other growing cooperative groups. Due to mutations new tag groups show up from time to time, and if these happen to be C players they start to increase. At each step there are a small number of mostly cooperative clusters present in the system, i.e., the average level of cooperation is high, but the prevailing tag values vary from time to time. \citet{jansen_n06} have considered the maintenance of cooperation through the so-called beard chromodynamics in a spatial system where the tag (color) and behavior (C or D strategy) are inherited via loosely coupled genes. Similar models for finite populations are analyzed by \citet{traulsen_plosone07} (further references therein).

A rather different tag-based model, similar in spirit to Public Good games (see Appendix \ref{app:g:pg}), was suggested by \citet{riolo_n01}. Each player $x$ is characterized by a tag value $\tau_x \in [0,1]$ and a tolerance level $\Delta_x>0$, which constitute the player's strategy in a continuous strategy space. Initially the tags and the tolerance levels are assigned to players at random. For a given step (generation) each player is paired with all others, but player $x$ only donates to player $y$ if $|\tau_x-\tau_y| \leq \Delta_x$. Agents with close enough tag values are forced to be altruistic.  Player $x$ incurs a cost $c$ and player $y$ receives a benefit $b$ ($b > c > 0$) if the tag-based condition of donation (cooperation) is satisfied. At the end of each step the players reproduce: the income of each player is compared with another player's income chosen at random, and the loser  adopts the winner's strategy $(\tau,\Delta)$. During the strategy adoption process mutations are allowed with a small probability. Using numerical simulations \citet{riolo_n01} demonstrated that tagging can again promote altruistic behavior (cooperation).

In the above model the strategies are characterized by two continuous parameters. \citet{traulsen_pre03} have shown that the main features can be reproduced with only four discrete strategies. The minimal model has two types of players (species), say Red and Blue, and both types can possess either minimum ($\Delta=0$) or maximum ($\Delta=1$) tolerance levels. Thus, (Red,1) and (Blue,1) donate to all others, while (Red,0) [resp., (Blue,0)] only donates to (Red,0) and (Red,1) [resp., (Blue,0) and (Blue,1)]. The corresponding payoff matrix is
\begin{equation}
\label{eq:mmtbcpo}
{\bf A}=\left( \matrix{b-c & b-c & b-c & -c  \cr
                       b-c & b-c & -c  & b-c \cr
                       b-c & b   & b-c & 0   \cr
                       b   & b-c & 0   & b-c \cr }\right) \;
\end{equation}
for the strategy order (Red,1), (Blue,1), (Red,0), and (Blue,0). This game has two pure Nash equilibria, (Red,0) and (Blue,0), and an evolutionary unstable mixed-strategy Nash equilibrium, where the two intolerant ($\Delta=0$) strategies are present with probability 1/2.

The most interesting features appear when mutations are introduced. For example, allowing that (Red,0) [resp., (Blue,0)] players mutate into (Red,1) [resp., (Blue,1)] players, the population of (Red,1) strategists can increase in a predominantly (Red,0) state, until the minority (Blue,0) players receive the highest income and begin to invade the system. In the resulting population the frequency of (Blue,1) increases until (Red,0) dominates, thus restarting the cycle. In short, this system exhibits spontaneous oscillations in the frequency of tags and tolerance levels, which were found to be one of the characteristic features in the more complicated model \citep{riolo_n01,sigmund_n01}. In the simplified version, however, many aspects of this dynamics can be investigated more quantitatively as was discussed by \citet{traulsen_pre03}.

The spatial (cellular automaton) version of the above discrete tag-tolerance model was studied by \citet{traulsen_pre04}. As was expected, the spatial version clearly demonstrated the formation of large Red and Blue domains. The spatial model allows for the occurrence of tolerant mutants in a natural way. In the cellular automaton rule they used, a player increased her tolerance level from $\Delta=0$ to 1, if she was surrounded by players with the same tag. In this case the cyclic dominance property of the strategies gives rise to a self-organizing pattern with rotating spirals and anti-spirals, which will be discussed in detail in the next Section. It is remarkable how these models sustain a poly-domain structure \citep{traulsen_pre03}. The coexistence of (Red,1) and (Blue,1) domains is stabilized by the appearance of intolerant strategies, (Red,0) and (Blue,0), along the peripheries of these regions.

The main feature of tag-based models can be recognized in the systems studied by \citet{ahmed_pa06} who considered what happens if the players are grouped, and their choices and/or evolutionary rules depend on the group tags. Evidently, tag-based models can be extended by allowing some variations of the tag, which is related to the evolving community structure on the network.

\section{Rock-Scissors-Paper games}
\label{sec:rsp}

Rock-Scissors-Paper games exemplify those systems from ecology or non-equilibrium physics where three states (strategies or species) cyclically dominate each other \citep{may_siam75,hofbauer_88,tainaka_lncs01}. Although cyclic dominance does not manifest itself directly in the payoff matrices of the Prisoner's Dilemma, that model can exhibit similar properties when three stochastic reactive strategies are considered \citep{nowak_jtb89,nowak_amc89}. Cyclic dominance can also be caused by spatial effects as observed in some spatial three-strategy evolutionary Public Good games \citep{hauert_s02,szabo_prl02} and Prisoner's dilemma games \citep{nowak_ijbc94,szabo_pre00a,szabo_pre02d}.

Rock-Scissors-Paper-type cycles can be observed directly in nature. One of the best known examples, described by \citet{sinervo_n96} are the three different mating strategies of the lizard species {\it Uta stansburiana}. Similarly, the interaction networks of many marine ecological communities involve three-species cycles \citep{johnson_prsb02}. Successional three-state systems such as space-grass-trees \citep{durrett_tpb98} or forest fire models with states: ``green tree", ``burning tree", and ``ash" \citep{bak_pla90,drossel_prl92,grassberger_jpa93} may also have similar dynamics. For many spatial predator-prey (Lotka-Volterra) models the ``empty site", ``prey", and ``predator" states also follow cyclically each other \citep{bascompte_98,tilman_97}. The synchronized versions of these phenomena have been studied in the framework of cellular automaton models by several authors \citep{greenberg_siam78,fisch_pd90,silvertown_je92,fuks_ddns01}.

Cyclic dominance occur in the so-called epidemiological \emph{SIRS models}. The mean-field version of the simpler SIR model was formulated by Lowell Reed and Wade Hampton Frost [their work was never published but is cited in \citep{newman_pre02}] and by \citet{kermack_prsa27}. The abbreviation SIR refers to the three possible states of individuals during the spreading of an infectious disease: ``susceptible", ``infected", and ``removed". In this context ``removed" means to be either recovered and become immune to further infection or being dead. For the more general SIRS model susceptible individuals replace removed ones with a certain rate representing the birth of susceptible offsprings or loosing immunity and regaining susceptible status after a while.

Another biological system demonstrates how cyclic dominance occurs in the biochemical warfare among three types of microbes. Bacteria extract toxic substances that are very effective against strains of their microbial cousins who do not produce the corresponding resistance factor \citep{durrett_jtb97,nakamaru_tpb00,czaran_pnas02,kerr_n02,kirkup_n04,neumann_jmb07}. For a certain toxin three types of bacteria can be distinguished: the Killer type produces both the toxin and the corresponding resistance factor (to prevent its suicide); the Resistant produces only the resistance factor; and the Sensitive produces neither. Colonies of Sensitive can always be invaded and replaced by Killers. At the same time Killer colonies can be invaded by Resistants, because the latter are immune to the toxin but do not carry the metabolic burden of synthesizing the toxic substance. Due to their faster reproduction rate they achieve competitive dominance over the Killer type. Finally the cycle is closed: Sensitives have metabolic advantage over Resistants, because they do not even have to pay the cost of producing the resistance factor.
In general cyclic dominance in biological systems can play two fundamental roles: supporting bio-diversity \citep{may_siam75,kerr_n02} and providing protection against external invaders \citep{boerlijst_pd91,szabo_pre01a}. 

It is worth mentioning, though, that cyclic invasion processes and the resultant spiral wave structures have been investigated in some physical, chemical, or physiological contexts for a long time. The typical chemical example is the Belousov-Zhabotinski reaction \citep{field_jcp74}. However, similar phenomena were found for the Rayleigh-Bernard convection in fluid layers discussed by \citep{busse_s80,toral_pa00}, as well as in many other excitable media, e.g., cardiac muscle \citep{wiener_aic46} and neural systems \citep{hempel_prl99}. Recently \citet{prager_pa03} have introduced a similar three-state model to consider stochastic excitable systems.

In the following we first survey the picture that can be deduced from Monte Carlo simulations of the simplest spatial evolutionary Rock-Scissors-Paper games. The results will be compared with the general mean-field approximation at different levels. Then the consequences of some modifications (extensions) will be discussed. We will consider what happens if we modify the dynamics, the payoff matrix, the background, or the number of strategies.

\subsection{Simulations on the square lattice}
\label{sec:rspsim}

In the Rock-Scissors-Paper model introduced by \citet{tainaka_jpsj88,tainaka_prl89,tainaka_pre94}, agents are located on the sites of a square lattice (with periodic boundary condition), and follow one of three possible strategies to be denoted shortly as $R$, $S$, and $P$. The evolution of the population is governed by random sequential invasion events between randomly chosen nearest neighbors. Nothing happens if the two agents follow the same strategy. For different strategies, however, the direction of invasion is determined by the rule of the Rock-Scissors-Paper game, that is, $R$ invades $S$ invades $P$ invades $R$.

Notice that this dynamical rule corresponds to the $K\to 0$ limit of Smoothed Imitation in Eqs.\ (\ref{gen_imit}) with (\ref{gen_imit2}), if the players' utility comes from a single game with one of their neighbors. For the sake of simplicity we choose the following payoff matrix:
\begin{equation}
{\bf A}=\left( \matrix{ 0 &  1 & -1 \cr
                       -1 &  0 &  1 \cr
                        1 & -1 &  0 \cr}\right).
\label{eq:rsppom}
\end{equation}

The above strategy adoption mechanism mimics a simple biological interaction rule between predators and their prey: the predator eats the prey and the predator's offspring occupies the empty site. This is why the present model is also called a cyclic predator-prey (or cyclic spatial Lotka-Volterra) model. The payoff matrix (\ref{eq:rsppom}) is equivalent to the adjacency matrix \citep{bollobas_98} of the three-species cyclic food web represented by a directed graph.

\begin{figure}[ht]
\centerline{\epsfig{file=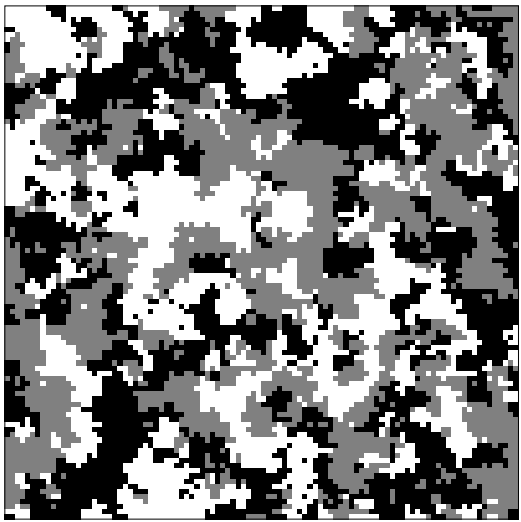,width=7cm}}
\caption{\label{fig:rsppatt}Typical spatial distribution of the three strategies on a block of $100 \times 100$ sites of a square lattice.}
\end{figure}

Due to its simplicity the present model can be studied accurately by Monte Carlo simulations on a finite $L \times L$ box with periodic boundary conditions. Independently of the initial state the system evolves into a self-organizing pattern, where each species is present with a probability $1/3$ on average (for a snapshot see Fig.~\ref{fig:rsppatt}). This pattern is characterized by perpetual cyclic invasions and the average velocity of the invasion fronts is approximately one lattice unit per Monte Carlo step.

\begin{figure}[ht]
\centerline{\epsfig{file=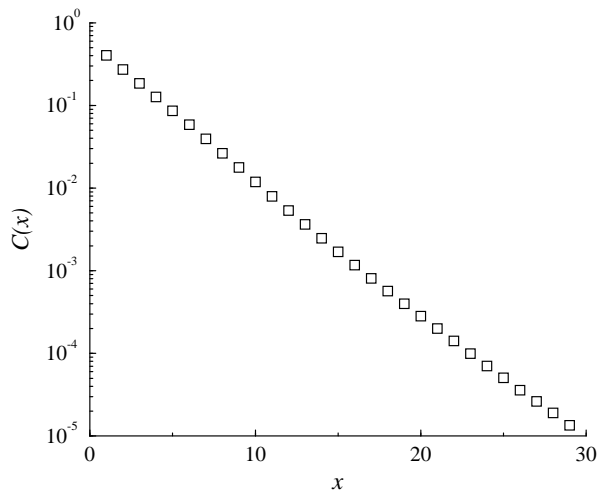,width=8cm}}
\caption{\label{fig:rspcfv}Correlation function {\it vs.} horizontal (or vertical) distance $x$ for the spatial distribution plotted in Fig.~\ref{fig:rsppatt}.}
\end{figure}

Some features of the self-organizing pattern can be described by traditional concepts. For example, the equal-time two-site correlation function is defined as
\begin{equation}
    C({\bf x}) = {1 \over L^2} \sum_{\bf y} \delta(s({\bf y}),s({\bf y+x})) -1/3,
    \label{eq:rspcf}
\end{equation}
where $\delta (s_1,s_2)$ is the Kronecker delta, and the summation runs over all the ${\bf y}$ sites. Figure~\ref{fig:rspcfv} shows $C({\bf x})$ as obtained by averaging over $2\cdot 10^4$ independent patterns with a linear size $L=4000$. The log-lin plot clearly indicates that the correlation function vanishes exponentially in the asymptotic region, $C(x) \propto e^{-x/\xi}$, where the numerical fit yields $\xi = 2.8(2)$. Above $x\approx 30$ the statistical error becomes comparable to $|C(x)|$.

Each species on the snapshot in Fig.~\ref{fig:rsppatt} will be invaded sooner or later by its predator. Using Monte Carlo simulations one can easily determine the average \emph{survival probability of individuals} $S_i(t)$ that defines those portion of sites where the individuals remain alive from a time $t_0$ to $t_0+t$ in the stationary state. The numerical results are consistent with an exponential decrease in the asymptotic region, $S_i(t) \simeq e^{-t/\tau_i}$ with $\tau_i = 1.9(1)$.

\citet{ravasz_pre04} studied a case when each newborn predator inherits the family name of its parent. One can study how the number of surviving families decreases while their average size increases. The \emph{survival probability of families} $S_f(t)$ gives the portion of surviving families (with at least one representative) at time $t+t_0$ if at time $t_0$ all the species had different family names. 
The numerical simulation revealed \citep{ravasz_pre04} that the survival probability of families varies as $S_f(t) \simeq \ln (t)/t$ in the asymptotic limit $t >> \tau_i$, which is a typical behavior for the two-dimensional voter model \citep{ben-naim_pre96}. This behavior is a direct consequence of the fact that for large times the confronting species (along the family boundaries) face their predator or prey with equal probability like in the voter model [for a textbook see \citet{liggett_85}].

Starting from an initial state of parallel stripes or concentric rings, the interfacial roughening of invasion fronts was studied by \citet{provata_pre03}. As expected, the propagating fronts show dynamical characteristics that are similar to those of the Eden growth model. This means that smooth interfaces become more and more irregular, and finally the pattern develops into a random domain structure as plotted in Fig.~\ref{fig:rsppatt}.

\begin{figure}[ht]
\centerline{\epsfig{file=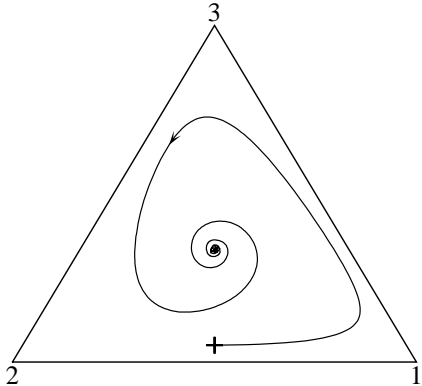,width=6cm}}
\caption{\label{fig:rspspiri}Trajectory of evolution during a Monte Carlo  simulation for $N=1000^2$ agents when the system is started from an asymmetric initial state indicated by a plus symbol. The seemingly smooth trajectory is decorated with noise, but the noise amplitude is comparable to the line width at the given system size.}
\end{figure}

Due to cyclic dominance, at any site of the lattice the species follow cyclically each other. Short-range interactions with noisy dynamics are not able to fully synchronize these oscillations. On the square lattice Monte Carlo simulations show damping oscillations of the strategy concentrations. If the system is started from an asymmetric initial state it spirals towards the symmetric state as is illustrated on the strategy simplex (ternary diagram) in Fig.~\ref{fig:rspspiri}. The symmetric state is asymptotically stable.

For small system sizes (e.g., $L=10$) the simulations show significantly different behavior, because occasionally one of the species becomes extinct. Then the system evolves into one of the three homogeneous states, and further evolution halts. The probability to reach one of these absorbing states decreases very fast as the system size increases.

This model was also investigated on a one-dimensional chain \citep{tainaka_jpsj88,frachebourg_prl96} and on a three-dimensional cubic lattice \citep{tainaka_prl89,tainaka_pre94}. The corresponding results will be discussed later on. First, however, we survey the prediction of the traditional mean-field approximation, whose result is independent of the spatial dimension $d$. Subsequently we will consider the predictions of the pair and the four-site approximations.

\subsection{Mean-field approximation}
\label{sec:rspmf}

Within the framework of the traditional mean-field approximation the system is characterized by the concentration of the three strategies. These quantities are directly related to the one-site configurational probabilities $p_1(s)$, where for later convenience we use the notation $s=1$, $2$, and $3$ for the three strategies $R$, $S$,and $P$, respectively. (We neglect the explicit notation of time dependence.) These time-dependent quantities satisfy the corresponding approximate mean value equation, Eq.\ (\ref{amveq}), i.e.,
\begin{eqnarray}
    \dot{p}_1(1)&=& p_1(1)[p_1(2)-p_1(3)] \;, \nonumber \\
    \dot{p}_1(2)&=& p_1(2)[p_1(3)-p_1(1)] \;, \nonumber \\
    \dot{p}_1(3)&=& p_1(3)[p_1(1)-p_1(2)] \;.
    \label{eq:rspmf}
\end{eqnarray}
The summation of these equations yields $\dot{p}_1(1)+\dot{p}_1(2)+\dot{p}_1(3)=0$, i.e., the present evolutionary rule leaves the sum of the concentrations unchanged, in agreement with the condition of normalization $\sum_s p_1(s) = 1$. It is also easy to see that Eq.\ (\ref{eq:rspmf}) gives $d \sum_s \ln[p_1(s)] / dt = \sum_s \dot{p}_1(s)/p_1(s) = 0$, hence
\begin{equation}
    m=p_1(1) p_1(2) p_1(3) = \mbox{constant} \;
    \label{eq:rspmfcm}
\end{equation}
is another constant of motion.

The equations of motion (\ref{eq:rspmf}) have some trivial stationary solutions. The symmetric (central) distribution $\varrho_1=\varrho_2=\varrho_3=1/3$ is invariant. Furthermore, Eqs.\ (\ref{eq:rspmf}) have three additional homogeneous solutions: $p_1(1)=1$, $p_1(2)=p_1(3)=0$, and two others obtained by cyclic permutations of the indices.

\begin{figure}[ht]
\centerline{\epsfig{file=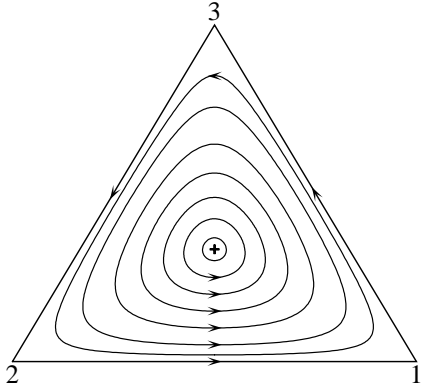,width=6cm}}
\caption{\label{fig:rspmf}Concentric orbits predicted by the mean-field approximation around the central point.}
\end{figure}

The homogeneous stationary solutions are unstable. Small perturbations are not damped, except for the unlikely situation when only prey are added to a homogeneous state of their predators. Under generic perturbations the homogeneous state is completely invaded by the corresponding predator strategy. On the other hand, perturbing the central solution the system exhibits a periodic oscillation, that is $p_1(s)=1/3+\varepsilon \sin(\omega t+2 s \pi /3)$ with $\omega = 1/\sqrt{3}$ for $s=1$, 2, and 3 in the limit $\varepsilon \to 0$. This periodic oscillation becomes more and more anharmonic for increasing initial perturbation amplitude, and finally the trajectory approaches the edges of the triangle on the strategy simplex 
as shown in Fig.~\ref{fig:rspmf}. The symmetric solution is stable but not asymptotically stable.

Notice the contradiction between the results of simulations and the prediction of the mean-field approximation which lacks the asymptotically stable solution. Contrary to our naive expectation, this contradiction becomes even more explicit at the level of the pair approximation.

The stochastic version of the three-species cyclic Lotka-Volterra system is studied by \citet{reichenbach_pre06,reichenbach_bcp06} within the formalism of an urn model which allows the investigation of finite size effects. It is found that fluctuations around the deterministic trajectories grow in time until two species become extinct. The extinction probability depends on the rescaled time $t/N$.

\subsection{Pair approximation and beyond}
\label{sec:rsppa}

In the pair approximation \citep{tainaka_pre94,sato_mmit97}, the system is characterized by the probabilities $p_2(s_1,s_2)$ of all strategy pairs $(s_1,s_2)$ on two neighboring sites ($s_1, s_2=1, 2, 3$). These quantities are directly related to the one-site probabilities, $p_1(s_1)$, through the compatibility conditions discussed in Appendix \ref{app:gmfa}. Neglecting technical details, now we discuss the results on the $d$-dimensional hyper-cubic lattice.

The equations of motion for the one-site configuration probabilities can be expressed by $p_2(s_1,s_2)$ as
\begin{eqnarray}
\dot{p}_1(1)&=& p_2(1,2)-p_2(1,3) \;, \nonumber \\
\dot{p}_1(2)&=& p_2(2,3)-p_2(2,1) \;, \nonumber \\
\dot{p}_1(3)&=& p_2(3,1)-p_2(3,2) \;.
\label{eq:rsppa1}
\end{eqnarray}
Note that in this case the sum of the one-site configuration probabilities remains unchanged, while the conservation law for their product is no longer valid, in contrast with
the prediction of the mean-field approximation [see Eq.\ (\ref{eq:rspmfcm})].


At the pair approximation level the time derivatives of the nine pair-configuration probabilities, $p_2(s_1,s_2)$ ($s_1, s_2=1, 2, 3$), satisfy nine coupled first-order nonlinear differential equations \citep{tainaka_pre94,sato_mmit97}.
For $d>1$ numerical integration confirms that the dynamics tends to a state satisfying rotation and reflection symmetries, $p_2(s_1,s_2)=p_2(s_2,s_1)$, independently of the initial state.

The equations have four trivial stationary solutions, the same as obtained in the mean-field approximation. In one of these solutions $p_1(1)=1$ and $p_2(1,1)=1$, and the other one- and two-site configuration probabilities are zero. For the non-trivial symmetric solution $p_1(1)=p_1(2)=p_1(3)=1/3$, and the pair configuration probabilities turn out to be \citep{tainaka_pre94,sato_mmit97}
\begin{eqnarray}
p_2(1,1)&=& p_2(2,2) = p_2(3,3)= {2d+1 \over 9(2d-1)} \;, \nonumber \\
p_2(1,2)&=& \cdots = p_2(3,2)= {2d-2 \over 9(2d-1)} \;.
\label{eq:rsppass}
\end{eqnarray}
In the limit $d \to \infty$ this solution reproduces the prediction of the mean-field approximation, $p_2(s_1,s_2)=p_1(s_1)p_1(s_2)=1/9$. In contrast with this, in $d=1$ the pair approximation predicts a vanishing probability for finding two different species on neighboring sites. This is due to a domain growth process, to be discussed in Sec.\ \ref{sec:rsp1d}.

On the square lattice ($d=2$) the pair approximation predicts
$p_2^{\rm (2p)}(1,1)=5/27
\approx 0.185$,
which is significantly lower than the Monte Carlo result, $p_2^{\rm (MC)}(1,1)=0.24595$.
In Appendix \ref{app:gmfa} we show how to deduce a correlation length from the pair configuration probabilities for two-state systems [see Eq.~(\ref{eq:xi_q})]. Using this technique we can determine the correlation length for the corresponding autocorrelation function,
\begin{equation}
    \xi^{\rm (2p)}= - {1 \over \ln \left(\frac{9}{2}\, [p_2(1,1)-p_1(1) p_1(1)]\right)} \;.
    \label{eq:rspxi_q}
\end{equation}
Substituting the solution (\ref{eq:rsppass}) into Eq.\ (\ref{eq:rspxi_q}) we find $\xi^{\rm (2p)}=1 / \ln 3 = 0.91$ for $d=2$, which is again significantly less than the simulation result.

\begin{figure}[ht]
\centerline{\epsfig{file=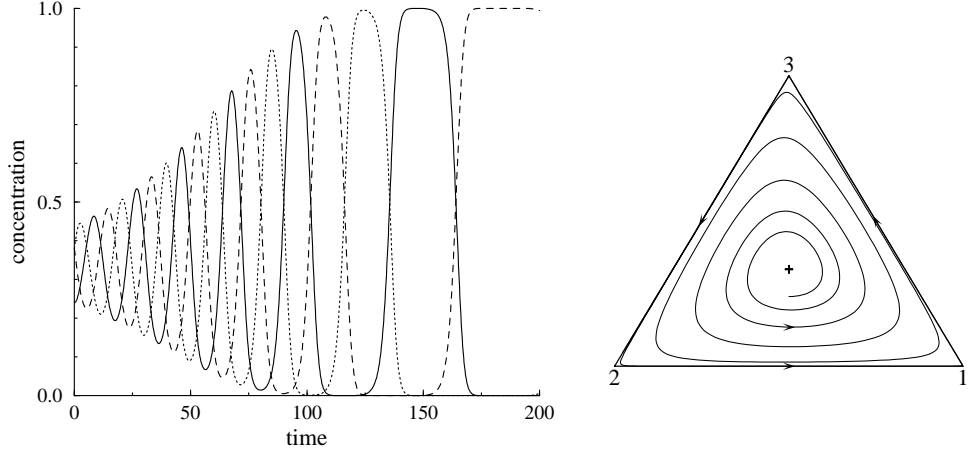,width=13cm}}
\caption{\label{fig:rsppair}Time-dependence of strategy concentrations (left) as predicted by the pair approximation for $d=2$. The system is started from an uncorrelated initial state with $p_1(1)=p_1(2)=0.36$ and $p_1(3)=0.24$. The corresponding trajectory spirals out and approaches the edges of the triplex (right).}
\end{figure}

For $d=2$ the numerical integration gives increasing oscillations in the strategy concentrations \citep{tainaka_pre94}. Figure \ref{fig:rsppair} illustrates that the amplitude of these oscillations approaches a saturation value with an exponentially increasing time period. In practice the theoretically endless oscillations are stopped by fluctuations (or rounding errors in the numerical integration), and the evolution terminates in one of the absorbing states. A similar behavior was found for higher dimensions too, with a decreasing pitch of the spiral for increasing $d$. The general properties of these so-called heteroclinic cycles and their relevance to ecological models were described and discussed in a more general context by \citet{may_siam75} and \citet{hofbauer_88}.

The above numerical investigations suggest that the stationary solutions of the pair approximation are unstable for small perturbations. This conjecture was confirmed analytically by \citet{sato_mmit97}.
This property of the pair approximation is qualitatively different from the mean-field or the Monte Carlo results, where the central solution is stable or asymptotically stable. This shortcoming of the pair approximation can be eliminated by using the more accurate four- and nine-site approximations \citep{szabo_jpa04}, which determine the probability of all possible configurations in the corresponding block. These approaches reproduce qualitatively well the spiral trajectories converging towards the symmetric stationary solution, as was found by the Monte Carlo simulations (see Fig.~\ref{fig:rspspiri}).

\subsection{One-dimensional models}
\label{sec:rsp1d}

\begin{figure}[ht]
\centerline{\epsfig{file=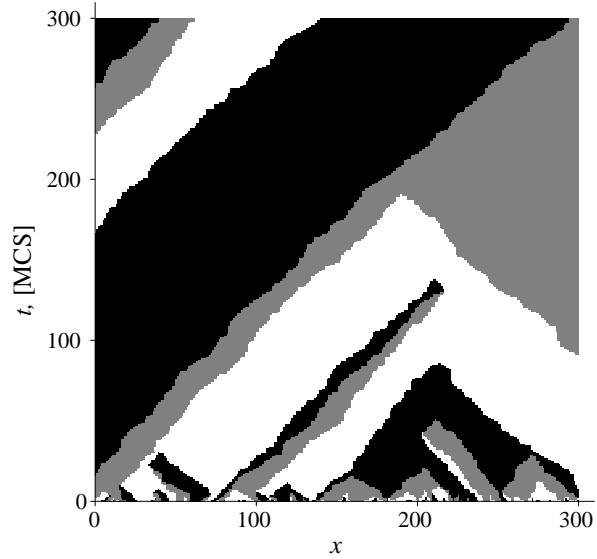,width=8cm}}
\caption{\label{fig:rspp1d}Time evolution of domains for random sequential update in the one-dimensional case.}
\end{figure}

In the one-dimensional case the boundaries separating homogeneous domains move left or right with the same average velocity. If two domain walls collide a domain disappears as is illustrated in Fig.~\ref{fig:rspp1d}. At the same time, the average size of the domains increases. Using Monte Carlo simulations \citet{tainaka_jpsj88}
found that the average number of domain walls decreases with time as $n_w \sim t^{-\alpha}$, where $\alpha \simeq 0.8$, contrary to the prediction of the pair approximation, which suggests $\alpha=1$. We will see that this discrepancy is related to the formation of ``superdomains" \citep{frachebourg_prl96,frachebourg_pre96}.

In this system one can distinguish six types of domain walls (or kinks), whose concentration is characterized by the pair configuration probabilities $p_2(s_1,s_2)$ ($s_1 \ne s_2$). The compatibility conditions, discussed in Appendix \ref{app:gmfa}, yields three constraints for the domain wall concentrations, $\sum_{s_2} p_2(s_1,s_2)=\sum_{s_2} p_2(s_2,s_1)$ ($s_1=1$, 2, and 3). Despite naive expectations these constraints allow the breaking of the spatial reflection symmetry in such a way that $p_2(1,2)-p_2(2,1) = p_2(2,3)-p_2(3,2) = p_2(3,1)-p_2(1,3)$. In particular cases the system can develop into a state, where all  interfaces move either to the right [$p_2(1,2)=p_2(2,3)=p_2(3,1)=0$] or to the left [$p_2(2,1)=p_2(3,2)=p_2(1,3)=0$]. Allowing this type of symmetry breaking, now we recall the time-dependent solution of the pair approximation for the one-dimensional case.

For the ``symmetric" solution we assume that $p_1(1)=p_1(2)=p_1(3)=1/3$, $p_2(1,2) = p_2(2,3) = p_2(3,1) = \rho_l/3$, and  $p_2(2,1) = p_2(3,2) = p_2(1,3) = \rho_r/3$, where $\rho_l$ and $\rho_r$ denote the total concentration of left and right moving interfaces. Under these conditions the pair approximation leads to 
\begin{eqnarray}
\dot{\rho}_l = -2 \rho_l^2 - 2 \rho_l \rho_r + \rho_r^2 \;, \nonumber \\
\dot{\rho}_r = -2 \rho_r^2 - 2 \rho_r \rho_l + \rho_l^2 \;.
\label{eq:rsprholr}
\end{eqnarray}
Starting from a random, uncorrelated initial state the time-dependence of the interface concentrations obeys the following form \citep{frachebourg_pre96}:
\begin{equation}
\rho_l(t)=\rho_r(t)= {1 \over 3+3t}\;.
\label{eq:rsp1dsol}
\end{equation}
We remind the reader that this prediction does not agree with the result of the Monte Carlo simulations, and the correct description requires a more careful analysis.

Figure \ref{fig:rspp1d} clearly shows that the evolution of the domain structure is governed by two basic types of elementary processes \citep{tainaka_prl89}. If two opposite moving interfaces collide they annihilate each other. On the other hand, the collision of two right (left) moving interfaces leads to their mutual annihilation, and simultaneously the creation of a left (right) moving interface. Equations (\ref{eq:rsprholr}) can be considered as the corresponding rate equations \citep{frachebourg_pre96}.

As a result of these elementary processes, one can observe the formation of superdomains (see Fig.~\ref{fig:rspp1d}), in which all the internal interfaces move to the same direction. The coarsening pattern can be characterized by two different length scales as illustrated by the following configuration:
\begin{equation}
P\overbrace{SSSPPPPRRRR\underbrace{SSSS}_{\cal \ell}PPRRRSSS}^{\cal L}R \;.
\label{eq:rsp1d2l}
\end{equation}
Here ${\cal L}$ denotes the size of the superdomain and ${\ell}$ is the size of a classical domain. Evidently, the average size of domains is related directly to the total number of interfaces, i.e., $L (\rho_l+\rho_r) = \langle {\ell} \rangle$.

Inside superdomains the interfaces perform biased random walks. Consequently, two parallel moving interfaces can meet and annihilate each other while a new, opposite moving interface is created. This in turn is annihilated by a suitable neighboring interface within a short time. This domain wall dynamics is similar to the one-dimensional ballistic annihilation process, when each particle has a fixed average velocity which may be either +1 or -1 with equal probability \citep{ben-naim_jpa96}. Using scaling arguments \citet{frachebourg_prl96,frachebourg_pre96} have shown that in these systems the superdomain and domain sizes growth respectively as
\begin{equation}
\langle {\cal L}(t) \rangle \sim t \;
\label{eq:rsp1dLt}
\end{equation}
and
\begin{equation}
\langle \ell(t) \rangle \sim t^{3/4} \;.
\label{eq:rsp1dLtb}
\end{equation}
Monte Carlo simulations 
have confirmed the above theoretical predictions. \citet{frachebourg_pre96} have observed that the average domain size increases approximately as $\langle \ell (t) \rangle^{\rm{(MC)}} \sim t^{\alpha}$ with $\alpha \simeq 0.79$, while the local slope $\alpha (t)= d \ln \ell (t) / d \ln t$ approaches the asymptotic value $\alpha = 3/4$.
This is an interesting result, because separately both the diffusion-controlled and the ballistic controlled annihilation processes yield the same coarsening exponent 1/2, whereas their combination gives a higher value, 3/4.


\subsection{Global oscillations on some structures}
\label{sec:rspgors}

\begin{figure}[ht]
\centerline{\epsfig{file=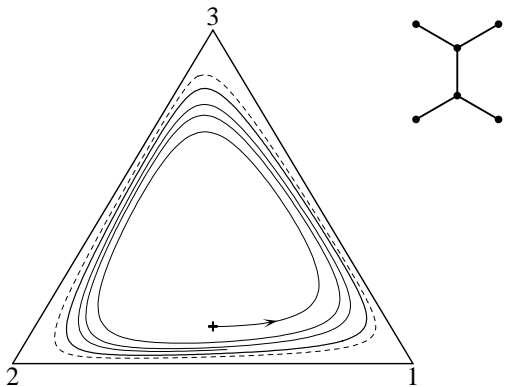,width=7cm}}
\caption{\label{fig:rsprrg3}Emergence of global oscillations in the strategy concentrations for the Rock-Scissors-Paper game on random regular graphs with $z=3$. A Monte Carlo simulation for $N=3\cdot 10^6$ sites shows that the system evolves from an uncorrelated initial state (denoted by the plus sign) towards a limit cycle (thick orbit). The dashed  orbit indicates the limit cycle predicted by the generalized mean-field method at the six-site approximation level. The corresponding six-site cluster is indicated on the right.}
\end{figure}

The pair approximation cannot handle short loops, which characterize all $d$-dimensional spatial structures. It is expected, however, that on tree-like structures, e.g., on the Bethe lattice, the qualitative prediction of the pair approximation can be valid. The Rock-Scissors-Paper game on the Bethe lattice with a coordination number $z$ was investigated by \citet{sato_mmit97} using the pair approximation. Their result is equivalent to those discussed above when substituting $z$ for $2d$ in Eq.~(\ref{eq:rsppass}). Unfortunately, the validity of these analytical results cannot be confirmed directly by Monte Carlo simulations, because of boundary effects present in any finite system. It is expected, however, that the growing spiral trajectories seen Fig.~\ref{fig:rsppair} can be observed on random regular graphs for sufficiently large $N$. The Monte Carlo simulations \citep{szolnoki_pre04a} have confirmed this analytical result on random regular graphs with $z=6$ (and similar behavior is expected for $z>6$, too). However, for $z=3$ and $4$ the  evolution tends towards a limit cycle as demonstrated in Fig.~\ref{fig:rsprrg3}.

For $z=3$ the generalized mean-field analysis can also be performed with four- and six-site clusters. The four-site approximation gives growing spiral trajectories, whereas the six-site approximation predicts a limit cycle with good quantitative agrement with the Monte Carlo simulation (see the thick solid and dashed lines in Fig.~\ref{fig:rsprrg3}). It is conjectured that the growth of the global oscillation is blocked by fluctuations whose role increases with decreasing $z$.

Different quantities can be introduced to characterize the extension of limit cycles. The simplest quantity is the amplitude, i.e., the difference between the maximum and minimum values of the strategy concentrations. However, this quantity is strongly affected by fluctuations, whose magnitude depends on the size of the system. There are other ways to quantify limit cycles, which are less sensitive to noise and to system size effects. For example, the limit cycle can be characterized by its average distance from the center. Besides it, one can define an order parameter $\Phi$ as the average relative area of the limit cycle on the ternary phase diagram (see Fig.~\ref{fig:rsprrg3}), compared to its maximum possible value (the area of the triangle). According to Monte Carlo results on random regular graphs, $\Phi^{\rm (RRG)}=0.750(5)$ and 0.980(1) for $z=3$ and 4, respectively \citep{szabo_jpa04,szolnoki_pre04b}.

An alternative definition of the order parameter can be based on the constant of motion in the mean-field approximation Eq.~(\ref{eq:rspmfcm}), i.e., $\Phi^{\prime}=1-3^3 \langle p_1(1) p_1(2) p_1(3) \rangle$. It was found that the product of strategy concentrations only exhibits a weak periodic oscillation along the limit cycle (the amplitude is about a few percent of the average value). The order parameters $\Phi$ and $\Phi^{\prime}$ can be easily calculated either in simulations or in the generalized mean-field approximation.

Evidently, both order parameters become zero on the square lattice and 1 on the random regular graph with a degree of $z=6$ and above. Thus, using these order parameters we can analyze quantitatively the transitions between the above mentioned three different sort of behaviors.

Now we present a simple model exhibiting two subsequent phase transitions when an adequate control parameter $r$ is tuned. The parameter $r$ characterizes the temporal randomness in the connectivity structure \citep{szabo_jpa04}. For $r=0$ the model is equivalent to the  spatial Rock-Scissors-Paper game on the square lattice. A randomly chosen player adopts one of the randomly chosen neighbor's strategy provided that the latter is dominant. With probability $0 < r < 1$ a standard (nearest) neighbor is replaced  by a randomly chosen other player in the system, and irrespectively of their distance a strategy adoption event occurs. When $r=1$, partners are chosen uniformly randomly, and the mean-field approximation becomes exact for sufficiently large $N$.

\begin{figure}[ht]
\centerline{\epsfig{file=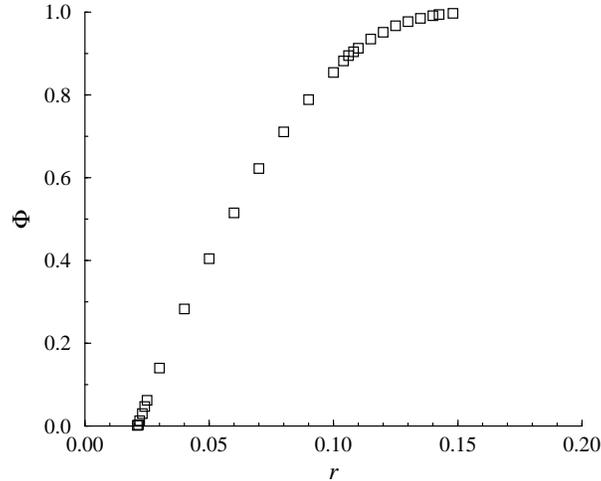,width=8cm}}
\caption{\label{fig:rspsqpmc}Monte Carlo results for $\Phi$, the relative area of the limit cycle, as a function of $r$, the probability of choosing random co-players instead of nearest neighbors on the square lattice.}
\end{figure}

The results of the Monte Carlo simulation in Fig.~\ref{fig:rspsqpmc} show three qualitatively different behaviors. For weak temporal randomness [$r< r_1=0.020(1)$] the spatial distribution of the strategies remains similar to the snapshots observed for $r=0$ (see Fig.~\ref{fig:rsppatt}). On the sites of the lattice the three strategies cyclically follow each other in a local sense. The short range interactions are not able to synchronize these local oscillations.

In the region $r_1<r<r_2=0.170(1)$ global oscillations occur. Finally for $r>r_2$, following a growing spiral trajectory, the (finite) system evolves into one of the three homogeneous states and remains there forever. The numerical values of $r_1$ and $r_2$ are surprisingly low, and indicate high sensitivity to the introduction of temporal randomness. Similar transitions can be observed for quenched randomness in random regular small-world structures. In this case, however, the corresponding threshold values are higher \citep{szolnoki_pre04b}.

The appearance of the global oscillation is a Hopf bifurcation \citep{hofbauer_98,kuznetsov_95}, as is indicated by the linear increase of $\Phi$ above $r_1$.\footnote{Note that in a Hopf bifurcation the amplitude increases by a square-root law above the transition point. However, in our case the order parameter is the area $\Phi$ which is quadratic in the amplitude.}

When $r \to r_2$ the order parameter $\Phi$ goes to 1 very smoothly. The Monte Carlo data are consistent with a power-law behavior: $1-\Phi \sim (r_2-r)^{\gamma}$ where $\gamma = 3.3(4)$. This power-law behavior seems to be a very robust feature of three-state systems with cyclic symmetries, because similar exponents were found for several other (regular) connectivity structures. It is worth noting that the generalized mean-field technique at the six-site approximation level (the corresponding cluster is shown in Fig.~\ref{fig:rsprrg3}) reproduces this behavior on the Bethe lattice with degree $z=3$ \cite{szolnoki_pre04b}.

The generalization of this model on other non-regular (quenched or temporal) connectivity structures is not straightforward. Different rules can be postulated to handle the variations in the degree. For example, within an elementary invasion process one can choose a site and one of its neighbors randomly, or one can select one of the edges of the connectivity graph with equal probability (the latter case favors the selection of sites with larger degree). \citet{masuda_pre06} have studied a case when the randomly chosen site (a potential predator) invades all possible prey sites in its neighborhood. The simulations indicated clearly that the system develops into the symmetric stationary state on both Erd{\H o}s-R\'enyi and scale-free (Barab\'asi-Albert) networks. In the light of these results it would be interesting to see what happens on other non-regular networks when using different evolutionary rules, and to clarify the relevant ingredients affecting the final stationary state.

\subsection{Rotating spiral arms}
\label{sec:rspspir}

The self-organizing spatiotemporal patterns of Rock-Scissors-Paper game on the square lattice are characterized by moving invasion fronts. Let us recall first the relevant topological difference between three- and two-color patterns. On the continuous plane two-color patterns are topologically equivalent to a structure of "islands in lakes in islands in lakes in ...", at least if four-edge vortices are neglected. In fact the probability of these four-edge vertices vanishes both in the continuum limit and also in the advanced states of the domain growth process for smooth interfaces. Furthermore, on the triangular lattice four-edges vortices are forbidden topologically. In contrast with this, three-edge vertices are distinctive points where three domains (or domain walls) meet on a three-color pattern. Due to cyclic dominance among the three states, the edges of these objects rotate clockwise or anti-clockwise. We will call these structures vortices and anti-vortices, respectively.

The three edges of the vortices and anti-vortices form spiral arms, because their average normal velocity is approximately constant. At the same time, the deterministic part of the interface motion is strongly decorated by noise, and the resultant irregularity is able to suppress the "deterministic" features of these rotating spiral arms.

\begin{figure}[ht]
\centerline{\epsfig{file=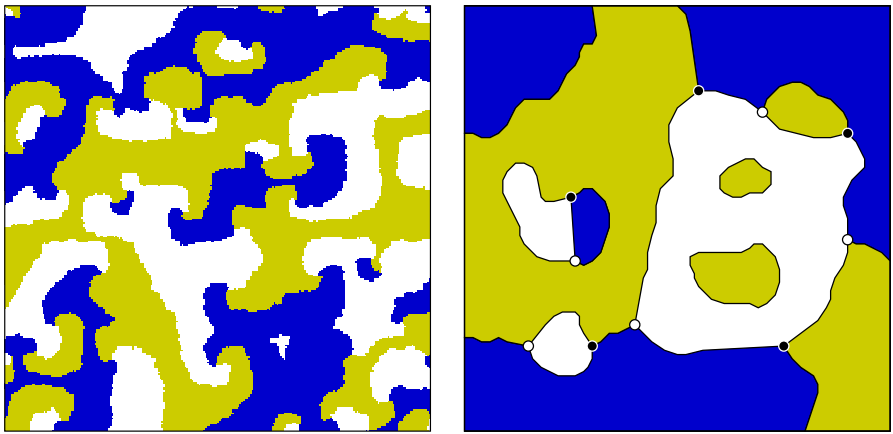,width=12cm}}
\caption{\label{fig:rspvavp}The left snapshot shows a typical rotating spiral structure of the three strategies on a block of $500 \times 500$ sites of a square lattice. The invasion rate in (\ref{eq:rsppmv}) is $\kappa=4/3$, and $\epsilon=0.05$. The schematic plot (right) illustrates how vortices (black dots) and anti-vortices (white dots) are positioned in the three-color model. A vortex and anti-vortex can be connected to each other by one, two, or three edges.}
\end{figure}

Rotating spiral arms become nicely visible in models, where the motion of the invasion fronts is affected by some surface tension. As the schematic three-color pattern in the right panel of Fig.~\ref{fig:rspvavp} indicates vortices and anti-vortices alternate each other along interfaces. In other words, a vortex is linked to three (not necessary different) anti-vortices by its arms, and vice versa. It also implies that vortices and anti-vortices are created or annihilated in pairs during interface motion \citep{szabo_pre99}.

Smooth interfaces can be observed for dynamical rules that suppress the probability of elementary processes trying to increase the length of an interface. This requirement can be fulfilled by a simple model \citep{szabo_pre02a}, where the probability of nearest neighbor ($y \in \Omega_x$) invasions is
\begin{equation}
    P (s_x \to s_y) =
    {1 \over 1+\exp[\kappa\, \Delta E_P + \varepsilon\, A (s_x,s_y) ]} \;.
    \label{eq:rsppmv}
\end{equation}
Cyclic dominance is built into the model through the payoff matrix (\ref{eq:rsppom}) with a magnitude denoted by $\varepsilon$ and $\kappa$ characterizing how interfacial energy is taken into account. The total length of the interface is defined by the Potts energy \citep{wu_rmp82},
\begin{equation}
    E_P = \sum_{\langle x,y \rangle} [1-\delta(s_x,s_y)]\;,
    \label{eq:pottse}
\end{equation}
where the summation runs over all nearest neighbor pairs and $\delta(s,s^{\prime})$ is the Kronecker delta. $\Delta E_P$ denotes the difference in the interface length between the final and the initial states. Notice that this rule enhances (suppresses) those elementary invasion events which decrease (increase) the total length of the interface. It also inhibits the creation of new islands inside homogeneous domains. For $\kappa>2$ the formation of horizontal and vertical interfaces is favored (as for the Potts model at low temperatures during the domain growth process) due to the anisotropic interfacial energy on the square lattice. However, this undesired effect becomes irrelevant in the region $\kappa \le 4/3$ (see left snapshot of Fig.~\ref{fig:rspvavp}), where our analysis mostly concentrated.

For $\kappa=0$, the interfacial energy does not play a role, and the model reproduces a version of the Rock-Scissors-Paper game, in which the probability of the direct invasion process is reduced, $P=[1+\tanh (\varepsilon /2)]/2<1$, and the inverse process is also allowed with probability $1-P$. \citet{tainaka_epl91} demonstrated that when $P \to 1/2$ (or $\varepsilon \to 0$) this model becomes equivalent to the \emph{three-state voter model} \citep{liggett_85,clifford_bm73,holley_ap75}. The behavior of the voter model depends on the spatial dimension $d$ \citep{ben-naim_pre96}. In two dimensions a very slow (logarithmic) domain coarsening occurs for $P=1/2$. In the presence of cyclic dominance ($\varepsilon > 0$ or $P>1/2$), however, the domain growth stops at a "typical size" that diverges as $P\to 1/2$. The mean-field aspects of this system were considered by \citet{ifti_epje03}.

\begin{figure}[ht]
\centerline{\epsfig{file=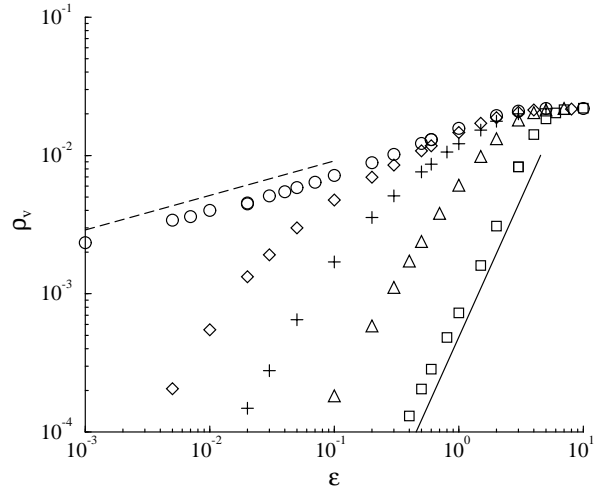,width=8cm}}
\caption{\label{fig:rsppmrv}Log-log plot of the concentration of vortices as a function of $\varepsilon$ for $\kappa=1$ (squares), 1/4 (triangles), 1/16 (pluses), 1/64 (diamonds), and 0 (circles). The solid (dashed) line represents a power-law behavior with an exponent of 2 (resp., 1/4).}
\end{figure}

In order to quantify the divergence of the typical length scale \citet{tainaka_epl91}  considered the average concentration of vortices and found a power-law behavior, $\rho_v \sim |P-1/2|^{\beta} \sim |\varepsilon|^{\beta}$. The numerical fit to the Monte Carlo data predicted $\beta =0.4$. Simulations performed by \citet{szabo_pre02a} on larger systems indicate a decrease in $\beta$ for smaller $\varepsilon$. Apparently, the Monte Carlo data in Fig.~\ref{fig:rsppmrv} are consistent with $\beta=1/4$. Unfortunately this conjecture is not yet supported by theoretical arguments, and a more rigorous numerical confirmation is hindered by the fast increase of the relaxation time as $\varepsilon \to 0$.

Figure \ref{fig:rsppmrv} shows a strikingly different behavior when the Potts energy is switched on, i.e., $\kappa>0$. The Monte Carlo data indicates a faster decrease in the concentration of vertices, tending towards a quadratic behavior, $ \rho_v \sim \varepsilon^2$, in the limit $\varepsilon \to 0$ for any $0 < \kappa \le 1$.

The geometrical features of the interfaces in Fig.~\ref{fig:rspvavp} demonstrate that these self-organizing patterns cannot be characterized by a single length scale [e.g., the correlation length, the average (horizontal or vertical) distance between the interfaces, or the inverse of $E_P/2N$] as it happens for traditional domain growing processes [for a survey on ordering phenomena see the review by \citet{bray_ap94}]. In the present case additional length scales can be introduced to characterize the average distance between the connected vortex--anti-vortex pairs or to describe the geometrical features of the (rotating) spiral arms.

On the square lattice the interfaces are polygons consisted of unit length elements whose relative tangential rotation can be $\Delta \theta= 0$ or $\pm \pi/2$. The tangential rotation of a vortex edge can be determined by summing up the values of $\Delta \theta$ along the given edge step by step from the vortex to the connected anti-vortex \citep{szabo_pre02a}. One can also determine the total length of the interface connecting a vortex--anti-vortex pair, as well as the average curvature for the given polygon. The average value of these quantities characterize the pattern. Beside vortex--anti-vortex edges one can observe islands, whose total perimeter can be evaluated in the knowledge of the Potts energy Eq.\ (\ref{eq:pottse}) and the total length of the vortex edges. Furthermore, vortices can also be classified by the number of anti-vortices, to which the given vortex is linked by vortex edges \citep{szolnoki_pre04a}. For example, the collision of two different islands (e.g., states $1$ and $2$ in the sea of state $0$) creates a vortex--anti-vortex pair linked to each other by three common vortex edges (see Fig.~\ref{fig:rspvavp}). If the evolution of the interfaces is affected by some surface tension reducing the total length, then this process will decrease the portion of vortex--anti-vortex pairs linked to each other by more than one edge.

Unfortunately, a geometrical analysis of the vortex edges on the square lattice requires  removal of all four-edge vertices for the given pattern. During this manipulation the four-edge vertex is considered as an instantaneous object which is about to transform into a vortex--anti-vortex pair or into two non-crossing interfaces. In many cases the effect of this manipulation is negligible and the subsequent geometrical analysis gives useful information about the three-color pattern \citep{szabo_pre02a}. It turns out, for example, that the average tangential rotation $\theta$ of the vortex edges increases as $\epsilon$ decreases, if the surface tension is switched on.
However, this quantity goes smoothly to zero when $\epsilon \to 0$ for $\kappa=0$ (voter model limit). In this latter case the vortex edges do not show the characteristic features of "spiral arms". The contribution of islands to the total length of interfaces is practically negligible for $\kappa > 0$. Consequently, pattern formation is governed dominantly by the confronting rotating spiral arms.

In the absence of surface tension ($\kappa=0$) \emph{interfacial roughening} \citep{provata_pre03,dornic_prl01} plays a crucial role as is illustrated by Fig.~\ref{fig:rsppatt}. The interfaces become more and more irregular, and the occasional overhanging brings about a mechanism that creates new islands. Moreover, the confronting fragments can also enclose a homogeneous territory and create islands. Due to these mechanisms the total perimeter of the islands becomes comparable to the total length of the vortex edges \citep{szabo_pre02a}. The resultant patterns and dynamics differ drastically from those appearing for smooth interfaces.

Despite the huge theoretical efforts aimed to clarify the main features of spiral patterns [for a review see \citet{cross_rmp93}], our current knowledge is still rather poor about the relationship between geometrical and dynamical characteristics. Using a geometrical approach suggested by \citet{brower_pra84}, the time evolution of an invasion front between a fixed vortex--anti-vortex pair was studied numerically by \citet{meron_prl88}. Unfortunately, the deterministic behavior of a single rotating spiral front has not yet been compared to the geometrical parameters averaged over vortex edges in the self-organizing patterns. Possible geometrical parameters of interest can be the  concentration of vortices ($\rho_v$), the total length of interfaces (or Potts energy), the total length and average tangential rotation (or curvature) of the vortex edges, whereas the dynamical parameters are those that describe the average normal velocity of invasion fronts and quantify interfacial roughening.

On the simple cubic lattice the above evolutionary Rock-Scissors-Paper game exhibits a self-organizing pattern, whose two-dimensional cross-section is similar to the one illustrated in Fig.~\ref{fig:rsppatt} \citep{tainaka_pre94}. The core of the vortices form directed strings in three dimensions, whose direction refers to being a vortex or an anti-vortex on the 2D cross-section. The core string typically closes in a ring, resembling a long thin toroid. The possible topological features of how the loops of these vortex strings are formed and linked together are discussed theoretically by \citet{winfree_pd84}. The three-dimensional simulations of the Rock-Scissors-paper game by \citet{tainaka_pre94} can serve as an illustration of the basic properties. Here it is worth mentioning that the investigation of evolving vortex string networks has a tradition in different areas of physics from topological defects in solids to cosmic strings \citep{vilenkin_94}. Many aspects of the associated dynamical rules and the generation of entangled string networks have also been extensively studied for superfluid turbulence [for surveys see \citet{schwarz_prl82,martins_prl04,bradley_prl05}]. The predicted richness of possible behaviors has not yet been fully identified in cyclically dominated three-state games.

\subsection{Cyclic dominance with different invasion rates}
\label{sec:rspur}

The Rock-Scissors-Paper game described above remains invariant for a cyclic permutation of species (strategies). However, some basic features of this system can also be observed when the exact symmetry does not hold. In order to demonstrate the distortion of the solution, first we study the effect of unequal invasion rates within the mean-field approximation. In this case the equations of motion [whose symmetric version was given in Eqs.~(\ref{eq:rspmf})] take the form:
\begin{eqnarray}
\dot{p}_1(1)&=& p_1(1)[w_{12}p_1(2)-w_{31}p_1(3)] \;, \nonumber \\
\dot{p}_1(2)&=& p_1(2)[w_{23}p_1(3)-w_{12}p_1(1)] \;, \nonumber \\
\dot{p}_1(3)&=& p_1(3)[w_{31}p_1(1)-w_{23}p_1(2)] \;,
\label{eq:rspmfas}
\end{eqnarray}
where $w_{s\hat{s}}>0$ defines the invasion rate between the predator $s$ and its prey $\hat{s}$. As for the symmetric case, these equations have three trivial homogeneous solutions,
\begin{eqnarray}
p_1(1)&=& 1 \;, \quad p_1(2)= 0, \quad p_1(3)=0 \;; \nonumber \\
p_1(1)&=& 0 \;, \quad p_1(2)= 1, \quad p_1(3)=0 \;; \nonumber \\
p_1(1)&=& 0 \;, \quad p_1(2)= 0, \quad p_1(3)=1 \;,
\label{eq:rspcsol}
\end{eqnarray}
which are unstable against the attack of the corresponding predators as discussed in Section \ref{sec:rspmf}. Furthermore, the system has
another stationary solution which describes the coexistence of all three species with  concentrations
\begin{equation}
p_1(1) = {w_{23} \over W} \;, \quad
p_1(2) = {w_{31} \over W} \;, \quad
p_1(3) = {w_{12} \over W} \;,
\label{eq:rspcsolb}
\end{equation}
where $W= w_{12}+w_{23}+w_{31}$. It should be emphasized that all species survive with the above concentrations, i.e., cyclic dominance provides a way for their coexistence in a wide range of parameters \citep{gilpin_an75,may_siam75,tainaka_pla93,durrett_jtb97}.

Notice, however, the counterintuitive response to the variation of the invasion rates. Naively, one may expect that the predator of the largest invasion rate should be in the most beneficial position. Contrary to this, Eq.~(\ref{eq:rspcsolb}) claims that the equilibrium concentration of a given species is proportional to the invasion rate of its prey. In fact, this unusual behavior is a direct consequence of cyclic dominance in all three-state systems.

In order to clarify this unexpected response, let us consider a simple example. Suppose that the concentration of species 1 is increased externally. In this case species 1 consumes more of species 2, whose concentration decreases. As a result, species 2 consumes less of species 3. Finally, species 3 gets the advantage in the dynamical balance, because it has more prey and less predators. In this example the external support can be realized by either increasing the corresponding invasion rate (here $w_{12}$), or converting randomly chosen individuals into species 1. The latter situation was analyzed by \citet{tainaka_pla93}, who studied what happens in a three-candidate (cyclic) voter model when one of the candidates is supported by the mass media during an election campaign. Further biological examples are discussed in \citet{frean_prsb01} and \citet{durrett_02}. The rigorous mathematical treatment of these equations of motion was given by \citet{hofbauer_88,gao_pla99,gao_pla00}.

\begin{figure}[ht]
\centerline{\epsfig{file=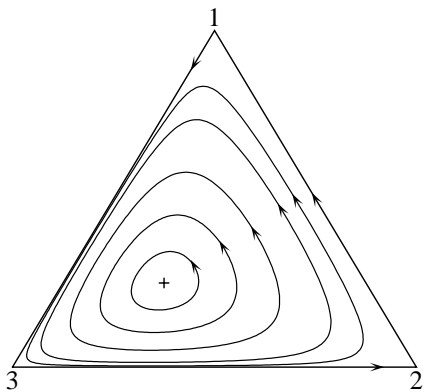,width=6cm}}
\caption{\label{fig:rspmfas}
Concentric orbits on the ternary phase diagram if the invasion rates are $w_{12}=2$ and $w_{23}=w_{31}=1$. The strategy concentrations in the central stationary state (indicated by the plus symbol) are shifted towards the predator of the  seeming beneficiary of the largest invasion rate.}
\end{figure}

In analogy to the symmetric case, a constant of motion can be constructed for Eqs.~(\ref{eq:rspmfas}). It is easy to check that
\begin{equation}
    m=p_1^{w_{23}}(1) p_1^{w_{31}}(2) p_1^{w_{12}}(3) = \mbox{constant} \;.
    \label{eq:rspmfascm}
\end{equation}
This means that the system evolves along closed orbits, i.e., the concentrations return periodically to their initial values as illustrated in Fig~\ref{fig:rspmfas}.

Shortly, according to the mean-field analysis the qualitative behavior of such asymmetric Rock-Scissors-Paper systems remain unchanged, despite the fact that different invasion rates break the cyclic symmetry \citep{ifti_epje03}. A special case of the above system with $w_{12}=w_{23} \ne w_{31}$ was analyzed by \citet{tainaka_pla95} using the pair approximation, which confirmed the above predictions. Furthermore, the pair approximation predicts that trajectories spiral out (as discussed for the symmetric case) if the system starts from a random initial state. Evidently, due to the absence of cyclic symmetry the three absorbing states are reached with unequal probabilities.

Simulations have confirmed that most of the relevant features are hardly affected by the loss of cyclic symmetry on spatial structures either.
The appearance of self-organizing patterns (sometimes with rotating spiral arms) were reported for many three-state models. Examples are the forest-fire models \citep{bak_pla90,drossel_prl92} and ecological models \citep{durrett_jtb97} mentioned already. Several aspects of phase transitions, spatial correlations, fluctuations, and finite size effects are studied by \citet{antal_pre01,mobilia_pre06,mobilia_jsp06} who considered lattice Lotka-Volterra models.

For the SIRS models the transitions $I \to R$ and $R \to S$ are not affected by the neighbors, and this may be the reason why the mean-field and pair approximations predict damping oscillations on hypercubic lattices for a wide range of parameters \citep{joo_pre04}. At the same time, emerging global oscillations were reported by \citet{kuperman_prl01} on small-world structures when they tuned the ratio of rewired links. For the latter system the amplitude of global oscillations never reached the saturation value (excluding finite size effects), i.e., a transition from the limit cycle to one of the absorbing states was not observed. Further analysis is needed to clarify the possible conditions (if any) for two subsequent phase transitions in these systems.

The unusual response for invasion rate asymmetries can also be observed in three-strategy evolutionary Prisoner's Dilemma games (discussed in Sec.\ \ref{sec:pd3s}) where cyclic dominance occurs. For example, if the original strategy adoption mechanism among defectors ($D$), cooperators ($C$), and "Tit-for-Tat" strategists ($T$) (recall that $D$ beats $C$ beats $T$ beats $D$) is superimposed externally by an additional probability of replacing randomly chosen $T$ strategists by $C$ strategists, then it is $D$ who eventually benefits \citep{szabo_pre00a}. In voluntary Public Good \citep{hauert_s02,szabo_prl02} or Prisoner's Dilemma \citep{szabo_pre02d} games the defector, cooperator, and loner strategies dominate cyclically each other. For all these models, the increase of the defector's income [i.e., the ``temptation to defect" parameter $b$ in Eq.\ (\ref{eq:vpdpo})] yields a decrease in the concentration of defectors, whereas the concentration of the corresponding predator ("Tit-for-Tat" or loner) increases, as was shown in Fig. \ref{fig:vpdsqmc}.

\subsection{Cyclic dominance for $Q > 3$ states}
\label{sec:rspns}

The three-state Rock-Scissors-Paper game can be generalized straightforwardly to $Q > 3$ states. For instance, cyclic dominance for four strategies was discussed by \citet{nowak_s04}. They considered a game where unconditional defectors (AllD) get replaced by the more successful Tit-for-Tat players, who are transformed into Generous Tit-for-Tat strategists due to environmental noise. They, in turn, get conquered by unconditional cooperators (AllC) not incurring the cost of inspection, but who are finally invaded by AllD strategists again. \citet{traulsen_pre03} faced a conceptually similar case when studied a simplified four-strategy version of the tag-based cooperation model studied earlier by \citet{riolo_n01} and \citet{nowak_n98}.

In the general case $Q$ different states (types, strategies, species) are allowed for each site ($s_i=1$, 2, ..., $Q$), and these states dominate each other cyclically (i.e., 1 beats 2, 2 beats 3, ...., and finally $Q$ beats 1). In the simplest case predator invasion occurs on two (randomly chosen) neighboring sites if these are occupied by a predator-prey pair, otherwise nothing happens. The interface is blocked between neutral pairs [e.g., (1,3) or (2,4)]. Evidently, the possible types of frozen (blocked) interfaces increase with $Q$.

Considering this model on the $d$-dimensional hyper-cubic lattice \citet{frachebourg_jpa98} showed that the spatial distribution evolves toward a frozen domain pattern if $Q$ exceeds a $d$-dependent threshold value.  In the one-dimensional case growing domains can be found if $Q<5$, and a finite system develops into one of the homogeneous states \citep{bramson_ap89,fisch_pd90}. During these domain growing processes (for $Q=3$ and 4) the formation of super-domains were studied by \citet{frachebourg_prl96}. They found that the concentrations of the moving and standing interfaces vanish algebraically with different exponents.

\begin{figure}[ht]
\centerline{\epsfig{file=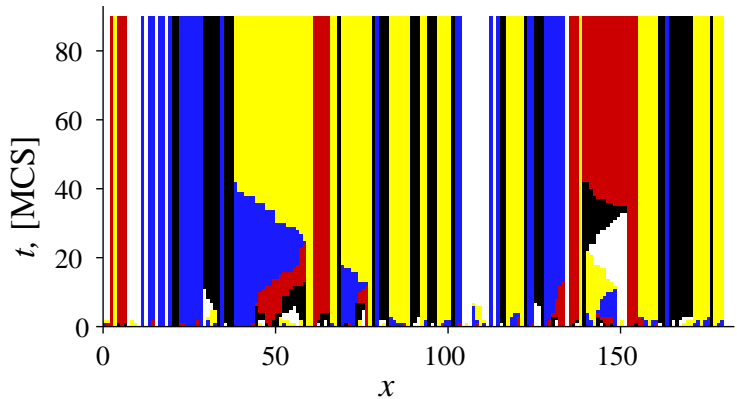,width=10cm}}
\caption{\label{fig:rsp5s1d}Time evolution of domain structure for five-species cyclic dominance in the one-dimensional system.}
\end{figure}

On the one-dimensional lattice for $Q \ge 5$, the invasion front moves until it collides with another interface. Evidently, if two oppositely moving interfaces meet, they annihilate each other. At the same time, the collision of a standing and a moving interface creates a third type of interface, which is either an oppositely moving or a standing one, as is shown in Fig.~\ref{fig:rsp5s1d}. Finally, all moving interfaces vanish and domain evolution halts.

Similar fixation occurs for higher spatial dimension $d$ if $Q \ge Q_{th}(d)$. Using the pair approximation \citet{frachebourg_jpa98} found that $Q_{th}(2)=14$ on the square lattice, and $Q_{th}(3)=23$ on the simple cubic lattice. In these calculations the numerical solution of the corresponding equations of motion were restricted to random uncorrelated initial distribution, and the following symmetries were assumed:
\begin{equation}
p_1(1)=p_1(2)= \cdots = p_1(Q)={1 \over Q} \;,
\label{eq:nsp1}
\end{equation}
\begin{equation}
p_2(s_1,s_2)=p_2(s_2,s_1) \;,
\label{eq:nsp2a}
\end{equation}
and
\begin{equation}
p_2(s_1,s_2)=p_2(s_1+\alpha,s_2+\alpha) \;,
\label{eq:nsp2b}
\end{equation}
where the state variables (e.g., $s_1+\alpha$; $1 \le \alpha < Q$) are taken modulo $Q$. The above predictions of the pair approximation for the threshold values of $Q$ were confirmed by extensive Monte Carlo simulations \citep{frachebourg_jpa98}.

For $Q=4$ the mean-field approximation leads to the following equations of motion (mean-field equations):
\begin{eqnarray}
\dot{p}_1(1)&=& p_1(1)p_1(2)-p_1(1)p_1(4) \;, \nonumber \\
\dot{p}_1(2)&=& p_1(2)p_1(3)-p_1(2)p_1(1) \;, \nonumber \\
\dot{p}_1(3)&=& p_1(3)p_1(4)-p_1(3)p_1(2) \;, \nonumber \\
\dot{p}_1(4)&=& p_1(4)p_1(1)-p_1(4)p_1(3) \;.
\label{eq:rsp4pa1}
\end{eqnarray}
These differential equations conserve the quantities
\begin{equation}
\sum_{s=1}^{Q} p_1(s) = 1 \;
\label{eq:rsp4n}
\end{equation}
and
\begin{equation}
\prod_{s=1}^{Q} p_1(s) = m \;.
\label{eq:rsp4p4}
\end{equation}
One can also derive two other (not independent) constants of motion,
\begin{eqnarray}
p_1(1)p_1(3) &=& m_{13} \;, \nonumber \\
p_1(2)p_1(4) &=& m_{24} \;,
\label{eq:rsp4p13}
\end{eqnarray}
where $m=m_{13} m_{24}$. These constraints assure that the system evolves along closed orbits, i.e., the concentrations return periodically to the initial state.

Notice that the equations of motion (\ref{eq:rsp4pa1}) are satisfied by the usual trivial stationary solutions: the central solution $p_1(1)=p_1(2)=p_1(3)=p_1(4)=1/4$, and by the four homogeneous absorbing solutions, e.g., $p_1(1)=1$ and $p_1(2)=p_1(3)=p_1(4)=0$. Beside these, Eqs. (\ref{eq:rsp4pa1}) have three continuous sets of stationary solutions, which can be parameterized by a single continuous parameter $\rho$. The first two with $0 < \rho < 1$ are trivial solutions,
\begin{equation}
p_1(1) = \rho , \;\; p_1(2)=0 , \;\; p_1(3)=1-\rho , \;\; p_1(4)=0 ,
\label{eq:rsp4s2sa}
\end{equation}
and
\begin{equation}
p_1(1) = 0 , \;\; p_1(2)=\rho , \;\; p_1(3)=0 , \;\; p_1(4)=1-\rho .
\label{eq:rsp4s2sb}
\end{equation}
The third with $0 \ge \rho \ge 1/2$ is non-trivial \citep{sato_amc02}:
\begin{eqnarray}
p_1(1) &=& p_1(3) = \rho \;, \nonumber \\
p_1(2) &=& p_1(4) = {1 \over 2} - \rho \;.
\label{eq:rsp4pnts}
\end{eqnarray}

The generalization of the mean-field equations (\ref{eq:rsp4pa1}) is straightforward for $Q > 4$. In these cases the sum and the product of the strategy concentrations are trivial constants of motion. Evidently, the corresponding homogeneous and central stationary solutions exist, as well as the mixed states involving neutral species in arbitrary concentrations [see Eqs. (\ref{eq:rsp4s2sa}) and (\ref{eq:rsp4s2sb})]. For even $Q$, the suitable version of (\ref{eq:rsp4pnts}) also remains valid. The difference between odd and even $Q$ shows up in the generalization of the constraints (\ref{eq:rsp4p13}). This generalization is only possible for $Q$ even, and the corresponding formulae express that the product of concentrations for even (or odd) label species remains constant during evolution.

In fact, all the above stationary solutions can be realized in lattice models, independently of the spatial dimension. The stability of these phases and their spatial competition will be studied later in Sec.\ \ref{sec:compass}.

Many aspects of cyclic $Q$-species predator-prey systems have not been investigated yet. Like for the Rock-Scissors-Paper game, it would be important to understand, how the topological features of the background graph can effect the emergence of global oscillations and the phenomenon of fixation in these more general systems.

The emergence of rotating spirals in a four-strategy evolutionary Prisoner's Dilemma game was studied by \citet{traulsen_pre04} using both synchronous and asynchronous updates on the square lattice. The four-color domain structure can be characterized by the distribution of (three- and four-edge) vortices. In this case, however, we have to distinguish a large number of different vortex types, not like in the three-state system where there are only two types, the vortex and anti-vortex. For the classification of vortices \citet{traulsen_pre04} used the concept of a topological charge. The spatial distribution of the vortices indicated that generally two cyclically dominated four-edge vortices split into three-edge vortices [with an edge separating the odd (even) label states]. Similar splitting of four-edge vortices was observed by \citet{lin_pre00}, when they studied the four-phase pattern of a reaction-diffusion system driven externally by a periodic perturbation.

\begin{figure}[ht]
\centerline{\epsfig{file=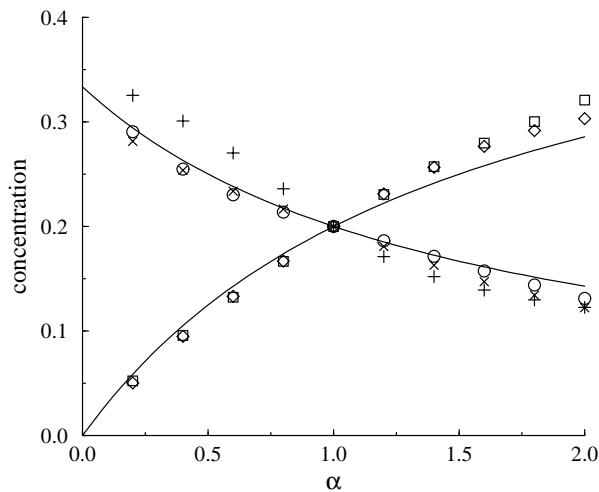,width=8cm}}
\caption{\label{fig:lv5s1a}Concentration of states as a function of $\alpha$ in the five-state model introduced by \citet{sato_amc02}. The Monte Carlo data are denoted by circles, pluses, squares, Xs, and diamonds for the states from 1 to 5. The solid lines show the prediction of the mean-field theory given by Eq.~(\ref{eq:rsp5pmfc}).}
\end{figure}

The lack of additional constants of motion for odd number of species implies a parity effect that becomes more striking when we consider the effect of different invasion rates. \citet{sato_amc02} studied what happens if the invasion rate from species 1 to 2 is varied [$w_{12}=\alpha$ in the notation of Eq. (\ref{eq:rspmfas})], while others are chosen to be unity [$w_{23}=w_{34}= \dots = w_{Q1}=1$ for $Q>3$]. Monte Carlo simulations were performed on a square lattice for $Q=3$, 4, 5, and 6. For $Q=5$ the results were similar to those described above for $Q=3$, i.e., the modification of $\alpha$ is beneficial for the predator of the seemingly favored species. In other words, although invasion rates with $\alpha >1$ seem to provide faster spreading for species 1, its ultimate concentration in the stationary state is the lowest among the species, whereas its predator (species 5) is present with the highest concentration. At the same time the concentration of species 3 is also enhanced, and this variation is accompanied by a reduction in concentration for the remaining species. The magnitude of the concentration variations is approximately proportional to $\alpha-1$, implying reversed tendencies for $\alpha < 1$. The prediction of the mean-field approximation for the central stationary solution,
\begin{eqnarray}
p_1(1)&=& p_1(2) = p_1(4) = {1 \over 3 + 2 \alpha} \;, \nonumber \\
p_1(3)&=& p_1(5) = {\alpha \over 3 + 2 \alpha} \;,
\label{eq:rsp5pmfc}
\end{eqnarray}
agrees qualitatively well with the Monte Carlo data as demonstrated in Fig.~\ref{fig:lv5s1a}. The more general solution for different invasion rates is given in the paper of \citet{sato_amc02}.

\begin{figure}[ht]
\centerline{\epsfig{file=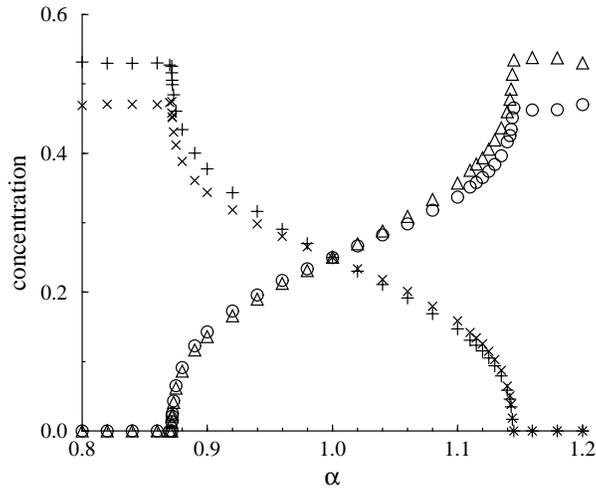,width=8cm}}
\caption{\label{fig:lv4s1a}Concentration of states {\it vs.}\ $\alpha$ characterizing the modified invasion rate from state 1 to 2 in the four-state cyclic predator-prey model. The Monte Carlo data are denoted by circles, pluses, triangles, and crosses for the states labeled from 1 to 4.}
\end{figure}

A drastically different behavior is predicted by the mean-field approximation for even $Q$ \citep{sato_amc02}. When $\alpha > 1$, the even label species die out after some transient phenomenon and finally the odd label species form a frozen mixed state. Conversely, if $\alpha < 1$ all odd label species become extinct. Thus, according to the mean-field approximation, the "central" solution $p_1(1)= \dots = p_1(Q)=1/Q$ only exists if $\alpha =1$ for even $Q$. Early Monte Carlo simulations, performed for the discrete $\alpha$ values, $\alpha = 1 \pm 0.2k$ ($k=0, 1, 2, 3$), seemed to confirm this prediction \citep{sato_amc02}.
However, repeating the simulations on a finer grid around $\alpha = 1$, we can observe the coexistence of all species for $Q=4$ as is shown in Fig.~\ref{fig:lv4s1a}. The stationary concentration of species 1 and 3 increase monotonously above $\alpha_1 \simeq 0.86$, until even label species die out simultaneously around the second threshold value $\alpha_2 \simeq 1.18$. According to our preliminary Monte Carlo results both extinction processes exhibit a similar power law behavior, $p_1(1) \simeq p_1(3) \simeq (\alpha - \alpha_1)^{\beta}$ and $p_1(2) \simeq p_1(4) \simeq (\alpha_2 - \alpha)^{\beta}$, where $\beta \simeq 0.55(5)$ in the close vicinity of the thresholds. A rigorous investigation of the universal features of these transitions is in progress.

For $Q=6$ simulations indicate a very similar picture to that in Fig.~\ref{fig:lv4s1a}. A remarkable difference is due to the fact that small islands of odd label species can remain frozen in large domains of even label species (e.g., species 1 survives forever within domains of species 4, and vice versa), therefore the value of $p_1(1)$, $p_1(3)$, and $p_1(5)$ remains finite below the first threshold value.

A similar parity effect was described by \citet{kobayashi_jpsj97} who considered a lattice version of the Lotka-Volterra model with a linear food web of $Q$ species. In this predator-prey model species 1 invades 2, 2 invades 3, ..., and $(Q-1)$ invades $Q$ with all rates chosen to be 1 except for the last one parameterized by $r$. On the square lattice  invasions between randomly chosen neighboring sites govern evolution. It is assumed, furthermore, that the top predator (species 1) dies randomly with a certain probability and its body is directly transformed into species $Q$. This model was investigated by using Monte Carlo simulations and mean-field approximations. Although the spatial effects are capable to stabilize the coexistence of species through the formation of a self-organizing pattern, \citet{kobayashi_jpsj97} have discovered fundamental differences depending on the parity of the number of species in these ecosystems.

The most relevant feature of the observed parity law can be interpreted as an interference phenomenon between direct and indirect effects \citep{kobayashi_jpsj97}. Assume that the concentration of one of the species is increased or decreased directly. This variation results in an opposite effect in the concentration of the corresponding prey and this indirect effects propagates through the cyclic food web. For even $Q$ the indirect effect will strengthen the direct modification, and the corresponding positive feedback can even eliminate the minority species. In contrast with this, for odd $Q$ the ecosystem exhibits an unusual response similar to the one described for the Rock-Scissors-Paper game with different invasion rates.

\section{Competing Associations}
\label{sec:compass}

We know that in classical game theory the number of Nash equilibria can be larger than one. In general,  the number of Nash equilibria (and ESSs in evolutionary games) rapidly increases with the number of pure strategies $Q$ for an arbitrary payoff matrix \citep{broom_mb00}. The situation becomes more complicated in  spatial evolutionary games, where each player follows one of the pure strategies, and interacts with a limited number of co-players. Particularly for large $Q$, the players can only experience a limited portion of all the possible constellations, and this intrinsic constraint affects the evolutionary process. Monte Carlo simulations demonstrated that the dynamical rule (as a way of optimization) can lead to many different stationary states in small systems. For large spatial models, these phases occur locally in the initial transient regime, and the neighboring phases compete with each other along the boundary separating them. The long-run state of the system develops as a result of random and/or deterministic invasion processes, and may have very complex spatio-temporal structure.

The large number of possible stationary states in these models was already seen in the mean-field description. In many of these solutions several strategies are missing, and these states can be considered as solutions to sub-games, in which the missing strategies are excluded by default. Henceforth, the possible stationary solutions will be considered as  ``associations of species".

For ecological systems the relevance of complex spatio-temporal structures \citep{watt_je47} and associations of species have been studied for a long time [for a survey and further references see \citet{johnson_tree02}]. In the subsequent sections we will discuss several simple models to demonstrate the surprisingly rich variety of behavior that can occur for predator-prey interactions. In these systems the survival of a species is strongly related to the existence of a suitable association that can provide a higher stability comparing to other possibilities.

\subsection{A four-species cyclic predator-prey model with local mixing}
\label{sec:lv4sx}

Let us first study the effect of local mixing in the four-species cyclic predator-prey model discussed in Section \ref{sec:rspns}. In this model each site $x$ of a square lattice can be occupied by an individual belonging to one of four species ($s_x=1, \ldots , 4$). The cyclic predator-prey relation is defined as above, i.e., 1 invades 2 invades 3 invades 4 invades 1. The evolutionary process repeats the following steps:
\begin{itemize}
\item[1)] Choose two neighboring sites at random;
\item[2)] If these sites are occupied by a predator--prey pair, the offspring of the predator occupies the prey's site [e.g., $(1,2) \to (1,1)$];
\item[3)] If the sites are occupied by a neutral pair of species, they exchange their sites with probability $\mu < 1$ [e.g., $(1,3) \to (3,1)$];
\item[4)] Nothing happens if the states are identical.
\end{itemize}

Starting from a random initial state these steps are repeated many times, and after a suitable thermalization time we determine the one- and two-site configuration probabilities, $p_1(s)$ and $p_2(s,s^{\prime})$; $s, s^{\prime}=1, \ldots , 4$, respectively, by averaging over a sufficiently long sampling time.

\begin{figure}[ht]
\centerline{\epsfig{file=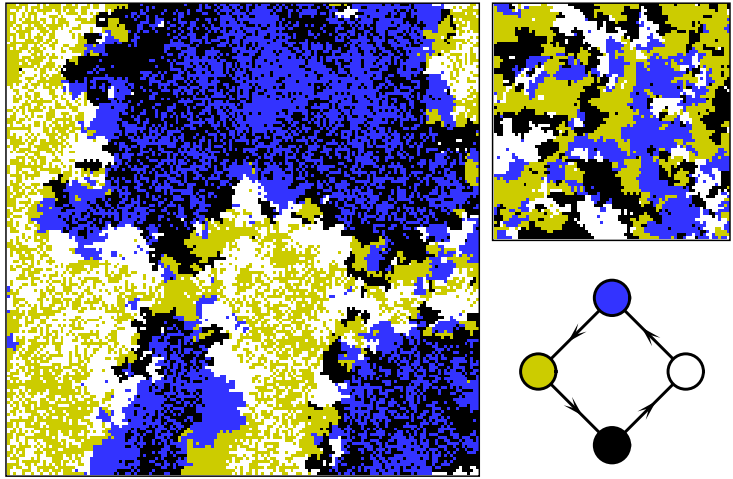,width=10cm}}
\caption{\label{fig:lv4sxpat}The large snapshot ($L=160$) shows the formation of the associations of neutral pairs during the domain growth process for $\mu=0.05$. Cyclic dominance in the food web is indicated in the lower right plot. The smaller snapshot (upper right) illustrates a typical distribution of species in the absence of site exchange, $\mu=0$.}
\end{figure}

The value of $\mu$ characterizes the strength of local mixing. If $\mu=0$, the model is equivalent to those discussed in Section \ref{sec:rspns}, and a typical snapshot is shown in the upper right part of Fig.~\ref{fig:lv4sxpat}.  As described above, this self-organizing pattern is sustained by traveling invasion fronts. Similar spatio-temporal pattern can be observed below a threshold value of mixing, $\mu<\mu_{cr}=0.02662(2)$ \citep{szabo_jpa05}. As $\mu$ approaches the threshold value, the irregularity of the interfaces separating neutral species increases, and one can observe the formation of small islands occupied by only two (neutral) species within the well-mixed state.

For $\mu<\mu_{cr}$ all the four species survive with the same concentration, $p_1(s)=1/4$, while the probability
\begin{equation}
p_{\rm np}=p_2(1,3)+p_2(3,1)+p_2(2,4)+p_2(4,2)
\label{eq:lv4sxpnp}
\end{equation}
of finding neutral pairs (np) on neighboring sites decreases as $\mu$ increases. At the same time, the Monte Carlo simulations indicate a decrease in the frequency of invasions. This quantity is nicely characterized by the probability of predator-prey pairs (ppp), defined as
\begin{eqnarray}
p_{\rm ppp}&=& p_2(1,2)+p_2(2,3)+p_2(3,4)+p_2(4,1)+ \nonumber \\
            && p_2(2,1)+p_2(3,2)+p_2(4,3)+p_2(1,4) \;.
\label{eq:lv4sxppp}
\end{eqnarray}
The Monte Carlo data indicates a first-order phase transition at $\mu=\mu_{cr}$.

Above the threshold value of $\mu$, simulations display a domain growth process, which is analogous to the one described by the kinetic Ising model below its critical temperature \citep{glauber_jmp63}. As illustrated in Fig.~\ref{fig:lv4sxpat}, there are two types of growing domains, formed from odd and even label species. These domains are separated by a boundary layer where the four species invades cyclically each other. This boundary layer serves as a "species reservoir", sustaining the symmetric composition within domains. In agreement with theoretical expectations \citep{bray_ap94}, the typical domain size increases with time as $l(t)\sim \sqrt{t}$. Thus, for any finite size this system develops into a mono-domain state, consisting of two neutral species with equal concentrations. Since there are no invasion events within this state, $p_{\rm ppp}$ can be considered as an order parameter that becomes zero in the stationary state for $\mu > \mu_{cr}$. Due to the site exchange mechanism the distribution of the two neutral species evolves into an uncorrelated (symmetric) state [e.g., $p_1(1)=p_1(3)=1/2$ and $p_2(1,1) =p_2(3,3) =p_2(1,3)= p_2(3,1)=1/4]$.

The traditional mean-field approximation cannot account for site exchange, therefore the corresponding equations of motion are the same as those in Eqs. (\ref{eq:rsp4pa1}) of Sec.~\ref{sec:rspns}. The two- and four-site approximations also fail, despite the fact that they directly involve the effect of site exchange. A qualitatively correct prediction can be achieved when the generalized mean-field approximation is performed on a $3 \times 3$ cluster. This discrepancy may indicate the strong relevance of complex processes (represented by several consecutive elementary steps in time and/or strongly correlated local structures), whose correct description would require larger cluster sizes beyond those analyzed thus far.

Let us recall that this system has four homogeneous stationary states, two continuous sets with two-species, and one symmetric stationary solution with four-species. Despite the large number of possible solutions, simulations only realize the symmetric four-species solution for $\mu<\mu_{cr}$, excepting situations when the system is started from some strongly correlated (almost homogeneous) initial states. The preference for two-species states for $\mu > \mu_{cr}$ is due to a particular feature of the food web. Namely, participants of these spatial associations mutually protect each other against all possible external invaders. For example, if an individual of species 1, who belongs to the well-mixed association of species 1 and 3, is invaded by an external invader of species 4, then one of its neighbors of species 3 will recapture the lost site within a short time. The individuals of species 1 will guard their partners of species 3 in the same way against attacks by species 2. For this reason, such states will be called "defensive alliances" in the sequel.

The present model has two equivalent defensive alliances, because members of the well-mixed association of species 2 and 4 can also protect each other against their respective predators using the above mechanism. Such defensive alliances are preferred to all other states when the mixing $\mu$ exceeds a threshold value.

The spatial competition of different associations can be visualized directly by Monte Carlo simulations, if the initial state is artificially set to contain large domains of two different associations. For example, one can study a strip-like initial structure, where parallel domains are occupied alternately by four-species (cyclic) and two-species (e.g., 1+3) associations. The linear variation of $p_{\rm ppp}$ with time measures the speed of the invasion process along the boundary, and we can determine the average velocity $v_{inv}$ of the invasion front. For $\mu < \mu_{cr}$ the four-species association expands, but its average invasion velocity vanishes linearly as the critical mixing rate is approached \citep{szabo_pre04a},
\begin{equation}
    v_{inv} \sim  (\mu_{cr}-\mu).
    \label{eq:assvinv}
\end{equation}

For $\mu>\mu_{cr}$ the application of this approach is less reliable. It seems to be strongly limited to those regions, where the spontaneous nucleation rate is very slow. The linear scaling of the average invasion velocity is accompanied by a divergence in the relaxation time as $\mu \to \mu_{cr}$, and we need very long run times in the vicinity of the threshold. On the other hand, Eq. (\ref{eq:assvinv}) can be used to improve the accuracy of $\mu_{cr}$ by deducing $v_{inv}$ from short time runs further away from $\mu_{cr}$. This approach becomes particularly efficient for situations, when the stationary state of the competing associations is not affected by the parameters (e.g., the mixed state is independent of $\mu$).

The emergence of defensive alliances seems to be very robust, and is not related to the simplicity of the present model. Similar phenomena were found for another model, where local mixing was built in in the form of random (nearest-neighbor) jumps to vacant sites \citep{szabo_pre04a}. Using similar dynamical rules \citet{he_ijmpc05} reported the formation of two defensive alliances (consisting of species with odd or even labels) in a cyclic six-species spatial predator-prey model. In a more realistic computer simulation \citet{sznaider_03} observed the spatial formation of defensive alliances on a continuous plane, where four species (with cyclic dominance) moved randomly, and created offsprings after they had catched and consumed a prey within a given distance.

\subsection{Defensive alliances}
\label{sec:defall}

In the previous section we discussed a spatial four-species cyclic predator-prey model, where two equivalent defensive alliances can exist. Their stability is supported by local mixing, i.e., site exchange between neutral pairs. The robustness of the formation of such types of defensive alliances was also confirmed for the six- and eight-species versions of the model. Evidently, larger $Q$ gives rise to more stationary states that consist of mutually neutral species. Nevertheless, Monte Carlo simulations indicate that only two types of states (associations) can survive after an evolutionary competition. If the mixing probability $\mu$ is smaller then a $Q$-dependent threshold $\mu_{cr}(Q)$, all species can coexist by forming a self-organizing pattern characterized by perpetual cyclic invasions. In the opposite case, $\mu>\mu_{cr}(Q)$, one of the two equivalent defensive alliances conquer the system in the end of a domain growth process. The two equivalent defensive alliances are composed of species with only odd and even labels, respectively. In the well-mixed phase, above $\mu_{cr}$, their stability is provided by a similar mechanism as for $Q=4$. In fact these are the only associations, which can protect themselves against attacks from the rest of the species. The threshold value $\mu_{cr}$ decreases for increasing $Q$. According to the Monte Carlo simulations, $\mu_{cr}(Q=6)=0.0065(1)$ and $\mu_{cr}(Q=8)=0.0028(1)$. Numerical studies face difficulties for $Q > 8$, because the domain growth velocity decreases as $Q$ increases.

The formation of defensive alliances seems to require local mixing. This is the reason why these states become dominant if $\mu$ exceeds a suitable threshold value. Cyclic dominance, however, may inherently imply another mechanism for the emergence of defensive alliances which is not necessarily based on local mixing \citep{boerlijst_pd91,szabo_pre01a,szabo_pre01b}.

\begin{figure}[ht]
\centerline{\epsfig{file=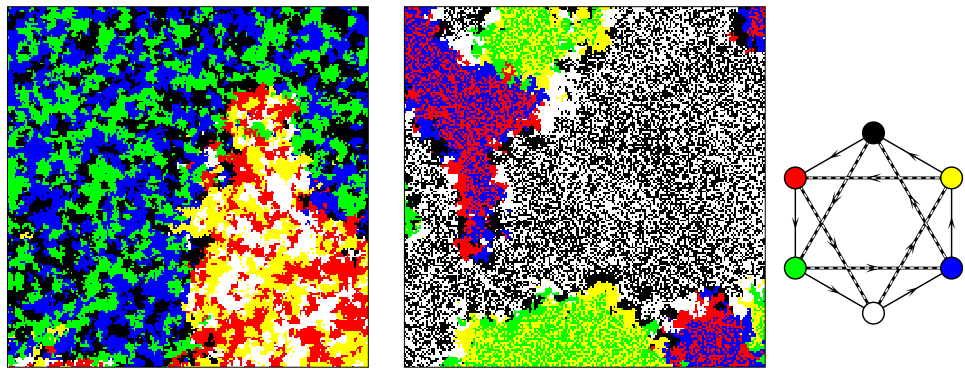,width=13cm}}
\caption{\label{fig:lv6s2ax} A six-species predator-pray model defined by the food web on the right. Members of the two possible cyclic defensive alliances are connected by thick dotted lines. Left panels show snapshots of the domain growth process for two different rates of mixing (left: $\mu=0$, right: $\mu=0.08$). }
\end{figure}

In order to see this, notice that within a self-organizing spatial pattern each individual has a finite life-time, because it will be consumed by its predator sooner or later.
Assume now that there is an association composed of cyclically dominating species (to be called a \emph{cyclic defensive alliance}). An external invader $s_{ex}$, outside the association, who is a predator of the internal species $s$ is eliminated from the system within the life-time, if the internal predator of $s$ also consumes $s_{ex}$. As an example Fig.\ \ref{fig:lv6s2ax} shows a six-species predator-prey system. The food web gives rise to two cyclic defensive alliances, whose spatial competitions for different mixing rates is also depicted.

In this six-species model each species has two predators and two prey in a way defined by the food web in Fig.~\ref{fig:lv6s2ax} \citep{szabo_jpa05}. The odd (even) label species can build up a self-organizing structure controlled by cyclic invasions. The members of these associations protect cyclically each other against external invaders as mentioned above. In the absence of local mixing, the Monte Carlo simulations on the square lattice visualize how domains of these defensive alliances grow (for an intermediate state see the upper snapshot in Fig.~\ref{fig:lv6s2ax}). Eventually the system develops into one of the "mono-domain" (mono-alliance) states.

The spatio-temporal structure of these associations seems to play a decisive role in the protection mechanism, because the corresponding mean-field approach cannot describe the extinction of external invaders. Contrary to the simulation results, the numerical solution of the mean-field equations predicts periodic oscillations. Very recently the relevance of the spatial structure was confirmed by \citet{kim_pre05}, who demonstrated that the formation of these defensive alliances can be impeded on complex networks, where a sufficiently large portion of the links is rewired.

Simulations on the square lattice show a strikingly different behavior, when we allow site exchange between neutral pairs (e.g., the black and white species in Fig.~\ref{fig:lv6s2ax}). This process favors the formation of two-species domains composed from a neutral pair of species. Surprisingly, these mixed states can also be considered as defensive alliances. Consequently, for sufficiently large mixing rate $\mu$, one can observe the formation of domains of all three two-species associations (see the right snapshot in Fig.~\ref{fig:lv6s2ax}).

According to the simulations a finite system evolves into one of the two cyclic three-species defensive alliances if $\mu < \mu_{cr}=0.0559(1)$. Due to the absence of neutral species the final self-organizing spatio-temporal structure is independent of $\mu$, and is equivalent to the one discussed in Section \ref{sec:rspsim}. In the opposite case, $\mu > \mu_{cr}$, the finite lattice evolves into an uncorrelated distribution of two neutral species, e.g., $p_1(1)=p_1(4)=1/2$ and $p_2(1,1)=p_2(4,4)=p_2(1,4)=p_2(4,1)=1/4$ \citep{szabo_pre04a}.

In this predator-prey system, either below or above $\mu_{cr}$, the domain growth process is controlled by the random motion of boundaries separating equivalent associations. In the next section we will consider what happens if the coexisting associations are different and/or cyclically dominate each other.

\subsection{Cyclic dominance between associations in a six-species predator-prey model}
\label{sec:lv6s2}

\begin{figure}[ht]
\centerline{\epsfig{file=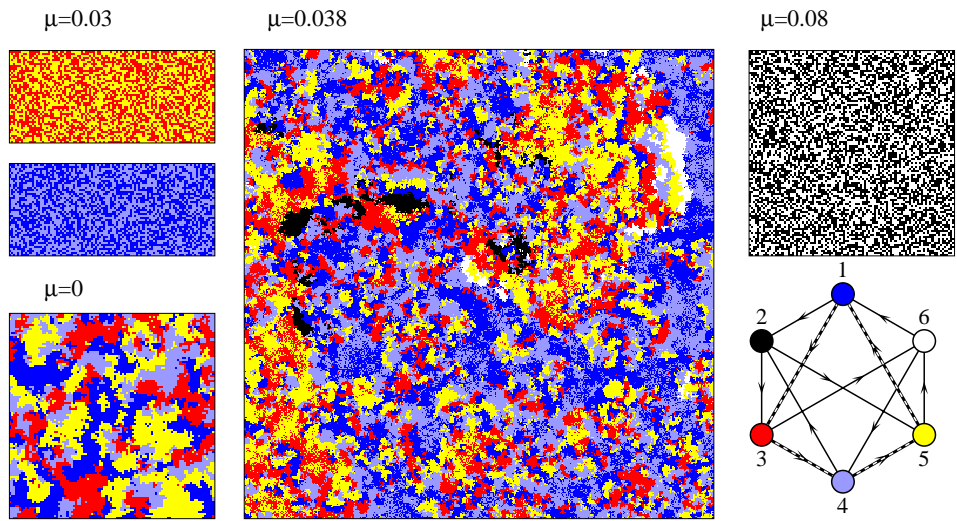,width=13cm}}
\caption{\label{fig:lv6dpts}Distribution of species in four typical stationary states obtained by simulations for values of $\mu$ given for each snapshot. The color of the labeled species and the predator-prey relations are defined by the food web.}
\end{figure}

In order to demonstrate the rich variety of possible complex behaviors for associations, now we consider in detail a six-species spatial predator-prey model embedding the features of the previous models \citep{szabo_jpa05}. The present system has six species, whose individuals are distributed on a square lattice in such a way that each site is occupied by one of the species. Each species has two predators and two prey according to the food web plotted in Fig.~\ref{fig:lv6dpts}. The invasion rates between any predator-prey pairs are chosen to be unity. Choosing sufficiently large system sizes with periodic boundary condition the system is started from a random initial state. Evolution is controlled by sequential invasion events if two neighboring sites, chosen at random, are occupied by a predator-prey pair. If the  sites are occupied by a neutral pair, they will exchange their positions with a probability $\mu$ characterizing the strength of local mixing. After a suitable thermalization time the system evolves into a state that we consider.

Before reporting the results of the Monte Carlo simulations it will be illuminating to recall the possible stationary states. This system has six homogeneous states which are unstable against invasions of the corresponding predators. There exist three two-species states, consisting of neutral pairs of species like 1+4, 2+6, and 3+5, with fixed composition, whose spatial distribution is frozen (becomes uncorrelated) for $\mu=0$ ($\mu>0$). Two subsystems, composed of species 2+3+4 or 1+5+6, are equivalent to the spatial Rock-Scissors-Paper game. One of the four-species subsystem (1+3+4+5) is analogous to those discussed in Section \ref{sec:lv4sx}. The existence of these states are confirmed by the mean-field approximation, which also allows the coexistence of all six species with the same fixed concentration, $p_1(s)= 1/6$ for $s=1, \ldots, 6$, or with oscillating concentrations.

According to the simulations performed on small systems ($L \simeq 10$), evolution ends in one of the one- or two-species absorbing states, and selection is hardly affected by the presence of local mixing. Unfortunately, the effect of system size on the transition from a random initial state into one of the absorbing (or even meta-stable) states is not yet considered rigorously. However, this type of investigation could be useful for understanding the specialization (differentiation) of biological cells.

On sufficiently large systems the prevalence of the homogeneous state can never be observed. The ultimate behavior is determined uniquely by the value of $\mu$ (for snapshots see Fig.~\ref{fig:lv6dpts}). In the absence of local mixing ($\mu=0$) two species (2 and 6 or black and white on the snapshot) die out within a short time and the surviving species form a self-organizing pattern maintained by cyclic invasions. We have to emphasize that this subsystem can be considered as a defensive alliance, whose stability is supported by its self-organizing pattern in the present six-species model. For example, if an external intruder of species 2 attacks a member of species 3 (or 5) then the offsprings will be eliminated by their common predator 1 (or 4).

For small $\mu$ values local mixing is not able to prevent the extinction of species 2 and 6, and the behavior of the resultant subsystem becomes equivalent to the cyclic four-species predator-prey model discussed above. This means that the self-organizing structure is eroded by the local mixing and the well-mixed associations of the neutral species (1+4 or 3+5) become favored above the first threshold value, $\mu_{c1}=0.0265$, given in Section \ref{sec:lv4sx}. Thus, for the region $\mu_{c1}<\mu<\mu_{c2}$ the final state of the system consists of only two neutral species, as shown by the snapshots in Fig.~\ref{fig:lv6dpts} for $\mu=0.03$.

The velocity of the segregation process increases with $\mu$. As a result, for $\mu > \mu_{c2}$ small (well-mixed) domains of species 1 and 4 (or 3 and 5) can form before the rest of the species (2 and 6) become extinct. The occurrence of these domains help minority species to survive, because both species 1 and 4 (3 and 5) are prey for species 6 (2). The largest snapshot, consisting of $400 \times 400$ lattice sites in Fig.~\ref{fig:lv6dpts} depicts the successive events controlling the evolution of this self-organizing pattern. In this snapshot there are only a few black and white sites (species 2 and 6) who invade (without hindrance) the territory occupied only by their two prey. Behind these invasion fronts the occupied territory is re-invaded by their predators, whose territory is again attacked by their respective predators, etc.

Although, as discussed above, the food web allows the formation of many other three- and four-species "cyclic" patterns, the present dynamics prefers four-species cyclic defensive alliance (1+3+4+5) to any other cyclic associations. On longer time scales, however, the strong mixing transforms these (cyclic) four-species domains into two-species associations consisting of neutral species, which can be invaded again by one of the minority species.

\begin{figure}[ht]
\centerline{\epsfig{file=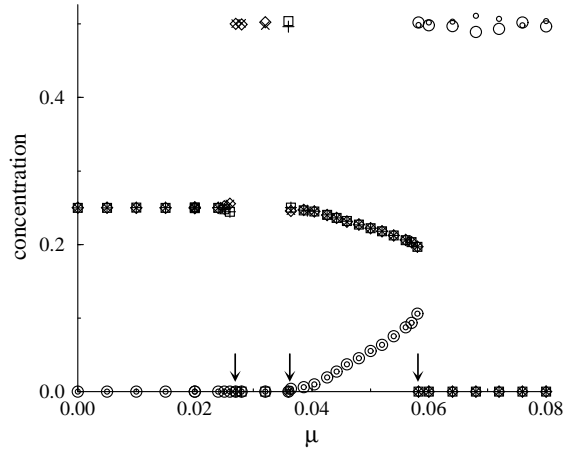,width=7.5cm}}
\caption{\label{fig:rho6d}Concentration of species {\it vs}.
$\mu$. Monte Carlo data are denoted by diamonds, large circles,
squares, crosses, pluses, and small circles for the six species labeled from 1 to 6, respectively. Arrows indicate the three transition points.}
\end{figure}

Notice that the above successive cyclic process is much more complicated than the simple invasion process modeled by the spatial Rock-Scissors-Paper game or even by the forest-fire model. In the present system two types of cycles, $(1+3+4+5) \to (1+4) \to 6 \to (1+2+3+4+5+6) \to (1+3+4+5)$ and $(1+3+4+5) \to (3+5) \to 2 \to (1+2+3+4+5+6) \to (1+3+4+5)$, are entangled. The boundaries between the territories of associations are fuzzy. Furthermore, transitions from an association into another have different time scales and characteristics depending on $\mu$. As a result, the concentration of the six species varies with $\mu$ if $\mu_{c2}<\mu<\mu_{c3}=0.0581(1)$. Within this region the concentrations of the species 2 and 6 increase with $\mu$  monotonously (as illustrated by the Monte Carlo data in Fig.~\ref{fig:rho6d}), and above the third threshold value they can form a mixed state prevailing the whole system.
In fact, the well-mixed distribution of species 2 and 6 also satisfies the criteria of defensive alliances, because they mutually protect each other against the rest of the species.

Determination of the second transition point is made difficult by increasing fluctuations in the concentrations when $\mu_{c2}$ is approached from above. Contrary to many critical phase transitions, where fluctuation of the vanishing order parameter diverges, here the largest concentration fluctuations are found for the majority species. Visualization of the spatial evolution indicates that the invasions of minority species (2 and 6) give rise to fast variations with large fluctuations and increasing correlation lengths if $\mu \to \mu_{c2}+$. Due to the mentioned technical difficulties the characteristic features of this transition are not yet fully clarified.

It is interesting that in the given model the coexistence of all species is only possible within a limited range of mixing, $\mu_{c2}<\mu<\mu_{c3}$. The steady-state concentrations are determined by the cyclic invasion rates between the (multi-species) associations. As discussed in Section \ref{sec:rspns}, this type of self-organizing patterns can be maintained for various invasion rates and mechanism, depending on the strength of mixing. This phenomenon is not unique as similar features were observed for another six-species model \citep{szabo_jpa05}. These results suggest that the stability of very complex ecological and catalytic chemical systems can be enhanced by cycles in the spatio-temporal pattern [for an early survey see the book by \citet{eigen_79}].

The above six-species states can be viewed as composite associations, which consist of simpler associations dominating cyclically each other. Following this procedure one can construct a hierarchy of associations, which is able to sustain bio-diversity at a very high level. In the last sections the analysis has been restricted to those systems, where only predator-prey interactions and local mixing were permitted. We expect, however, that suitable extensions of the evolutionary dynamics may induce an even larger variety of behaviors.

It is widely accepted that self-organization plays a crucial role in pre-biotical evolution. Self-organizing processes can provide a way for transforming nonliving matter to living organisms [for a brief survey of recent research see  \citet{rasmussen_s04}]. The systematic investigation of simplified evolutionary games as toy models can serve as a bottom-up approach that helps us clarifying theoretically the fundamental phenomena, mechanisms, and environmental requirements needed for creating living systems.

\section{Conclusions and outlook}
\label{sec:co}

In this article we reviewed our current understanding of evolutionary games that have become of increased interest to biologists, economists, and social scientists in recent years. Following a pedagogical way we surveyed the most relevant elements of the classic theory of games, as well as the main components of evolutionary game theory. One of the principal goals was to demonstrate the applicability of concepts and approaches originally developed in statistical physics to study many-particle systems. For this purpose we investigated in detail some representative games like the Prisoner's Dilemma and the Rock-Scissors-Paper game for different connectivity structures and evolutionary rules. The analysis of these games nicely illustrated  the possible richness of behavior. Evolutionary games can exhibit different stationary states and phase transitions when fundamental parameters like the payoffs, the level of noise, the dimensionality or the connectivity structure are varied.

In comparison with physical systems, evolutionary games can show more complicated behaviors which depend sensitively on the applied dynamic rules and the topological features of the underlying social graph. Usually the behavior of a physical system can be reproduced qualitatively well by the traditional mean-field approximation. However, the appropriate description of an evolutionary game requires more sophisticated techniques, taking the fine details appearing both in the model and in the configuration space into account. Although, extended versions of the mean-field technique are useful and are capable of providing correct results on a high enough level, they should frequently be complemented by time-consuming numerical calculations.

Most of our discussion focused on infinitely large systems where the concept of phase transitions is applicable. We have to emphasize, however, that for many realistic systems finite-size models and approaches (e.g., Moran process, fixation) can give more appropriate descriptions. At the same time the corresponding techniques are capable of deducing simple analytical relations [see \citep{nowak_s06}]. The clarification of finite-size effects is especially needed for those systems which have two (or more) time and/or spatial scales.

Unfortunately, systematic investigations are strongly limited by the wide variety of microscopic rules and the large number of parameters. For example, at the beginning many questions in social dilemmas were only investigated  within the framework of the Prisoner's Dilemma, although similar problems and questions naturally arise for the Hawk-Dove and some other games too [see the papers by \citet{nowak_ijbc93,killingback_prslb96,hauert_n04,doebeli_el05,
sysiaho_epjb05,tomassini_pre06}]. The extension of research towards these models in recent years has started to provide a more complete picture about the emergence of cooperation in social dilemmas in general \citep{santos_pnas06,hauert_jtb06a,ohtsuki_n06}).

One of the most striking features of evolutionary games is the occurrence of self-organizing patterns. These patterns can be characterized by additional (internal) parameters (e.g., average time period of local oscillations, geometrical properties of the spatial patterns, {\it etc}). We hope that future efforts will reveal the detailed relationships between these internal properties. The appearance of self-organization is strongly related to the absence of detailed balance in the microscopic states. This feature can be described by introducing a "probability current" between microscopic states. The investigation of the distribution of probability currents (the "vorticity structure") \citep{schnakenberg_rmp76,zia_cm06} can provide further information about the possible relations between microscopic mechanisms and the resulting spatial structures.

The present review is far from being complete. Missing related topics include minority games, games with random payoffs, quantum games, and those systems where the payoffs come from multi-person (e.g. Public Good) games. About minority games, as mentioned previously, the reader can find detailed discussions in the books by \citet{challet_04} and by \citet{coolen_05}. Methods developed originally within spin glass theory [for reviews see \citep{mezard_87,gyorgyi_pr01}] are used successfully by \citet{berg_prl98,berg_epl99}, who studied the properties of two-player games with random payoffs in the limit $Q \to \infty$.

Many aspects of evolutionary game theory are utilized directly in the developments of computational learning algorithms [for a survey see the paper by \citet{wolpert_pre04} and further references therein]. In these systems the players represent algorithms, and the Darwinian selection among the different mutants are controlled by some utility function (payoff). Thus, this research area combines three fields: game theory, machine learning, and optimization theory.

Another perspective research topic is quantum game theory which involves the quantum superposition of pure strategies [for a brief overview we suggest \citet{lee_n01,lee_pw02} and further references therein]. Utilizing the quantum entanglement this extension may offer strategies more favorable than classical ones (for a nice example see the description of the Quantum Penny Flip in Appendix \ref{app:g:qpf}). As quantum entanglement enforces common interest, the Pareto optimum can be easily achieved by using quantum strategies even for the Prisoner's Dilemma, as was discussed by \citet{eisert_prl99}. Players can avoid the dilemma if they both follow quantum strategies. This two-person quantum game was experimentally realized by \citet{du_prl02} on a nuclear magnetic resonance quantum computer. It is believed that the deeper understanding of quantum games can inspire the development of efficient evolutionary rules for achieving the optimal global payoff in social dilemmas.

\section*{Acknowledgments}

Enlightening discussions with Tibor Antal, Ulf Dieckmann, Attila Szolnoki, Arne Traulsen, and Jeromos Vukov are gratefully acknowledged. This work was supported by the Hungarian National Research Fund (OTKA T-47003).

\appendix

\section{Games}
\label{sec:app:g}

\subsection{Battle of the Sexes}
\label{app:g:bos}

This game is a typical (asymmetric) Coordination game. Imagine a wife and a husband who forget whether they have agreed to attend an opera production or a sports event in the evening. Both events are unique and they should decide simultaneously without communication where to go. The husband prefers the sports event, his wife the opera, but both prefer being together rather than alone. The payoff matrix is:
\begin{equation}
\begin{tabular}{lc|crccc}
  & & & \multicolumn{4}{c}{\emph{Wife}} \\
  & & & Opera &    & & Sports \\[0.5ex] \hline
  & &  &  & &  &    \\[-2.5ex]
  &Opera\phantom{a}&  & (1,2) & &  & (0,0)  \\[-1.2ex]
\emph{Husband}\phantom{a} &   & &  & &   &     \\[-1.2ex]
  &Sports\phantom{a}& & (0,0) & &   & (2,1)  \\
\end{tabular}
\end{equation}

Note that Battle of the Sexes is a completely different game in the biology literature \citep{ hofbauer_98}. That game is structurally similar to Matching Pennies.

\subsection{Chicken game}
\label{app:g:ch}

Bertrand Russell, in his book ``Common Sense and Nuclear Warfare" \citep{russell_59}, describes the game as follows: ``... a sport which, I am told, is practised by some youthful degenerates. This sport is called "Chicken!" It is played by choosing a long straight road with a white line down the middle and starting two very fast cars towards each other from opposite ends. Each car is expected to keep the wheels of one side on the white line. As they approach each other, mutual destruction becomes more and more imminent. If one of them swerves from the white line before the other, the other, as he passes, shouts "Chicken!" and the one who has swerved becomes an object of contempt...."

Chicken is mathematically equivalent to the Hawk-Dove game and to the Snowdrift game.

\subsection{Coordination game}
\label{app:g:co}

Whenever players should choose an identical action, whatever it is, to receive high payoff, we speak about a Coordination game. As a typical example think about competing technologies like video standards (VHS vs Betamax) or storage formats (Blue-ray vs HD DVD). When firms are capable of agreeing on the technology to apply, market sales and thus profits are high. However, in the lack of a common technology standard the market is full of compatibility problems, and buyers are reluctant to purchase. Sales and profits are low for all producers. Ironically, the actual technology chosen has less importance, and it may well happen that the market coordinates on an inferior product, as was undoubtedly the case for the QWERTY keyboard system.

The Coordination game is also the adequate mathematical metaphor behind social conventions such as our collective choice of the side of the road we drive on, the time zones we define, or the signs we associate with given meanings in our linguistic communication.

In general, the players can have $Q > 2$ options, and the Nash equilibria can be achieved by agreeing on which one to use. For different payoffs (see Battle of the Sexes) the players can prefer different Nash equilibria.

\subsection{Hawk-Dove game}
\label{app:g:hd}

Assume a population of animals (birds or others) where individuals are equal in almost all biological properties except one: their aggressiveness in interactions with others. This behavioral attribute is genetically coded, and animals exist in two forms: the aggressive type, to be called Hawk, and the cooperative type, to be called Dove. Assume that each time two animals meet, they compete for a resource $R$ that can represent food. When two Doves meet they simply share the resource. When two Hawks meet they fight, and one of them (randomly) gets seriously injured, while the other takes the resource. Finally if a Dove meets a Hawk, the Dove escapes without fighting, and the Hawk takes the full resource without injury. The payoff matrix is
\begin{equation}\label{eq:HDG}
\begin{tabular}{lc|ccccc}
  & & & \multicolumn{4}{c}{\emph{Animal 2}} \\
  & & & Hawk &    & & Dove \\[0.5ex] \hline
  & &  &  & &  &    \\[-2.5ex]
  &Hawk\phantom{a}&  & $\left(\displaystyle\frac{V-C}{2},\frac{V-C}{2}\right)$ & &  & $(V,0)$  \\[-1.2ex]
\emph{Animal 1} &   & &  & &   &     \\[-1.2ex]
  &Dove\phantom{a}& & $(0,V)$ & &   & $\left(\displaystyle\frac{V}{2},\frac{V}{2}\right)$  \\
\end{tabular} \qquad
\end{equation}
where $C>V$ is the cost of injury. The game assumes darwinian evolution, in which fitness is associated with average payoff in the population.

The Hawk-Dove game is mathematically equivalent to economists' Chicken and Snowdrift games.

\subsection{Matching Pennies}
\label{app:g:mp}

This is a $2 \times 2$ zero-sum matrix game. The players first determine who will be the winner for an outcome with the "same" or "different" sides of the coins. Then, each player conceals in her palm a penny either with its face up or down. The players reveal their choices simultaneously. If the pennies match (both are head or both are tail), the player "Same" receives one dollar from player "Different". Otherwise, player "Different" wins and receives one dollar from the other player. The payoff matrix is
\begin{equation}
\begin{tabular}{lc|ccccc}
  & & & \multicolumn{4}{c}{\emph{Different}} \\
  & & & Head &    & & Tail \\[0.5ex] \hline
  & &  &  & &  &    \\[-2.5ex]
  &Head\phantom{a}&  & $\left(1,-1\right)$ & &  & $(-1,1)$  \\[-1.2ex]
\emph{Same}  &   & &  & &   &     \\[-1.2ex]
  &Tail\phantom{a}& & $(-1,1)$ & &   & $\left(1,-1\right)$  \\
\end{tabular} \qquad
\end{equation}

The Nash equilibrium of this game is a mixed strategy: each player chooses heads or tails with equal probability.

This game is equivalent to "Throwing Fingers" and "Odds or Evens".

\subsection{Minority game}
\label{app:g:min}

In this $N$-person game ($N$ is odd) the players should choose between two options simultaneously and only that action is successful which is chosen by the minority. Such a situation can occur in financial markets where agents choose between "selling" and "buying", or in traffic when drivers choose from two possible roads to take. The original version of the game, suggested by \citet{arthur_aer94}, used as an example the El Farol bar in Santa Fe, where Irish music is only enjoyable if the bar is not too crowded, and agents should decide whether to visit the bar or stay at home. The evolutionary version was introduced by \citet{challet_pa97}. Many aspects of the game are discussed in two recent books by \citet{challet_04} and by \citet{coolen_05}.

\subsection{Prisoner's Dilemma}
\label{app:g:pd}

The story for this game was invented by the mathematician Albert Tucker in 1950, when he wished to illustrate the difficulty of analyzing certain kinds of games studied previously by Melvin Dresher and Merill Flood (scientist at RAND Corporation, Santa Monica, California). Tucker's paradox has since given rise to an enormous literature in areas as diverse as philosophy, biology, economics, behavioral and political sciences, as well as game theory itself. The story of the "Prisoner's Dilemma" is the following:

Two burglars are arrested after their joint burglary and held separately by the police. However, the police does not have sufficient proof in order to have them convicted, therefore the prosecutor visits each of them and offers the same deal: if one confesses (called defection in the context of game theory) and the other remains silent (cooperation -- with the other prisoner), the silent accomplice receives a three-year sentence and the confessor goes free. If both stay silent then the police can only give both burglars one year for a minor charge. If both confess, each burglar receives a two-year sentence.

According to the traditional notation the payoff matrix is
\begin{equation}\label{eq:app_pd}
\begin{tabular}{lc|ccccc}
  & & & \multicolumn{4}{c}{\emph{Prisoner 2}} \\
  & & & Defect &    & & Cooperate \\[0.5ex] \hline
  & &  &  & &  &    \\[-2.5ex]
  &Defect\phantom{a}&  & $(P,P)$ & &  & $(T,S)$  \\[-1.2ex]
\emph{Prisoner 1}  &   & &  & &   &     \\[-1.2ex]
  &Cooperate\phantom{a}& & $(S,T)$ & &   & $(R,R)$  \\
\end{tabular} \qquad
\end{equation}
where $P$ means "{\bf P}unishment for mutual defection", $T$ "{\bf T}emptation to defect", $S$ "{\bf S}ucker's payoff", and $R$ "{\bf R}eward for mutual cooperation". The matrix elements satisfy the following rank ordering: $S<P<R<T$. For the repeated version (Iterated Prisoner's Dilemma) usually the additional constraint, $T+S<2R$, is also assumed. This ensures the long-run advantage of mutual cooperation against a strategy profile where players cooperate and defect alternatively in opposite phase.

The reader can find further details about the history of this game in the book by \citet{poundstone_92}, together with many important applications. Experimental investigations of human behavior in Prisoner's Dilemma situations have been expanding since the works of \citet{trivers_qrb71} and \citet{wedekind_pnas96}.

\subsection{Public Good game}
\label{app:g:pg}

In an experimental realization of the Public Good game \citep{ledyard_95,hauert_s02}, an experimenter gives some money $c$ to each of $N$ players. The players decide independently and simultaneously how much to invest (if any) to a common pool. The collected sum is multiplied by a factor $r$ ( $1 < r < N-1$) and is redistributed to the $N$ players equally, independently of their individual contributions. The maximum total income is achieved if all players contribute maximally. In this case each player receives $rc$, thus the final payoff is $(r-1)c$. Players are faced with the temptation of being free-riders, i.e., to take advantage of the common pool without contributing to it. In other words, any individual investment is a loss for the player because only a portion $r/N<1$ will be repaid. Consequently,  rational players invest nothing and the corresponding state is known as the "Tragedy of the Commons" \citep{hardin_s68}. In the game theory literature this game is frequently referred to as the Tragedy of the Commons, the Free Rider problem, the Social Dilemma, or the Multi-person Prisoner's Dilemma. The large variety of names reflects the large number of situations when members of a society can benefit from the efforts of others, while having a temptation not to pay the cost of these efforts \citep{baumol_52,samuelson_res54,hardin_bs71}.

Evidently, if there are only two players whose choices are restricted to two options (to invest nothing or all) then this game becomes a Prisoner's Dilemma. Conversely, the $N$-person round robin Prisoner's Dilemma game is equivalent to a Public Good game.

\subsection{Quantum Penny Flip}
\label{app:g:qpf}

This game was played by Captain Picard (P) and Q (two characters in the American TV series Star Trek: The Next Generation) on the bridge of the Starship Enterprise. Q offers his help to the crew provided that P can beat him. In this game first P puts a penny head up into a box, thereafter they can reverse the penny without seeing it (first Q, then P, and finally Q again). In this two-person zero-sum game Q wins if the penny is head up when opening the box. Between two classical players this game has a Nash equilibrium when each player selects one of the two possibilities randomly in subsequent steps. In the present situation, however, Q possesses a curious capability: he can create a quantum superposition state of the penny (we can think of the quantum coin as the spin of an electron). \citet{meyer_prl99} shows that due to this advantage Q can always win this game, if he puts the coin into a quantum state, i.e., an equal mixture of head $(1,0)^T$ and tail $(0,1)^T$ [i.e., $a(1,0)^T +b(0,1)^T$, where $a$ and $b$ are C numbers satisfying the conditions $a \bar{a}+b \bar{b}=1$ and $|a|=|b|=1/\sqrt{2}$]. In the next step P can only perform a classical spin reversal or leave the quantum coin unchanged. Q wins because the second quantum operation allows him to rotate the coin into its original (head up) state independently of the action of P.

In the quantum mechanical context \citep{lee_pw02} the $1/2$ spin has two pure states: pointing up or down along the $z$ axis. The classical spin flip made (or not) by P rotates the spin along the $x$ axis. In his first action Q performs a Hadamard operation creating a spin state pointing along the $+x$ axis. This state cannot be affected by P so Q can easily restore the initial (spin up) state by the second (inverse Hadamard) operation.

\subsection{Rock-Scissors-Paper game}
\label{app:g:rsp}

The Rock-Scissors-Paper game is a two-person, zero-sum game with three pure strategies (items) named ``rock", ``scissors", and ``paper". After a synchronization procedure depending on culture, the two players have to choose a strategy simultaneously and show it by their hands. If the players form the same item then it is a tie, and the round is repeated once again. In this game each item is superior to one other, and inferior to the third one: rock beats scissors beat paper beats rock. In other words, rock (indicated by keeping the hand in a fist) crushes scissors, scissors (by extending the first two fingers and holding them apart) cut paper, and paper (by holding the hand flat) covers rock.

Several extensions of this game have been invented and are played occasionally word wide. These may allow further choices. For example, in the Rock-Scissors-Paper-Spock-Lizard game each item beats two others and is beaten by the remaining two ones, that is, scissors cut paper covers rock crushes lizard poisons Spock smashes scissors decapitate lizard eats paper disproves Spock vaporizes rock crushes scissors.

\subsection{Snowdrift game}
\label{app:g:sd}

Two drivers are trapped on opposite sides of a snowdrift. They have to choose between two options: (1) to get out and start shoveling (cooperate); (2) to remain in the car (defect). If both drivers are willing to shovel then each one has the benefit $b$ of getting home and they share the cost $c$ of the work, i.e., each receives the reward of mutual cooperation $R=b-c/2$.
If both drivers choose defection then they do not get home and obtain zero benefit ($P=0$). If only one of them shovels then both get home, however, the defector income ($T=b$) is not reduced by the cost of shoveling, whereas the cooperator gets $S=b-c$. This $2 \times 2$ matrix game becomes equivalent to the Prisoner's Dilemma in Eq.\ (\ref{eq:app_pd}) when $2b>c>b>0$. However, for $b>c>0$ the payoffs generate the Snowdrift game, which is in fact equivalent to the Hawk-Dove game. Two different versions of the spatial evolutionary Snowdrift game were very recently introduced and studied by \citet{hauert_n04} and \citet{sysiaho_epjb05}.

\subsection{Stag Hunt game}
\label{app:g:sh}

The story of the Stag Hunt game was briefly described by Rousseau in {\it A Discourse on Inequality} (1755). In Maurice Cranston's translation deer means stag:

{\it If it was a matter of hunting a deer, everyone well realized that he must remain faithfully at his post; but if a hare happened to pass within the reach of one of them, we cannot doubt that he would have gone off in pursuit of it without scruple and, having caught his own prey, he would have cared very little about having caused his companions to lose theirs.}

Each hunter prefers stag to hare and hare to nothing. In the context of game theory this means that the highest income is reached if each hunter chooses hunting deer. The chance of a successful deer hunt increases with the number of hunters, and practically there is no chance of bagging a deer by oneself. At the same time the chance of getting a hare is independent of what others do. Consequently, for two hunters the payoffs can be given by a bi-matrix as
\begin{equation}\label{eq:app_sh}
\begin{tabular}{lc|ccccc}
  & & & \multicolumn{4}{c}{\emph{Hunter 2}} \\
  & & & Hare &    & & Stag \\[0.5ex] \hline
  & &  &  & &  &    \\[-2.5ex]
  &Hare\phantom{a}&  & $(1,1)$ & &  & $(2,0)$  \\[-1.2ex]
\emph{Hunter 1}  &   & &  & &   &     \\[-1.2ex]
  &Stag\phantom{a}& & $(0,2)$ & &   & $(3,3)$  \\
\end{tabular} \qquad
\end{equation}

The Stag Hunt game is a prototype of the social contract, it is in fact a special case of Coordination games.

\subsection{Ultimatum game}
\label{app:g:ultim}

In the Ultimatum game two players have to agree on how to share a sum of money. One of the randomly chosen players (called proposer) makes an
offer and the other (responder) can either accept or reject it. If the offer is accepted, the money is shared accordingly; if rejected, both players receive nothing. In the one-shot game rational responders
should accept any positive offer, while the best choice for the proposer is to offer the minimum (positive) sum \citep{gintis_00}. In human experiments \citep{guth_jebo82,thaler_jep88,henrich_aer01} the majority of proposers offer 40 to 50\% of the total sum, and the responders frequently reject offers below 30\%. This human behavior can be reproduced by several evolutionary versions of the Ultimatum game \citep{nowak_s00,page_jtb00,sanchez_jtb05}. A spatial version of the game was studied by \citet{page_prsb00}.

\section{Strategies}
\label{app:str}

\subsection{Pavlovian strategies}
\label{app:str:pav}

The Pavlov strategy was introduced in the context of the Iterated Prisoner's Dilemma. By definition Pavlov works according to the following algorithm: ``repeat your latest action if that produced one of the two highest possible payoffs, and switch to the other possible action, if your last round payoff was one of the two lowest possible payoffs". As such Pavlov belongs to the more general class of Win-Stay-Lose-Shift strategies, which define a direct payoff criterium (aspiration level) for strategy change. An alternative definition frequently appearing in the literature is ``cooperate if and only if you and your opponent used the same move in the previous round". Of course, this translates into the same rule for the Prisoner's Dilemma.

For a general discussion of Pavlovian strategies see \citet{kraines_td89}.

\subsection{Tit-for-Tat}
\label{app:str:tft}

The Tit-for-Tat strategy suggested by Anatol Rapoport become world-wide known after winning the computer tournaments conducted by \citet{axelrod_84}. For the Iterated Prisoner's Dilemma this strategy starts by cooperating in the first step and afterwards repeats the previous decision of the opponent.

Tit-for-Tat cooperates mutually with all so-called ``nice" strategies, and it is never the first to defect. On long time scales the Tit-for-Tat strategy cannot be exploited, because any defection is retaliated by playing defection until the co-player chooses cooperation again. Then the opponent's extra income gained at the first defection is returned to Tit-for-Tat. At the same time, it is a forgiving strategy, because it is willing to cooperate again until the next defection of its opponent. Against a Tit-for-Tat the best choice is mutual cooperation. In multi-agent evolutionary Prisoner's Dilemma games the Tit-for-Tat strategy helps effectively to maintain cooperative behavior. \citet{axelrod_84} concluded that individuals should follow this strategy in a declarative way in all repeated decision situations analogous to the Prisoner's Dilemma.

This deterministic strategy, however, has a drawback in noisy environments, where two Tit-for-Tat strategists may easily end up alternating cooperation and defection in opposite phase without real hope to get out from this deadlock. Most of the modified versions of Tit-for-Tat were introduced to eliminates or reduce this shortcoming. For example, \emph{Tit for Two Tats} only defects if its opponent has defected twice in a row; \emph{Generous (Forgiving) Tit-for-Tat} cooperates with some probability even if the co-player has defected previously, etc.

\subsection{Win stay lose shift}
\label{app:str:wsls}

It seems that the Win-Stay-Lose-Shift (WSLS) idea as a learning rule was originally introduced as early as 1911 by \citet{thorndike_11}. WSLS strategies use a heuristic  update rule which depends on a direct payoff criterium, a so-called aspiration level, dictating the agent when to change her intended action. If the payoff average of recent rounds is above the aspiration level the agent keeps her original action, otherwise she switches to a new one. As such the aspiration level distinguishes between winning and losing situations. In the case when there are more than one alternatives to switch for, the choice can be random. The simplest example of a WSLS-type strategy is Pavlov, introduced in the context of the Iterated Prisoner's Dilemma. Pavlov was demonstrated to be able to defeat Tit-for-Tat in noisy environments \citep{nowak_n93,kraines_td93}, thanks to its ability to correct mistakes and exploit unconditional cooperators better than Tit-for-Tat.

\subsection{Stochastic reactive strategies}
\label{app:str:pq}

Decision in reactive strategies only depend on the previous move of the opponent. As suggested by \citet{nowak_jtb89,nowak_amc89} in the context of the Prisoner's Dilemma game, these strategies are characterized by two parameters, $p$ and $q$ ($0 \le p, q \le 1$), which denote the probability to cooperate after the opponent has cooperated or defected. The definition of these strategies is made complete by introducing a third parameter $u$ characterizing the probability of cooperation in the first step. For many evolutionary rules the value of $u$ becomes irrelevant in the stationary population of reactive strategies $(u,p,q)$. Evidently, for $p=q$ the decision is independent of the opponent's choice and the strategy $(0.5,0.5,0.5)$ represents a completely random decision. The strategies $(0,0,0)$ and $(1,1,1)$ are equivalent to unconditional defection (AllD) and unconditional cooperation (AllC), respectively. Tit-for-Tat and Suspicious Tit-for-Tat can be represented as $(1,1,0)$ and $(0,1,0)$.

The above class of reactive strategies was extended by \citet{nowak_geb95} in a way that allows different cooperation probabilities for all possible outcomes realized in the previous step. This strategy can be represented as $(p_1,p_2,p_3,p_4)$ where $p_1$, $p_2$, $p_3$, and $p_4$ denote the cooperation probability after the players' previous decisions $(C,C)$, $(C,D)$, $(D,C)$, and $(D,D)$, respectively. (For noisy systems we can ignore the probability $u$ of cooperating in the first step.) Evidently, some elements of this wider set of strategies can also be related to other strategies. For example, Tit-for-Tat corresponds to $(1,0,1,0)$.

\section{Generalized mean-field approximations}
\label{app:gmfa}

Now we give a simple and concise introduction to the use of the generalized mean-field technique on such lattice systems where the spatial distribution of $Q$ different states is described by a set of site variables denoted shortly as $s_x= 1, \ldots , Q$. For the sake of simplicity the present description will be formulated on a square lattice at the levels of one-, two-, and four-site approximations.

In this approach the translation and rotation invariant states of the system are described by a set configuration probabilities on compact clusters of sites with different sizes (and forms). The time-dependence of these configuration probabilities is not denoted. Thus, the one-site configuration probability $p_1(s_1)$ characterizes the probability of finding state $s_1$ for any site $x$ of the lattice. Similarly, $p_2(s_1,s_2)$ indicates the configuration probability of a pair of states $(s_1,s_2)$ on two neighboring sites independently of both the position and direction of the two-site cluster. These quantities satisfy the compatibility conditions that obey simple forms if translation invariance holds, i.e.,
\begin{eqnarray}
\label{eq:2s_c}
p_1(s_1)&=&\sum_{s_2} p_2(s_1,s_2) \nonumber \\
              &=&\sum_{s_2} p_2(s_2,s_1) \;.
\end{eqnarray}
On the square lattices the quantities $p_4(s_1,s_2,s_3,s_4)$ describe all the possible four-site configuration probabilities on the $2 \times 2$ clusters
of sites. In this notation the horizontal and vertical pairs are $(s_1,s_2)$, $(s_3,s_4)$, and  $(s_1,s_3)$, $(s_2,s_4)$, respectively. Evidently, these quantities are related to the pair configuration probabilities via the following compatibility conditions:
\begin{eqnarray}
p_2(s_1,s_2)
&=&\sum_{s_3,s_4} p_4(s_1,s_2,s_3,s_4) \nonumber \\
&=&\sum_{s_3,s_4} p_4(s_3,s_4,s_1,s_2) \nonumber \\
&=&\sum_{s_3,s_4} p_4(s_1,s_3,s_2,s_4) \nonumber \\
&=&\sum_{s_3,s_4} p_4(s_3,s_4,s_1,s_2) \;.
\label{eq:4s_c}
\end{eqnarray}
The above compatibility conditions [including the normalization $\sum_{s_1} p_1(s_1)=1$] allow us to reduce the number of parameters in the description of all the possible configuration probabilities. For example, assuming translation and rotation symmetry in a two-state ($Q=2$) system, the one- and two-site configuration probabilities can be characterized by two parameters,
\begin{eqnarray}
p_1(1)&=& \varrho \;, \nonumber \\
p_1(2)&=& 1-\varrho \;, \label{eq:2s_p1} \\
p_2(1,1)&=& \varrho^2 +q \;,\nonumber \\
p_2(1,2)&=& \varrho (1-\varrho) -q \;, \nonumber \\
p_2(2,1)&=& \varrho (1-\varrho) -q \;, \nonumber \\
p_2(2,2)&=& (1 - \varrho)^2 +q \;, \label{eq:2s_p2}
\end{eqnarray}
where $0 \le \varrho \le 1$ means the average concentration of state $1$, and $q$ denotes the deviation from the prediction of the mean-field approximation [$ - \min (\varrho^2, (1-\varrho)^2) \le q \le \varrho (1-\varrho)$]. The reader can easily check that under the same conditions the four-site configuration probabilities can be described by introducing three additional parameters.
Notice that this type of parametrization reduces the number of variables we have to determine by solving a suitable set of equations of motion. Evidently, significantly more parameters are necessary for $Q > 2$ despite the possible additional symmetries (e.g., cyclic symmetry in the Rock-Scissors-Paper game) reducing the number of independent configuration probabilities. The choice of an appropriate parametrization can significantly simplify the calculations.

Within the framework of the traditional mean-field theory the configuration probabilities on a given cluster are approximated by a product of one-site configuration probabilities, i.e., $p_2(s_1,s_2) \simeq p_1(s_1) p_1(s_2)$ [corresponding to $q=0$ in Eqs.~(\ref{eq:2s_p2})] and $p_4(s_1,s_2,s_3,s_4) \simeq p_1(s_1) p_1(s_2) p_1(s_3) p_1(s_4)$. For this approach the compatibility conditions remain self-consistent and the normalization is conserved.

Adopting the Bayesian extension process \citep{gutowitz_pd87} we can construct a better approximation for the configuration probabilities on large clusters by building them from configuration probabilities on smaller clusters. For example, on three subsequent sites the configuration probability can be constructed from two two-site configuration probabilities as
\begin{equation}
\label{eq:3p_2p}
p_3(s_1,s_2,s_3)\simeq {p_2(s_1,s_2) p_2(s_2,s_3) \over p_1 (s_2)} \;.
\end{equation}
At the level of the pair approximation on $k$ ($k>2$) subsequent sites (positioned linearly) the corresponding expression obeys the following form:
\begin{equation}
\label{eq:3p_kp}
p_k(s_1,\ldots,s_k)\simeq p_2(s_1,s_2)\prod_{n=2}^{k-1} {p_2(s_n,s_{n+1}) \over p_1 (s_n)} \;.
\end{equation}
This equation predicts exponentially decreasing configuration probabilities for
a homogeneous distribution on a $k$-site block. For example,
\begin{equation}
\label{eq:3p_k1}
p_k(1,\ldots,1)\simeq p_1(1)\left[ {p_2(1,1) \over p_1 (1)}\right]^k =
p_1(1) e^{-k/\xi_1} \;,
\end{equation}
where $\xi_1=-1/\ln(p_2(1,1)/p_1(1))$ characterizes the typical size of the homogeneous domain.

Knowing the pair configuration probabilities the autocorrelation functions can be expressed as
\begin{equation}
C_{s}(x) = \sum_{s_2, \ldots, s_{x-1}} p_{x+1}(s,s_2,\ldots,s_{x-1},s) - p_1(s)p_1(s) \;.
\label{eq:corrfn}
\end{equation}
For the parametrization given by Eqs.~(\ref{eq:2s_p1}) and (\ref{eq:2s_p2}) in a two-state system this autocorrelation function obeys a simple form, $C_s(x=1)=q$. Using successive iterations one can easily derive the following relation:
\begin{equation}
C_{s}(x) = \varrho (1-\varrho ) \left[{q \over \varrho (1-\varrho ) }\right]^x
= \varrho (1-\varrho ) e^{-x/ \xi} \;,
\label{eq:cfq}
\end{equation}
where the correlation length is
\begin{equation}
\xi = -{1 \over \ln{q \over \varrho (1-\varrho)}} \;.
\label{eq:xi_q}
\end{equation}

\begin{figure}[ht]
\centerline{\epsfig{file=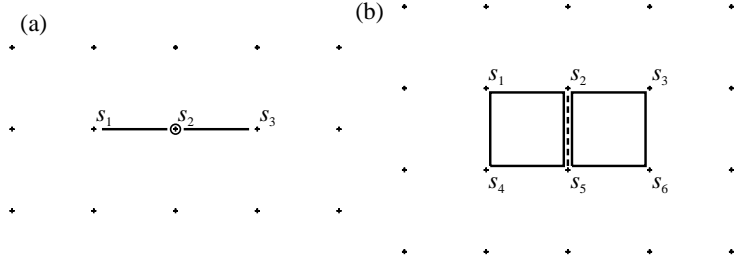,width=10cm}}
\caption{\label{fig:graphrep}Graphical representation of the construction of a configuration probability from configuration probabilities on smaller clusters for a three-site (a) and a six-site cluster (b).}
\end{figure}

On a $3 \times 2$ cluster the six-site configuration probabilities can be approximated as a product of two four-site configuration probabilities,
\begin{equation}
\label{eq:6p_4p}
p_6(s_1, \ldots ,s_6)\simeq
{p_4(s_1,s_2,s_4,s_5) p_4(s_2,s_3,s_5,s_6) \over p_2 (s_2,s_5)},
\end{equation}
where the configuration probability on the overlapping region of the two clusters appears in the denominator. The above expressions can be represented graphically as shown in Fig.~\ref{fig:graphrep}. In this representation the solid lines and squares represent two- and four-site configuration probabilities (in the nominator) with site variables at the connected sites (indicated by pluses), while the same quantities in the denominator are plotted by dashed lines.
At the same time the quantities $p_1(s_2)$ are denoted by closed (open) circles at the suitable sites if they appear in the nominator (denominator) of the corresponding products. In these constructions the compatibility conditions are not satisfied completely, although the normalization is conserved.
For example, the compatibility condition is broken for the second-neighbor pair configuration probabilities because $\sum_{s_2} p_3(s_1,s_2,s_3)$ deviates from those predicted by Eq. (\ref{eq:3p_2p}), $\sum_{s_2} p_2(s_1,s_2) p_2(s_2,s_3)$], while it remains valid for the nearest-neighbor pair configurations.

Each elementary step in the variation of the spatial distribution of states will modify the above configuration probabilities. In the knowledge of the dynamical rules we can derive a set of the equations of motion that summarizes the contributions of all types of elementary steps with the suitable weight. In order to demonstrate the essence of the derivation of these equations now we choose a very simple model.

Let us consider a three-species, cyclic predator-prey model (also called spatial Rock-Scissors-Paper game) on a square lattice where species 1 invades species 2 invades species 3 invades species 1 with the same invasion rates chosen to be unity. We assume that the evolution is governed by random sequential updates. More precisely, a randomly chosen species can be invaded by one of the four neighbors chosen randomly with a probability of $1/4$.

\begin{figure}[ht]
\centerline{\epsfig{file=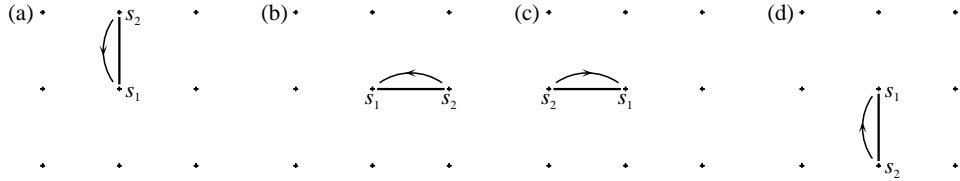,width=13cm}}
\caption{\label{fig:invsq1}The possible invasion processes from $s_2$ to $s_1$ which modify the one-site configuration probabilities.}
\end{figure}

When a predator $s_2$ invades the site of its neighboring prey $s_1$ then this process decreases the value of $p_1(s_1)$ and simultaneously increases $p_1(s_2)$ with the same magnitude whose rate becomes unity with a suitable choice of the time scale. Due to the symmetries, all the four possible invasion processes (see Fig.~\ref{fig:invsq1}) give the same contribution to the time derivative of the one-site configuration probabilities. Consequently, the derivative of $p_1(s_1)$ with respect to time can be expressed as
\begin{eqnarray}
\dot{p}_1(s_1)=
&-&\sum_{s_x}p_2(s_1,s_x) \Gamma_{\rm rsp}(s_1 \to s_x) \nonumber \\
&+&\sum_{s_x}p_2(s_x,s_1) \Gamma_{\rm rsp}(s_x \to s_1) \label{eq:inv1gen}
\end{eqnarray}
where $\Gamma_{\rm rsp} (s_x \to s_y)=1$ if the species $s_y$ is the predator of species $s_y$ and $\Gamma_{\rm rsp} (s_x \to s_y)=0$ otherwise.

In fact, the above general mathematical formulae hide the simplicity of both the calculation and results because the corresponding three equations for $s_1=1, 2,$ and $3$ obey the following forms:
\begin{eqnarray}
\dot{p}_1(1)&=&p_2(1,3)-p_2(1,2) \;, \nonumber \\
\dot{p}_1(2)&=&p_2(2,1)-p_2(2,3) \;, \nonumber \\
\dot{p}_1(3)&=&p_2(3,2)-p_2(3,1) \;.
\label{eq:inv1s}
\end{eqnarray}
Notice that the time-derivative of the functions $p_1(s_1)$ depend on the
two-site configuration probabilities $p_2(s_1,s_2)$. For the traditional mean-field approximations the two-site configuration probabilities are approximated as $p_2(s_1,s_2)=p_1(s_1)p_1(s_2)$. This approximation yields Eqs.~(\ref{eq:rspmf}) and the solutions are discussed in the section \ref{sec:rspmf}.

\begin{figure}[ht]
\centerline{\epsfig{file=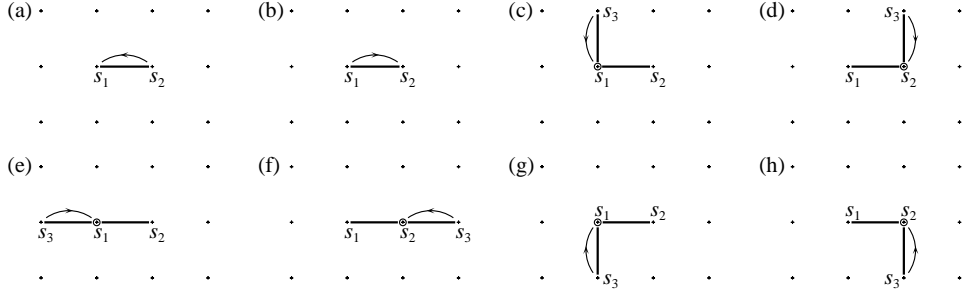,width=13cm}}
\caption{\label{fig:invsq2}The horizontal two-site configuration probabilities $p_2(s_1,s_2)$ are modified by the possible invasion processes indicated by arrows.}
\end{figure}

More accurate results can be obtained by deriving further set of equations of motion for $p_2(s_1,s_2)$ to improve the present approach. In analogy with the derivation of Eqs. (\ref{eq:inv1gen}) we can sum up the contributions of all the elementary processes (see Fig.~\ref{fig:invsq2}).

Within a horizontal pair there are two types of internal invasions [processes (a) and (b) in Fig.~\ref{fig:invsq2}] affecting the configuration probability $p_2(s_1,s_2)$. Their contribution to the function $\dot{p}_2(s_1,s_2)$ is not influenced by the surrounding states, and therefore it is proportional to the probability of the initial pair configuration $p_2(s_1,s_2)$. The contribution of the external invasions [processes (c) - (h) in Fig.~\ref{fig:invsq2}], however, are proportional to the corresponding three-site configuration probabilities that can be constructed from pair configuration probabilities as described above. This approximation neglects the shape differences between the three-site clusters. In this case the approximative equation of motions for $p_2(s_1,s_2)$ can be given as
\begin{eqnarray}
\dot{p}_2(s_1,s_2)=
&-&{1 \over 4}p_2(s_1,s_2)\Gamma_{\rm rsp}(s_2 \to s_1) \nonumber \\
&-&{1 \over 4}p_2(s_1,s_2)\Gamma_{\rm rsp}(s_1 \to s_2) \nonumber \\
&+&{1 \over 4}\delta(s_1,s_2)\sum_{s_3}p_2(s_1,s_3)\Gamma_{\rm rsp}(s_3 \to s_1) \nonumber \\
&+&{1 \over 4}\delta(s_1,s_2)\sum_{s_3}p_2(s_3,s_2)\Gamma_{\rm rsp}(s_3 \to s_2) \nonumber \\
&-&{3 \over 4}\sum_{s_3}{p_2(s_3,s_1)p_2(s_1,s_2)\over p_1(s_1)} \Gamma_{\rm rsp}(s_1 \to s_3)  \nonumber \\
&-&{3 \over 4}\sum_{s_3}{p_2(s_1,s_2)p_2(s_2,s_3)\over p_1(s_2)} \Gamma_{\rm rsp}(s_2 \to s_3)  \nonumber \\
&+&{3 \over 4}\sum_{s_3}{p_2(s_1,s_3)p_2(s_3,s_2)\over p_1(s_3)} \Gamma_{\rm rsp}(s_x \to s_1)  \nonumber \\
&+&{3 \over 4}\sum_{s_3}{p_2(s_1,s_3)p_2(s_3,s_2)\over p_1(s_3)} \Gamma_{\rm rsp}(s_x \to s_2) \; ,
\label{eq:emp2}
\end{eqnarray}
where $\delta (s_1,s_2)$ denotes the Kronecker delta. Here it is worth mentioning that the right hand side of Eq.\ (\ref{eq:emp2}) only depends on the pair configuration probabilities if the compatibility conditions Eq.\ (\ref{eq:2s_c}) are taken into consideration.

\begin{figure}[ht]
\centerline{\epsfig{file=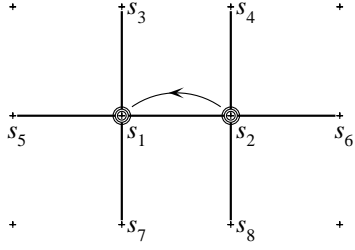,width=5cm}}
\caption{\label{fig:invnni2}The horizontal two-site configuration probabilities $p_2(s_1,s_2)$ are modified by the internal invasion process (indicated by the arrow) with a probability depending on all the labeled sites. The graphical representation of the construction of the corresponding eight-site configuration probabilities from pair configuration probabilities is illustrated as above.}
\end{figure}

A more complicated set of equations of motion can be derived for those systems where the probability of the elementary invasion (strategy adoption) processes depends not only on the given two site variables but on their surroundings too. Such a situation is illustrated in Fig.~\ref{fig:invnni2} where the probability of strategy adoption from one of the randomly chosen neighbors (e.g., $s_2$ is substituted for $s_1$) depends on the neighborhood ($s_3, \ldots , s_8$).

In this case the negative $\Delta_{s_1 \to s_2}$ contribution of this elementary process to $\dot{p}_2(s_1,s_2)$ is
\begin{equation}
\Delta_{s_1 \to s_2} = -\sum_{s_3,s_4,s_5, \atop s_6,s_7,s_8} p_8(s_1, \ldots , s_8) \Gamma (s_1 \to s_2),
\end{equation}
where some possible definitions for $\Gamma (s_1 \to s_2)$ are discussed in Section \ref{sec:mur} [for a definite expression see Eq.\ (\ref{eq:smoothedimit})]. Simultaneously, these processes give opposite ($-\Delta_{s_1 \to s_2}$) contribution to $\dot{p}_2(s_2,s_2)$. According to the above graphical representation, at the pair approximation level the eight-site configuration probabilities can be approximated as
\begin{eqnarray}
p_8(s_1, \ldots , s_8) &\simeq&
   p_2(s_1,s_2){p_2(s_1,s_3)p_2(s_1,s_5)p_2(s_1,s_7) \over p_1^3(s_1)} \nonumber \\
&& \cdot {p_2(s_2,s_4)p_2(s_2,s_6)p_2(s_2,s_8)  \over p_1^3(s_2)},
\label{eq:8p}
\end{eqnarray}
where the contributions of the denominator are indicated by the three concentric circles at the sites $s_1$ and $s_2$ in Fig.~\ref{fig:invnni2}. Similar terms describe the contributions of the opposite strategy adoption ($s_2 \to s_1$,
as well as of those processes where an external strategy will be substituted for either $s_1$ or $s_2$ (e.g., $s_1 \to s_3$). Thus, the complete set of the equations of motion for $p_2(s_1,s_2)$ can be obtained by summing up the contributions of each possible elementary process. Despite the simplicity of the derivation of the corresponding differential equations, the final formulae become very lengthy even with the use of sophisticated notations. This is the reason why we do not display the whole formulae. It is emphasized, however, that this difficulty becomes irrelevant in the numerical solutions, because efficient algorithms can be developed to derive the equations of motion.

For a small number of independent configuration probabilities we always have some chance to find analytical solutions. Besides this two methods are used to solve numerically the resultant set of equations of motions. In the first case the differential equations are integrated numerically (with respect to time) starting from an initial state with configuration probabilities satisfying the symmetry conditions. This is the only way if the system tends to a limit cycle. In general, the initial state can be any uncorrelated state with arbitrary concentration of states. Special initial states can be chosen if we wish to study the competition between two types of ordered domains which do not have common configurations. Using this approach, we can study the average velocity of interfaces separating two different homogeneous states. For example, at the level of the pair approximation in a one-dimensional, two-state system the evolution can be started from a state: $p_2(1,0)=p_2(0,1)=\epsilon$ and $p_2(0,0)=p_2(1,1)=1/2-\epsilon$, where $0<\epsilon << 1$ means the density of $(0,1$ and $(1,0)$ domain walls. After some transient events the quantity $\dot{p}_1(1)/2\epsilon$ characterizes the average velocity of the growth of large domains of state 1.

In the second case the stationary solution(s) ($\dot{p}_2(s_1,s_2)$=0) can be found by using the standard Newton-Raphson method \citep{ralston_65}. The efficiency of this method can be increased by reducing the number of independent configuration probabilities by taking all symmetries into consideration. For this iteration technique the solution algorithm should be started from a state staying within the region of convergence. Sometimes it is convenient to find a solution by using the above mentioned numerical integration for a given values of parameters, and afterward we can repeat the Newton-Raphson method whereas parameters are varied very slowly. It is emphasized that this method can find the unstable solutions too.

At the level of two-site approximation the construction of the configuration probabilities on large clusters neglects several pair correlations that may be important for some models (e.g., the construction represented graphically in Fig.~\ref{fig:invnni2} does not involve explicitly the pair correlations between the states $s_3$ and $s_4$). The best way to overcome this difficulty is to extend further the method. The larger the cluster size we use in this technique, the higher the accuracy we can achieve. Sometimes the gradual increase of the cluster size results in qualitative improvement in the prediction as discussed in Secs.\ \ref{sec:spdrsu} and \ref{sec:rsppa}. The investigation of the configuration probabilities on $2 \times 2$ (or larger) clusters becomes important for models where the local structure of distribution, the motion of invasion fronts, and/or the long range correlations play relevant roles. The corresponding set of the equations of motion can be derived via the straightforward generalization of the method described above.
That is, we sum the contributions to the time-derivative of configuration probabilities coming from all the possible elementary processes.

\begin{figure}[ht]
\centerline{\epsfig{file=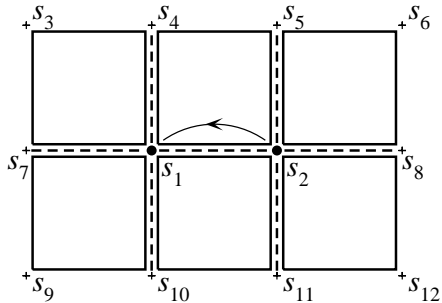,width=6cm}}
\caption{\label{fig:invnni4}Strategy adoption from $s_2$ to $s_1$ will modify all the four four-site configuration probabilities on the $2 \times 2$ blocks including the sites $s_1$.}
\end{figure}

Evidently, the contributions to $\dot{p}_4(s_1,s_2,s_3,s_4)$ are determined by the strategy distributions on larger clusters. However, the corresponding configuration probabilities can be constructed from four-site configuration probabilities as shown in Fig.~\ref{fig:invnni4}, where we assumed that the invasion rates are affected by the surrounding sites. Using this trick the quantities $\dot{p}_4(s_1,s_2,s_3,s_4)$ can be expressed as nonlinear functions of $p_4(s_1,s_2,s_3,s_4)$. The above symmetric construction takes explicitly into account all the relevant four-site correlations, while the condition of normalization for the large clusters are no longer valid. Generally this shortcoming of the present construction does not cause difficulties in the calculations.

The applicability of this method is limited by the large number of different configuration probabilities which increases exponentially with the cluster size and also by the increasing number of terms in the corresponding equations of motion. Current computer facilities allow us to perform such numerical investigations on even $3 \times 3$ clusters if the dynamical rule is not very complicated \citep{szolnoki_pre05a}. For the one-dimensional systems this method is used successfully up to cluster size 11 \citep{dickman_pre02} for the investigation of a three-state stochastic sandpile model, and even longer clusters could be studied for some two-state systems \citep{szolnoki_pre02b}.

Knowing the configuration probabilities for the different phases one can give an estimation for the velocity of interface separating two stationary domains. The method suggested by \citet{ellner_jmb98} is based on the pair correlations in the confronting phases. This approach proved to be successful for an evolutionary prisoner's Dilemma on the one-dimensional lattice \citep{szabo_pre00a} and also for a driven lattice gas model \citep{dickman_pre01}.

Disregarding technical difficulties this method can be easily adapted either for other (spatial) lattice structures with arbitrary dimensions or for Bethe lattices \citep{szabo_pre00b}. In fact, the pair approximation (with a construction represented in Fig.~\ref{fig:invnni2}) is expected to give a better prediction on Bethe lattices, because here the correlation between two sites can be mediated along only a single path.

The above description was only concentrated on the effect of asynchronous nearest-neighbor invasions (strategy adoptions). With suitable modifications, however, this method can be adjusted to consider other local dynamical rules, such as the appearance of mutants and the strategy exchange or diffusion when two site variables change simultaneously.

Variants of this method can be used to study stationary states (evolved from a random initial state) for deterministic or stochastic cellular automata \citep{gutowitz_pd87,atman_pre03}. In this case one should derive a set of recursion equations for the configuration probabilities at a discrete time $t+1$ as a function of the configuration probabilities at time $t$. Using the above described construction of configuration probabilities we can obtain a finite set of equations whose stationary solution can be found analytically or numerically. In this case difficulties can arise for those constructions which do not conserve normalization (such an example is shown in Fig.~\ref{fig:invnni4}).


\begin{thebibliography}{427}
\expandafter\ifx\csname natexlab\endcsname\relax\def\natexlab#1{#1}\fi
\expandafter\ifx\csname url\endcsname\relax
  \def\url#1{\texttt{#1}}\fi
\expandafter\ifx\csname urlprefix\endcsname\relax\def\urlprefix{URL }\fi

\bibitem[{Abramson and Kuperman(2001)}]{abramson_pre01}
Abramson, G., Kuperman, M., 2001. Social games in a social network. Phys. Rev.
  E 63, 030901(R).

\bibitem[{Ahmed et~al.(2006)Ahmed, Hegazi, Elettreby, and Askar}]{ahmed_pa06}
Ahmed, E., Hegazi, A.~S., Elettreby, M.~F., Askar, S.~S., 2006. On multi-team
  games. Physica A 369, 809--816.

\bibitem[{Aktipis(2004)}]{aktipis_jtb04}
Aktipis, C.~A., 2004. Known when to walk away: contingent movement and the
  evolution of cooperation. J. Theor. Biol. 231, 249--260.

\bibitem[{Albert and Barab{\'a}si(2002)}]{albert_rmp02}
Albert, R., Barab{\'a}si, A.-L., 2002. Statistical mechanics of complex
  networks. Rev. Mod. Phys. 74, 47--97.

\bibitem[{Alexander(1987)}]{alexander_87}
Alexander, R.~D., 1987. The biology of moral systems. Aldinee de Gruyter, New
  York.

\bibitem[{Alonso-Sanz et~al.(2001)Alonso-Sanz, Mart{\'{\i}}n, and
  Mart{\'{\i}}n}]{alonsosanz_ijbc01b}
Alonso-Sanz, R., Mart{\'{\i}}n, C., Mart{\'{\i}}n, M., 2001. The effect of
  memory in the spatial continuous-valued prisoner's dilemma. Int. J. Bifurcat.
  Chaos 11, 2061--2083.

\bibitem[{Amaral et~al.(2000)Amaral, Scala, Barth{\'e}l{\'e}my, and
  Stanley}]{amaral_pnas00}
Amaral, L. N.~A., Scala, A., Barth{\'e}l{\'e}my, M., Stanley, H.~E., 2000.
  Classes of small-world networks. Proc. Natl. Acad. Sci. USA 97, 11149--11152.

\bibitem[{Antal and Droz(2001)}]{antal_pre01}
Antal, T., Droz, M., 2001. Phase transitions and oscillations in a lattice
  prey-predator model. Phys. Rev. E 63, 056119.

\bibitem[{Antal and Scheuring(2006)}]{antal_bmb06}
Antal, T., Scheuring, I., 2006. Fixation of strategies for an evolutionary game
  in finite populations. Bull. Math. Biol. 68, 1923--1944.

\bibitem[{Arthur(1994)}]{arthur_aer94}
Arthur, W.~B., 1994. Inductive reasoning and bounded rationality. Am. Econ.
  Rev. 84, 406--411.

\bibitem[{Ashlock et~al.(1996)Ashlock, Smucker, Stanley, and
  Tesfatsion}]{ashlock_bs96}
Ashlock, D., Smucker, M.~D., Stanley, E.~A., Tesfatsion, L., 1996. Preferential
  partner selection in an evolutionary study of prisoner's dilemma. BioSystems
  37, 99--125.

\bibitem[{Atman et~al.(2003)Atman, Dickman, and Moreira}]{atman_pre03}
Atman, A. P.~F., Dickman, R., Moreira, J.~G., 2003. Phase diagram of a
  probabilistic cellular automaton with three-site interactions. Phys. Rev. E
  67, 016107.

\bibitem[{Aumann(1992)}]{aumann_92}
Aumann, R.~J., 1992. Irrationality in game theory. In: Dasgupta, P., Gale, D.,
  Hart, O., Maskin, E. (Eds.), Economic Analysis of Markets and Games: Essays
  in Honor of Frank Hahn. MIT Press, Cambridge, MA, p. 214–227.

\bibitem[{Axelrod(1980{\natexlab{a}})}]{axelrod_jcr80a}
Axelrod, R., 1980{\natexlab{a}}. Effective choice in the prisoner's dilemma. J.
  Confl. Resol. 24, 3--25.

\bibitem[{Axelrod(1980{\natexlab{b}})}]{axelrod_jcr80b}
Axelrod, R., 1980{\natexlab{b}}. More effective choice in the prisoner's
  dilemma. J. Confl. Resol. 24, 379--403.

\bibitem[{Axelrod(1984)}]{axelrod_84}
Axelrod, R., 1984. The Evolution of Cooperation. Basic Books, New York.

\bibitem[{Axelrod(1986)}]{axelrod_apsr86}
Axelrod, R., 1986. An evolutionary approach to norms. Am. Polit. Sci. Rev. 80,
  1095--1111.

\bibitem[{Axelrod et~al.(2006)Axelrod, Axelrod, and Pienta}]{axelrod_pnas06}
Axelrod, R., Axelrod, D.~E., Pienta, K.~J., 2006. Evolution of cooperation
  among tumor cells. Proc. Natl. Acad. Sci. 103, 13474--13479.

\bibitem[{Axelrod and Dion(1988)}]{axelrod_s88}
Axelrod, R., Dion, D., 1988. The further evolution of cooperation. Science 242,
  1385--1390.

\bibitem[{Axelrod and Hamilton(1981)}]{axelrod_s81}
Axelrod, R., Hamilton, W.~D., 1981. The evolution of cooperation. Science 211,
  1390--1396.

\bibitem[{Bak et~al.(1990)Bak, Chen, and Tang}]{bak_pla90}
Bak, P., Chen, K., Tang, C., 1990. A forest-fire model and some thoughts on
  turbulence. Phys. Lett. A 147, 297--300.

\bibitem[{Bala and Goyal(2000)}]{bala_em00}
Bala, V., Goyal, S., 2000. A noncooperative model of network formation.
  Econometrica 68, 1181--1229.

\bibitem[{Balkenborg and Schlag(1995)}]{balkenborg_95b}
Balkenborg, D., Schlag, K.~H., 1995. Evolutionary stability in asymmetric
  population games.
\newline\urlprefix\url{http://www.wiwi.uni-bonn.de/sfb/papers/1995/b/bonnsfb31%
4.ps}

\bibitem[{Ball(2004)}]{ball_04}
Ball, P., 2004. Critical mass: how one thing leads another. William Heinemann,
  London.

\bibitem[{Barab{\'a}si and Albert(1999)}]{barabasi_s99}
Barab{\'a}si, A.-L., Albert, R., 1999. Emergence of scaling in random networks.
  Science 286, 509--512.

\bibitem[{Bascompte and Sol{\'e}(1998)}]{bascompte_98}
Bascompte, J., Sol{\'e}, G. (Eds.), 1998. Modeling Spatiotemporal Dynamics in
  Ecology. Springer, New York.

\bibitem[{Baumol(1952)}]{baumol_52}
Baumol, W.~J., 1952. Welfare Economics and the Theory of the State. Harward
  University Press, Cambridge, MA.

\bibitem[{Ben-Naim et~al.(1996{\natexlab{a}})Ben-Naim, Frachebourg, and
  Krapivsky}]{ben-naim_pre96}
Ben-Naim, E., Frachebourg, L., Krapivsky, P.~L., 1996{\natexlab{a}}. Coarsening
  and persistence in the voter model. Phys. Rev. E 53, 3078--3087.

\bibitem[{Ben-Naim et~al.(1996{\natexlab{b}})Ben-Naim, Redner, and
  Krapivsky}]{ben-naim_jpa96}
Ben-Naim, E., Redner, S., Krapivsky, P.~L., 1996{\natexlab{b}}. Two scales in
  asynchnonous ballistic annihilation. J. Phys. A: Math. Gen. 29, L561--L568.

\bibitem[{Benaim and Weibull(2003)}]{benaim_e03}
Benaim, M., Weibull, J., 2003. Deterministic approximation of stochastic
  evolution in games. Econometrica 71, 873--903.

\bibitem[{Benzi et~al.(1982)Benzi, Parisi, Sutera, and Vulpiani}]{benzi_t82}
Benzi, R., Parisi, G., Sutera, A., Vulpiani, A., 1982. Stochastic resonance in
  climatic change. TELLUS 34, 10--16.

\bibitem[{Berg and Engel(1998)}]{berg_prl98}
Berg, J., Engel, M., 1998. Matrix games, mixed strategies, and statistical
  mechanics. Phys. Rev. Lett. 81, 4999--5002.

\bibitem[{Berg and Weigt(1999)}]{berg_epl99}
Berg, J., Weigt, M., 1999. Entropy and typical properties of \protect{Nash}
  equilibria in two-player games. Europhys. Lett. 48, 129--135.

\bibitem[{Bidaux et~al.(1989)Bidaux, Boccara, and Chat{\'e}}]{bidaux_pra89}
Bidaux, R., Boccara, N., Chat{\'e}, H., 1989. Order of the transition versus
  space dimension in a family of cellular automata. Phys. Rev. A 39,
  3094--3105.

\bibitem[{Biely et~al.(2007)Biely, Dragosits, and Thurner}]{biely_cm05}
Biely, C., Dragosits, K., Thurner, S., 2007. The prisoner's dilemma on
  co-evolving networks under perfect rationality. Physica D 228, 40--48.

\bibitem[{Binder and M{\"u}ller-Krumbhaar(1974)}]{binder_prb74}
Binder, K., M{\"u}ller-Krumbhaar, H., 1974. Investigations of metastable states
  and nucleation in the kinetic \protect{Ising} model. Phys. Rev. B 9,
  2328--2353.

\bibitem[{Binmore(1994)}]{binmore_94}
Binmore, K.~G., 1994. Playing fair: game theory and the social contract. MIT
  Press, Cambridge, MA.

\bibitem[{Binmore and Samuelson(1992)}]{binmore_jet92}
Binmore, K.~G., Samuelson, L., 1992. Evolutionary stability in repeated games
  played by finite automata. J. Econ. Theory 57, 278--305.

\bibitem[{Binmore et~al.(1995)Binmore, Samuelson, and Vaughan}]{binmore_geb95}
Binmore, K.~G., Samuelson, L., Vaughan, R., 1995. Musical chairs: Modeling
  noisy evolution. Games Econ. Behav. 11, 1--35.

\bibitem[{Bishop and Cannings(1978)}]{bishop_jtb78}
Bishop, T., Cannings, C., 1978. A generalized war of attrition. J. Theor. Biol.
  70, 85--124.

\bibitem[{Blarer and Doebeli(1999)}]{blarer_el99}
Blarer, A., Doebeli, M., 1999. Resonance effects and outbreaks in ecological
  time series. Ecol. Lett. 2, 167--177.

\bibitem[{Blume and Durlauf(2003)}]{blume_igtr03}
Blume, L., Durlauf, S., 2003. Equilibrium concepts for social interaction
  models. International Game Theory Review 5, 193--209.

\bibitem[{Blume(1993)}]{blume_geb93}
Blume, L.~E., 1993. The statistical-mechanics of strategic interactions. Games
  Econ. Behav. 5, 387--424.

\bibitem[{Blume(1995)}]{blume_geb95}
Blume, L.~E., 1995. The statistical-mechanics of best-response strategy
  revision. Games Econ. Behav. 11, 111--145.

\bibitem[{Blume(1998)}]{blume_98}
Blume, L.~E., 1998. Population games. In: Arthur, W.~B., Lane, D., Durlauf, S.
  (Eds.), The Economy as a Complex Adaptive System \protect{II}.
  Addison-Wesley.

\bibitem[{Blume(2003)}]{blume_geb03}
Blume, L.~E., 2003. How noise matters. Games Econ. Behav. 44, 251--271.

\bibitem[{Boccaletti et~al.(2006)Boccaletti, Latora, Moreno, Chavez, and
  Hwang}]{boccaletti_pr06}
Boccaletti, S., Latora, V., Moreno, Y., Chavez, M., Hwang, D., 2006. Complex
  networks: Structure and dynamics. Phys. Rep. 424, 175--308.

\bibitem[{Boerlijst and Hogeweg(1991)}]{boerlijst_pd91}
Boerlijst, M.~C., Hogeweg, P., 1991. Spiral wave structure in pre-biotic
  evolution: Hypercycles stable against parasites. Physica D 48, 17--28.

\bibitem[{Boerlijst et~al.(1997)Boerlijst, Nowak, and
  Sigmund}]{boerlijst_jtb97}
Boerlijst, M.~C., Nowak, M.~A., Sigmund, K., 1997. The logic of contrition. J.
  Theor. Biol. 185, 281--293.

\bibitem[{Bollob\'{a}s(1985)}]{bollobas_85}
Bollob\'{a}s, B., 1985. Random Graphs. Academic Press, New York.

\bibitem[{Bollob{\'a}s(1998)}]{bollobas_98}
Bollob{\'a}s, B., 1998. Modern Graph Theory. Springer, New York.

\bibitem[{Bomze(1983)}]{bomze_bc83}
Bomze, I., 1983. Lotka-volterra equations and replicator dynamics: A
  two-dimensional classification. Biol. Cybern. 48, 201--211.

\bibitem[{Bomze(1995)}]{bomze_bc95}
Bomze, I.~M., 1995. \protect{Lotka-Volterra} equation and replicator dynamics:
  New issues in classification. Biol. Cybern. 72, 447--453.

\bibitem[{Bradley et~al.(2005)Bradley, Clubb, Fisher, Gu{\'e}nault, Haley,
  Matthews, Pickett, Tsepelin, and Zaki}]{bradley_prl05}
Bradley, D.~I., Clubb, D.~O., Fisher, S.~N., Gu{\'e}nault, A.~M., Haley, R.~P.,
  Matthews, C.~J., Pickett, G.~R., Tsepelin, V., Zaki, K., 2005. Emission of
  discerete vortex rings by vibrating grid in superfluid \protect{$^3$He-B}: A
  precursor to quantum turbulence. Phys. Rev. Lett. 95, 035302.

\bibitem[{Bramson and Griffeath(1989)}]{bramson_ap89}
Bramson, M., Griffeath, D., 1989. Flux and fixation in cyclic particle systems.
  Ann. Prob. 17, 26--45.

\bibitem[{Brauchli et~al.(1999)Brauchli, Killingback, and
  Doebeli}]{brauchli_jtb99}
Brauchli, K., Killingback, T., Doebeli, M., 1999. Evolution of cooperation in
  spatially structured populations. J. Theor. Biol. 200, 405--417.

\bibitem[{Bray(1994)}]{bray_ap94}
Bray, A.~J., 1994. Theory of phase ordering kinetics. Adv. Phys. 43, 357--459.

\bibitem[{Broom(2000)}]{broom_mb00}
Broom, M., 2000. Bounds on the number of esss of a matrix game. Math. Biosci.
  167, 163--175.

\bibitem[{Brosig(2002)}]{brosig_jebo02}
Brosig, J., 2002. Identifying cooperative behavior: some experimental results
  in a prisoner's dilemma game. J. Econ. Behav. Org. 47, 275--290.

\bibitem[{Brower et~al.(1984)Brower, Kessler, Koplik, and Levin}]{brower_pra84}
Brower, R.~C., Kessler, D.~A., Koplik, J., Levin, H., 1984. Geometrical models
  of interface evolution. Phys. Rev. A 29, 1335--1342.

\bibitem[{Brown(1951)}]{brown_51}
Brown, G.~W., 1951. Iterative solution of games by fictious play. In: Koopmans,
  T.~C. (Ed.), Activity Analysis of Production and Allocation. Wiley, New York,
  pp. 373--376.

\bibitem[{Busse and Heikes(1980)}]{busse_s80}
Busse, F.~H., Heikes, K.~E., 1980. Convection in a rotating layer: a simple
  case of turbulence. Science 208, 173--175.

\bibitem[{Camerer and Thaler(1995)}]{camerer_jep95}
Camerer, C.~F., Thaler, R.~H., 1995. Anomalies: Ultimatums, dictators and
  manners. J. Econ. Persp. 9~(2), 209--219.

\bibitem[{Cardy and T{\"a}uber(1996)}]{cardy_prl96}
Cardy, J.~L., T{\"a}uber, U.~C., 1996. Theory of branching and annihilating
  random walks. Phys. Rev. Lett. 77, 4780--4783.

\bibitem[{Cardy and T{\"a}uber(1998)}]{cardy_jsp98}
Cardy, J.~L., T{\"a}uber, U.~C., 1998. Field theory of branching and
  annihilating random walks. J. Stat. Phys. 90, 1--56.

\bibitem[{Challet et~al.(2004)Challet, Marsili, and Zhang}]{challet_04}
Challet, D., Marsili, M., Zhang, Y.-C., 2004. Minority Games: Interacting
  Agents in Financial Markets. Oxford Univ. Press, Oxford.

\bibitem[{Challet and Zhang(1997)}]{challet_pa97}
Challet, D., Zhang, Y.-C., 1997. Emergence of cooperation and organization in
  an evolutionary game. Physica A 246, 407--418.

\bibitem[{Chandler(1987)}]{chandler_87}
Chandler, D., 1987. Introduction to Modern Statistical Mechanics. Oxford Univ.
  Press, Oxford.

\bibitem[{Cheng et~al.(2004)Cheng, Reeves, Vorobeychik, and
  Wellman}]{cheng_proc04}
Cheng, S.-F., Reeves, D.~M., Vorobeychik, Y., Wellman, M.~P., 2004. Notes on
  equilibria in symmetric games. In: AAMAS-04 Workshop on Game-Theoretic and
  Decision-Theoretic Agents. New York.

\bibitem[{Chiappin and de~Oliveira(1999)}]{chiappin_pre99}
Chiappin, J. R.~N., de~Oliveira, M.~J., 1999. Emergence of cooperation among
  interacting individuals. Phys. Rev. E 59, 6419--6421.

\bibitem[{Clifford and Sudbury(1973)}]{clifford_bm73}
Clifford, P., Sudbury, A., 1973. A model for spatial conflict. Biometrika 60,
  581--588.

\bibitem[{Colman(1995)}]{colman_95}
Colman, A., 1995. Game theory and its application. Butterworth--Heinemann,
  Oxford, UK.

\bibitem[{Conlisk(1996)}]{conlisk_jel96}
Conlisk, J., 1996. Why bounded rationality? J. Econ. Lit. XXXIV, 669--700.

\bibitem[{Coolen(2005)}]{coolen_05}
Coolen, A. C.~C., 2005. The Mathematical Theory of Minority Games: Statistical
  Mechanics of Interacting Agents. Oxford Univ. Press, Oxford.

\bibitem[{Coricelli et~al.(2004)Coricelli, Fehr, and Fellner}]{coricelli_jcr04}
Coricelli, G., Fehr, D., Fellner, G., 2004. Partner selection in public goods
  experiments. J. Conflict Resol. 48, 356--378.

\bibitem[{Cressman(2003)}]{cressman_03}
Cressman, R., 2003. Evolutionary Dynamics and Extensive Form Games. MIT Press,
  Cambridge, MA.

\bibitem[{Cross and Hohenberg(1993)}]{cross_rmp93}
Cross, M.~C., Hohenberg, P.~C., 1993. Pattern formation outside of equilibrium.
  Rev. Mod. Phys. 65, 851--1112.

\bibitem[{Cz{\'a}r{\'a}n et~al.(2002)Cz{\'a}r{\'a}n, Hoekstra, and
  Pagie}]{czaran_pnas02}
Cz{\'a}r{\'a}n, T.~L., Hoekstra, R.~F., Pagie, L., 2002. Chemical warfare
  between microbes promotes biodiversity. Proc. Natl. Acad. Sci. USA 99,
  786--790.

\bibitem[{Dawkins(1976)}]{dawkins_76}
Dawkins, R., 1976. The Selfish Gene. Oxford Univ. Press, Oxford.

\bibitem[{Der{\'e}nyi et~al.(2005)Der{\'e}nyi, Palla, and
  Vicsek}]{derenyi_prl05}
Der{\'e}nyi, I., Palla, G., Vicsek, T., 2005. Clique percolation in random
  networks. Phys. Rev. Lett. 94, 160202.

\bibitem[{Dickman(2001)}]{dickman_pre01}
Dickman, R., 2001. First- and second-order phase transitions in a driven
  lattice gas with nearest-neighbor exclusion. Phys. Rev. E 64, 016124.

\bibitem[{Dickman(2002)}]{dickman_pre02}
Dickman, R., 2002. n-site approximations and coherent-anomaly-method analysis
  for a stochastic sandpile. Phys. Rev. E 66, 036122.

\bibitem[{Dieckmann and Metz(1996)}]{dieckmann_jmb96}
Dieckmann, U., Metz, J. A.~J., 1996. The dynamical theory of coevolution: a
  derivation from stochastic ecological processes. J. Math. Biol. 34, 579--612.

\bibitem[{Doebeli and Hauert(2005)}]{doebeli_el05}
Doebeli, M., Hauert, C., 2005. Models of cooperation based on prisoner's
  dilemma and snowdrift game. Ecol. Lett. 8, 748--766.

\bibitem[{Domany and Kinzel(1984)}]{domany_prl84}
Domany, E., Kinzel, W., 1984. Equivalence of cellular automata to
  \protect{Ising} models and directed percolation. Phys. Rev. Lett. 53,
  311--314.

\bibitem[{Dornic et~al.(2001)Dornic, Chat{\'e}, Chave, and
  Hinrichsen}]{dornic_prl01}
Dornic, I., Chat{\'e}, H., Chave, J., Hinrichsen, H., 2001. Critical coarsening
  without surface tension: The universality class of the voter model. Phys.
  Rev. Lett. 87, 045701.

\bibitem[{Dorogovtsev and Mendes(2003)}]{dorogovtsev_03}
Dorogovtsev, S.~N., Mendes, J. F.~F., 2003. Evolution of Networks: From
  Biological Nets to the Internet and WWW. Oxford Univ. Press, Oxford.

\bibitem[{Dorogovtsev et~al.(2001)Dorogovtsev, Mendes, and
  Samukhin}]{dorogovtsev_pre01}
Dorogovtsev, S.~N., Mendes, J. F.~F., Samukhin, A.~N., 2001. Size-dependent
  degree distribution of a scale-free growing network. Phys. Rev. E 63, 062101.

\bibitem[{Douglass et~al.(1993)Douglass, Wilkens, Pantazelou, and
  Moss}]{douglass_n93}
Douglass, J.~K., Wilkens, L., Pantazelou, E., Moss, F., 1993. Noise enhancement
  of information transfer in crayfish mechanoreceptors by stochastic resonance.
  Nature 365, 337--340.

\bibitem[{Drossel(2001)}]{drossel_ap01}
Drossel, B., 2001. Biological evolution and statistical physics. Adv. Phys. 50,
  209--295.

\bibitem[{Drossel and Schwabl(1992)}]{drossel_prl92}
Drossel, B., Schwabl, F., 1992. Self-organized critical forest-fire model.
  Phys. Rev. Lett. 69, 1629--1732.

\bibitem[{Du et~al.(2002)Du, Li, Xu, Shi, Wu, Zhou, and Han}]{du_prl02}
Du, J., Li, H., Xu, X., Shi, M., Wu, J., Zhou, X., Han, R., 2002. Experimental
  realization of quantum games on a quantum computer. Phys. Rev. Lett. 88,
  137902.

\bibitem[{Dugatkin and Mesterton-Gibbons(1996)}]{dugatkin_bs96}
Dugatkin, L.~A., Mesterton-Gibbons, M., 1996. Cooperation among unrelated
  individuals: reciprocial altruism, by-product mutualism and group selection
  in fishes. BioSystems 37, 19--30.

\bibitem[{Duran and Mulet(2005)}]{duran_pd05}
Duran, O., Mulet, R., 2005. Evolutionary prisoner's dilemma in random graphs.
  Physica D 208, 257--265.

\bibitem[{Durrett(2002)}]{durrett_02}
Durrett, R., 2002. Mutual invadability implies coexistence in spatial models.
  American Mathematical Society, Providence, RI.

\bibitem[{Durrett and Levin(1997)}]{durrett_jtb97}
Durrett, R., Levin, S., 1997. Allelopathy in spatial distributed populations.
  J. Theor. Biol. 185, 165--171.

\bibitem[{Durrett and Levin(1998)}]{durrett_tpb98}
Durrett, R., Levin, S., 1998. Spatial aspects of interspecific competition.
  Theor. Pop. Biol. 53, 30--43.

\bibitem[{Dutta and Jackson(2003)}]{dutta_03}
Dutta, B., Jackson, M.~O., 2003. Models of the Formation of Networks and
  Groups. Springer-Verlag, Berlin.

\bibitem[{Ebel and Bornholdt(2002)}]{ebel_pre02}
Ebel, H., Bornholdt, S., 2002. Coevolutionary games on networks. Phys. Rev. E
  66, 056118.

\bibitem[{Eigen and Schuster(1979)}]{eigen_79}
Eigen, M., Schuster, P., 1979. The Hypercycle, A Principle of Natural
  Self-Organization. Springer-Verlag, Berlin.

\bibitem[{Eisert et~al.(1999)Eisert, Wilkens, and Lewenstein}]{eisert_prl99}
Eisert, J., Wilkens, M., Lewenstein, M., 1999. Quantum games and quantum
  strategies. Phys. Rev. Lett. 83, 3077--3080.

\bibitem[{Ellner et~al.(1998)Ellner, Sasaki, Haraguchi, and
  Matsuda}]{ellner_jmb98}
Ellner, S.~P., Sasaki, A., Haraguchi, Y., Matsuda, H., 1998. Word-of-mouth
  communication and social learning. J. Math. Biol. 36, 469--484.

\bibitem[{Equ{\'{\i}}luz et~al.(2005)Equ{\'{\i}}luz, Zimmermann, Cela-Conde,
  and San~Miguel}]{eguiluz_ajs05}
Equ{\'{\i}}luz, V.~M., Zimmermann, M.~G., Cela-Conde, C.~J., San~Miguel, M.,
  2005. Cooperation and the emergence of role differentiation in the dynamics
  of social networks. Am. J. Sociology 110, 977--1008.

\bibitem[{Erd{\H{o}}s and R{\'e}nyi(1959)}]{erdos_pmd59}
Erd{\H{o}}s, P., R{\'e}nyi, A., 1959. On random graphs. Publ. Math. Debrecen 6,
  290--297.

\bibitem[{Fehr and Fischbacher(2003)}]{fehr_n03}
Fehr, E., Fischbacher, U., 2003. The nature of human altruism. Nature 425,
  785--791.

\bibitem[{Field and Noyes(1974)}]{field_jcp74}
Field, R.~J., Noyes, R.~M., 1974. Oscillations in chemical systems.
  \protect{IV}. limit cycle behavior in a model of real chemical reaction. J.
  Chem. Phys. 60, 1877--1884.

\bibitem[{Fisch(1990)}]{fisch_pd90}
Fisch, R., 1990. Cyclic cellular automata and related processes. Physica D 45,
  19--25.

\bibitem[{Fisher(1930)}]{fisher_30}
Fisher, R.~A., 1930. The genetical theory of natural selection. Clarendon
  Press, Oxford.

\bibitem[{F{\"o}llmer(1974)}]{follmer_jme74}
F{\"o}llmer, H., 1974. Random economics with noisy interacting agents. J. Math.
  Econ. 1, 51--62.

\bibitem[{Forsythe et~al.(1994)Forsythe, Horowitz, Savin, and
  Sefton}]{forsythe_geb94}
Forsythe, R., Horowitz, J.~L., Savin, N.~E., Sefton, M., 1994. Fairness in
  simple bargaining experiments. Games Econ. Behav. 6, 347--369.

\bibitem[{Fort and Viola(2005)}]{fort_jsm05}
Fort, H., Viola, S., 2005. Spatial patterns and scale freedom in prisoner's
  dilemma cellular automata with \protect{Pavlovian} strategies. J. Stat. Mech.
  Theor. Exp. 2, P01010.

\bibitem[{Foster and Young(1990)}]{foster_tpb90}
Foster, D., Young, P., 1990. Stochastic evolutionary game dynamics. Theor. Pop.
  Biol. 38, 219--232.

\bibitem[{Frachebourg and Krapivsky(1998)}]{frachebourg_jpa98}
Frachebourg, L., Krapivsky, P.~L., 1998. Fixation in a cyclic
  \protect{Lotka-Volterra} model. J. Phys. A 31, L287--L293.

\bibitem[{Frachebourg et~al.(1996{\natexlab{a}})Frachebourg, Krapivsky, and
  Ben-Naim}]{frachebourg_prl96}
Frachebourg, L., Krapivsky, P.~L., Ben-Naim, E., 1996{\natexlab{a}}.
  Segregation in a one-dimensional model of interacting species. Phys. Rev.
  Lett. 77, 2125--2128.

\bibitem[{Frachebourg et~al.(1996{\natexlab{b}})Frachebourg, Krapivsky, and
  Ben-Naim}]{frachebourg_pre96}
Frachebourg, L., Krapivsky, P.~L., Ben-Naim, E., 1996{\natexlab{b}}. Spatial
  organization in cyclic \protect{Lotka-Volterra} systems. Phys. Rev. E 54,
  6186--6200.

\bibitem[{Frean and Abraham(2001)}]{frean_prsb01}
Frean, M., Abraham, E.~D., 2001. Rock-scissors-paper and the survival of the
  weakest. Proc. R. Soc. Lond. B 268, 1--5.

\bibitem[{Freidlin and Wentzell(1984)}]{freidlin_84}
Freidlin, M.~I., Wentzell, A.~D., 1984. Random Perturbations of Dynamical
  Systems. Springer-Verlag, New York.

\bibitem[{Frick and Schuster(2003)}]{frick_n03}
Frick, T., Schuster, S., 2003. An example of the prisoner's dilemma in
  biochemistry. Naturwissenschaften 90, 327--331.

\bibitem[{Friedman(1971)}]{friedman_res71}
Friedman, J., 1971. A non-cooperative equilibrium for supergames. Review of
  Economic Studies 38, 1--12.

\bibitem[{Fudenberg and Levine(1998)}]{fudenberg_98}
Fudenberg, D., Levine, D., 1998. The Theory of Learning in Games. MIT Press,
  Cambridge MA.

\bibitem[{Fudenberg and Maskin(1986)}]{fudenberg_e86}
Fudenberg, D., Maskin, E., 1986. The folk theorem in repeated games with
  discounting and incomplete information. Econometrica 54, 533--554.

\bibitem[{Fudenberg and Tirole(1991)}]{fudenberg_91}
Fudenberg, D., Tirole, J., 1991. Game Theory. MIT Press, Cambridge, MA.

\bibitem[{Fuks and Lawniczak(2001)}]{fuks_ddns01}
Fuks, H., Lawniczak, A.~T., 2001. Individual-based lattice model for spatial
  spread of epidemics. Discr. Dyn. Nat. Soc. 6, 191--200.

\bibitem[{Gambarelli and Owen(2004)}]{gambarelli_td04}
Gambarelli, G., Owen, G., 2004. The coming of game theory. Theor. Decis. 56,
  1--18.

\bibitem[{Gammaitoni et~al.(1998)Gammaitoni, H{\"a}nggi, Jung, and
  Marchasoni}]{gammaitoni_rmp98}
Gammaitoni, L., H{\"a}nggi, P., Jung, P., Marchasoni, F., 1998. Stochastic
  resonance. Rev. Mod. Phys. 70, 223--287.

\bibitem[{Gao(1999)}]{gao_pla99}
Gao, P., 1999. Direct integration method and first integrals for
  three-dimensional \protect{Lotka-Volterra} systems. Phys. Lett. A 255,
  253--258.

\bibitem[{Gao(2000)}]{gao_pla00}
Gao, P., 2000. Hamiltonian structure and first integrals for the
  \protect{Lotka-Volterra} systems. Phys. Lett. A 273, 85--96.

\bibitem[{Gardiner(2004)}]{gardiner_04}
Gardiner, C.~W., 2004. Handbook of Stochastic Methods, 3rd Edition. Springer.

\bibitem[{Gardner(1970)}]{gardner_sa70}
Gardner, M., 1970. Mathematical games: the fantastic combination of
  \protect{John Conway's} new solitary game 'life'. Sci. Am. 223, 120--123.

\bibitem[{Gatenby and Maini(2003)}]{gatenby_n03}
Gatenby, R.~A., Maini, P.~K., 2003. Cancer summed up. Nature 421, 321--321.

\bibitem[{Geritz et~al.(1997)Geritz, Metz, Kisdi, and
  Mesz{\'e}na}]{geritz_prl97}
Geritz, S. A.~H., Metz, J. A.~J., Kisdi, E., Mesz{\'e}na, G., 1997. Dynamics of
  adaptation and evolutionary branching. Phys. Rev. Lett. 78, 2024--2027.

\bibitem[{Gibbons(1992)}]{gibbons_92}
Gibbons, R., 1992. Game Theory for Applied Economists. Princeton University
  Press, Princeton.

\bibitem[{Gilpin(1975)}]{gilpin_an75}
Gilpin, M.~E., 1975. Limit cycles in competition communities. Am. Nat. 108,
  207--228.

\bibitem[{Gintis(2000)}]{gintis_00}
Gintis, H., 2000. Game Theory Evolving. Princeton University Press, Princeton.

\bibitem[{Glauber(1963)}]{glauber_jmp63}
Glauber, R.~J., 1963. Time-dependent statistics of the \protect{Ising} model.
  J. Math. Phys 4, 294--307.

\bibitem[{G{\'o}mez-Garde{\~ n}ez et~al.(2007)G{\'o}mez-Garde{\~ n}ez,
  Campillo, Flor{\'\i}a, and Moreno}]{gomez_prl07}
G{\'o}mez-Garde{\~ n}ez, J., Campillo, M., Flor{\'\i}a, L.~M., Moreno, Y.,
  2007. Dynamical organization of cooperation in complex topologies. Phys. Rev.
  Lett. 98, 108103.

\bibitem[{Gould and Eldredge(1993)}]{gould_n93}
Gould, S.~J., Eldredge, N., 1993. Punctuated equilibrium comes of age. Nature
  366, 223--227.

\bibitem[{Grassberger(1982)}]{grassberger_zpb82}
Grassberger, P., 1982. On phase-transitions in \protect{Schl{\"o}gl} 2nd model.
  Z. Phys. B 47, 365--374.

\bibitem[{Grassberger(1993)}]{grassberger_jpa93}
Grassberger, P., 1993. On a self-organized critical forest-fire model. J. Phys.
  A 26, 2081--2089.

\bibitem[{Greenberg and Hastings(1978)}]{greenberg_siam78}
Greenberg, J.~M., Hastings, S.~P., 1978. Spatial patterns for discrete models
  of diffusion in exitable media. SIAM J. Appl. Math. 34, 515--523.

\bibitem[{Grim(1995)}]{grim_jtb95}
Grim, P., 1995. The greater generosity of the spatialized prisoner's dilemma.
  J. Theor. Biol. 173, 353--359.

\bibitem[{Grim(1996)}]{grim_bs96}
Grim, P., 1996. Spatialization and greater generosity in stochastic prisoner's
  dilemma. BioSystems 37, 3--17.

\bibitem[{Guan et~al.(2006)Guan, Wu, Huang, Xu, and Wang}]{guan_epl06}
Guan, J.-Y., Wu, Z.-X., Huang, Z.-G., Xu, X.-J., Wang, Y.-H., 2006. Promotion
  of cooperation induced by nonlinear attractive effect in spatial prisoner's
  dilemma game. Europhys. Lett. 76, 1214--1220.

\bibitem[{G{\"u}th et~al.(1982)G{\"u}th, Schmittberger, and
  Schwarze}]{guth_jebo82}
G{\"u}th, W., Schmittberger, R., Schwarze, B., 1982. An experimental analysis
  of ultimatum bargaining. J. Econ. Behav. Org. 24, 153--172.

\bibitem[{Gutowitz et~al.(1987)Gutowitz, Victor, and Knight}]{gutowitz_pd87}
Gutowitz, H.~A., Victor, J.~D., Knight, B.~W., 1987. Local structure theory for
  cellular automata. Physica D 28, 18--48.

\bibitem[{Gy{\"o}rgyi(2001)}]{gyorgyi_pr01}
Gy{\"o}rgyi, G., 2001. Techniques of replica symmetry breaking and the storage
  problem of the \protect{McCulloch-Pitts} neuron. Phys. Rep. 342, 263--392.

\bibitem[{Haken(1988)}]{haken_88}
Haken, H., 1988. Information and Self-organization. Springer-Verlag, Berlin.

\bibitem[{Hales(2000)}]{hales_lnai00}
Hales, D., 2000. Cooperation without space or memory: Tags, groups and the
  prisoner's dilemma. In: Moss, S., Davidsson, P. (Eds.), Multi-Agent-Based
  Simulation. Vol. 1979 of Lecture Notes in Artificial Intelligence.
  Springer-Verlag, Berlin, pp. 157--166.

\bibitem[{Hamilton(1964{\natexlab{a}})}]{hamilton_jtb64a}
Hamilton, W.~D., 1964{\natexlab{a}}. Genetical evolution of social behavior
  \protect{I}. J. Theor. Biol. 7, 1--16.

\bibitem[{Hamilton(1964{\natexlab{b}})}]{hamilton_jtb64b}
Hamilton, W.~D., 1964{\natexlab{b}}. Genetical evolution of social behavior
  \protect{II}. J. Theor. Biol. 7, 17--52.

\bibitem[{Hanaki et~al.(2006)Hanaki, Peterhansl, Dodds, and
  Watts}]{hanaki_ms06}
Hanaki, N., Peterhansl, A., Dodds, P.~S., Watts, D.~J., 2006. Cooperation in
  evolving social networks. Management Science (in press).

\bibitem[{Hardin(1968)}]{hardin_s68}
Hardin, G., 1968. The tragedy of the commons. Science 162, 1243--1248.

\bibitem[{Hardin(1971)}]{hardin_bs71}
Hardin, R., 1971. Collective action as an aggregeable n-prisoner's dilemma.
  Behav. Sci. 16, 472--481.

\bibitem[{Harris(1974)}]{harris_ap74}
Harris, T.~E., 1974. Contect interactions on a lattice. Ann. Prob. 2, 969--988.

\bibitem[{Harsanyi and Selten(1988)}]{harsanyi_88}
Harsanyi, J.~C., Selten, R., 1988. A General Theory of Equilibrium Selection in
  Games. MIT Press, Cambridge, MA.

\bibitem[{Hauert(2001)}]{hauert_prslb01}
Hauert, C., 2001. Fundamental clusters in spatial $2\times 2$ games. Proc. R.
  Soc. Lond. B 268, 761--769.

\bibitem[{Hauert(2002)}]{hauert_ijbc02}
Hauert, C., 2002. Effects of space in $2\times 2$ games. Int. J. Bifurc. Chaos
  12, 1531--1548.

\bibitem[{Hauert(2006)}]{hauert_jtb06b}
Hauert, C., 2006. Spatial effects in social dilemmas. J. Theor. Biol 240,
  627--636.

\bibitem[{Hauert et~al.(2002)Hauert, De~Monte, Hofbauer, and
  Sigmund}]{hauert_s02}
Hauert, C., De~Monte, S., Hofbauer, J., Sigmund, K., 2002. Volunteering as
  \protect{Red Queen} mechanism for cooperation in public goods game. Science
  296, 1129--1132.

\bibitem[{Hauert and Doebeli(2004)}]{hauert_n04}
Hauert, C., Doebeli, M., 2004. Spatial structure often inhibits the evolution
  of cooperation in the snowdrift game. Nature 428, 643--646.

\bibitem[{Hauert et~al.(2006)Hauert, Michor, Nowak, and
  Doebeli}]{hauert_jtb06a}
Hauert, C., Michor, F., Nowak, M.~A., Doebeli, M., 2006. Synergy and
  discounting of cooperation in social dilemmas. J. Theor. Biol 239, 195--202.

\bibitem[{Hauk and Nagel(2001)}]{hauk_jcr01}
Hauk, E., Nagel, R., 2001. Choice of partners in multiply two-person prisoner's
  dilemma games: Experimental study. J. Conflict Resol. 45, 770--793.

\bibitem[{He et~al.(2005)He, Cai, Wang, and Pan}]{he_ijmpc05}
He, M., Cai, Y., Wang, Z., Pan, Q.-H., 2005. The influence of species' number
  and the density of vacant sites on the defensive alliance. Int. J. Mod. Phys.
  C 16, 1861--1868.

\bibitem[{Helbing(1996)}]{helbing_td96}
Helbing, D., 1996. A stochastic behavioral model and a microscopic foundation
  of evolutionary game theory. Theor. Decis. 40, 149--179.

\bibitem[{Helbing(1998)}]{helbing_incoll98}
Helbing, D., 1998. Microscopic foundation of stochastic game dynamical
  equations. In: Leinfellner, W., K{\"o}hler, E. (Eds.), Game Theory,
  Experience, Rationality. Kluwer Academic, Dordrecht, pp. 211--224,
  cond-mat/9805395.

\bibitem[{Helbing et~al.(2005)Helbing, Sch{\"o}nhof, Stark, and
  Holyst}]{helbing_acs05}
Helbing, D., Sch{\"o}nhof, M., Stark, H.-U., Holyst, J.~A., 2005. How
  individuals learn to take turns: Emergence of alternating cooperation in a
  congestion game and the prisoner's dilemma. Adv. Complex Syst. 8, 87--116.

\bibitem[{Hempel et~al.(1999)Hempel, Schimansky-Geier, and
  Garcia-Ojalvo}]{hempel_prl99}
Hempel, H., Schimansky-Geier, L., Garcia-Ojalvo, J., 1999. Noise-sustained
  pulsating patterns and global oscillations in subexitable media. Phys. Rev.
  Lett. 82, 3713--3716.

\bibitem[{Henrich et~al.(2001)Henrich, Boyd, Bowles, Camerer, Fehr, Gintis, and
  McElreath}]{henrich_aer01}
Henrich, J., Boyd, R., Bowles, S., Camerer, C., Fehr, E., Gintis, H.,
  McElreath, R., 2001. In search of homo economicus: behavioral experiments in
  15 small-scale societies. Am. Econ. Rev. 91, 73--78.

\bibitem[{Hinrichsen(2000)}]{hinrichsen_ap00}
Hinrichsen, H., 2000. Non-equilibrium critical phenomena and phase transitions
  into absorbing states. Adv. Phys. 49, 815--958.

\bibitem[{Hofbauer et~al.(1997)Hofbauer, Hutson, and Vickers}]{hofbauer_na97}
Hofbauer, J., Hutson, V., Vickers, G.~T., 1997. Travelling waves for games in
  economics and biology. Nonlin. Anal. 30, 1235--1244.

\bibitem[{Hofbauer and Sigmund(1988)}]{hofbauer_88}
Hofbauer, J., Sigmund, K., 1988. The Theory of Evolution and Dynamical Systems.
  Cambridge University Press, Cambridge.

\bibitem[{Hofbauer and Sigmund(1998)}]{hofbauer_98}
Hofbauer, J., Sigmund, K., 1998. Evolutionary Games and Population Dynamics.
  Cambridge University Press, Cambridge.

\bibitem[{Hofbauer and Sigmund(2003)}]{hofbauer_bams03}
Hofbauer, J., Sigmund, K., 2003. Evolutionary game dynamics. Bull. Am. Math.
  Soc. 40, 479--519.

\bibitem[{Holland(1995)}]{holland_95}
Holland, J.~H., 1995. Hidden order: How adaption builds complexity. Addison
  Wesley, Reading, MA.

\bibitem[{Holley and Liggett(1975)}]{holley_ap75}
Holley, R., Liggett, T.~M., 1975. Ergodic theorems for weakly interacting
  systems and the voter model. Ann. Probab. 3, 643--663.

\bibitem[{Holme et~al.(2003)Holme, Trusina, Kim, and Minnhagen}]{holme_pre03}
Holme, P., Trusina, A., Kim, B.~J., Minnhagen, P., 2003. Prisoner's dilemma in
  real-world acquintance networks: Spikes and quasiequilibria induced by the
  interplay between structure and dynamics. Phys. Rev. E 68, 030901.

\bibitem[{Huberman and Glance(1993)}]{huberman_pnas93}
Huberman, B., Glance, N., 1993. Evolutionary games and computer simulations.
  Proc. Natl. Acad. Sci. USA 90, 7716--7718.

\bibitem[{Ifti and Bergersen(2003)}]{ifti_epje03}
Ifti, M., Bergersen, B., 2003. Survival and extension in cyclic and neutral
  three-species systems. Eur. Phys. J. E 10, 241--248.

\bibitem[{Ifti et~al.(2004)Ifti, Killingback, and Doebeli}]{ifti_jtb04}
Ifti, M., Killingback, T., Doebeli, M., 2004. Effects of neighbourhood size and
  connectivity on the spatial prisoner's dilemma. J. Theor. Biol. 231, 97--106.

\bibitem[{Imhof et~al.(2005)Imhof, Fudenberg, and Nowak}]{imhof_pnas05}
Imhof, L.~A., Fudenberg, D., Nowak, M.~A., 2005. Evolutionary cycles of
  cooperation and defection. Proc. Natl. Acad. Sci. USA 102, 10797--10800.

\bibitem[{Jackson(2005)}]{jackson_05}
Jackson, M.~O., 2005. A survey of models of network formation: Stability and
  efficiency. In: Demange, G., Wooders, M. (Eds.), Group formation in economics
  : networks, clubs and coalitions. Cambridge University Press, Cambridge, UK,
  pp. 11--57.

\bibitem[{Jansen and Baalen(2006)}]{jansen_n06}
Jansen, V. A.~A., Baalen, M., 2006. Altruism through beard chromodynamics.
  Nature 440, 663--666.

\bibitem[{Janssen(1981)}]{janssen_zpb81}
Janssen, H.~K., 1981. On the non-equilibrium phase-transition in
  reaction-diffusion systems with an absorbing state. Z. Phys. B 42, 151--154.

\bibitem[{Jensen(1991)}]{jensen_pra91}
Jensen, I., 1991. Universality class of a one-dimensional cellular automaton.
  Phys. Rev. A 43, 3187--3189.

\bibitem[{Johnson and Boerlijst(2002)}]{johnson_tree02}
Johnson, C.~R., Boerlijst, M.~C., 2002. Selection at the level of the
  community: the importance of spatial structures. Trends Ecol. Evol. 17,
  83--90.

\bibitem[{Johnson and Seinen(2002)}]{johnson_prsb02}
Johnson, C.~R., Seinen, I., 2002. Selection for restraint in competitive
  ability in spatial competition systems. Proc. Roy. Soc. Lond. B 269,
  655--663.

\bibitem[{Joo and Lebowitz(2004)}]{joo_pre04}
Joo, J., Lebowitz, J.~L., 2004. Pair approximation of the stochastic
  susteptible-infected-recovered-susceptible epidemic node on the hypercubic
  lattice. Phys. Rev. E 70, 036114.

\bibitem[{Jung and Mayer-Kress(1995)}]{jung_prl95}
Jung, P., Mayer-Kress, G., 1995. Spatiotemporal stochastic resonance in
  excitable media. Phys. Rev. Lett. 74, 2130--2133.

\bibitem[{Kandori et~al.(1993)Kandori, Mailath, and Rob}]{kandori_e93}
Kandori, M., Mailath, G.~J., Rob, R., 1993. Learning, mutation, and long-run
  equilibria in games. Econometrica 61, 29--56.

\bibitem[{Katz et~al.(1983)Katz, Lebowitz, and Spohn}]{katz_prb83}
Katz, S., Lebowitz, J.~L., Spohn, H., 1983. Phase transitions in stationary
  nonequilibrium states of model lattice systems. Phys. Rev. B 28, 1655--1658.

\bibitem[{Katz et~al.(1984)Katz, Lebowitz, and Spohn}]{katz_jsp84}
Katz, S., Lebowitz, J.~L., Spohn, H., 1984. Nonequilibrium steady states of
  stochastic lattice gas models of fast ionic conductors. J. Stat. Phys. 34,
  497--537.

\bibitem[{Kawasaki(1972)}]{kawasaki_72}
Kawasaki, K., 1972. Kinetics of \protect{Ising} models. In: Domb, C., Green,
  M.~S. (Eds.), Phase Transitions and Critical Phenomena, Vol. 2. Academic
  Press, London, pp. 443--501.

\bibitem[{Kelly(1979)}]{kelly_79}
Kelly, F.~P., 1979. Reversibility and Stochastic Networks. John Wiley, London.

\bibitem[{Kermack and McKendrick(1927)}]{kermack_prsa27}
Kermack, W.~O., McKendrick, A.~G., 1927. Contribution to the mathematical
  theory of epidemics. Proc. Roy. Soc. Lond. A 115, 700--721.

\bibitem[{Kerr et~al.(2002)Kerr, Riley, Feldman, and Bohannan}]{kerr_n02}
Kerr, B., Riley, M.~A., Feldman, M.~W., Bohannan, B. J.~M., 2002. Local
  dispersal promotes biodiversity in a real-life game of rock-paper-scissors.
  Nature 418, 171--174.

\bibitem[{Killingback and Doebeli(1996)}]{killingback_prslb96}
Killingback, T., Doebeli, M., 1996. Spatial evolutionary game theory: Hawks and
  doves revisited. Proc. R. Soc. Lond. B 263, 1135--1144.

\bibitem[{Killingback and Doebeli(1998)}]{killingback_jtb98}
Killingback, T., Doebeli, M., 1998. Self-organized criticality in spatial
  evolutionary game theory. J. Theor. Biol. 191, 335--340.

\bibitem[{Killingback et~al.(1999)Killingback, Doebeli, and
  Knowlton}]{killingback_prslb99}
Killingback, T., Doebeli, M., Knowlton, N., 1999. Variable investment, the
  continuous prisoner's dilemma, and the origin of cooperation. Proc. R. Soc.
  Lond. B 266, 1723--1728.

\bibitem[{Kim et~al.(2005)Kim, Liu, Um, and Lee}]{kim_pre05}
Kim, B.~J., Liu, J., Um, J., Lee, S.-I., 2005. Instability of defensive
  alliances in the predator-prey model on complex networks. Phys. Rev. E 72,
  041906.

\bibitem[{Kim et~al.(2002)Kim, Trusina, Holme, Minnhagen, Chung, and
  Choi}]{kim_pre02}
Kim, B.~J., Trusina, A., Holme, P., Minnhagen, P., Chung, J.~S., Choi, M.~Y.,
  2002. Dynamic instabilities induced by asymmetric influence: Prisoner's
  dilemma game in small-world networks. Phys. Rev. E 66, 021907.

\bibitem[{Kinzel(1985)}]{kinzel_zpb85}
Kinzel, W., 1985. Phase transitions of cellular automata. Z. Phys. B 58,
  229--244.

\bibitem[{Kirchkamp(2000)}]{kirchkamp_jebo00}
Kirchkamp, O., 2000. Spatial evolution of automata in the prisoner's dilemma.
  J. Econ. Behav. Org. 43, 239--262.

\bibitem[{Kirkup and Riley(2004)}]{kirkup_n04}
Kirkup, B.~C., Riley, M.~A., 2004. Antibiotic-mediated antagonism leads to a
  bacterial game of rock-paper-scissors in vivo. Nature 428, 412--414.

\bibitem[{Kittel(2004)}]{kittel_04}
Kittel, C., 2004. Introduction to Solid State Physics, 8th edition. John Wiley
  \& Sons, Chichester.

\bibitem[{Kobayashi and Tainaka(1997)}]{kobayashi_jpsj97}
Kobayashi, K., Tainaka, K., 1997. Critical phenomena in cyclic ecosystems:
  Parity law and selfstructuring extinction pattern. J. Phys. Soc. Jpn. 66,
  38--41.

\bibitem[{Kraines and Kraines(1989)}]{kraines_td89}
Kraines, D., Kraines, V., 1989. Pavlov and the prisoner's dilemma. Theor.
  Decis. 26, 47--79.

\bibitem[{Kraines and Kraines(1993)}]{kraines_td93}
Kraines, D., Kraines, V., 1993. Learning to cooperate with pavlov an adaptive
  strategy for the iterated prisoner's dilemma with noise. Theor. Decis. 35,
  107--150.

\bibitem[{Krapivsky et~al.(2000)Krapivsky, Redner, and
  Leyvraz}]{krapivsky_prl00}
Krapivsky, P.~L., Redner, S., Leyvraz, F., 2000. Connectivity of growing random
  networks. Phys. Rev. Lett. 85, 4629--4632.

\bibitem[{Kreft(2004)}]{kreft_mb04}
Kreft, J.-U., 2004. Biofilms promote altruism. Microbiology 150, 2751--2760.

\bibitem[{Kreps et~al.(1982)Kreps, Milgrom, Roberts, and Wilson}]{kreps_jet82}
Kreps, D.~M., Milgrom, P., Roberts, J., Wilson, R., 1982. Rational cooperation
  in the finitely repeated prisoners-dilemma. J. Econ. Theor. 27, 245--252.

\bibitem[{Kuperman and Abramson(2001)}]{kuperman_prl01}
Kuperman, M., Abramson, G., 2001. Small world effects in an epidemiological
  model. Phys. Rev. Lett. 86, 2909--2912.

\bibitem[{Kuznetsov(1995)}]{kuznetsov_95}
Kuznetsov, Y.~A., 1995. Elements of Applied Bifurcation Theory.
  Springer-Verlag, New York.

\bibitem[{Ledyard(1995)}]{ledyard_95}
Ledyard, J.~O., 1995. Public goods: A survey of experimental research. In:
  Kagel, J.~H., Roth, A.~E. (Eds.), The Handbook of Experimental Economics.
  Princeton University Press, Princeton, NJ, pp. 111--194.

\bibitem[{Lee and Johnson(2001)}]{lee_n01}
Lee, C.~F., Johnson, N.~F., 2001. Playing by quantum rules. Nature 414,
  244--245.

\bibitem[{Lee and Johnson(2002)}]{lee_pw02}
Lee, C.~F., Johnson, N.~F., 2002. Let the quantum games begin. Phys. World 15,
  25--29.

\bibitem[{Lee and Valentinyi(2000)}]{lee_res00}
Lee, I.~H., Valentinyi, {\'A}., 2000. Interactive contagion. Rev. Econ. Stud.
  67, 47--66.

\bibitem[{Lewontin(1961)}]{lewontin_jtb61}
Lewontin, 1961. Evolution and the theory of games. J. Theor. Biol. 1, 382--403.

\bibitem[{Lieberman et~al.(2005)Lieberman, Hauert, and Nowak}]{lieberman_n05}
Lieberman, E., Hauert, C., Nowak, M.~A., 2005. Evolutionary dynamics on graphs.
  Nature 433, 312--316.

\bibitem[{Liggett(1985)}]{liggett_85}
Liggett, T.~M., 1985. Interacting Particle Systems. Springer-Verlag, New York.

\bibitem[{Lim et~al.(2002)Lim, Chen, and Jayaprakash}]{lim_pre02}
Lim, Y.~M., Chen, K., Jayaprakash, C., 2002. Scale-invariant behavior in a
  spatial game of prisoner's dilemma. Phys. Rev. E 65, 026134.

\bibitem[{Lin et~al.(2000)Lin, Hagberg, Ardelea, Bertram, Swinney, and
  Meron}]{lin_pre00}
Lin, A.~I., Hagberg, A., Ardelea, A., Bertram, M., Swinney, H.~L., Meron, E.,
  2000. Four-phase patterns in forced oscillatory systems. Phys. Rev. E 62,
  3790--3798.

\bibitem[{Lindgren(1997)}]{lindgren_proc97}
Lindgren, K., 1997. Evolutionary dynamics in game-theoretic models. In: Arthur,
  W.~B., Durlauf, S., Lane, D.~A. (Eds.), The economy as an evolving complex
  system \protect{II}. Santa Fe Institute, Perseus Books, Reading, Mass.

\bibitem[{Lindgren and Nordahl(1994)}]{lindgren_pd94}
Lindgren, K., Nordahl, M.~G., 1994. Evolutionary dynamics of spatial games.
  Physica D 75, 292--309.

\bibitem[{MacLean and Gudelj(2006)}]{maclean_n06}
MacLean, R.~C., Gudelj, I., 2006. resource competition and social conflict in
  experimental populations of yeast. Nature 441, 498--501.

\bibitem[{Macy and Flache(2002)}]{macy_pnas02}
Macy, M.~W., Flache, A., 2002. Learning dynamics in social dilemmas. Proc.
  Natl. Acad. Sci. USA 99, 7229--7236.

\bibitem[{Marro and Dickman(1999)}]{marro_99}
Marro, J., Dickman, R., 1999. Nonequilibrium Phase Transitions in Lattice
  Models. Cambridge University Press, Cambridge.

\bibitem[{Marsili and Zhang(1997)}]{marsili_pa97}
Marsili, M., Zhang, Y.-C., 1997. Fluctuations around \protect{Nash} equilibria
  in game theory. Physica A 245, 181--188.

\bibitem[{Martins et~al.(2004)Martins, Moore, and Shellard}]{martins_prl04}
Martins, C. J. A.~P., Moore, J.~N., Shellard, E. P.~S., 2004. Unified model for
  vortex-string network evolution. Phys. Rev. Lett. 92, 251601.

\bibitem[{Masuda and Aihara(2003)}]{masuda_pla03}
Masuda, N., Aihara, K., 2003. Spatial prisoner's dilemma optimally played in
  small-world networks. Phys. Lett. A 313, 55--61.

\bibitem[{Masuda and Konno(2006)}]{masuda_pre06}
Masuda, N., Konno, N., 2006. Networks with dispersed degrees save stable
  coexistence of species in cyclic competition. Phys. Rev. E 74, 066102.

\bibitem[{May and Leonard(1975)}]{may_siam75}
May, R.~M., Leonard, W.~J., 1975. Nonlinear aspects of competition between
  three species. SIAM J. Appl. Math. 29, 243--253.

\bibitem[{Maynard~Smith(1978)}]{maynard_sa78}
Maynard~Smith, J., 1978. The evolution of behaviors. Sci. Am. 239, 176--192.

\bibitem[{Maynard~Smith(1982)}]{maynard_82}
Maynard~Smith, J., 1982. Evolution and the theory of games. Cambridge
  University Press, Cambridge.

\bibitem[{Maynard~Smith and Price(1973)}]{maynard_n73}
Maynard~Smith, J., Price, G.~R., 1973. The logic of animal conflict. Nature
  246, 15--18.

\bibitem[{Meron and Pelc{\'e}(1988)}]{meron_prl88}
Meron, E., Pelc{\'e}, P., 1988. Model for spiral wave formation in exitable
  media. Phys. Rev. Lett. 60, 1880--1883.

\bibitem[{Metz et~al.(1996)Metz, Geritz, Mesz{\'e}na, Jacobs, and van
  Heerwaarden}]{metz_96}
Metz, J. A.~J., Geritz, S. A.~H., Mesz{\'e}na, G., Jacobs, F. J.~A., van
  Heerwaarden, J.~S., 1996. Adaptive dynamics, a geometrical study of the
  consequences of nearly faithful reproduction. In: van Strien, S.~J., Lunel,
  S. M.~V. (Eds.), Stochastic and spatial structures of dynamical systems.
  North Holland, p. 183–231.

\bibitem[{Meyer(1999)}]{meyer_prl99}
Meyer, D.~A., 1999. Quantum strategies. Phys. Rev. Lett. 82, 1052--1055.

\bibitem[{M{\'e}zard et~al.(1987)M{\'e}zard, Parisi, and Virasoro}]{mezard_87}
M{\'e}zard, M., Parisi, G., Virasoro, M.~A., 1987. Spin Glass Theory and
  Beyond. World Scientific, Singapore.

\bibitem[{Mi{\c e}kisz(2004{\natexlab{a}})}]{miekisz_jpa04}
Mi{\c e}kisz, J., 2004{\natexlab{a}}. Statistical mechanics of spatial
  evolutionary games. J. Phys. A: Math. Gen. 37, 9891--9906.

\bibitem[{Mi{\c e}kisz(2004{\natexlab{b}})}]{miekisz_jsp04}
Mi{\c e}kisz, J., 2004{\natexlab{b}}. Stochastic stability in spatial games. J.
  Stat. Phys. 117, 99--110.

\bibitem[{Mi{\c e}kisz(2004{\natexlab{c}})}]{miekisz_pa04}
Mi{\c e}kisz, J., 2004{\natexlab{c}}. Stochastic stability in spatial
  three-player games. Physica A 343, 175--184.

\bibitem[{Milgram(1967)}]{milgram_pt67}
Milgram, S., 1967. The small world problem. Psychol. Today 2, 60--67.

\bibitem[{Mobilia et~al.(2006{\natexlab{a}})Mobilia, Georgiev, and
  T{\"a}uber}]{mobilia_pre06}
Mobilia, M., Georgiev, I.~T., T{\"a}uber, U.~C., 2006{\natexlab{a}}.
  Fluctuations and correlations in lattice models for predator-prey
  interaction. Phys. Rev. E 73, 040903(R).

\bibitem[{Mobilia et~al.(2006{\natexlab{b}})Mobilia, Georgiev, and
  T{\"a}uber}]{mobilia_jsp06}
Mobilia, M., Georgiev, I.~T., T{\"a}uber, U.~C., 2006{\natexlab{b}}. Phase
  transitions and spatio-temporal fluctuations in stochastic lattice
  \protect{Lotka-Volterra} models. J. Stat. Phys. in press.

\bibitem[{Molander(1985)}]{molander_jcr85}
Molander, P., 1985. The optimal level generosity in a selfish, uncertain
  environment. J. Conflict Resolut. 29, 611--618.

\bibitem[{Monderer and Shapley(1996)}]{Monderer_geb96}
Monderer, D., Shapley, L.~S., 1996. Potential games. Games Econ. Behav. 14,
  124--143.

\bibitem[{Moran(1962)}]{moran_62}
Moran, P. A.~P., 1962. The Statistical Processes of Evolutionary Theory.
  Clarendon, Oxford, UK.

\bibitem[{Mukherji et~al.(1996)Mukherji, Rajan, and Slagle}]{mukherji_n96}
Mukherji, A., Rajan, V., Slagle, J.~R., 1996. Robustness of cooperation. Nature
  125, 125--126.

\bibitem[{Nakamaru and Iwasa(2000)}]{nakamaru_tpb00}
Nakamaru, M., Iwasa, Y., 2000. Competition by allelopathy proceeds in traveling
  waves: Colicin-immune strain aids collicin-sensitive strain. Theor. Pop.
  Biol. 57, 131--144.

\bibitem[{Nakamaru et~al.(1997)Nakamaru, Matsuda, and Iwasa}]{nakamaru_jtb97}
Nakamaru, M., Matsuda, H., Iwasa, Y., 1997. The evolution of cooperation in a
  lattice-structured population. J. Theor. Biol. 184, 65--81.

\bibitem[{Nash(1950)}]{nash_pnas50}
Nash, J., 1950. Equilibrium points in n-person games. Proc. Nat. Acad. Sci. USA
  36, 48--49.

\bibitem[{Neumann and Schuster(2007)}]{neumann_jmb07}
Neumann, G., Schuster, S., 2007. Continuous model for the rock-scissors-paper
  game between bacteriocin producing bacteria. J. Math. Biol. in press.

\bibitem[{Newman(2002)}]{newman_pre02}
Newman, M. E.~J., 2002. Spread of epidemic disease on networks. Phys. Rev. E
  66, 016128.

\bibitem[{Newman(2003)}]{newman_siamr03}
Newman, M. E.~J., 2003. The structure and function of complex networks. SIAM
  Review 45, 167--256.

\bibitem[{Newman and Watts(1999)}]{newman_pre99}
Newman, M. E.~J., Watts, D.~J., 1999. Scaling and percolation in the
  small-world network model. Phys. Rev. E 60, 7332--7342.

\bibitem[{Nowak(1990{\natexlab{a}})}]{nowak_jtb90}
Nowak, M., 1990{\natexlab{a}}. An evolutionary stable strategy may be
  inaccessible. J. Theor. Biol. 142, 237--241.

\bibitem[{Nowak and Sigmund(1989{\natexlab{a}})}]{nowak_amc89}
Nowak, M., Sigmund, K., 1989{\natexlab{a}}. Game-dynamical aspects of the
  prisoner's dilemma. Appl. Math. Comp. 30, 191--213.

\bibitem[{Nowak and Sigmund(1989{\natexlab{b}})}]{nowak_jtb89}
Nowak, M., Sigmund, K., 1989{\natexlab{b}}. Oscillation in the evolutionary
  reciprocity. J. Theor. Biol. 137, 21--26.

\bibitem[{Nowak and Sigmund(1990)}]{nowak_aam90}
Nowak, M., Sigmund, K., 1990. The evolution of stochastic strategies in the
  prisoner's dilemma. Acta Appl. Math. 20, 247--265.

\bibitem[{Nowak(1990{\natexlab{b}})}]{nowak_tpb90}
Nowak, M.~A., 1990{\natexlab{b}}. Stochastic strategies in the prisoner's
  dilemma. Theor. Pop. Biol. 38, 93--112.

\bibitem[{Nowak(2006{\natexlab{a}})}]{nowak_06}
Nowak, M.~A., 2006{\natexlab{a}}. Evolutionary Dynamics. Harvard University
  Press, Cambridge, MA.

\bibitem[{Nowak(2006{\natexlab{b}})}]{nowak_s06}
Nowak, M.~A., 2006{\natexlab{b}}. Five rules for the evolution of cooperation.
  Science 314, 1560--1563.

\bibitem[{Nowak et~al.(1994{\natexlab{a}})Nowak, Bonhoeffer, and
  May}]{nowak_ijbc94}
Nowak, M.~A., Bonhoeffer, S., May, R.~M., 1994{\natexlab{a}}. More spatial
  games. Int. J. Bifurcat. Chaos 4, 33--56.

\bibitem[{Nowak et~al.(1994{\natexlab{b}})Nowak, Bonhoeffer, and
  May}]{nowak_pnas94}
Nowak, M.~A., Bonhoeffer, S., May, R.~M., 1994{\natexlab{b}}. Spatial games and
  the maintenance of cooperation. Proc. Natl. Acad. Sci. USA 91, 4877--4881.

\bibitem[{Nowak and May(1992)}]{nowak_n92b}
Nowak, M.~A., May, R.~M., 1992. Evolutionary games and spatial chaos. Nature
  359, 826--829.

\bibitem[{Nowak and May(1993)}]{nowak_ijbc93}
Nowak, M.~A., May, R.~M., 1993. The spatial dilemmas of evolution. Int. J.
  Bifurcat. Chaos 3, 35--78.

\bibitem[{Nowak et~al.(2000)Nowak, Page, and Sigmund}]{nowak_s00}
Nowak, M.~A., Page, K.~M., Sigmund, K., 2000. Fairness versus reason in the
  ultimatum game. Science 289, 1773--1775.

\bibitem[{Nowak et~al.(2004)Nowak, Sasaki, Taylor, and Fudenberg}]{nowak_n04}
Nowak, M.~A., Sasaki, A., Taylor, C., Fudenberg, D., 2004. Emergence of
  cooperation and evolutionary stability in finite populations. Nature 428,
  646--650.

\bibitem[{Nowak and Sigmund(1992)}]{nowak_n92a}
Nowak, M.~A., Sigmund, K., 1992. Tit for tat in heterogeneous population.
  Nature 355, 250--253.

\bibitem[{Nowak and Sigmund(1993)}]{nowak_n93}
Nowak, M.~A., Sigmund, K., 1993. A strategy of win-stay, lose-shift that
  outperforms tit-for-tat in the prisoner's dilemma game. Nature 364, 56--58.

\bibitem[{Nowak and Sigmund(1995)}]{nowak_geb95}
Nowak, M.~A., Sigmund, K., 1995. Invasion dynamics of the finitely repeated
  prisoner's dilemma. Games Econ. Behav. 11, 364--390.

\bibitem[{Nowak and Sigmund(1998)}]{nowak_n98}
Nowak, M.~A., Sigmund, K., 1998. Evolution of indirect reciprocity by image
  scoring. Nature 393, 573--577.

\bibitem[{Nowak and Sigmund(1999)}]{nowak_n99}
Nowak, M.~A., Sigmund, K., 1999. Phage-lift for game theory. Nature 399,
  367--368.

\bibitem[{Nowak and Sigmund(2004)}]{nowak_s04}
Nowak, M.~A., Sigmund, K., 2004. Evolutionary dynamics of biological games.
  Science 303, 793--799.

\bibitem[{Nowak and Sigmund(2005)}]{nowak_n05}
Nowak, M.~A., Sigmund, K., 2005. Evolution of indirect reciprocity. Nature 437,
  1291--1298.

\bibitem[{Ohta(2002)}]{ohta_pnas02}
Ohta, T., 2002. Near-neutrality in evolution of genes and gene regulation.
  Proc. Natl. Acad. Sci. USA 99, 16134--16137.

\bibitem[{Ohtsuki et~al.(2006)Ohtsuki, Hauert, Lieberman, and
  Nowak}]{ohtsuki_n06}
Ohtsuki, H., Hauert, C., Lieberman, E., Nowak, M.~A., 2006. A simple rule for
  the evolution of cooperation on graphs and social networks. Nature 441,
  502--505.

\bibitem[{Ohtsuki et~al.(2007{\natexlab{a}})Ohtsuki, Nowak, and
  Pacheco}]{ohtsuki_prl07}
Ohtsuki, H., Nowak, M.~A., Pacheco, J.~M., 2007{\natexlab{a}}. Breaking the
  symmetry between interaction and replacement in evolutionary graph theory.
  Phys. Rev. Lett. 98, 108106.

\bibitem[{Ohtsuki et~al.(2007{\natexlab{b}})Ohtsuki, Pacheco, and
  Nowak}]{ohtsuki_jtb07b}
Ohtsuki, H., Pacheco, J.~M., Nowak, M.~A., 2007{\natexlab{b}}. Evolutionary
  graph theory: breaking the symmetry between interaction and replacement. J.
  Theor. Biol. 246, 681--694.

\bibitem[{Pacheco and Santos(2005)}]{pacheco_aipcp05}
Pacheco, J.~M., Santos, F.~C., 2005. Network dependence of the dilemmas of
  cooperation. In: Mendes, J. F.~F. (Ed.), Science of Complex Networks: From
  Biology to the Intertnet and WWW, AIP Conf. Proc. No. 776. AIP, Melville, NY,
  pp. 90--100.

\bibitem[{Pacheco et~al.(2006{\natexlab{a}})Pacheco, Traulsen, and
  Nowak}]{pacheco_jtb06}
Pacheco, J.~M., Traulsen, A., Nowak, M.~A., 2006{\natexlab{a}}. Active linking
  in evolutionary games. J. Theor. Biol. 243, 437--443.

\bibitem[{Pacheco et~al.(2006{\natexlab{b}})Pacheco, Traulsen, and
  Nowak}]{pacheco_prl06}
Pacheco, J.~M., Traulsen, A., Nowak, M.~A., 2006{\natexlab{b}}. Coevolution of
  strategy and structure in complex networks with dynamical linking. Phys. Rev.
  Lett. 97, 258103.

\bibitem[{Page and Nowak(2000)}]{page_jtb00}
Page, K.~M., Nowak, M.~A., 2000. A generalized adaptive dynamics framework can
  descibe the evolutionary ultimatum game. J. Theor. Biol. 209, 173--179.

\bibitem[{Page et~al.(2000)Page, Nowak, and Sigmund}]{page_prsb00}
Page, K.~M., Nowak, M.~A., Sigmund, K., 2000. The spatial ultimatum game. Proc.
  Roy. Soc. Lond. B 267, 2177--2182.

\bibitem[{Palla et~al.(2005)Palla, Der{\'e}nyi, Farkas, and Vicsek}]{palla_n05}
Palla, G., Der{\'e}nyi, I., Farkas, I., Vicsek, T., 2005. Uncovering the
  overlapping community structure of complex networks in nature and society.
  Nature 435, 814--818.

\bibitem[{Panchanathan and Boyd(2004)}]{panchanathan_n04}
Panchanathan, K., Boyd, R., 2004. Indirect reciprocity can stabilize
  cooperation without the second-order free rider problem. Nature 432,
  499--502.

\bibitem[{Perc(2005)}]{perc_pre05}
Perc, M., 2005. Spatial coherence resonance in excitable media. Phys. Rev. E
  72, 016207.

\bibitem[{Perc(2006)}]{perc_njp06a}
Perc, M., 2006. Coherence resonance in spatial prisoner's dilemma game. New J.
  Phys. 8, 22.

\bibitem[{Perc(2007)}]{perc_njp07}
Perc, M., 2007. Premature seizure of traffic flow due to the introduction of
  evolutionary games. New J. Phys. 9, 3.

\bibitem[{Perc and Marhl(2006)}]{perc_njp06b}
Perc, M., Marhl, M., 2006. Evolutionary and dynamical coherence resonance in
  the pair approximated prisoner's dilemma game. New J. Phys. 8, 142.

\bibitem[{Pettit and Sugden(1989)}]{pettit_jp89}
Pettit, P., Sugden, R., 1989. The backward induction paradox. J. Philos. 86,
  169--182.

\bibitem[{Pfeiffer and Schuster(2005)}]{pfeiffer_tbs05}
Pfeiffer, T., Schuster, S., 2005. Game-theoretical approaches to studying the
  evolution of biochemical systems. TRENDS Biochem. Sci. 30, 20--25.

\bibitem[{Pfeiffer et~al.(2001)Pfeiffer, Schuster, and
  Bonhoeffer}]{pfeiffer_s01}
Pfeiffer, T., Schuster, S., Bonhoeffer, S., 2001. Cooperation and competition
  in the evolution of \protect{ATP}-producing pathways. Science 292, 504--507.

\bibitem[{Pikovsky and Kurths(1997)}]{pikovsky_prl97}
Pikovsky, A.~S., Kurths, J., 1997. Coherence resonance in a noise-driven
  excitable system. Phys. Rev. Lett. 78, 775--778.

\bibitem[{Posch(1999)}]{posch_jtb99}
Posch, M., 1999. Win-stay, lose-shift strategies for repeated games--memory
  length, aspiration levels and noise. J. Theor. Biol. 198, 183--195.

\bibitem[{Posch et~al.(1999)Posch, Pichler, and Sigmund}]{posch_prslb99}
Posch, M., Pichler, A., Sigmund, K., 1999. The efficiency of adapting
  aspiration levels. Proc. R. Soc. Lond. B 266, 1427--1435.

\bibitem[{Poundstone(1992)}]{poundstone_92}
Poundstone, W., 1992. Prisoner's Dilemma. Doubleday, New York.

\bibitem[{Prager et~al.(2003)Prager, Naundorf, and
  Schimansky-Geier}]{prager_pa03}
Prager, T., Naundorf, B., Schimansky-Geier, L., 2003. Coupled three-state
  oscillators. Physica A 325, 176--185.

\bibitem[{Provata and Tsekouras(2003)}]{provata_pre03}
Provata, A., Tsekouras, G.~A., 2003. Spontaneous formation of dynamical
  patterns with fractal fronts in the cyclic lattice \protect{Lotka-Volterra}
  model. Phys. Rev. E 67, 056602.

\bibitem[{Ralston(1965)}]{ralston_65}
Ralston, A., 1965. A first course in numerical analysis. McGraw Hill, New York.

\bibitem[{Rapoport and Guyer(1966)}]{rapoport_gs66}
Rapoport, A., Guyer, M., 1966. A taxonomy of $2 \times 2$ games. Yearbook of
  the Society for General Systems 11, 203--214.

\bibitem[{Rasmussen et~al.(2004)Rasmussen, Chen, Deamer, Krakauer, Packard,
  Stadler, and Bedau}]{rasmussen_s04}
Rasmussen, S., Chen, L., Deamer, D., Krakauer, D.~C., Packard, N.~H., Stadler,
  P.~F., Bedau, M.~A., 2004. Transition from nonliving to living matter.
  Science 303, 963--965.

\bibitem[{Ravasz et~al.(2004)Ravasz, Szab{\'o}, and Szolnoki}]{ravasz_pre04}
Ravasz, M., Szab{\'o}, G., Szolnoki, A., 2004. Spreading of families in cyclic
  predator-prey models. Phys. Rev. E 70, 012901.

\bibitem[{Reichenbach et~al.(2006{\natexlab{a}})Reichenbach, Mobilia, and
  Frey}]{reichenbach_pre06}
Reichenbach, T., Mobilia, M., Frey, E., 2006{\natexlab{a}}. Coexistence versus
  extinction in the stochastic cyclic \protect{Lotka-Volterra} model. Phys.
  Rev. E 74, 011907.

\bibitem[{Reichenbach et~al.(2006{\natexlab{b}})Reichenbach, Mobilia, and
  Frey}]{reichenbach_bcp06}
Reichenbach, T., Mobilia, M., Frey, E., 2006{\natexlab{b}}. Stochastic effects
  on biodiversity in cyclic coevolutionary dynamics. Banach Center
  Publications, in press.

\bibitem[{Ren et~al.(2006)Ren, Wang, Yan, and Wang}]{ren_cm06a}
Ren, J., Wang, W.-X., Yan, G., Wang, B.-H., 2006. Emergence of cooperation
  induced by preferential learning. arXiv:physics/0603007.

\bibitem[{Riolo et~al.(2001)Riolo, Cohen, and Axelrod}]{riolo_n01}
Riolo, R.~L., Cohen, M.~D., Axelrod, R., 2001. Evolution of cooperation without
  reciprocity. Nature 414, 441--443.

\bibitem[{Robson(1990)}]{robson_jtb90}
Robson, A.~J., 1990. Efficiency in evolutionary games: \protect{Darwin},
  \protect{Nash} and the secret handshake. J. Theor. Biol. 144, 379--396.

\bibitem[{Roca et~al.(2006)Roca, Cuesta, and S{\'a}nchez}]{roca_prl06}
Roca, C.~P., Cuesta, J.~A., S{\'a}nchez, A., 2006. Time scales in evolutionary
  dynamics. Phys. Rev. Lett. 97, 158701.

\bibitem[{Russell(1959)}]{russell_59}
Russell, B., 1959. Common Sense of Nuclear Warfare. George Allen and Unwin
  Ltd., London.

\bibitem[{Saijo and Yamato(1999)}]{saijo_jet99}
Saijo, T., Yamato, T., 1999. Voluntary participation game with a non-exludable
  public good. J. Econ. Theory 84, 227--242.

\bibitem[{Samuelson(1997)}]{samuelson_97}
Samuelson, L., 1997. Evolutionary Games and Equilibrium Selection. MIT Press,
  Cambridge, Mass.

\bibitem[{Samuelson(1954)}]{samuelson_res54}
Samuelson, P.~A., 1954. The pure theory of public expenditure. Rev. Econ. Stat.
  36, 387--389.

\bibitem[{S{\'a}nchez and Cuesta(2005)}]{sanchez_jtb05}
S{\'a}nchez, A., Cuesta, J.~A., 2005. Altruism may arise from individual
  selection. J, Theor. Biol. 235, 233--240.

\bibitem[{Sandholm and Dokumaci(2006)}]{sandholm_dynamo}
Sandholm, W.~H., Dokumaci, E., 2006. Dynamo, version 1.3.3, 5/9/06. freeware
  software.

\bibitem[{Santos and Pacheco(2005)}]{santos_prl05}
Santos, F.~C., Pacheco, J.~M., 2005. Scale-free networks provide a unifying
  framework for the emergence of cooperation. Phys. Rev. Lett. 95, 098104.

\bibitem[{Santos and Pacheco(2006)}]{santos_jeb06}
Santos, F.~C., Pacheco, J.~M., 2006. A new route to the evolution of
  cooperation. J. Evol. Biol. 19, 726--733.

\bibitem[{Santos et~al.(2006{\natexlab{a}})Santos, Pacheco, and
  Lenaerts}]{santos_ploscb06}
Santos, F.~C., Pacheco, J.~M., Lenaerts, T., 2006{\natexlab{a}}. Cooperation
  prevails when individuals adjust their social ties. PLoS. Comput. Biol. 2,
  1284--1290.

\bibitem[{Santos et~al.(2006{\natexlab{b}})Santos, Pacheco, and
  Lenaerts}]{santos_pnas06}
Santos, F.~C., Pacheco, J.~M., Lenaerts, T., 2006{\natexlab{b}}. Evolutionary
  dynamics of social dilemmas in structured heterogeneous populations. Proc.
  Natl. Acad. Sci. USA 103, 3490--3494.

\bibitem[{Santos et~al.(2005)Santos, Rodrigues, and Pacheco}]{santos_pre05}
Santos, F.~C., Rodrigues, J.~F., Pacheco, J.~M., 2005. Epidemic spreading and
  cooperation dynamics on homogeneous small-world networks. Phys. Rev. E 72,
  056128.

\bibitem[{Santos et~al.(2006{\natexlab{c}})Santos, Rodrigues, and
  Pacheco}]{santos_prslb06}
Santos, F.~C., Rodrigues, J.~F., Pacheco, J.~M., 2006{\natexlab{c}}. Graph
  topology plays a determinant role in the evolution of cooperation. Proc. Roy.
  Soc. Lond. B 273, 51--55.

\bibitem[{Sato et~al.(1997)Sato, Konno, and Yamaguchi}]{sato_mmit97}
Sato, K., Konno, N., Yamaguchi, T., 1997. Paper-scissors-stone game on trees.
  Mem. Muroran Inst. Tech. 47, 109--114.

\bibitem[{Sato et~al.(2002)Sato, Yoshida, and Konno}]{sato_amc02}
Sato, K., Yoshida, N., Konno, N., 2002. Parity law for population dynamics of
  n-species with cyclic advantage competition. Appl. Math. Comp. 126, 255--270.

\bibitem[{Schlag(1998)}]{schlag_jet98}
Schlag, K.~H., 1998. Why imitate, and if so, how? a bounded rational approach
  to multi-armed bandits. J. Econ. Theory 78, 130--156.

\bibitem[{Schlag(1999)}]{schlag_jme99}
Schlag, K.~H., 1999. Which one should \protect{I} imitate? J. Math. Econ. 31,
  493--522.

\bibitem[{Schmittmann and Zia(1995)}]{schmittmann_95}
Schmittmann, B., Zia, R. K.~P., 1995. Statistical mechanics of driven diffusive
  systems. In: Domb, C., Lebowitz, J.~L. (Eds.), Phase Transitions and Critical
  Phenomena, Vol. 17. Academic Press, London.

\bibitem[{Schnakenberg(1976)}]{schnakenberg_rmp76}
Schnakenberg, J., 1976. Network theory of microscopic and macroscopic behavior
  of master equation systems. Rev. Mod. Phys. 48, 571--585.

\bibitem[{Schwarz(1982)}]{schwarz_prl82}
Schwarz, K.~W., 1982. Generation of superfluid turbulence deduced from simple
  dynamical rules. Phys. Rev. Lett. 49, 282--285.

\bibitem[{Schweitzer et~al.(2002)Schweitzer, Behera, and
  M{\"u}hlenbein}]{schweitzer_acs02}
Schweitzer, F., Behera, L., M{\"u}hlenbein, H., 2002. Evolution of cooperation
  in a spatial prisoner's dilemma. Adv. Complex Systems 5, 269--299.

\bibitem[{Schweitzer et~al.(2005)Schweitzer, Mach, and
  M{\"u}hlebein}]{schweitzer_lnems05}
Schweitzer, F., Mach, R., M{\"u}hlebein, H., 2005. Agents with heterogeneous
  strategies interacting in a spatial \protect{IPD}. In: Lux, T., Reitz, S.,
  Samanidou, E. (Eds.), Nonlinear Dynamics and Heterogeneous Interacting
  Agents, Lecture Notes in Economics and Mathematical Systems, Vol. 550.
  Springer, Berlin, pp. 87--102.

\bibitem[{Selten(1965)}]{selten_zgs65}
Selten, R., 1965. Spieltheoretische behandlung eines oligopolmodells mit
  nachfragetragheit. Z. Gesamte Staatswiss. 121, 301--324.

\bibitem[{Selten(1980)}]{selten_jtb80}
Selten, R., 1980. A note on evolutionarily stable strategies in asymmetric
  animal conflict. J. Theor. Biol. 84, 93--101.

\bibitem[{Semmann et~al.(2003)Semmann, Krambeck, and Milinski}]{semmann_n03}
Semmann, D., Krambeck, H.-J., Milinski, M., 2003. Volunteering leads to
  rock-paper-scissors dynamics in a public goods game. Nature 425, 390--393.

\bibitem[{Shapley(1964)}]{shapley_ams64}
Shapley, L., 1964. Some topics in two person games. Ann. Math. Studies 5,
  1--28.

\bibitem[{Sigmund(1993)}]{sigmund_93}
Sigmund, K., 1993. Games of Life: Exploration in Ecology, Evolution and
  Behavior. Oxford University Press, Oxford, UK.

\bibitem[{Sigmund and Nowak(2001)}]{sigmund_n01}
Sigmund, K., Nowak, M.~A., 2001. Tides of tolerance. Nature 414, 403--405.

\bibitem[{Silvertown et~al.(1992)Silvertown, Holtier, Johnson, and
  Dale}]{silvertown_je92}
Silvertown, J., Holtier, S., Johnson, J., Dale, P., 1992. Cellular automaton
  models of interspecific competition of space -- the effect of pattern on
  process. J. Ecol. 80, 527--534.

\bibitem[{Sinervo and Lively(1996)}]{sinervo_n96}
Sinervo, B., Lively, C.~M., 1996. The rock-paper-scissors game and the
  evolution of alternative male strategies. Nature 380, 240--243.

\bibitem[{Skyrms(2003)}]{skyrms_03}
Skyrms, B., 2003. Stag-Hunt Game and the Evolution of Social Structure.
  Cambridge University Press, Cambridge.

\bibitem[{Skyrms and Pemantle(2000)}]{skyrms_pnas00}
Skyrms, B., Pemantle, R., 2000. A dynamic model of social network formation.
  Proc. Natl. Acad. Sci. USA 97, 9340--9346.

\bibitem[{Stanley(1971)}]{stanley_71}
Stanley, H.~E., 1971. Introduction to Phase Transitions and Critical Phenomena.
  Clarendon Press, Oxford.

\bibitem[{Sysi-Aho et~al.(2005)Sysi-Aho, Saram{\"a}ki, Kert{\'e}sz, and
  Kaski}]{sysiaho_epjb05}
Sysi-Aho, M., Saram{\"a}ki, J., Kert{\'e}sz, J., Kaski, K., 2005. Spatial
  snowdrift game with myopic agents. Eur. Phys. J. B 44, 129--135.

\bibitem[{Szab{\'o}(2000)}]{szabo_pre00b}
Szab{\'o}, G., 2000. Branching annihilating random walk on random regular
  graphs. Phys. Rev. E 62, 7474--7477.

\bibitem[{Szab{\'o}(2005)}]{szabo_jpa05}
Szab{\'o}, G., 2005. Competing associations in six-species predator-prey
  models. J. Phys. A: Math. Gen. 38, 6689--6702.

\bibitem[{Szab{\'o} et~al.(2000)Szab{\'o}, Antal, Szab{\'o}, and
  Droz}]{szabo_pre00a}
Szab{\'o}, G., Antal, T., Szab{\'o}, P., Droz, M., 2000. Spatial evolutionary
  prisoner's dilemma game with three strategies and external constraints. Phys.
  Rev. E 62, 1095--1103.

\bibitem[{Szab{\'o} and Cz{\'a}r{\'a}n(2001{\natexlab{a}})}]{szabo_pre01b}
Szab{\'o}, G., Cz{\'a}r{\'a}n, T., 2001{\natexlab{a}}. Defensive alliances in
  spatial models of cyclical population interactions. Phys. Rev. E 64, 042902.

\bibitem[{Szab{\'o} and Cz{\'a}r{\'a}n(2001{\natexlab{b}})}]{szabo_pre01a}
Szab{\'o}, G., Cz{\'a}r{\'a}n, T., 2001{\natexlab{b}}. Phase transition in a
  spatial \protect{Lotka-Volterra} model. Phys. Rev. E 63, 061904.

\bibitem[{Szab{\'o} and Hauert(2002{\natexlab{a}})}]{szabo_pre02d}
Szab{\'o}, G., Hauert, C., 2002{\natexlab{a}}. Evolutionary prisoner's dilemma
  games with voluntary participation. Phys. Rev. E 66, 062903.

\bibitem[{Szab{\'o} and Hauert(2002{\natexlab{b}})}]{szabo_prl02}
Szab{\'o}, G., Hauert, C., 2002{\natexlab{b}}. Phase transitions and
  volunteering in spatial public goods games. Phys. Rev. Lett. 89, 118101.

\bibitem[{Szab{\'o} et~al.(1999)Szab{\'o}, Santos, and Mendes}]{szabo_pre99}
Szab{\'o}, G., Santos, M.~A., Mendes, J. F.~F., 1999. Vortex dynamics in a
  three-state model under cyclic dominance. Phys. Rev. E 60, 3776--3780.

\bibitem[{Szab{\'o} and Sznaider(2004)}]{szabo_pre04a}
Szab{\'o}, G., Sznaider, G.~A., 2004. Phase transition and selection in a
  four-species predator-prey model. Phys. Rev. E 69, 031911.

\bibitem[{Szab{\'o} and Szolnoki(2002)}]{szabo_pre02a}
Szab{\'o}, G., Szolnoki, A., 2002. Three-state cyclic voter model extended with
  \protect{Potts} energy. Phys. Rev. E 65, 036115.

\bibitem[{Szab{\'o} et~al.(2004)Szab{\'o}, Szolnoki, and
  Izs{\'a}k}]{szabo_jpa04}
Szab{\'o}, G., Szolnoki, A., Izs{\'a}k, R., 2004. Rock-scissors-paper game on
  regular small-world networks. J. Phys. A: Math. Gen. 37, 2599--2609.

\bibitem[{Szab{\'o} and T{\H{o}}ke(1998)}]{szabo_pre98}
Szab{\'o}, G., T{\H{o}}ke, C., 1998. Evolutionary prisoner's dilemma game on a
  square lattice. Phys. Rev. E 58, 69--73.

\bibitem[{Szab{\'o} and Vukov(2004)}]{szabo_pre04b}
Szab{\'o}, G., Vukov, J., 2004. Cooperation for volunteering and partially
  random partnerships. Phys. Rev. E 69, 036107.

\bibitem[{Szab{\'o} et~al.(2005)Szab{\'o}, Vukov, and Szolnoki}]{szabo_pre05}
Szab{\'o}, G., Vukov, J., Szolnoki, A., 2005. Phase diagrams for an
  evolutionary prisoner's dilemma game on two-dimensional lattices. Phys. Rev.
  E 72, 047107.

\bibitem[{Sznaider(2003)}]{sznaider_03}
Sznaider, G.~A., 2003. Unpublished results.

\bibitem[{Szolnoki(2002)}]{szolnoki_pre02b}
Szolnoki, A., 2002. Dynamical mean-field approximation for a pair contact
  process with a particle source. Phys. Rev. E 66, 057102.

\bibitem[{Szolnoki and Szab{\'o}(2004{\natexlab{a}})}]{szolnoki_pre04b}
Szolnoki, A., Szab{\'o}, G., 2004{\natexlab{a}}. Phase transitions for
  rock-scissors-paper game on different networks. Phys. Rev. E 70, 037102.

\bibitem[{Szolnoki and Szab{\'o}(2004{\natexlab{b}})}]{szolnoki_pre04a}
Szolnoki, A., Szab{\'o}, G., 2004{\natexlab{b}}. Vertex dynamics during domain
  growth in three-state models. Phys. Rev. E 70, 027101.

\bibitem[{Szolnoki and Szab{\'o}(2005)}]{szolnoki_pre05a}
Szolnoki, A., Szab{\'o}, G., 2005. Three-state potts model in combination with
  the rock-scissors-paper game. Phys. Rev. E 71, 027102.

\bibitem[{Szolnoki and Szab{\'o}(2007)}]{szolnoki_epl07}
Szolnoki, A., Szab{\'o}, G., 2007. Cooperation enhanced by inhomogeneous
  activity of teaching for evolutionary prisoner's dilemma games. Europhys.
  Lett. 77, 30004.

\bibitem[{Tainaka(1988)}]{tainaka_jpsj88}
Tainaka, K., 1988. Lattice model for the \protect{Lotka-Volterra} system. J.
  Phys. Soc. Jpn. 57, 2588--2590.

\bibitem[{Tainaka(1989)}]{tainaka_prl89}
Tainaka, K., 1989. Stationary pattern of vortices or strings in biological
  systems: lattice version of the \protect{Lotka-Volterra} model. Phys. Rev.
  Lett. 63, 2688--2691.

\bibitem[{Tainaka(1993)}]{tainaka_pla93}
Tainaka, K., 1993. Paradoxial effect in a three-candidate voter model. Phys.
  Lett. A 176, 303--306.

\bibitem[{Tainaka(1994)}]{tainaka_pre94}
Tainaka, K., 1994. Vortices in a model ecosystem. Phys. Rev. E 50, 3401--3409.

\bibitem[{Tainaka(1995)}]{tainaka_pla95}
Tainaka, K., 1995. Indirect effect in cyclic voter models. Phys. Lett. A 207,
  53--57.

\bibitem[{Tainaka(2001)}]{tainaka_lncs01}
Tainaka, K., 2001. Physics and ecology of rock-paper-scissors game. In:
  Marsland, T., Frank, I. (Eds.), Lecture Notes in Computer Science. Vol. 2063.
  Springer Verlag, Berlin, pp. 384--395.

\bibitem[{Tainaka and Itoh(1991)}]{tainaka_epl91}
Tainaka, K., Itoh, Y., 1991. Topological phase transition in biological
  ecosystems. Europhys. Lett. 15, 399--404.

\bibitem[{Tang et~al.(2006)Tang, Wang, Wu, and Wang}]{tang_epjb06}
Tang, C.-L., Wang, W.-X., Wu, X., Wang, B.-H., 2006. Effects of average degree
  on cooperation in networked evolutionary game. Eur. Phys. J. B 53, 411--415.

\bibitem[{Taylor et~al.(2004)Taylor, Fundenberg, Sasaki, and
  Nowak}]{taylor_bmb04}
Taylor, C., Fundenberg, D., Sasaki, A., Nowak, M.~M., 2004. Evolutionary game
  dynamics in finite population. Bull. Math. Biol. 66, 1621--1644.

\bibitem[{Taylor and Jonker(1978)}]{taylor_mb78}
Taylor, P., Jonker, L., 1978. Evolutionary stable strategies and game dynamics.
  Math. Biosci. 40, 145--156.

\bibitem[{Thaler(1988)}]{thaler_jep88}
Thaler, R.~H., 1988. Anomalies - the ultimatum game. J. Econ. Perspect. 2,
  195--206.

\bibitem[{Thorndike(1911)}]{thorndike_11}
Thorndike, E.~L., 1911. Animal Intelligence. Macmillan, New York.

\bibitem[{Tilman and Kareiva(1997)}]{tilman_97}
Tilman, D., Kareiva, P. (Eds.), 1997. Spatial Ecology. Princeton University
  Press, Princeton.

\bibitem[{Tomassini et~al.(2006)Tomassini, Luthi, and
  Giacobini}]{tomassini_pre06}
Tomassini, M., Luthi, L., Giacobini, M., 2006. Hawks and doves games on
  small-world networks. Phys. Rev. E 73, 016132.

\bibitem[{Tomochi and Kono(2002)}]{tomochi_pre02}
Tomochi, M., Kono, M., 2002. Spatial prisoner's dilemma games with dynamic
  payoff matrices. Phys. Rev. E 65, 026112.

\bibitem[{Toral et~al.(2000)Toral, San~Miguel, and Gallego}]{toral_pa00}
Toral, R., San~Miguel, M., Gallego, R., 2000. Period stabilization in the
  \protect{Busse-Heikes} model of the \protect{K{\"u}ppers-Lortz} instability.
  Physica A 280, 315--336.

\bibitem[{Traulsen and Claussen(2004)}]{traulsen_pre04}
Traulsen, A., Claussen, J.~C., 2004. Similarity based cooperation and spatial
  segregation. Phys. Rev. E 70, 046128.

\bibitem[{Traulsen et~al.(2005)Traulsen, Claussen, and Hauert}]{traulsen_prl05}
Traulsen, A., Claussen, J.~C., Hauert, C., 2005. Coevolutionary dynamics: From
  finite to infinite populations. Phys. Rev. Lett. 95, 0238701.

\bibitem[{Traulsen et~al.(2006{\natexlab{a}})Traulsen, Claussen, and
  Hauert}]{traulsen_pre06a}
Traulsen, A., Claussen, J.~C., Hauert, C., 2006{\natexlab{a}}. Coevolutionary
  dynamics in large, but finite populations. Phys. Rev. E 74, 011901.

\bibitem[{Traulsen and Nowak(2006)}]{traulsen_pnas06}
Traulsen, A., Nowak, M.~A., 2006. Evolution of cooperation by multilevel
  selection. Proc. Natl. Acad. Sci. USA 103, 10952--10955.

\bibitem[{Traulsen and Nowak(2007)}]{traulsen_plosone07}
Traulsen, A., Nowak, M.~A., 2007. Chromodynamics of cooperation in finite
  populations. PLoS ONE 2~(3), e270.

\bibitem[{Traulsen et~al.(2006{\natexlab{b}})Traulsen, Nowak, and
  Pacheco}]{traulsen_pre06c}
Traulsen, A., Nowak, M.~A., Pacheco, J.~M., 2006{\natexlab{b}}. Stochastic
  dybamics of invasion and fixation. Phys. Rev. E 74, 011909.

\bibitem[{Traulsen et~al.(2007{\natexlab{a}})Traulsen, Nowak, and
  Pacheco}]{traulsen_jtb07}
Traulsen, A., Nowak, M.~A., Pacheco, J.~M., 2007{\natexlab{a}}. Stochastic
  payoff evaluation increases the temperature of selection. J. Theor. Biol.
  244, 349--356.

\bibitem[{Traulsen et~al.(2006{\natexlab{c}})Traulsen, Pacheco, and
  Imhof}]{traulsen_pre06b}
Traulsen, A., Pacheco, J.~M., Imhof, L.~A., 2006{\natexlab{c}}. Stochasticity
  and evolutionary stability. Phys. Rev. E 74, 021905.

\bibitem[{Traulsen et~al.(2007{\natexlab{b}})Traulsen, Pacheco, and
  Nowak}]{traulsen_jtb07b}
Traulsen, A., Pacheco, J.~M., Nowak, M.~A., 2007{\natexlab{b}}. Pairwise
  comparison and selection temperature in evolutionary game dynamics. J. Theor.
  Biol. 246, 522--529.

\bibitem[{Traulsen et~al.(2004)Traulsen, R{\"o}hl, and
  Schuster}]{traulsen_prl04}
Traulsen, A., R{\"o}hl, T., Schuster, H.~G., 2004. Stochastic gain in
  population dynamics. Phys. Rev. Lett. 93, 028701.

\bibitem[{Traulsen and Schuster(2003)}]{traulsen_pre03}
Traulsen, A., Schuster, H.~G., 2003. A minimal model for tag-based cooperation.
  Phys. Rev. E 68, 046129.

\bibitem[{Trivers(1985)}]{trivers_85}
Trivers, R., 1985. Social Evolution. Benjamin Cummings, Menlo Park.

\bibitem[{Trivers(1971)}]{trivers_qrb71}
Trivers, R.~L., 1971. The evolution of reciprocal altruism. Q. Rev. Biol. 46,
  35--57.

\bibitem[{Turner and Chao(1999)}]{turner_n99}
Turner, P.~E., Chao, L., 1999. Prisoner's dilemma in an \protect{RNA} virus.
  Nature 398, 441--443.

\bibitem[{Vainstein and Arenzon(2001)}]{vainstein_pre01}
Vainstein, M.~H., Arenzon, J.~J., 2001. Disordered environments in spatial
  games. Phys. Rev. E 64, 051905.

\bibitem[{Vainstein et~al.(2007)Vainstein, Silva, and
  Arenzon}]{vainstein_jtb07}
Vainstein, M.~H., Silva, A. T.~C., Arenzon, J.~J., 2007. Does mobility decrease
  cooperation? J. Theor. Biol. 244, 722--728.

\bibitem[{Vilenkin and Shellard(1994)}]{vilenkin_94}
Vilenkin, A., Shellard, E. P.~S., 1994. Cosmic Strings and Other Topological
  Defects. Cambridge University Press, Cambridge, UK.

\bibitem[{von Neumann(1928)}]{neumann_ma28}
von Neumann, J., 1928. Zur theorie der gesellschaftsspiele. Mathematische
  Annalen 100, 295--320, english Translation Fin Tucker, A. W. and R. D. Luce,
  ed., Contributions to the Theory of Games IV, Annals of Mathematics Studies
  40, 1959.

\bibitem[{von Neumann and Morgenstern(1944)}]{neumann_44}
von Neumann, J., Morgenstern, O., 1944. Theory of Games and Economic Behaviour.
  Princeton University Press, Princeton.

\bibitem[{Vukov et~al.(2006)Vukov, Szab{\'o}, and Szolnoki}]{vukov_pre06}
Vukov, J., Szab{\'o}, G., Szolnoki, A., 2006. Cooperation in the noisy case:
  Prisoner's dilemma game on two types of regular random graphs. Phys. Rev. E
  73, 067103.

\bibitem[{Wakano(2006)}]{wakano_mb06}
Wakano, J.~Y., 2006. A mathematical analysis on public goods games in the
  continuous space. Math. Biosci 201, 72--89.

\bibitem[{Walker(1995)}]{walker_web95}
Walker, P., 1995. An outline of the history of game theory. Working Paper,
  Department of Economics, University of Canterbury, New Zealand.

\bibitem[{Watt(1947)}]{watt_je47}
Watt, A.~S., 1947. Pattern and process in plant community. J. Ecol. 35, 1--22.

\bibitem[{Watts and Strogatz(1998)}]{watts_n98}
Watts, D.~J., Strogatz, S.~H., 1998. Collective dynamics of 'small world'
  networks. Nature 393, 440--442.

\bibitem[{Wedekind and Milinski(1996)}]{wedekind_pnas96}
Wedekind, C., Milinski, M., 1996. Human cooperation in the simultaneous and
  alternating prisoner's dilemma: \protect{Pavlov} versus tit-for-tat
  strategies. Proc. Natl. Acad. Sci. U.S.A. 93, 2686--2698.

\bibitem[{Weibull(1995)}]{weibull_95}
Weibull, J.~W., 1995. Evolutionary Game Theory. MIT Press, Cambridge, MA.

\bibitem[{Weibull(2004)}]{weibull_bookch04}
Weibull, J.~W., 2004. Testing game theory, \protect{Boston University} working
  paper, available at
  http://www.bu.edu/econ/workingpapers/papers/Jorgen
12.pdf.

\bibitem[{Weidlich(1991)}]{weidlich_pr91}
Weidlich, W., 1991. Physics and social-science - the approach of synergetics.
  Phys. Rep. 204, 1--163.

\bibitem[{Wiener and Rosenblueth(1946)}]{wiener_aic46}
Wiener, N., Rosenblueth, A., 1946. Conduction of impulses in cardiac muscle.
  Arc. Inst. Cardiol. (Mexico) 16, 205--265.

\bibitem[{Wild and Taylor(2005)}]{wild_prslb04}
Wild, G., Taylor, P.~D., 2005. Fittness and evolutionary stability in game
  theoretic models of finite populations. Proc. Roy. Soc. Lond. B 271,
  2345--2349.

\bibitem[{Wilhelm and Baynes(1977)}]{wilhelm_50}
Wilhelm, R., Baynes, C.~F., 1977. The I Ching or Book of Changes, the R.
  Wilhelm translation from Chinese to German and rendered into English by C. F.
  Baynes. Princeton University Press, Princeton, NJ.

\bibitem[{Winfree and Strogatz(1984)}]{winfree_pd84}
Winfree, A.~T., Strogatz, S.~H., 1984. Singular filaments organize chemical
  waves in three dimensions \protect{IV}: Wave taxonomy. Physica D 13,
  221--233.

\bibitem[{Wolfram(1983)}]{wolfram_rmp83}
Wolfram, S., 1983. Statistical mechanics of cellular automata. Rev. Mod. Phys.
  55, 601--644.

\bibitem[{Wolfram(1984)}]{wolfram_pd84}
Wolfram, S., 1984. Universality and complexity in cellular automata. Physica D
  10, 1--35.

\bibitem[{Wolfram(2002)}]{wolfram_02}
Wolfram, S., 2002. A New Kind of Science. Wolfram Media Inc., Champaign.

\bibitem[{Wolpert et~al.(2004)Wolpert, Tumer, and Bandari}]{wolpert_pre04}
Wolpert, D., Tumer, K., Bandari, E., 2004. Improving search algorithms by using
  intelligent coordinates. Phys. Rev. E 69, 017701.

\bibitem[{Wormald(1981)}]{wormald_jctb81}
Wormald, N.~C., 1981. The asymptotic distribution of short cycles in random
  regular graphs. J. Combin. Theor. B 31, 168--182.

\bibitem[{Wormald(1999)}]{wormald_inc99}
Wormald, N.~C., 1999. Models of random regular graphs. In: Lamb, J.~D., Preece,
  D.~A. (Eds.), Surveys in Combinatorics, London Mathematical Society Lecture
  Note Series. Vol. 267. Cambridge Univ. Press, Cambridge, pp. 239--298.

\bibitem[{Wu(1982)}]{wu_rmp82}
Wu, F.~Y., 1982. The \protect{Potts} model. Rev. Mod. Phys. 54, 235--268.

\bibitem[{Wu et~al.(2005{\natexlab{a}})Wu, Xu, Chen, and Wang}]{wu_pre05}
Wu, Z.-X., Xu, X.-J., Chen, Y., Wang, Y.-H., 2005{\natexlab{a}}. Spatial
  prisoner's dilemma game with volunteering in \protect{Newmann-Watts}
  small-world networks. Phys. Rev. E 71, 037103.

\bibitem[{Wu et~al.(2005{\natexlab{b}})Wu, Xu, and Wang}]{wu_cm05}
Wu, Z.-X., Xu, X.-J., Wang, Y.-H., 2005{\natexlab{b}}. Does the scale-free
  topology favor the emergence of cooperation. arXiv:physics/0508220.

\bibitem[{Wu et~al.(2006)Wu, Xu, and Wang}]{wu_cpl06}
Wu, Z.-X., Xu, X.-J., Wang, Y.-H., 2006. Prisoner's dilemma game with
  heterogeneous influental effect on regular small-world networks. Chin. Phys.
  Lett. 23, 531--534.

\bibitem[{Young(1993)}]{young_e93}
Young, P., 1993. The evolution of conventions. Econometrica 61, 57--84.

\bibitem[{Zeeman(1980)}]{zeeman_80}
Zeeman, E.~C., 1980. Population dynamics from game theory. In: Lecture Notes in
  Mathematics, Vol. 819. Springer, New York, pp. 471--497.

\bibitem[{Zia and Schmittmann(2006)}]{zia_cm06}
Zia, R. K.~P., Schmittmann, B., 2006. A possible classification of
  nonequilibrium steady states. J. Phys. A: Math. Gen. 39, L407--L413.

\bibitem[{Zimmermann and Egu{\'{\i}}luz(2005)}]{zimmermann_pre05}
Zimmermann, M.~G., Egu{\'{\i}}luz, V., 2005. Cooperation, social networks and
  the emergence of leadership in a prisoner's dilemma with local interactions.
  Phys. Rev. E 72, 056118.

\bibitem[{Zimmermann et~al.(2004)Zimmermann, Egu{\'{\i}}luz, and
  San~Miguel}]{zimmermann_pre04}
Zimmermann, M.~G., Egu{\'{\i}}luz, V., San~Miguel, M., 2004. Coevolution of
  dynamical states and interactions in dynamic networks. Phys. Rev. E 69,
  065102(R).

\bibitem[{Zimmermann et~al.(2001)Zimmermann, Egu\'{\i}luz, and
  San~Miguel}]{zimmermann_01}
Zimmermann, M.~G., Egu\'{\i}luz, V.~M., San~Miguel, M., 2001. Cooperation,
  adaption and the emergence of leadership. In: Zimmermann, J.~B., Kirman, A.
  (Eds.), Economics and Heterogeneous Interacting Agents. Springer Verlag,
  Berlin, pp. 73--86.

\bibitem[{Zimmermann et~al.(2000)Zimmermann, Egu{\'{\i}}luz, San~Miguel, and
  Spadaro}]{zimmermann_00}
Zimmermann, M.~G., Egu{\'{\i}}luz, V.~M., San~Miguel, M., Spadaro, A., 2000.
  Cooperation in adaptive network. In: Ballot, G., Weisbuch, G. (Eds.),
  Application of Simulations in Social Sciences. Hermes Science Publications,
  Paris, pp. 283--297.

\end{thebibliography}

\end{document}